\pdfoutput=1
\documentclass[10pt]{article}
\usepackage[margin=1in]{geometry}
\usepackage{amsmath}
\usepackage{amssymb}
\usepackage{caption}
\usepackage{graphicx}
\usepackage{booktabs}
\usepackage{float}
\usepackage{CJKutf8}
\usepackage[linktoc=all,hidelinks]{hyperref}
\graphicspath{{figs/}}
\newcommand{\figcont}[1]{\addtocounter{figure}{-1}\caption{\emph{Continued.} #1}}

\title{\vspace{-2em}SA-AI (Spalart--Allmaras with Autogenous Inception):\\
Technical Summary\vspace{-0.5em}}
\author{Qiqi Wang (Flexcompute, Inc.\ / MIT)}
\date{\today}

\begin{document}
\maketitle

\begin{center}
  \begin{minipage}{0.6\textwidth}\centering
  \begin{CJK}{UTF8}{bsmi}埏埴以為器，當其無，有器之用。\end{CJK}\\[3pt]
  \emph{Clay is shaped into a vessel:\\
  it is the emptiness within that makes it useful.}\\[3pt]
  ---\emph{Daodejing}, ch.~11
  \end{minipage}
\end{center}

\setcounter{tocdepth}{2}
\tableofcontents

\section*{Notation}
\begin{tabbing}
  $\sigma_P,\sigma_D;\;\tau$\quad\quad \= \kill
  $\tilde\nu$ \> SA working variable \\
  $\tilde S$ \> SA modified vorticity, Eq.~\eqref{w:closure} \\
  $\nu,\;\mu_t$ \> kinematic (laminar) viscosity and eddy viscosity, Eq.~\eqref{w:fv1bypass} \\
  $\chi,\;\chi_\infty$ \> $\tilde\nu/\nu$ and its freestream (seed) value \\
  $c_{v1},\;f_{v1}$ \> SA near-wall constant ($7.1$) and damping function \\
  $P,\;D$ \> production and destruction of $\tilde\nu$, Eq.~\eqref{w:blend} \\
  $P_\mathrm{SA},\;P_\mathrm{AI}$ \> turbulent (SA) and laminar-amplification production \\
  $c_{\nu,\mathrm{ai}}$ \> laminar-viscosity scale in the SA diffusion term only ($1/6$) \\
  $d$ \> distance to the nearest wall \\
  $\omega$ \> vorticity magnitude \\
  $Re_\Omega$ \> vorticity Reynolds number, $d^2\omega/\nu$ \\
  $X,\;Y,\;Z$ \> velocity, shear, and curvature indicators (velocity scale), Eq.~\eqref{w:ind} \\
  $\hat X,\;\hat Y,\;\hat Z$ \> the same, normalized by $U_\mathrm{loc}$: direction cosines on the unit sphere \\
  $U_\mathrm{loc}$ \> local velocity scale, $\sqrt{X^2+Y^2+Z^2}$, Eq.~\eqref{w:ind} \\
  $\hat\Omega$ \> vorticity fraction $Y/\sqrt{X^2+Y^2}$, Eq.~\eqref{w:omhat} \\
  $I$ \> accumulated inflection, Eq.~\eqref{w:integral} \\
  $\hat I$ \> nondimensionalized accumulated inflection, Eq.~\eqref{w:omhat} \\
  $a,\;a_{\max}$ \> amplification rate and its free-shear ceiling ($0.19$) \\
  $\sigma_{Re}$ \> onset gate on $Re_\Omega$, Eq.~\eqref{w:rate} \\
  $Re_\Omega^{\mathrm c}$ \> onset threshold in $Re_\Omega$, Eq.~\eqref{w:onset} \\
  $C,\;A,\;B,\;k$ \> onset-threshold shape constants and scale, Eq.~\eqref{w:onset} \\
  $w$ \> onset ramp width, Eq.~\eqref{w:rate} \\
  $\sigma_P,\sigma_D;\;\tau$ \> production and destruction handover weights and the handover width, Eq.~\eqref{w:sigma} \\
  $\sigma_{v1};\;q,\;g_\chi,\;g_q$ \> $f_{v1}$-bypass blend weight and its two gates, Eqs.~\eqref{w:fv1q}--\eqref{w:fv1bypass} \\
  $N,\;N_\mathrm{crit}$ \> amplification factor and its critical value ($e^N$ method), Eq.~\eqref{w:mack} \\
  $Tu$ \> freestream turbulence intensity \\
  $\mathbf u,\;\boldsymbol\omega$ \> velocity and vorticity vectors \\
  $\mathbf d,\;\boldsymbol\ell$ \> wall-distance vector $d\,\nabla d$ and curvature vector, Eq.~\eqref{w:gram} \\
  $H$ \> boundary-layer shape factor \\
  $\beta$ \> Falkner--Skan wedge parameter \\
  $\alpha$ \> angle of attack \\
  $c,\;x/c$ \> chord and chordwise coordinate \\
  $\eta$ \> fraction of semi-span \\
  $x_\mathrm{tr};\;x_{LS},\,x_R$ \> transition-onset station; laminar-separation and reattachment stations \\
  $Re,\;M$ \> chord Reynolds number and Mach number \\
  $Re_x,\;Re_\theta$ \> streamwise and momentum-thickness Reynolds numbers \\
  $Re_D,\;Re_L$ \> cylinder-diameter and spheroid-length Reynolds numbers \\
  $C_f,\;C_d,\;C_l$ \> skin-friction, drag, and lift coefficients \\
  L0, L1, L2 \> the three grid-refinement levels, coarse to fine \\
\end{tabbing}

\clearpage

\section{Model}

\subsection{Transport equation and blended sources}

The model enters the mean flow through the turbulent viscosity alone:
\begin{gather}
  \frac{\partial\rho}{\partial t} + \nabla\!\cdot\!(\rho\mathbf u) = 0,
  \qquad
  \frac{\partial(\rho\mathbf u)}{\partial t}
    + \nabla\!\cdot\!(\rho\,\mathbf u\!\otimes\!\mathbf u)
    = -\nabla p + \nabla\!\cdot\!\boldsymbol\tau, \notag\\[3pt]
  \boldsymbol\tau = (\mu+\mu_t)\big[\nabla\mathbf u
    + (\nabla\mathbf u)^{\!\top}
    - \tfrac23(\nabla\!\cdot\!\mathbf u)\,\mathbf I\big],
  \qquad
  \mu_t = \big[(1-\sigma_{v1})\,f_{v1}(\chi)+\sigma_{v1}\big]\,\rho\tilde\nu,
  \label{w:ns}
\end{gather}
with $\sigma_{v1}$ the lifted-layer bypass weight of Eq.~\eqref{w:fv1bypass}. The
working variable is transported by
\begin{equation}
  \frac{D\tilde\nu}{Dt} = P - D
  + \frac{1}{\sigma}\nabla\!\cdot\!\big[(c_{\nu,\mathrm{ai}}\,\nu+\tilde\nu)\,\nabla\tilde\nu\big]
  + \frac{c_{b2}}{\sigma}\,|\nabla\tilde\nu|^2,
  \label{w:sa}
\end{equation}
\begin{equation}
  P = \max\!\big[\,(1-\sigma_P)\,a\,\sigma_{Re}\,\omega\,\tilde\nu,\;\;\sigma_P\,P_\mathrm{SA}\,\big],
  \qquad
  D = \sigma_D\,D_\mathrm{SA},
  \label{w:blend}
\end{equation}
$a$ and $\sigma_{Re}$ are Eq.~\eqref{w:rate}; $\sigma_P,\sigma_D$ Eq.~\eqref{w:sigma}.
$\sigma_P\!\equiv\!1$ (and with it $\sigma_D\!=\!1$),
$c_{\nu,\mathrm{ai}}\!=\!1$, and $\sigma_{v1}\!\equiv\!0$ recover baseline SA
exactly.
\begin{gather}
  P_\mathrm{SA} = c_{b1}\tilde S\,\tilde\nu, \qquad
  D_\mathrm{SA} = c_{w1}f_w\!\left(\frac{\tilde\nu}{d}\right)^{\!2}, \qquad
  \tilde S = \omega + \frac{\tilde\nu}{\kappa^2 d^2}\,f_{v2}, \qquad
  f_{v2} = 1 - \frac{\chi}{1+\chi f_{v1}}, \notag\\[3pt]
  f_{v1} = \frac{\chi^3}{\chi^3+c_{v1}^3}, \qquad
  f_w = g\!\left(\frac{1+c_{w3}^6}{g^6+c_{w3}^6}\right)^{\!1/6}, \qquad
  g = r + c_{w2}(r^6-r), \qquad
  r = \min\!\left(\frac{\tilde\nu}{\tilde S\,\kappa^2 d^2},\,10\right),
  \label{w:closure}
\end{gather}
The
non-negative $\tilde S$ modification and the negative-$\tilde\nu$
safeguards of Allmaras et al.\ (2012) are retained, with one
consistency requirement: the negative-branch diffusion factor $f_n$ is
formed with the scaled coefficient $c_{\nu,\mathrm{ai}}\,\nu$, which
keeps the total diffusivity positive,
$c_{\nu,\mathrm{ai}}\nu+f_n\tilde\nu>0$.
\begin{equation}
  \sigma_P = \sigma_t(\chi;\tau), \qquad
  \sigma_t(\chi;\tau) = \max\!\left[\,1-e^{-(\chi-1)/\tau},\;0\right], \qquad
  \sigma_D = 1-\frac{c_{b1}}{\kappa^2c_{w1}}\,(1-\sigma_P).
  \label{w:sigma}
\end{equation}

\subsection{Local indicators}

\begin{figure}[H]\centering
\includegraphics[width=0.68\textwidth]{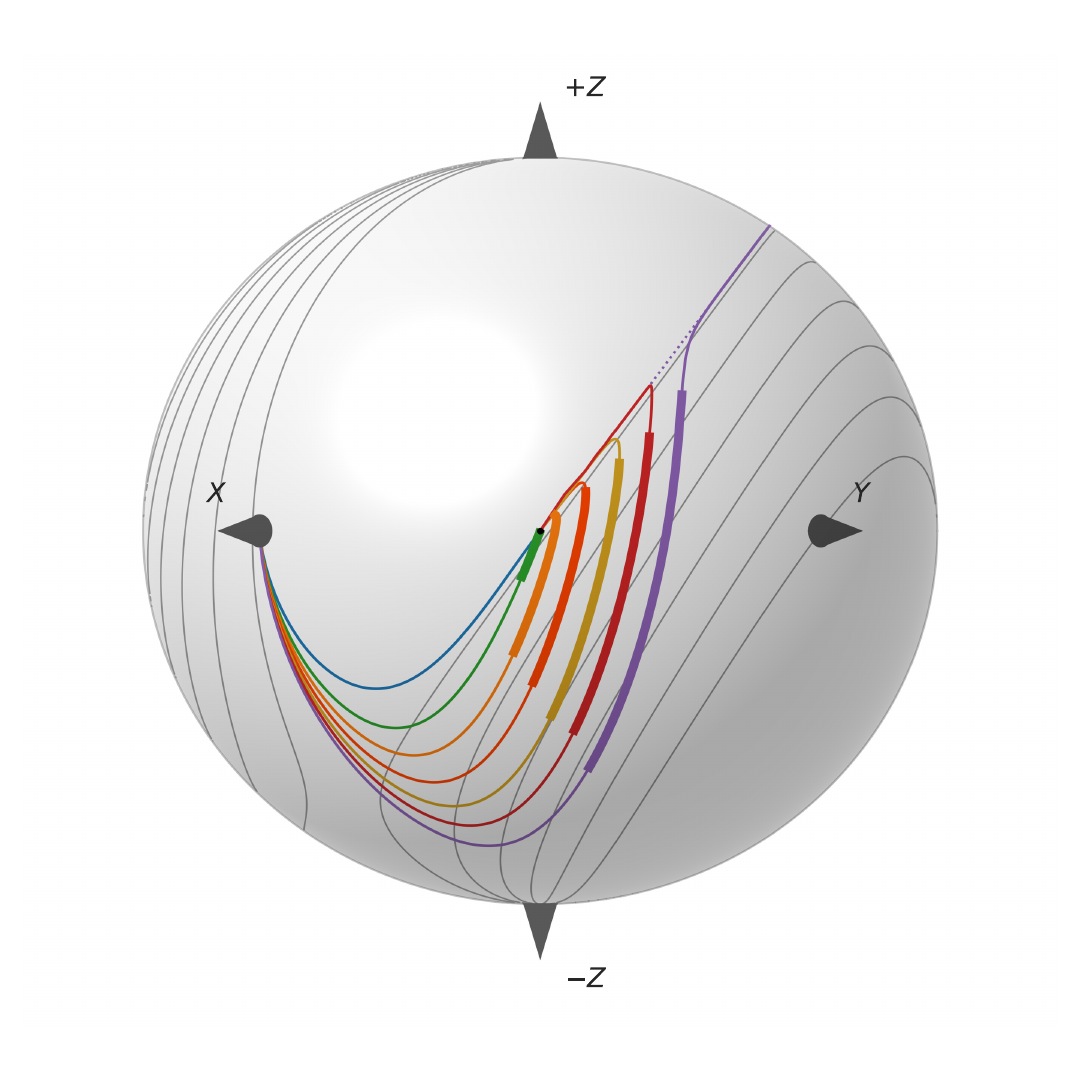}
\caption{Indicator sphere. Falkner--Skan profiles $\beta=+0.30$, $0$,
$-0.10$, $-0.16$, $-0.19$, $-0.1988$; separated Stewartson
$\beta=-0.19$, $H\approx4.9$; $\hat\Omega\hat I$ contours
$\{0.025,0.2,0.4,0.6,0.8,1.0\}$.}\label{f:sphere}
\end{figure}

In a thin quasi-two-dimensional layer the profile is captured to second
order about any wall-normal point by three velocity scales formed from
the wall distance $d$: the velocity $u$, the shear $d\,u'$, and the
curvature $\tfrac12 d^2u''$. Normalized by their common magnitude, the local
velocity scale $U_\mathrm{loc}=\sqrt{u^2+(du')^2+(\tfrac12 d^2u'')^2}$, they
place the local state
on the unit sphere of directions
\begin{equation}
  (X,Y,Z) = \big(u,\;d\,u',\;\tfrac12 d^2u''\big),
  \qquad
  (\hat X,\hat Y,\hat Z) = (X,Y,Z)\big/U_\mathrm{loc},
  \label{w:ind}
\end{equation}
so unhatted symbols carry the velocity scale and hatted ones are the
direction cosines on the unit sphere,
\begin{equation}
  \hat\Omega = \frac{\hat Y}{\sqrt{\hat X^2+\hat Y^2}}
  = \frac{Y}{\sqrt{X^2+Y^2}}
  \qquad\text{(vorticity fraction)},
  \qquad
  \hat I = \hat Y - \hat X - \hat Z
  \qquad\text{(accumulated inflection)}.
  \label{w:omhat}
\end{equation}
A near-wall parabola $u=c_1y+c_2y^2$ satisfies $\hat I=0$ identically.
\begin{equation}
  I(d) \equiv U_\mathrm{loc}\,\hat I = d\,u'-u-\tfrac12 d^2u''
  = -\int_0^d \tfrac12\,s^2\,u'''(s)\,\mathrm{d}s,
  \label{w:integral}
\end{equation}
Every profile leaves the wall with
$\hat I$ third-order small ($u'''_w=0$ on a steady wall). The vorticity
Reynolds number $Re_\Omega=d^2\omega/\nu$ is reserved for the onset
gate. Fig.~\ref{f:sphere}.

In the solver the triple is assembled from four local vector-calculus
objects: the velocity $\mathbf u$, the vorticity
$\boldsymbol\omega$, the wall-distance vector
$\mathbf d=d\,\nabla d$, and the curvature vector
$\boldsymbol\ell=\tfrac12\langle\mathbf d,\mathbf d\rangle\,\nabla^2\mathbf u$,
with $\nabla^2\mathbf u=-\nabla\times\boldsymbol\omega$ supplying the
second derivative.
\begin{equation}
  (X,\,Y,\,Z) \;=\;
  \Big(\,\langle\mathbf u,\mathbf u\rangle,\;
  \sqrt{\langle\mathbf u,\mathbf u\rangle\,
        \langle\mathbf d,\mathbf d\rangle\,
        \langle\boldsymbol\omega,\boldsymbol\omega\rangle},\;
  \langle\boldsymbol\ell,\mathbf u\rangle\,\Big),
  \qquad
  \hat\Omega = \frac{Y}{\sqrt{X^2+Y^2}}, \quad
  \hat I = \frac{Y-X-Z}{\sqrt{X^2+Y^2+Z^2}},
  \label{w:gram}
\end{equation}
The solver's equivalent \emph{ratio} realization,
\begin{equation}
  (X,\,Y,\,Z) \;=\;
  \big(\,|\mathbf u|,\;\; d\,|\boldsymbol\omega|,\;\;
  \tfrac12 d^2(\nabla^2\mathbf u)\!\cdot\!\hat{\mathbf u}\,\big),
  \label{w:ratio}
\end{equation}
coincides with Eq.~\eqref{w:gram} wherever $\mathbf u\neq0$.

\subsection{Amplification rate and onset threshold}

\begin{figure}[H]\centering
\includegraphics[width=0.62\textwidth]{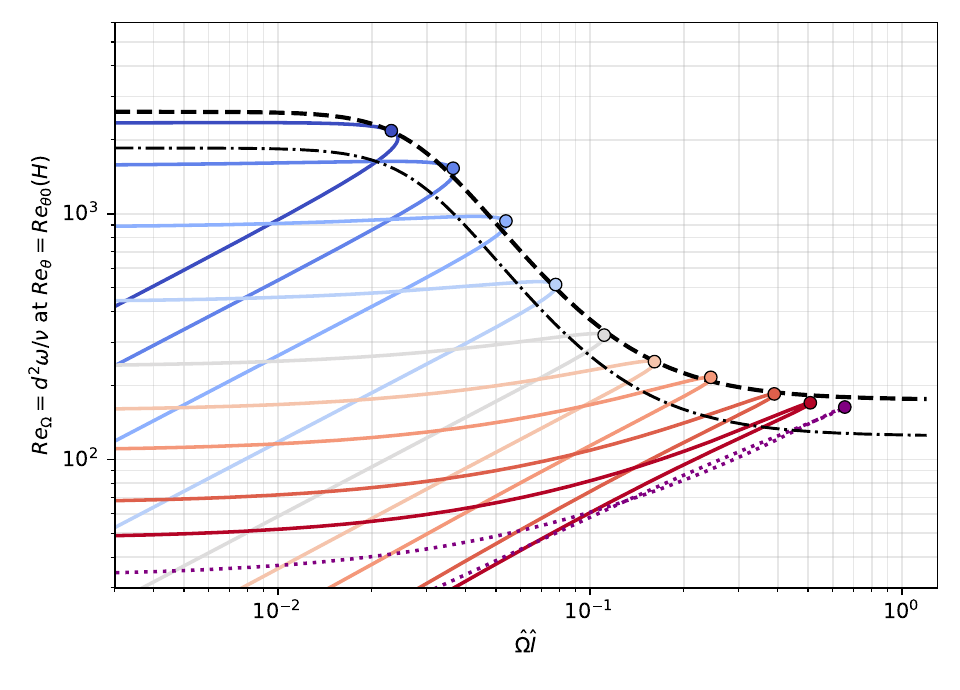}
\caption{Onset-threshold graze. Attached family $\beta=+0.15$,
$+0.10$, $+0.05$, $0$, $-0.05$, $-0.10$, $-0.15$, $-0.19$, $-0.1988$,
each at its Drela--Giles $Re_{\theta 0}(H)$; separated Stewartson
$\beta=-0.19$ ($H\approx4.9$, $Re_{\theta 0}\approx26$).
$\max_y\hat\Omega\hat I=0.162$/$0.078$/$0.037$ at $\beta=-0.10$/$0$/$+0.10$.
Envelope $k=1$; model threshold $k=0.712$.}\label{f:graze}
\end{figure}

\begin{figure}[H]\centering
\includegraphics[width=0.99\textwidth]{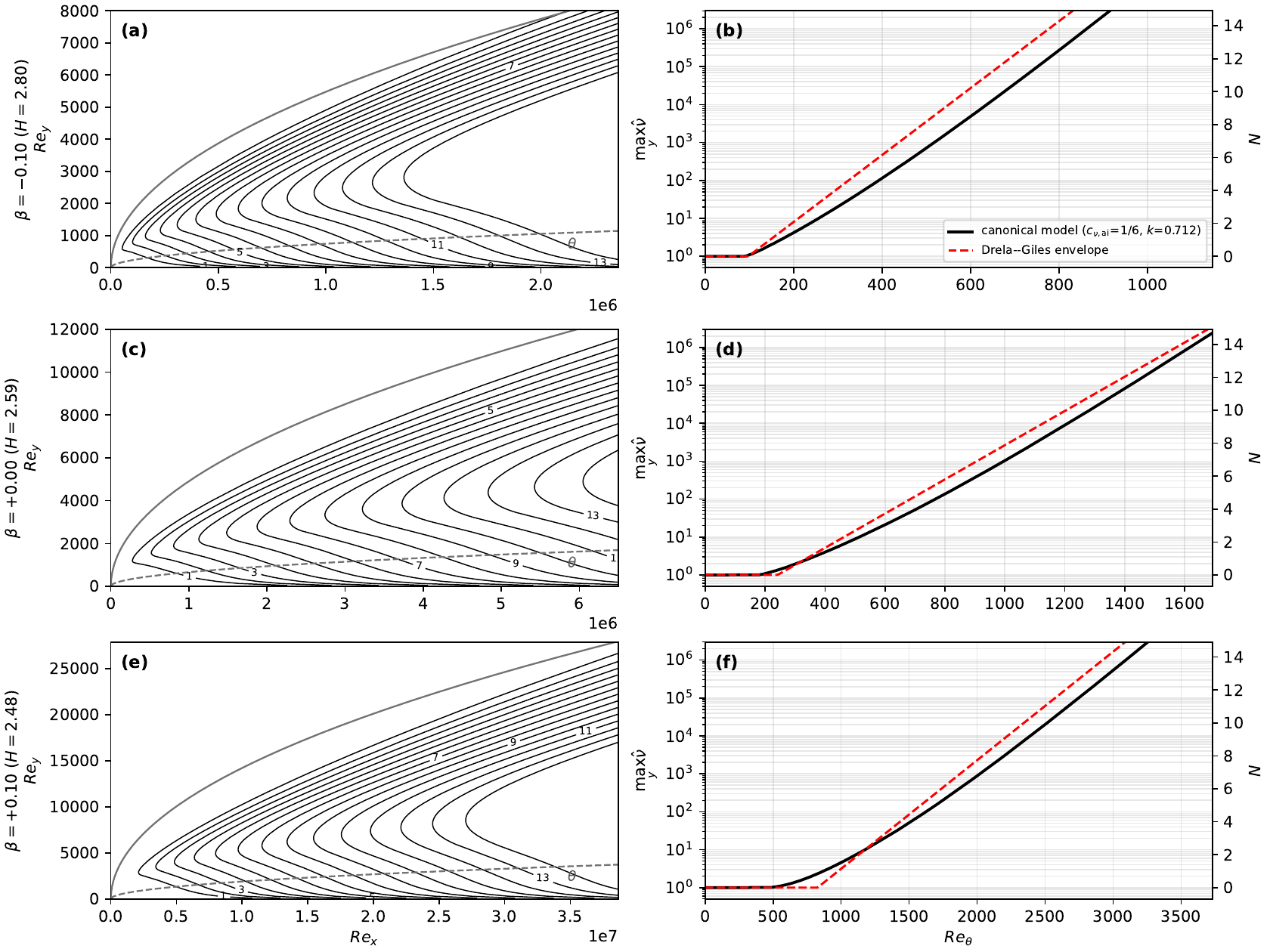}
\caption{Disturbance transport, three wedges. Adverse $\beta=-0.10$
($\max_y\hat\Omega\hat I=0.162$), Blasius $\beta=0$ ($0.078$),
favorable $\beta=+0.10$ ($0.037$); seeded $\hat\nu_\infty=1$,
$a_{\max}=0.19$, $(c_{\nu,\mathrm{ai}},k)=(\tfrac16,\,0.712)$. Anchor
$N\!=\!9$ crossing $7\%$ past Drela; favorable $N\!=\!1$ station
${\sim}11\%$ early.}\label{f:nuhat}
\end{figure}

\begin{figure}[H]\centering
\includegraphics[width=\textwidth]{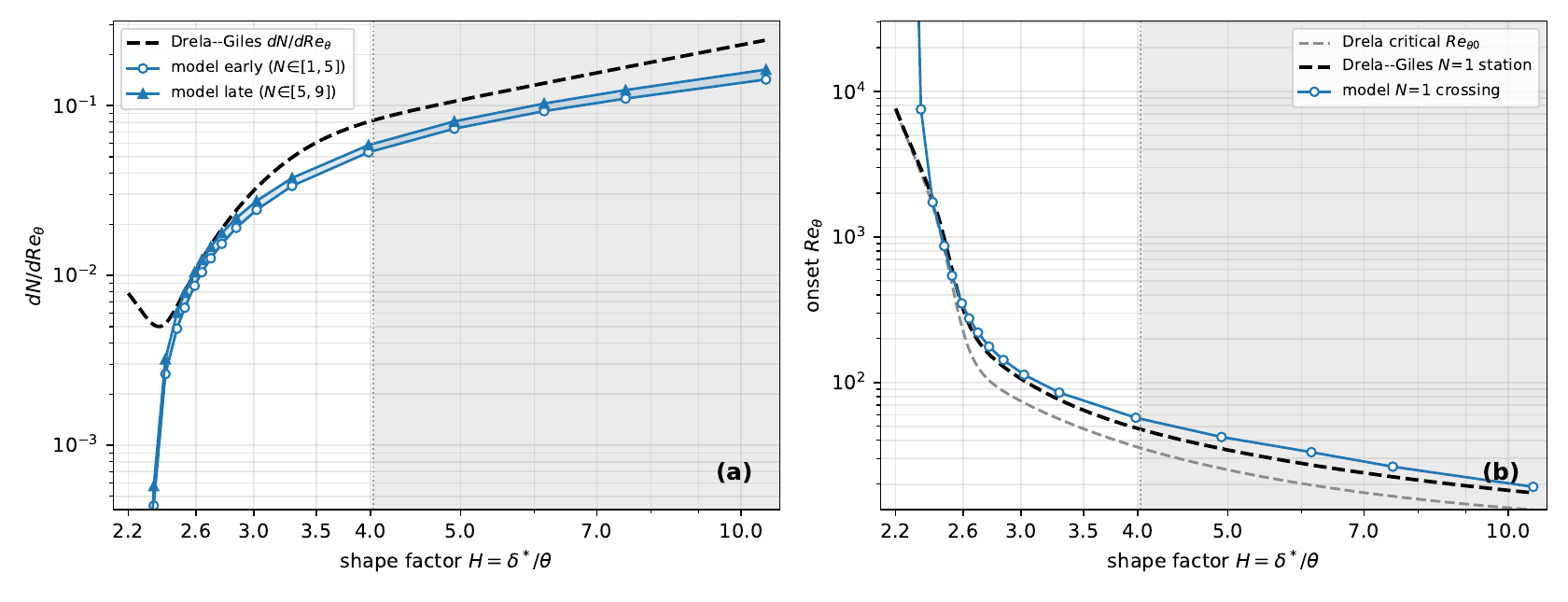}
\caption{Calibrated model vs Drela--Giles. Stagnation $H\approx2.2$,
Blasius $H\approx2.59$, separation limit $H=3.98$, reversed-flow
Stewartson $H\approx10.6$; $a_{\max}=0.19$, $c_{\nu,\mathrm{ai}}=1/6$.
(b)~onset tracks the Drela--Giles $N\!=\!1$ stations within
${\sim}\pm12\%$ from $H=2.44$ ($\beta=+0.15$) to
$H=3.98$.}\label{f:calibrate}
\end{figure}

The complete amplification source is
\begin{equation}
  P_{\mathrm{AI}} = a\,\sigma_{Re}\,\omega\,\tilde\nu, \qquad
  a = a_{\max}\,\operatorname{clip}\!\big\langle\,\hat\Omega\,\hat I\,\big\rangle_{0}^{1},
  \qquad
  \sigma_{Re} = \sigma_{Re}\!\big(Re_\Omega/Re_\Omega^{\mathrm c}(\hat\Omega\hat I)\big), \qquad
  \sigma_{Re}(\zeta)=\tfrac12\big[1+\tanh\!\big((\zeta-1)/w\big)\big],
  \label{w:rate}
\end{equation}
\begin{equation}
  Re_\Omega^{\mathrm c}(\hat\Omega\hat I) = k\,\operatorname{softmin}_2\!\Big(C,\;
      A + B\,\langle\hat\Omega\hat I\rangle_+^{-2}\Big),
  \qquad
  \operatorname{softmin}_2(x_1,x_2)=\frac{x_1x_2}{\sqrt{x_1^2+x_2^2}},
  \qquad \langle x\rangle_+=\max(x,0).
  \label{w:onset}
\end{equation}
Fig.~\ref{f:graze}: graze ratios $0.93$--$1.14$,
root-mean-square $\sim6\%$, from incipient separation through
$\beta=+0.15$. The marched Blasius $N\!=\!9$ crossing lands at
$Re_\theta=1181$, $7\%$ past Drela's $1108$; across the attached
family the $N\!=\!1$ stations land on the Drela--Giles stations to
$6\%$ root-mean-square (adverse wedges within $\pm4\%$, the favorable
pair $11$--$12\%$ early). By $\beta=+0.35$ the rate is
$\approx0.1\times$ Drela and the onset $\approx3\times$ its critical
$Re_\theta$; by $\beta\approx0.45$ transition has retreated beyond any
station of practical relevance, and stronger favorable wedges are
suppressed outright (Figs.~\ref{f:nuhat}--\ref{f:calibrate}).

\subsection{Reduced laminar diffusion}

The molecular coefficient in the diffusion term of Eq.~\eqref{w:sa}
only is scaled by $c_{\nu,\mathrm{ai}}<1$, leaving the mean-flow
viscosity, the SA production and destruction, and
$\chi=\tilde\nu/\nu$ (formed with the true $\nu$) untouched.

Marched on Blasius, against the Drela--Giles $N\!=\!1$ station
$Re_\theta=337.9$ and slope $1.039\times10^{-2}$:
\begin{center}\small
\begin{tabular}{ccc}
  \toprule
  $c_{\nu,\mathrm{ai}}$ & $N\!=\!1$ station & rate at $N\!=\!1$ \\
  \midrule
  $1$            & $556.7$ & $61\%$ \\
  $1/2$          & $450.7$ & $71\%$ \\
  $1/4$          & $387.5$ & $79\%$ \\
  $1/6$          & $365.2$ & $84\%$ \\
  $1/12$         & $337.4$ & $90\%$ \\
  $1/24$         & $320.8$ & $95\%$ \\
  $1/48$         & $312.7$ & $98\%$ \\
  $0$            & $304.6$ & $103\%$ \\
  \bottomrule
\end{tabular}
\end{center}
The rate is the $N\!=\!1$--$5$ secant as a fraction of the Drela--Giles
slope. The model uses
\begin{equation*}
  c_{\nu,\mathrm{ai}}=\tfrac16 .
\end{equation*}
The equilibrium disturbance-band thickness scales as
$\sqrt{c_{\nu,\mathrm{ai}}}$, so every halving thins the band the grid
must resolve by $\sqrt2$.

\subsection{Handover blend and destruction tie}

The destruction is tied, not gated off: $\sigma_D$ tracks $\sigma_P$ with
slope $c_{b1}/(\kappa^2c_{w1})\approx0.249$ from the floor
$(1+c_{b2})/(\sigma c_{w1})\approx0.751$, the two summing to one by the
definition of $c_{w1}$. On the constant-stress wall layer SA's solution is
the linear profile $\tilde\nu=\kappa u_\tau y$, with
$P_\mathrm{SA}=c_{b1}u_\tau^2$,
$D_\mathrm{SA}=c_{w1}\kappa^2u_\tau^2$ and self-diffusion
$(1+c_{b2})\kappa^2u_\tau^2/\sigma$ all height-independent; requiring the
gated balance to hold pointwise gives Eq.~\eqref{w:sigma}, under which that
profile is an exact solution at every height and for any $\tau$. In the
laminar range the destruction floor is inert: below $10^{-3}$ of the
amplification production up to the $\chi=1$ crossing, at most $1.4\%$ of the
$\sigma_P$-weighted SA production at $\chi=c_{v1}$, and worth less than $1\%$
on the anchor and prediction stations.

\subsection{Eddy-viscosity assembly}

The near-wall damping $f_{v1}$ enforces the attached-wall-layer
identity $\chi=\kappa y^+$. Writing $y^+:=\chi/\kappa$, the reconstruction
\begin{equation}
  q = \frac{|\mathbf u|\,d/\nu}{\,y^+\,U^+(y^+)\,},
  \qquad
  U^+(y^+)=\tfrac1\kappa\ln(1+\kappa y^+)
  +7.8\big(1-e^{-y^+/11}-\tfrac{y^+}{11}e^{-y^+/3}\big)
  \;\;\text{(Reichardt)},
  \label{w:fv1q}
\end{equation}
sets the bypass weight of Eq.~\eqref{w:ns},
\begin{equation}
  \sigma_{v1} = g_\chi(\chi)\,g_q(q), \qquad
  g_\chi=\mathrm{clip}(\chi-1,0,1), \qquad
  g_q=\mathrm{clip}((q-2)/2,0,1),
  \label{w:fv1bypass}
\end{equation}
inert for $\chi<1$ (negligible $\mu_t$ either way), for
$\chi\gtrsim30$ ($f_{v1}\to1$), and in the attached inner layer
($q=1$, $G$ shut). Controlled on/off batteries bound the net effect on
attached-layer outer edges below $1.5$ drag counts (flat plate and
NLF at their cold-budget protocol).

\subsection{Freestream seed}

A case seeded
with $\chi_\infty$ reaches the handover where its envelope crosses
$N_\mathrm{crit}=\ln(c_{v1}/\chi_\infty)$.
\begin{equation}
  N_\mathrm{crit}(Tu) = -8.43 - 2.4\,\ln(Tu_{\mathrm{frac}}),
  \qquad
  \chi_\infty = c_{v1}\,e^{-N_\mathrm{crit}},
  \label{w:mack}
\end{equation}
One
$e$-fold in $\chi_\infty$ is one unit of $N_\mathrm{crit}$, which on
the Blasius envelope ($dN/dRe_\theta\approx10^{-2}$) shifts onset by
$\Delta Re_\theta\approx100$. Strongly bypass-dominated conditions
($Tu\gtrsim1\%$) are out of scope: at $Tu=1\%$ the map leaves only
$N_\mathrm{crit}\approx2.6$, and by $Tu\approx3\%$ the seed exceeds
$c_{v1}$ outright.

\subsection{Constants}

Baseline SA, unchanged:
\begin{equation*}
  c_{b1}=0.1355,\quad \sigma=\tfrac{2}{3},\quad c_{b2}=0.622,\quad
  \kappa=0.41,\quad c_{w1}=\frac{c_{b1}}{\kappa^2}+\frac{1+c_{b2}}{\sigma},\quad
  c_{w2}=0.3,\quad c_{w3}=2,\quad c_{v1}=7.1 .
\end{equation*}

SA-AI:
\begin{table}[H]\centering\small
\begin{tabular}{llp{3.9in}}
\toprule
constant & value & determination \\
\midrule
$a_{\max}$ & $0.19$ & tanh-layer Kelvin--Helmholtz temporal eigenvalue, $\omega_{i,\max}/\omega_{\mathrm{peak}}=0.1897$ \\
$(C,A,B)$ & $(2600,\,175,\,2)$ & graze envelope of the attached Falkner--Skan family at its critical $Re_\theta$ (rms $\sim6\%$) \\
$k$ & $0.712$ & Blasius march crosses $N\!=\!1$ at the Drela--Giles station $Re_\theta=338$ at $c_{\nu,\mathrm{ai}}=\tfrac16$ \\
$w$ & $0.35$ & onset ramp width; $N\!=\!1$ stations move $<\!1\%$ halved, $<\!2.5\%$ doubled \\
$c_{\nu,\mathrm{ai}}$ & $1/6$ & frozen-profile growth eigenvalue reaches the Drela--Giles rate just past critical ($1.02$ at $Re_\theta\!=\!400$) \\
$\tau$ & $4$ & numerics bracket: handover $78\%$ complete at $\chi=c_{v1}$, $97\%$ at $\chi\approx14.8$ \\
$\sigma_D$ tie & Eq.~\eqref{w:sigma} & derived: constant-stress wall-layer balance; linear wall solution exact \\
bypass $(g_\chi,g_q)$ & Eq.~\eqref{w:fv1bypass} & parameter-free law-of-the-wall discriminator; $g_q$ ramps over $q\in(2,4)$ \\
$(A,B)_\mathrm{Mack}$ & $(-8.43,\,2.4)$ & Mack 1977 $e^N$ correlation, wholesale \\
\bottomrule
\end{tabular}
\end{table}

Free constants: baseline SA $7$ ($9$ with the retained Allmaras
modifications); SA-AI $+8$, $+2$ imported from Mack, $19$ total; Drela's
envelope $e^N$ in XFOIL $13$, plus $2$ compressibility constants and
$N_\mathrm{crit}$.

\subsection{Running the model}

\begin{figure}[H]\centering
\includegraphics[width=\textwidth]{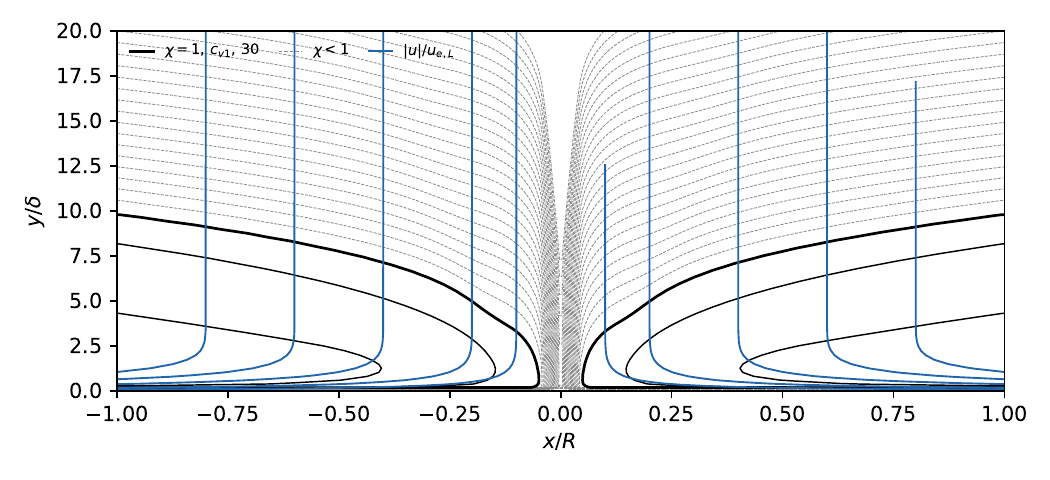}
\caption{Attachment-anchored turbulent branch, $L=3000$
($Re_r=9\times10^6$), $\max\chi=76$. Standard SA, frozen Hiemenz,
$\chi_\infty=0$. Abscissa $x/R$ (arc length in nose radii, $R/\delta=L$);
ordinate $y/\delta$. $\chi$ contours dashed per decade for $\chi<1$
($10^{-24}$--$10^{-1}$; the zero freestream leaves no floor, so $\chi$
decays smoothly over decades), solid at $1$, $c_{v1}$, $30$; velocity
magnitude
$|u|/u_{e,L}$ (blue, normalized by the $x\!=\!L$ edge speed) at $0.1$,
$0.2$, $0.4$, $0.6$, $0.8$. Branch data in
Table~\ref{t:stagbistab}.}\label{f:stagbistab}
\end{figure}

On the frozen Hiemenz stagnation field with the freestream
seed exactly zero, standard SA sustains the wedge above a critical
nose Reynolds number bisected to $Re_r=4.66$--$4.74\times10^5$
(Fig.~\ref{f:stagbistab}).

\begin{table}[H]
  \centering\small
  \caption{Attachment-anchored branch: steady $\max\chi$ versus nose
  Reynolds number $Re_r=L^2$ on the frozen Hiemenz wedge, standard SA,
  $\chi_\infty=0$. Below the bisected critical band the layer collapses
  to $\chi=0$.}
  \label{t:stagbistab}
  \begin{tabular}{ccc}
    \toprule
    $L=\sqrt{Re_r}$ & $Re_r$ & steady $\max\chi$ \\
    \midrule
    $\le 683$ & $\le 4.66\times10^5$ & $<10^{-3}$ (collapses to $0$) \\
    $683$--$689$ & $4.66$--$4.74\times10^5$ & critical band (bisected) \\
    $689$ & $4.74\times10^5$ & $2.19$ \\
    $700$ & $4.90\times10^5$ & $6.15$ \\
    $1000$ & $1.00\times10^6$ & $24.1$ \\
    $3000$ & $9.00\times10^6$ & $75.7$ \\
    \bottomrule
  \end{tabular}
\end{table}

The protocol used for every result in this paper: the
$\tilde\nu$-equation residual is scaled by $f^{\,1-g}$, where $g$
rises smoothly from $0$ (laminar) to $1$ (turbulent) with $\chi$ and
$f=10^{-2}$ (the solver uses $g=1-e^{-(\chi-c_{v1})/4}$ above
$\chi=c_{v1}$, zero below). Because
the freestream boundary condition flows through the same scaling, the
freestream viscosity ratio supplied to the solver is $\chi_\infty f$;
the seeds quoted throughout are the physical $\chi_\infty$, recovered
at the boundary-layer edge.

\section{Results}

Coupled RANS: Flow360, unstructured node-centered second-order
finite volume, Roe flux, $M=0.1$. Airfoil cases:
$\chi_\infty=c_{v1}e^{-9}\approx8.76\times10^{-4}$ (the
$N_\mathrm{crit}=9$ anchor; via Mack's map, $Tu\approx0.07\%$); two
independently generated mesh families---all-triangular unstructured
cavity meshes, constrained-Delaunay through the boundary layer, and
all-quadri\-lateral structured O-grids---at three refinement levels,
$1.5\times10^4$ to $2.6\times10^5$ nodes.
Transition locations are the solver's near-wall $\chi=1$ crossing
(the $\chi=c_{v1}$ crossing sits up to $0.05\,c$ aft where the front
grows slowly); bubble stations are signed-$C_f$ zero crossings. The
cross-solver replication is a cell-centered, pressure-based,
incompressible OpenFOAM implementation of the model on the same
flat-plate grid specification and the same structured airfoil grids
(the port omits the $f_{v1}$ bypass); thirty airfoil cases plus the
flat-plate sweep.

\subsection{The Blasius flat plate}

\begin{figure}[H]\centering
\includegraphics[width=0.90\textwidth]{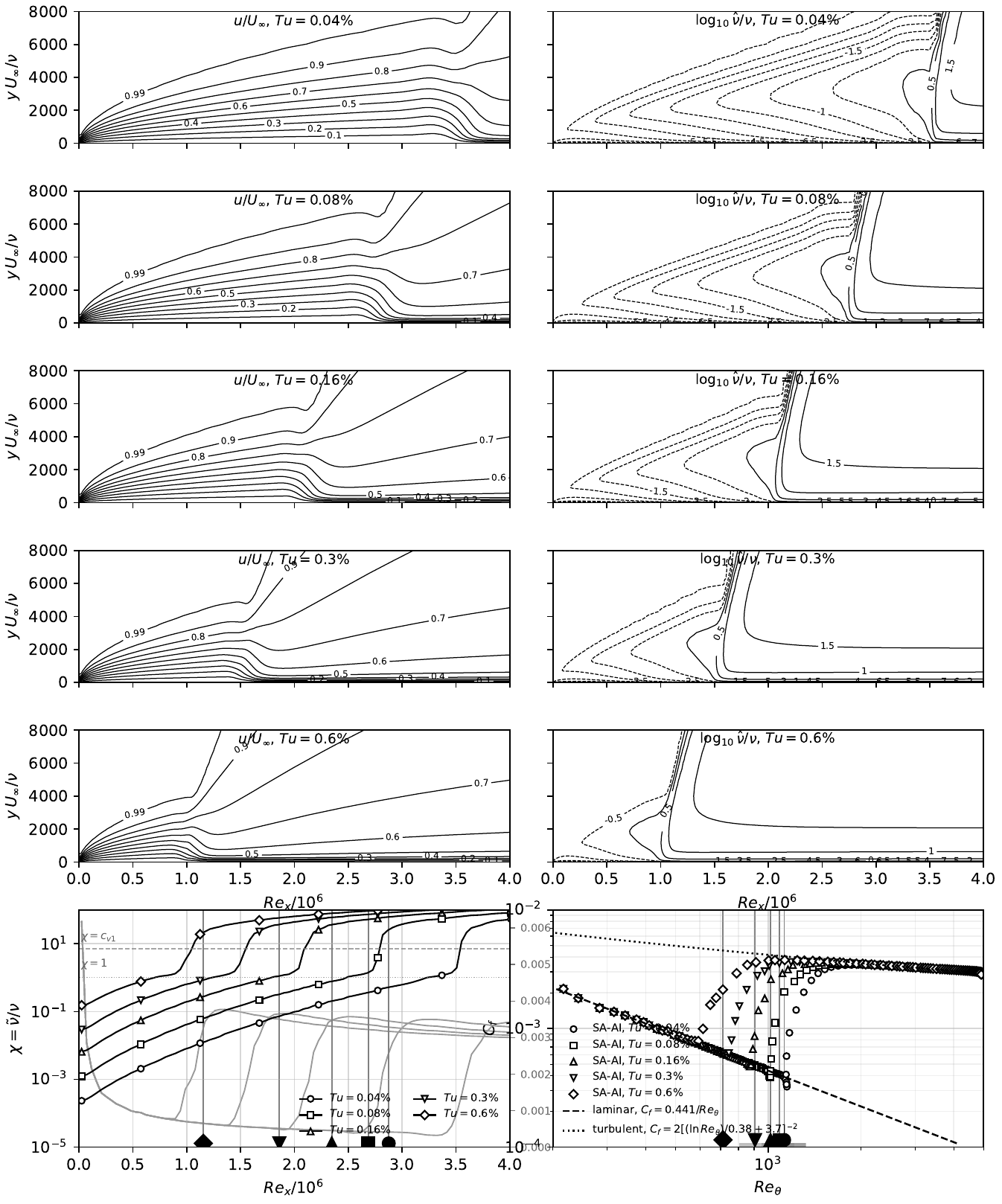}
\caption{Flat plate, Flow360. Five levels $Tu=0.04$--$0.60\%$;
references $\chi=1$ and $\chi=c_{v1}=7.1$; laminar
$C_f=0.441/Re_\theta$, turbulent
$C_f=2[(\ln Re_\theta)/0.38+3.7]^{-2}$; $\chi=c_{v1}$ onset brackets
Abu-Ghannam \& Shaw within $10\%$.}\label{f:flatplate}
\end{figure}

\begin{figure}[H]\centering
\includegraphics[width=0.90\textwidth]{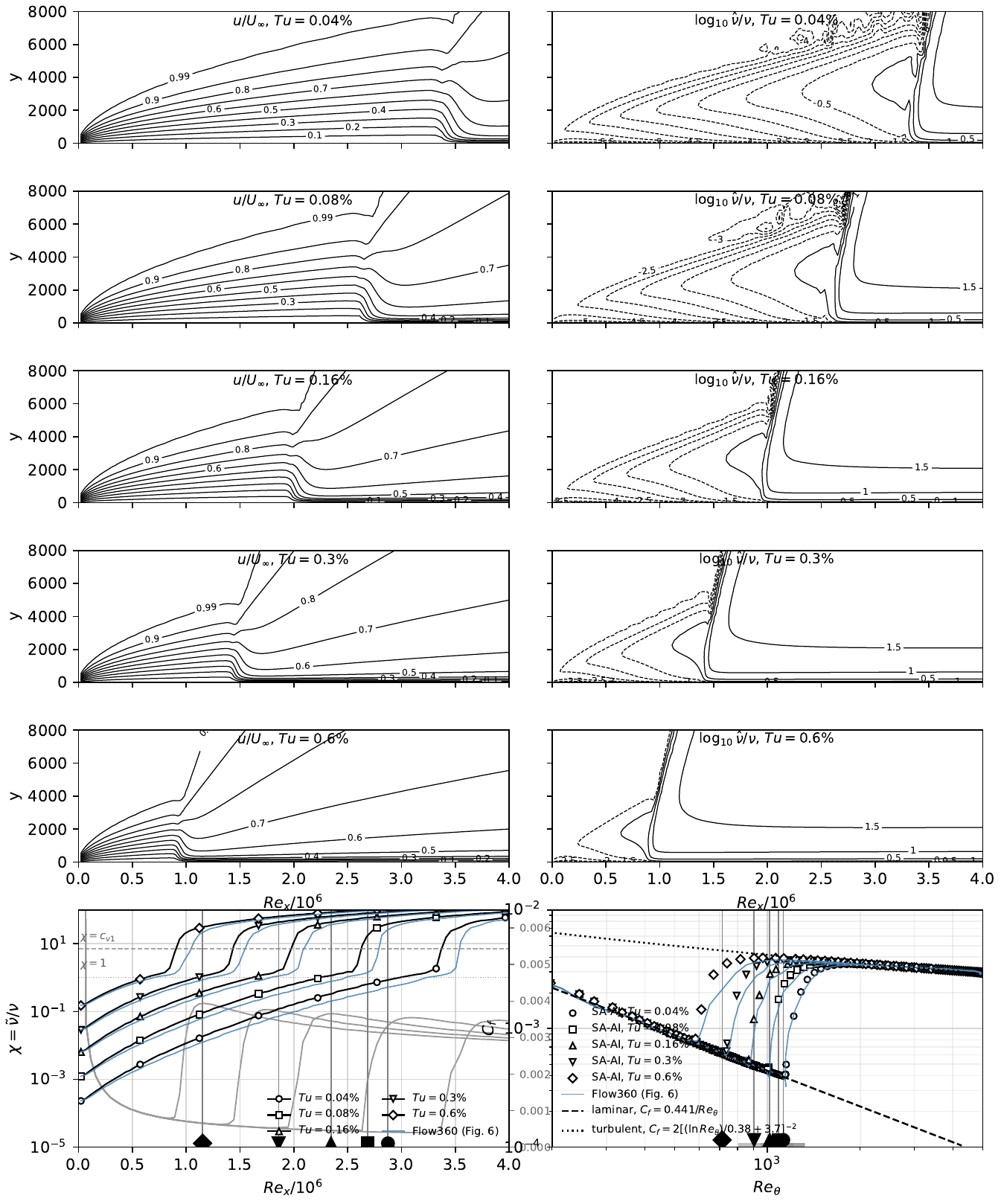}
\caption{Flat plate, OpenFOAM. Same sweep and layout as
Fig.~\ref{f:flatplate}, five levels $Tu=0.04$--$0.60\%$; thin steel-blue
curves in both bottom panels are the Flow360 result of
Fig.~\ref{f:flatplate}.}\label{f:flatplateof}
\end{figure}

Structured $320\times80$ grid, $Re_x\le6\times10^6$,
$y^+\lesssim0.3$; five seeds from Mack's map
($\chi_\infty\approx2.3\times10^{-4}$, $1.2\times10^{-3}$,
$6.3\times10^{-3}$, $2.9\times10^{-2}$, $1.5\times10^{-1}$).
Figure~\ref{f:flatplate}: at the $\chi=c_{v1}$
crossing the onset brackets the Abu-Ghannam \& Shaw correlation
within $10\%$ across the range. Figure~\ref{f:flatplateof}: $\chi=1$
crossings within $7\%$ in $Re_\theta$. Table~\ref{t:flatplate}: both
solvers' crossings.

\begin{table}[H]
  \centering\small
  \caption{Flat-plate transition-onset $Re_\theta$ (integrated from the
  computed profiles) at the $\chi\!=\!1$ and $\chi\!=\!c_{v1}$
  crossings, Flow360 and the OpenFOAM replication (OF), against the
  Abu-Ghannam \& Shaw correlation.}
  \label{t:flatplate}
  \begin{tabular}{cccccc}
    \toprule
    $Tu$ [\%] & AGS $Re_\theta$ & $\chi\!=\!1$ & $\chi\!=\!c_{v1}$ & OF $\chi\!=\!1$ & OF $\chi\!=\!c_{v1}$ \\
    \midrule
    0.04 & 1126 & 1149 & 1203 & 1098 & 1154 \\
    0.08 & 1088 & 1002 & 1069 & 949 & 1004 \\
    0.16 & 1017 & 845 & 934 & 798 & 887 \\
    0.3 & 905 & 700 & 815 & 656 & 739 \\
    0.6 & 713 & 524 & 685 & 487 & 589 \\
    \bottomrule
  \end{tabular}
\end{table}

\subsection{The NLF(1)-0416 airfoil}

At L2 every front sits within
$0.016\,c$ and the lift within $0.9\%$ of the SA-AI values.

At the four in-window
incidences both computed fronts lie inside the submittal band on both
surfaces (at $2^\circ$, $0.338$/$0.344\,c$ upper against
$0.040$--$0.453\,c$ and $0.604$/$0.607\,c$ lower against
$0.080$--$0.654\,c$).

Across the Reynolds sweep the upper-surface front moves forward monotonically and joins the
$4\times10^6$ benchmark without a step
($0.446\to0.418\to0.400\to0.387\,c$ at $0^\circ$;
$0.379\to0.320\to0.279\to0.254\,c$ at $4^\circ$).
Somers reports the last laminar and first turbulent orifice, so each lift
coefficient carries a
$0.05\,c$ bracket rather than a point; all four panels of his Fig.~9 are
digitized in \texttt{data/somers1981\_nlf0416\_transition\_by\_Re.json}
(\texttt{repro/cfd/digitize\_somers\_fig9.py}).
Table~\ref{t:extnlfsomers}: $20$ of the $27$ covered comparisons fall inside
the bracket, the misses being $0.012$--$0.029\,c$ aft on the upper surface at
$\alpha=4^\circ$, $Re=2$ and $3\times10^6$, and $0.020\,c$ ahead on the lower
surface at $\alpha=4^\circ$, $Re=10^6$. Digitization check: all $28$ points of
the same data as replotted in Coder's dissertation fall inside these brackets.

\begin{table}[H]
  \centering\small
  \caption{NLF(1)-0416 Reynolds sweep against the Somers TP-1861 Fig.~9 transition brackets at matched lift (last laminar to first turbulent orifice). $20$ of the $27$ covered comparisons fall inside the bracket. $^\dagger$Sec.~V benchmark rows.}
  \label{t:extnlfsomers}
  \begin{tabular}{cc c ccc ccc}
    \toprule
    & & & \multicolumn{3}{c}{upper surface $x_\mathrm{tr}$} & \multicolumn{3}{c}{lower surface $x_\mathrm{tr}$} \\
    \cmidrule(lr){4-6}\cmidrule(lr){7-9}
    $\alpha$ & $Re/10^6$ & $c_l$ & str & cav & measured & str & cav & measured \\
    \midrule
    $0$ & $1$ & 0.4898 & 0.446 & 0.456 & $0.421$--$0.472$ & 0.613 & 0.616 & $0.601$--$0.651$ \\
    $0$ & $2$ & 0.5053 & 0.418 & 0.414 & $0.401$--$0.451$ & 0.604 & 0.611 & $0.602$--$0.652$ \\
    $0$ & $3$ & 0.5109 & 0.400 & 0.400 & $0.367$--$0.417$ & 0.586 & 0.586 & $0.585$--$0.634$ \\
    $0$ & $4$\rlap{$^\dagger$} & 0.5139 & 0.387 & -- & $0.352$--$0.400$ & 0.566 & -- & $0.532$--$0.582$ \\
    $4$ & $1$ & 0.9584 & 0.379 & 0.379 & $0.350$--$0.400$ & 0.631 & 0.630 & $0.651$--$0.701$ \\
    $4$ & $2$ & 0.9744 & 0.320 & 0.323 & $0.258$--$0.308$ & 0.619 & 0.628 & $0.601$--$0.651$ \\
    $4$ & $3$ & 0.9786 & 0.279 & 0.287 & $0.211$--$0.260$ & 0.612 & 0.626 & $0.602$--$0.650$ \\
    $4$ & $4$\rlap{$^\dagger$} & 0.9805 & 0.254 & -- & -- & 0.610 & -- & $0.601$--$0.650$ \\
    \bottomrule
  \end{tabular}
\end{table}

\begin{table}[H]
  \centering\small
  \caption{NLF(1)-0416 Reynolds sweep on the L2 pair at the two incidence anchors, $M=0.1$ and the paper-wide seed.}
  \label{t:extnlfresweep}
  \begin{tabular}{cc cccc cccc}
    \toprule
    & & \multicolumn{4}{c}{structured} & \multicolumn{4}{c}{cavity} \\
    \cmidrule(lr){3-6}\cmidrule(lr){7-10}
    $\alpha$ & $Re/10^6$ & $C_L$ & $C_D$ & $x_\mathrm{tr}^\mathrm{up}$ & $x_\mathrm{tr}^\mathrm{lo}$
             & $C_L$ & $C_D$ & $x_\mathrm{tr}^\mathrm{up}$ & $x_\mathrm{tr}^\mathrm{lo}$ \\
    \midrule
    $0$ & $1$ & 0.4898 & 0.00777 & 0.446 & 0.613 & 0.4937 & 0.00825 & 0.456 & 0.616 \\
    $0$ & $2$ & 0.5053 & 0.00636 & 0.418 & 0.604 & 0.5112 & 0.00685 & 0.414 & 0.611 \\
    $0$ & $3$ & 0.5109 & 0.00588 & 0.400 & 0.586 & 0.5181 & 0.00633 & 0.400 & 0.586 \\
    $4$ & $1$ & 0.9584 & 0.00890 & 0.379 & 0.631 & 0.9574 & 0.00942 & 0.379 & 0.630 \\
    $4$ & $2$ & 0.9744 & 0.00766 & 0.320 & 0.619 & 0.9764 & 0.00814 & 0.323 & 0.628 \\
    $4$ & $3$ & 0.9786 & 0.00740 & 0.279 & 0.612 & 0.9819 & 0.00784 & 0.287 & 0.626 \\
    \bottomrule
  \end{tabular}
\end{table}

\begin{figure}[H]\centering
\includegraphics[width=\textwidth]{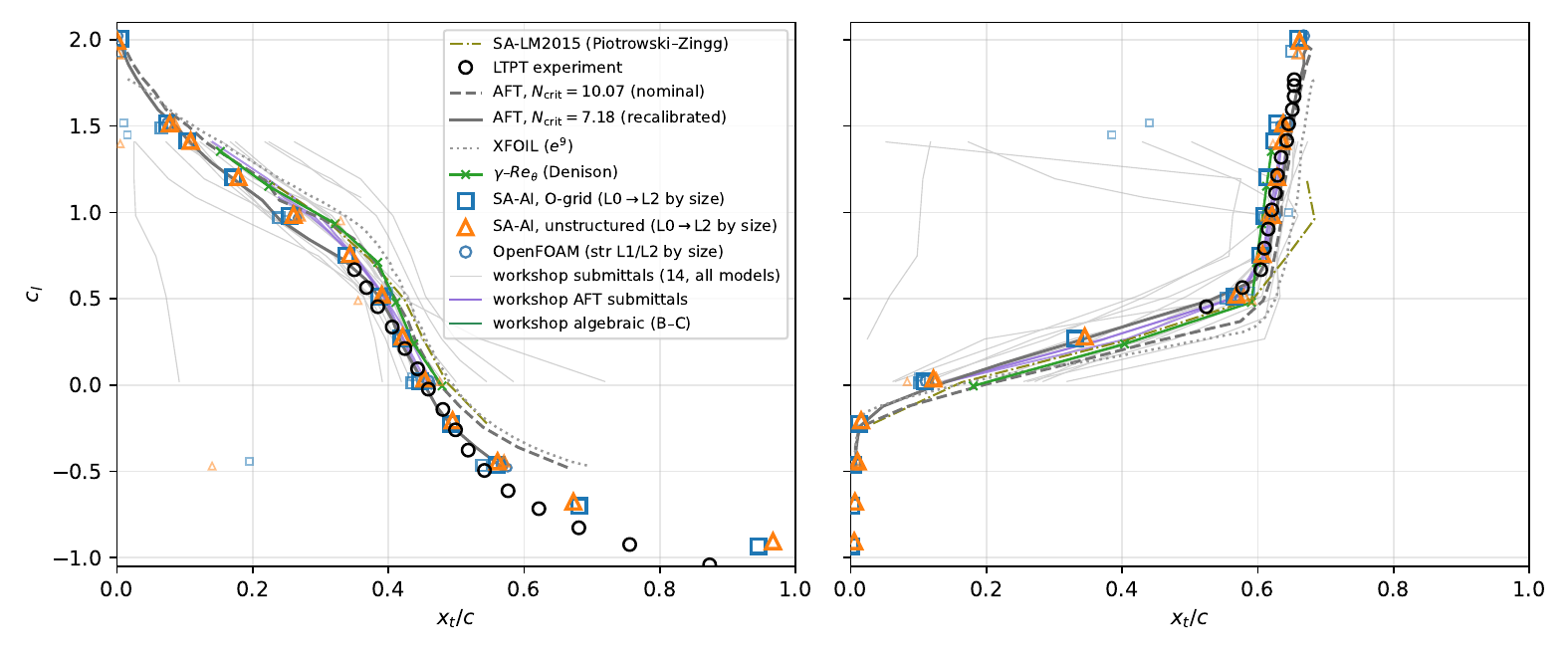}
\caption{NLF(1)-0416 transition fronts, $Re=4\times10^6$; near-wall
$\chi=1$ convention. SA-AI at
$\alpha=-8^\circ,-4^\circ,0^\circ,4^\circ,9^\circ,15^\circ$ on all six
grids, and at $-12^\circ$, $-10^\circ$, $-6^\circ$, $-2^\circ$, $2^\circ$,
$6^\circ$, $8^\circ$ on the L2 pair; workshop
sweep $\alpha=-4^\circ$--$8^\circ$.}\label{f:nlfaft}
\end{figure}

\begin{figure}[H]\centering
\includegraphics[width=\textwidth]{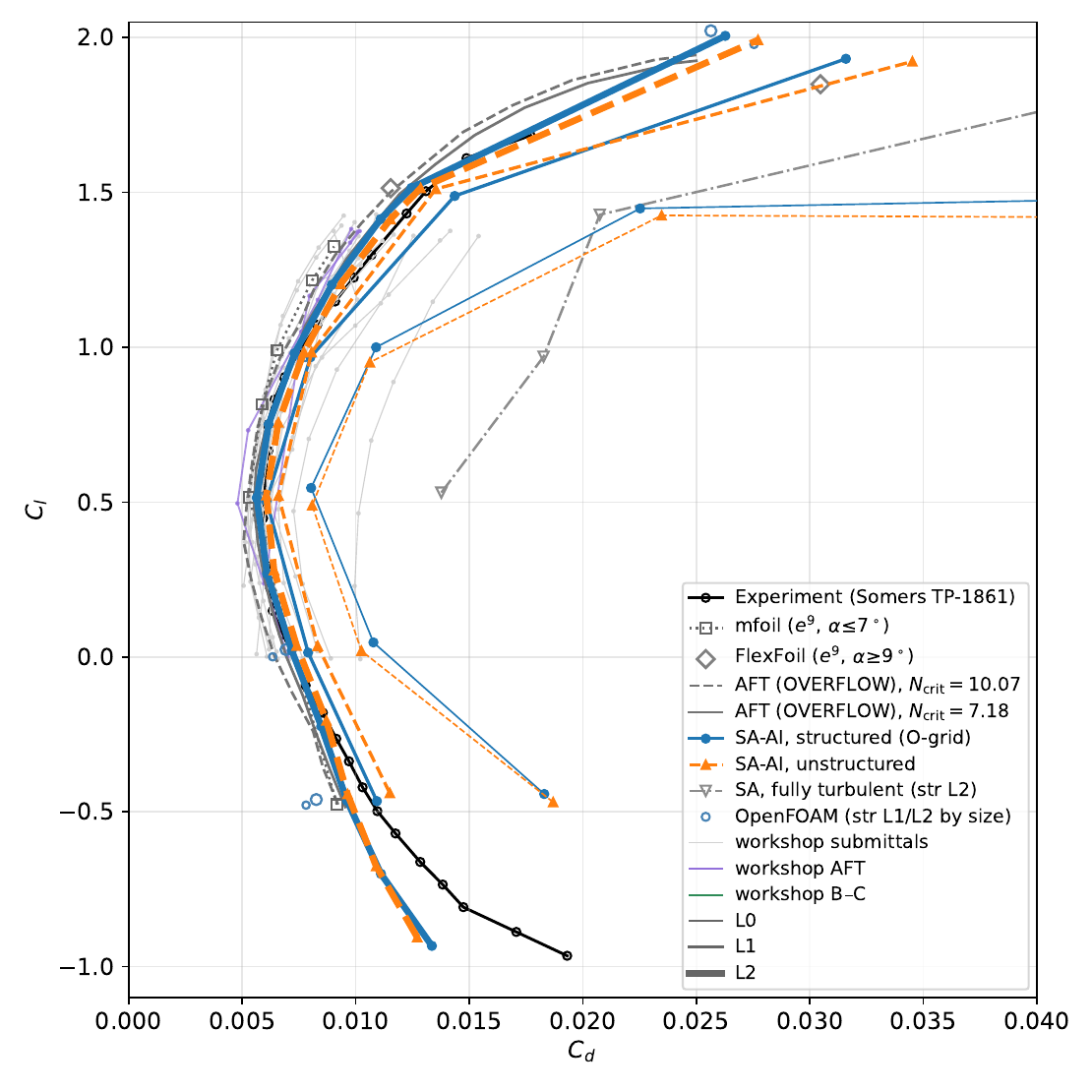}
\caption{NLF(1)-0416 drag polar, $Re=4\times10^6$; the L2 pair carries the
seven extension incidences as well, so it spans the measured polar down to
$\alpha=-12^\circ$ ($c_l=-0.93$). Fully turbulent SA
($\chi_\infty=3$): $1.6$--$2.5\times$ above the SA-AI drag,
$+144$/$+150\%$ in the bucket at $\alpha=0^\circ$/$4^\circ$,
$+67$/$+74\%$ at $9^\circ$/$15^\circ$.}\label{f:nlfpolar}
\end{figure}

\begin{figure}[H]\centering
\includegraphics[width=0.99\textwidth]{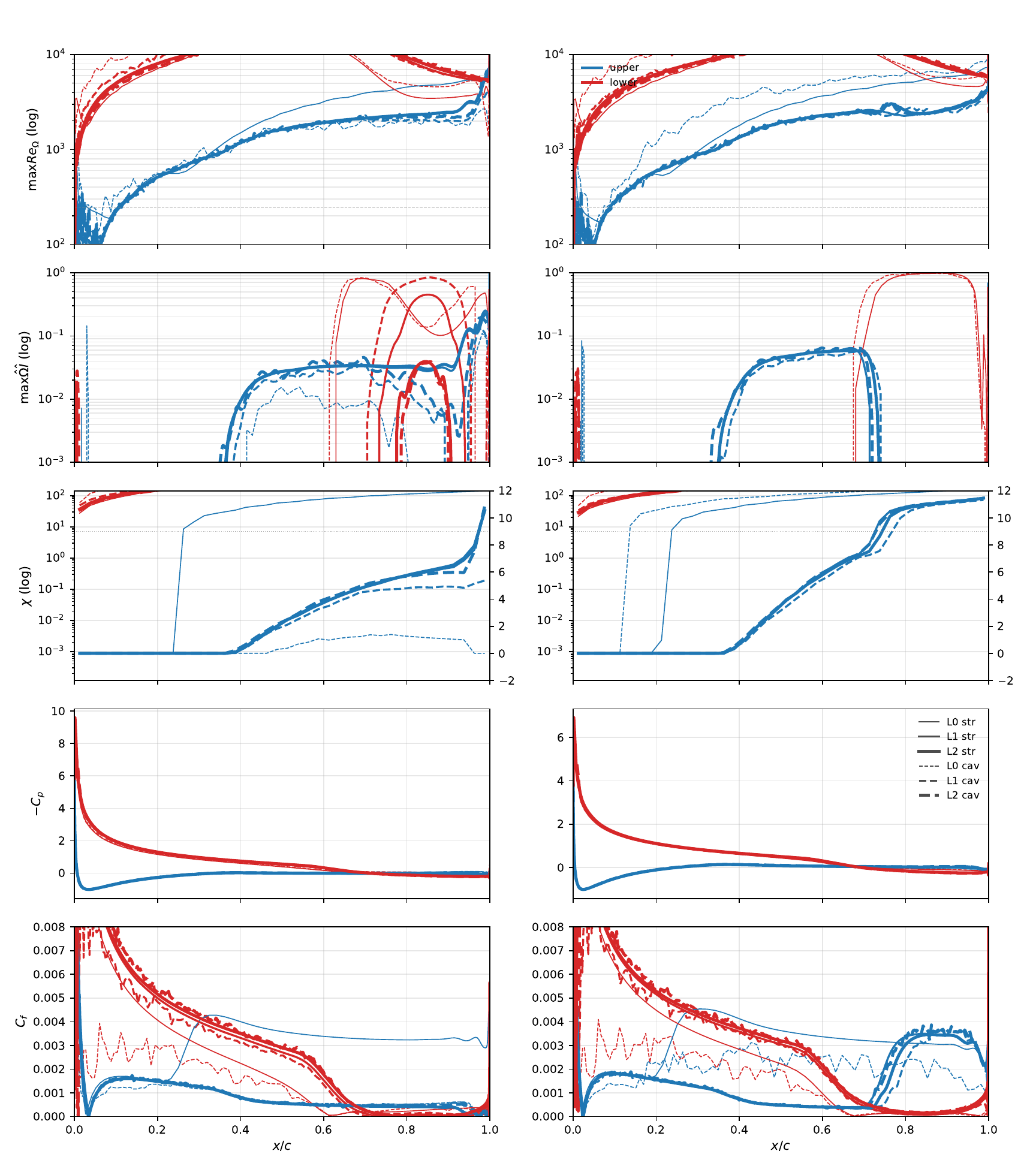}
\caption{NLF(1)-0416 surface distributions, $Re=4\times10^6$, all six grids; pages run in ascending incidence. $\alpha=-12^\circ$, $\alpha=-10^\circ$.}\label{f:nlfcflow}
\end{figure}

\begin{figure}[H]\centering
\includegraphics[width=0.99\textwidth]{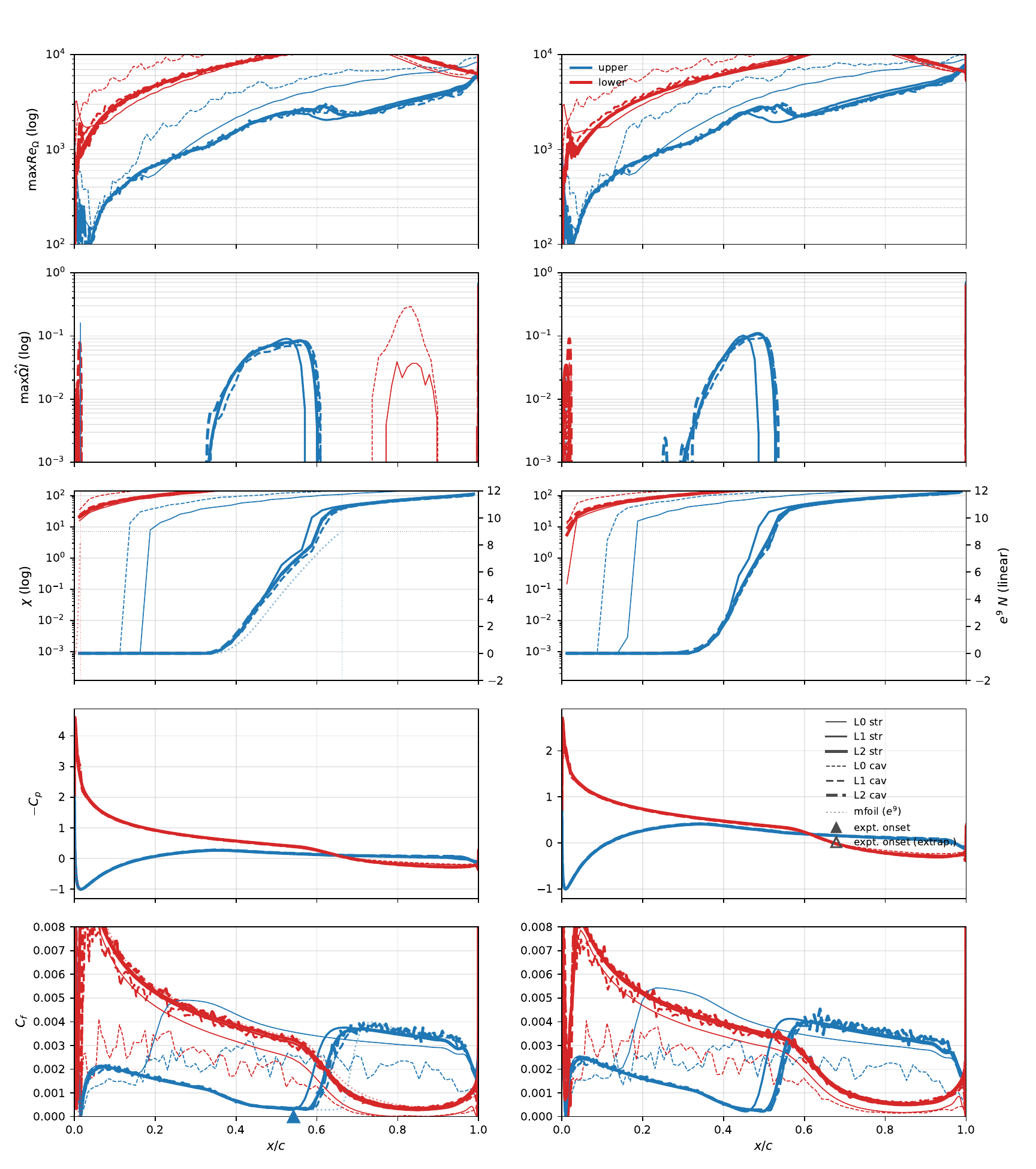}
\figcont{$\alpha=-8^\circ$, $\alpha=-6^\circ$.}
\end{figure}

\begin{figure}[H]\centering
\includegraphics[width=0.99\textwidth]{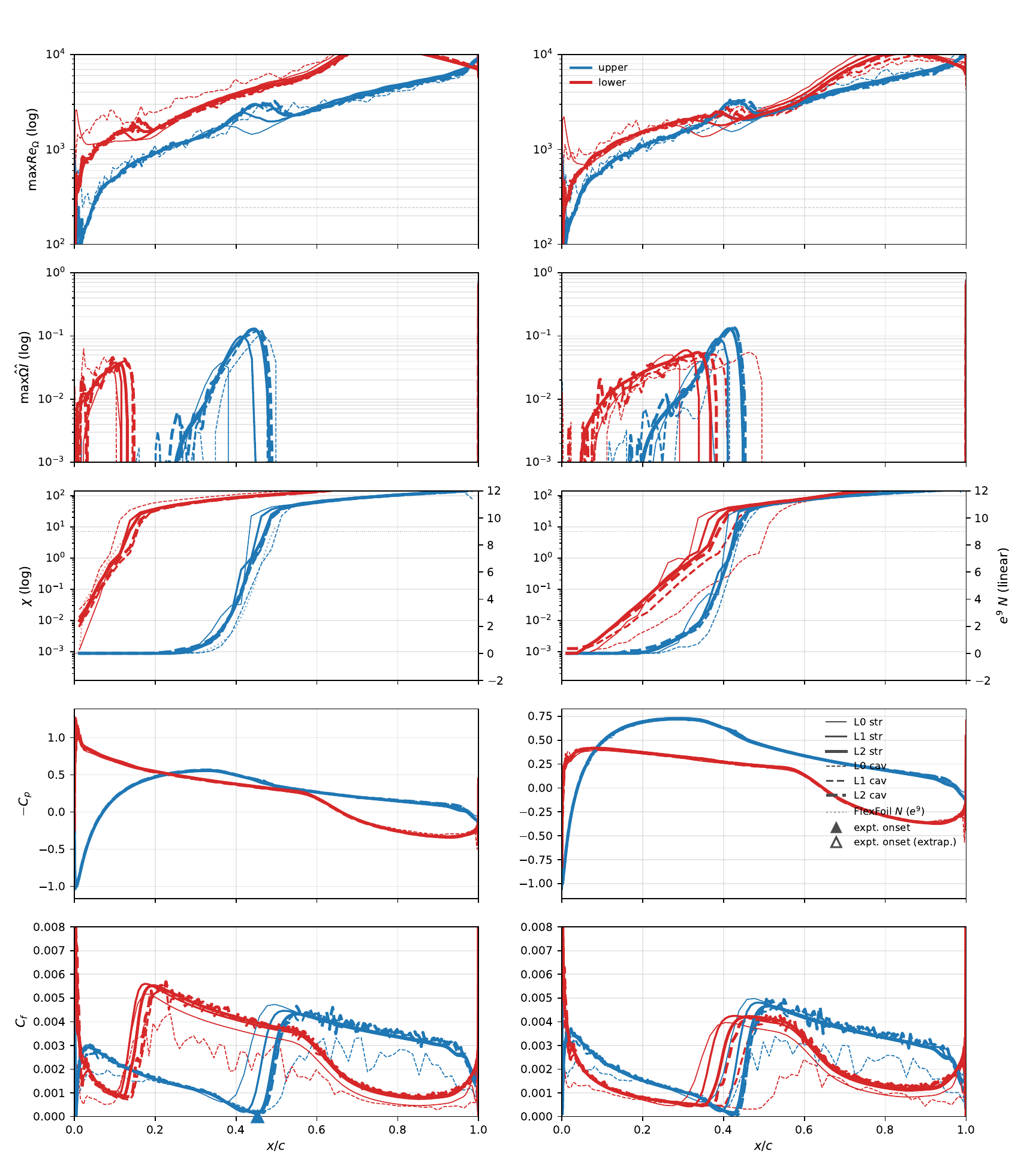}
\figcont{$\alpha=-4^\circ$, $\alpha=-2^\circ$.}
\end{figure}

\begin{figure}[H]\centering
\includegraphics[width=0.99\textwidth]{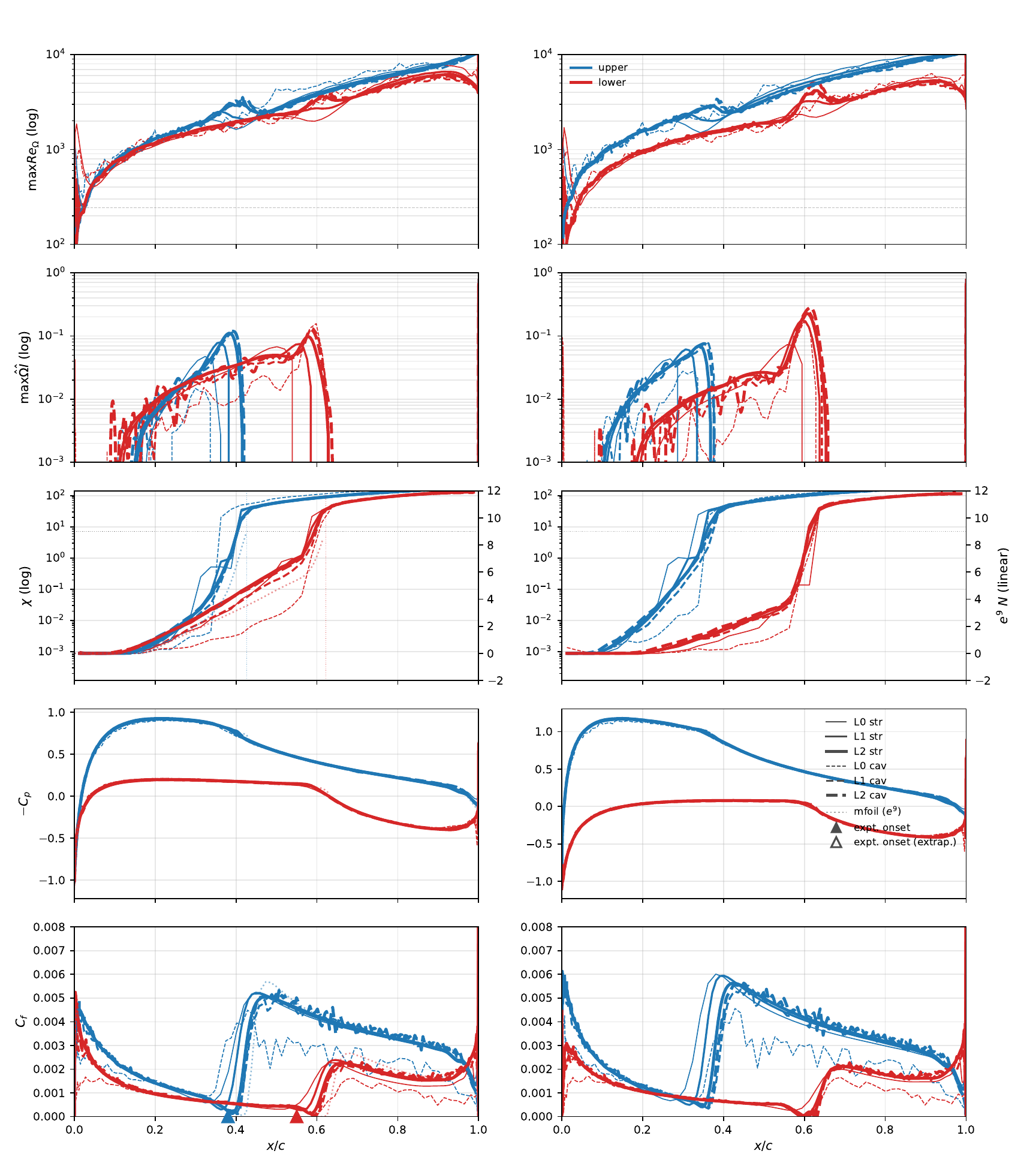}
\figcont{$\alpha=0^\circ$, $\alpha=2^\circ$.}
\end{figure}

\begin{figure}[H]\centering
\includegraphics[width=0.99\textwidth]{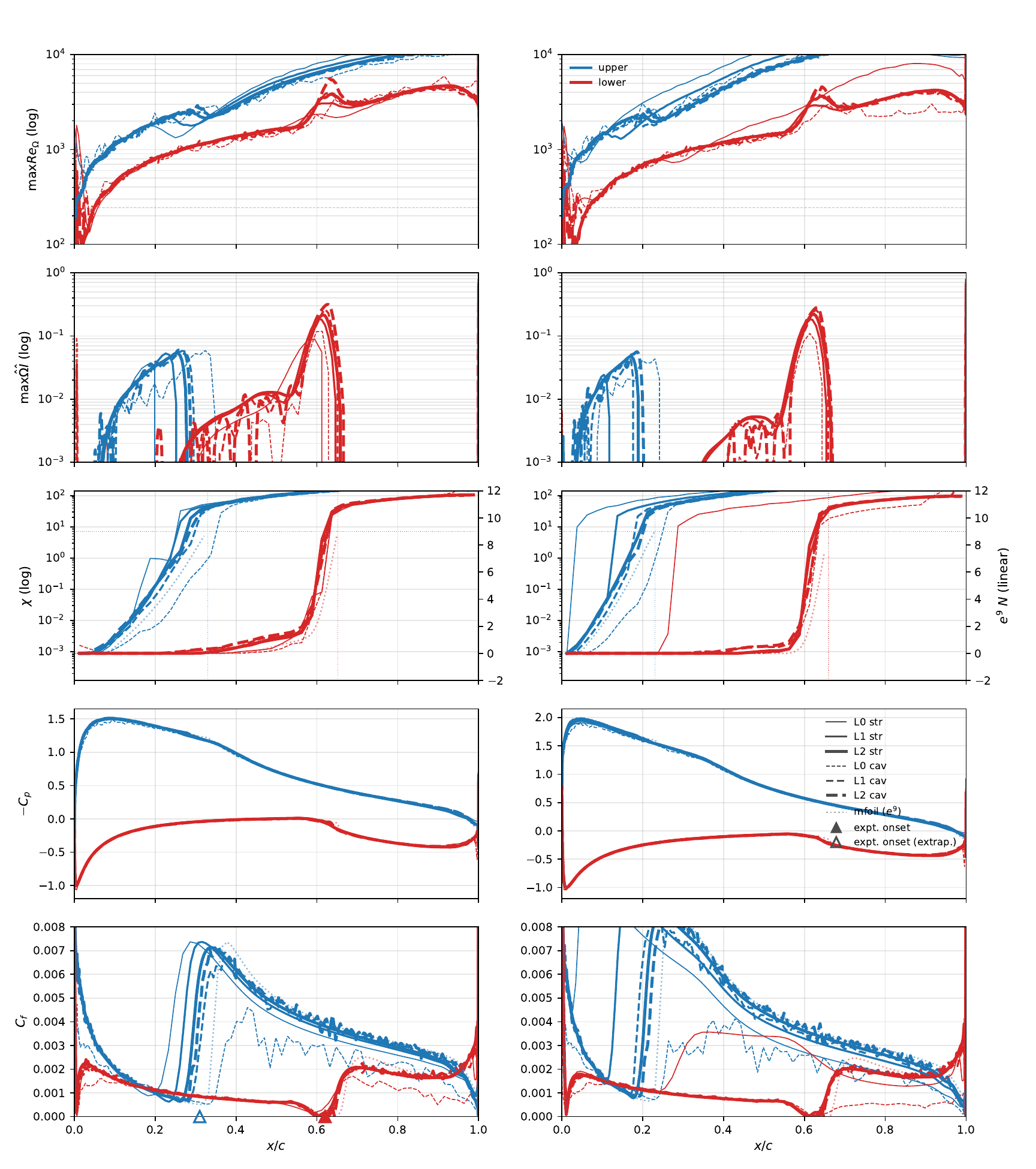}
\figcont{$\alpha=4^\circ$, $\alpha=6^\circ$.}
\end{figure}

\begin{figure}[H]\centering
\includegraphics[width=0.99\textwidth]{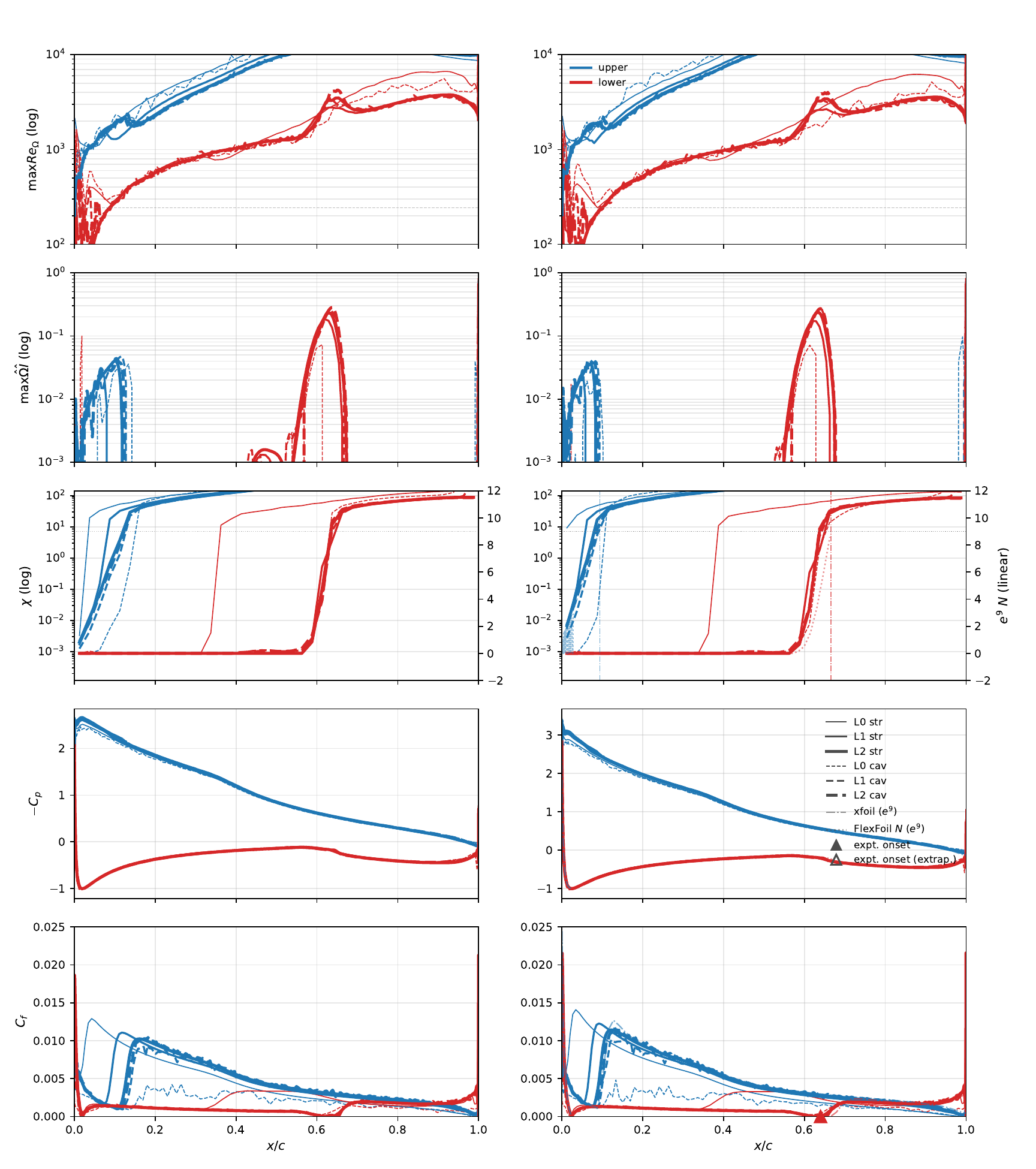}
\figcont{$\alpha=8^\circ$, $\alpha=9^\circ$.}
\end{figure}

\begin{figure}[H]\centering
\includegraphics[width=0.55\textwidth]{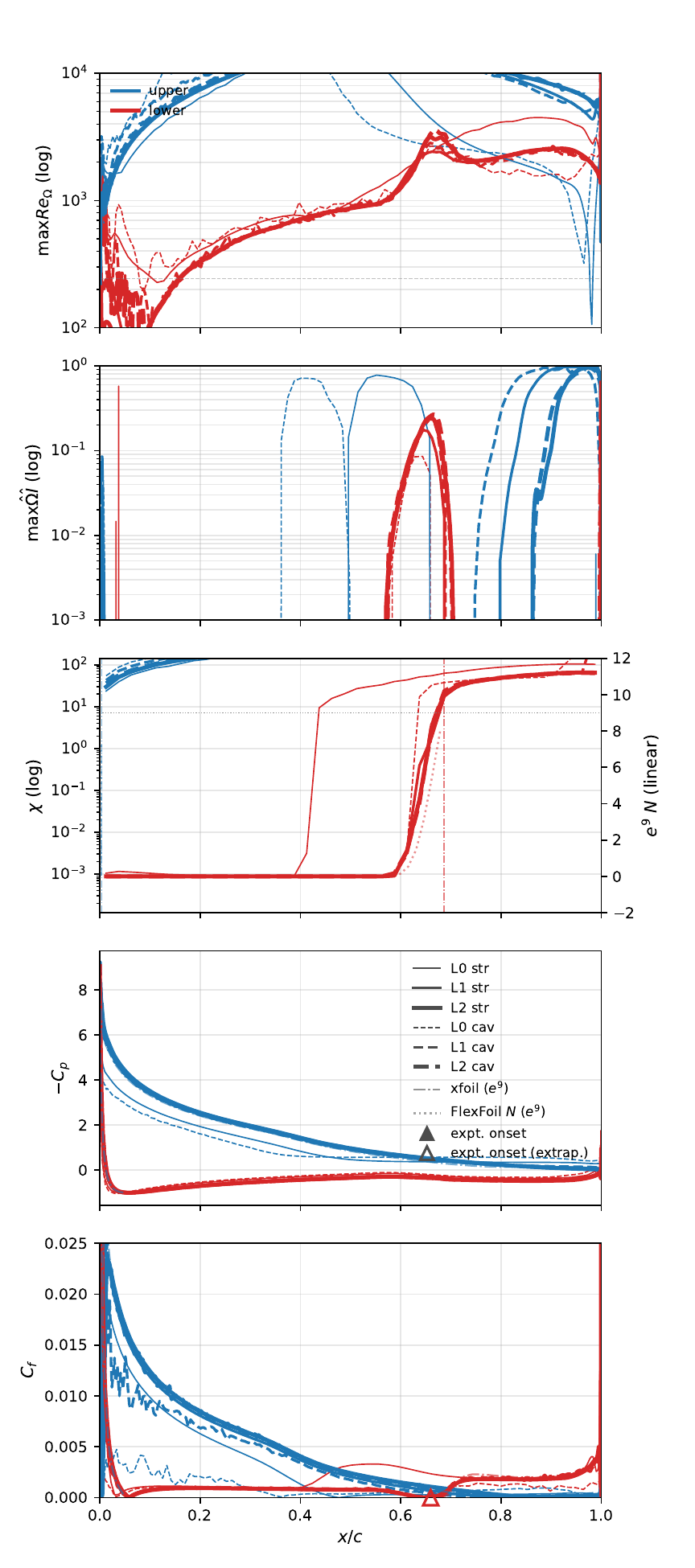}
\figcont{$\alpha=15^\circ$.}
\end{figure}

\begin{table}[H]
  \centering\scriptsize\setlength{\tabcolsep}{2pt}
  \caption{NLF(1)-0416, $Re\!=\!4\!\times\!10^6$: forces and
  transition locations (near-wall $\chi\!=\!1$ fronts). The OpenFOAM
  port is independent and omits the $f_{v1}$ bypass. The seven extension
  incidences exist on the L2 pair only. Experimental columns are LTPT
  measurements digitized from Coder's Figs.~5-22--5-25; drag and
  transition are read at the measured $C_L$, and blanks fall outside the
  digitized range or past the stall break.}
  \label{t:nlf}
  \begin{tabular}{ll cccc cccc cccc cccc}
    \toprule
    & & \multicolumn{4}{c}{Flow360, structured O-grid} & \multicolumn{4}{c}{Flow360, unstructured cavity} & \multicolumn{4}{c}{OpenFOAM, structured} & \multicolumn{4}{c}{Experiment (LTPT)} \\
    \cmidrule(lr){3-6}\cmidrule(lr){7-10}\cmidrule(lr){11-14}\cmidrule(lr){15-18}
    $\alpha$ & grid & $C_L$ & $C_D$ & $x_\mathrm{tr}^\mathrm{up}$ & $x_\mathrm{tr}^\mathrm{lo}$ & $C_L$ & $C_D$ & $x_\mathrm{tr}^\mathrm{up}$ & $x_\mathrm{tr}^\mathrm{lo}$ & $C_L$ & $C_D$ & $x_\mathrm{tr}^\mathrm{up}$ & $x_\mathrm{tr}^\mathrm{lo}$ & $C_L$ & $C_D$ & $x_\mathrm{tr}^\mathrm{up}$ & $x_\mathrm{tr}^\mathrm{lo}$ \\
    \midrule
    $-12$ & L2 & -0.9329 & 0.01335 & 0.945 & 0.001 & -0.9050 & 0.01271 & 0.967 & 0.005 & -- & -- & -- & -- & -- & -- & -- & -- \\
    \addlinespace
    $-10$ & L2 & -0.7004 & 0.01111 & 0.681 & 0.002 & -0.6756 & 0.01090 & 0.673 & 0.006 & -- & -- & -- & -- & -- & -- & -- & -- \\
    \addlinespace
    $-8$ & L0 & -0.4426 & 0.01830 & 0.195 & 0.011 & -0.4689 & 0.01869 & 0.140 & 0.012 & -0.4460 & 0.01241 & 0.556 & 0.007 & -0.4745 & 0.01052 & 0.538 & -- \\
    $-8$ & L1 & -0.4656 & 0.01093 & 0.538 & 0.006 & -0.4397 & 0.01150 & 0.571 & 0.011 & -0.4785 & 0.00781 & 0.576 & 0.000 &  &  &  &  \\
    $-8$ & L2 & -0.4646 & 0.00962 & 0.559 & 0.004 & -0.4426 & 0.00964 & 0.561 & 0.010 & -0.4606 & 0.00826 & 0.568 & 0.000 &  &  &  &  \\
    \addlinespace
    $-6$ & L2 & -0.2240 & 0.00848 & 0.492 & 0.012 & -0.2052 & 0.00870 & 0.495 & 0.016 & -- & -- & -- & -- & -0.2410 & 0.00881 & 0.496 & -- \\
    \addlinespace
    $-4$ & L0 & 0.0469 & 0.01078 & 0.435 & 0.100 & 0.0197 & 0.01025 & 0.476 & 0.083 & 0.0388 & 0.00748 & 0.424 & 0.018 & -0.0122 & 0.00695 & 0.457 & -- \\
    $-4$ & L1 & 0.0142 & 0.00790 & 0.435 & 0.102 & 0.0357 & 0.00832 & 0.452 & 0.126 & 0.0005 & 0.00634 & 0.464 & 0.117 &  &  &  &  \\
    $-4$ & L2 & 0.0208 & 0.00718 & 0.447 & 0.109 & 0.0363 & 0.00739 & 0.453 & 0.122 & 0.0239 & 0.00693 & 0.459 & 0.111 &  &  &  &  \\
    \addlinespace
    $-2$ & L2 & 0.2683 & 0.00609 & 0.419 & 0.330 & 0.2800 & 0.00640 & 0.421 & 0.345 & -- & -- & -- & -- & 0.2140 & 0.00595 & 0.424 & -- \\
    \addlinespace
    $0$ & L0 & 0.5458 & 0.00803 & 0.400 & 0.590 & 0.4896 & 0.00808 & 0.355 & 0.592 & 0.5473 & 0.00669 & 0.387 & 0.564 & 0.4553 & 0.00567 & 0.384 & 0.525 \\
    $0$ & L1 & 0.5048 & 0.00607 & 0.395 & 0.553 & 0.5209 & 0.00660 & 0.392 & 0.579 & 0.5034 & 0.00603 & 0.393 & 0.572 &  &  &  &  \\
    $0$ & L2 & 0.5139 & 0.00564 & 0.387 & 0.566 & 0.5219 & 0.00606 & 0.391 & 0.569 & 0.5153 & 0.00573 & 0.395 & 0.568 &  &  &  &  \\
    \addlinespace
    $2$ & L2 & 0.7510 & 0.00616 & 0.338 & 0.604 & 0.7573 & 0.00659 & 0.344 & 0.607 & -- & -- & -- & -- & 0.6694 & 0.00587 & -- & 0.605 \\
    \addlinespace
    $4$ & L0 & 1.0001 & 0.01090 & 0.250 & 0.645 & 0.9513 & 0.01062 & 0.330 & 0.621 & 1.0114 & 0.00955 & 0.271 & 0.618 & 0.8903 & 0.00639 & -- & 0.615 \\
    $4$ & L1 & 0.9678 & 0.00800 & 0.238 & 0.610 & 0.9839 & 0.00805 & 0.269 & 0.614 & 0.9670 & 0.00778 & 0.266 & 0.616 &  &  &  &  \\
    $4$ & L2 & 0.9805 & 0.00730 & 0.255 & 0.610 & 0.9848 & 0.00772 & 0.260 & 0.621 & 0.9793 & 0.00749 & 0.261 & 0.625 &  &  &  &  \\
    \addlinespace
    $6$ & L2 & 1.2019 & 0.00895 & 0.171 & 0.614 & 1.2042 & 0.00933 & 0.179 & 0.629 & -- & -- & -- & -- & 1.1049 & 0.00830 & -- & 0.626 \\
    \addlinespace
    $8$ & L2 & 1.4139 & 0.01109 & 0.104 & 0.624 & 1.4137 & 0.01153 & 0.109 & 0.636 & -- & -- & -- & -- & 1.2941 & 0.01001 & -- & 0.633 \\
    \addlinespace
    $9$ & L0 & 1.4483 & 0.02252 & 0.015 & 0.385 & 1.4251 & 0.02345 & 0.105 & 0.641 & 1.5277 & 0.01496 & 0.063 & 0.618 & 1.4026 & 0.01104 & -- & 0.642 \\
    $9$ & L1 & 1.4886 & 0.01436 & 0.065 & 0.625 & 1.5112 & 0.01353 & 0.085 & 0.635 & 1.5066 & 0.01273 & 0.076 & 0.633 &  &  &  &  \\
    $9$ & L2 & 1.5146 & 0.01245 & 0.073 & 0.628 & 1.5139 & 0.01286 & 0.078 & 0.638 & 1.5151 & 0.01270 & 0.075 & 0.644 &  &  &  &  \\
    \addlinespace
    $15$ & L0 & 1.5172 & 0.07109 & 0.010 & 0.440 & 1.3976 & 0.11602 & 0.005 & 0.623 & 1.7210 & 0.05409 & 0.000 & 0.634 & -- & -- & -- & -- \\
    $15$ & L1 & 1.9317 & 0.03159 & 0.004 & 0.650 & 1.9232 & 0.03454 & 0.001 & 0.659 & 1.9780 & 0.02754 & 0.000 & 0.665 &  &  &  &  \\
    $15$ & L2 & 2.0057 & 0.02628 & 0.005 & 0.659 & 1.9924 & 0.02772 & 0.000 & 0.662 & 2.0221 & 0.02564 & 0.000 & 0.667 &  &  &  &  \\
    \bottomrule
  \end{tabular}
\end{table}

\clearpage
\begin{figure}[H]\centering
\includegraphics[width=\textwidth,height=0.94\textheight,keepaspectratio]{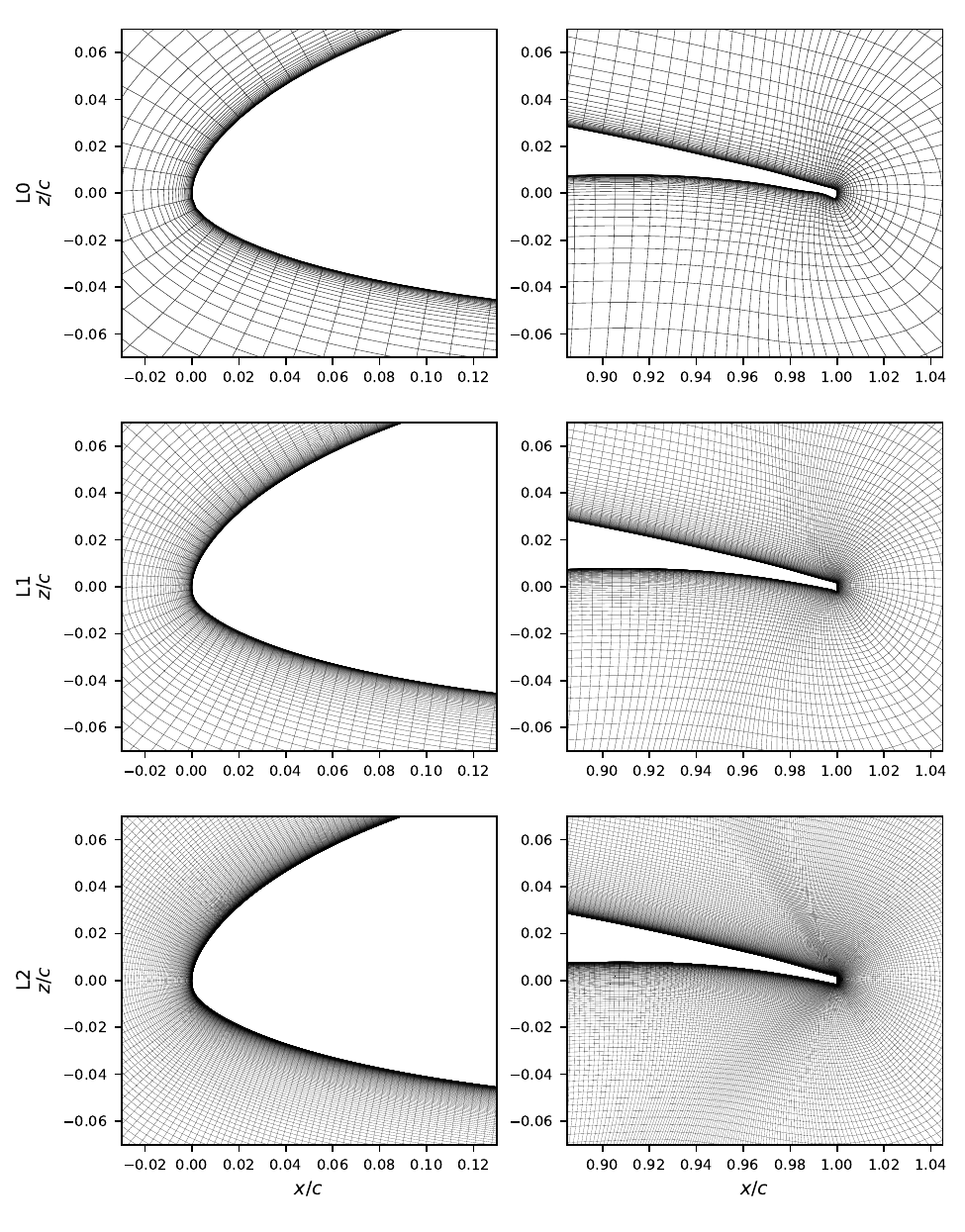}
\caption{NLF(1)-0416 structured O-grid: leading and trailing edge (columns),
levels L0--L2 (rows).}\label{f:nlfmeshstr}
\end{figure}
\begin{figure}[H]\centering
\includegraphics[width=\textwidth,height=0.94\textheight,keepaspectratio]{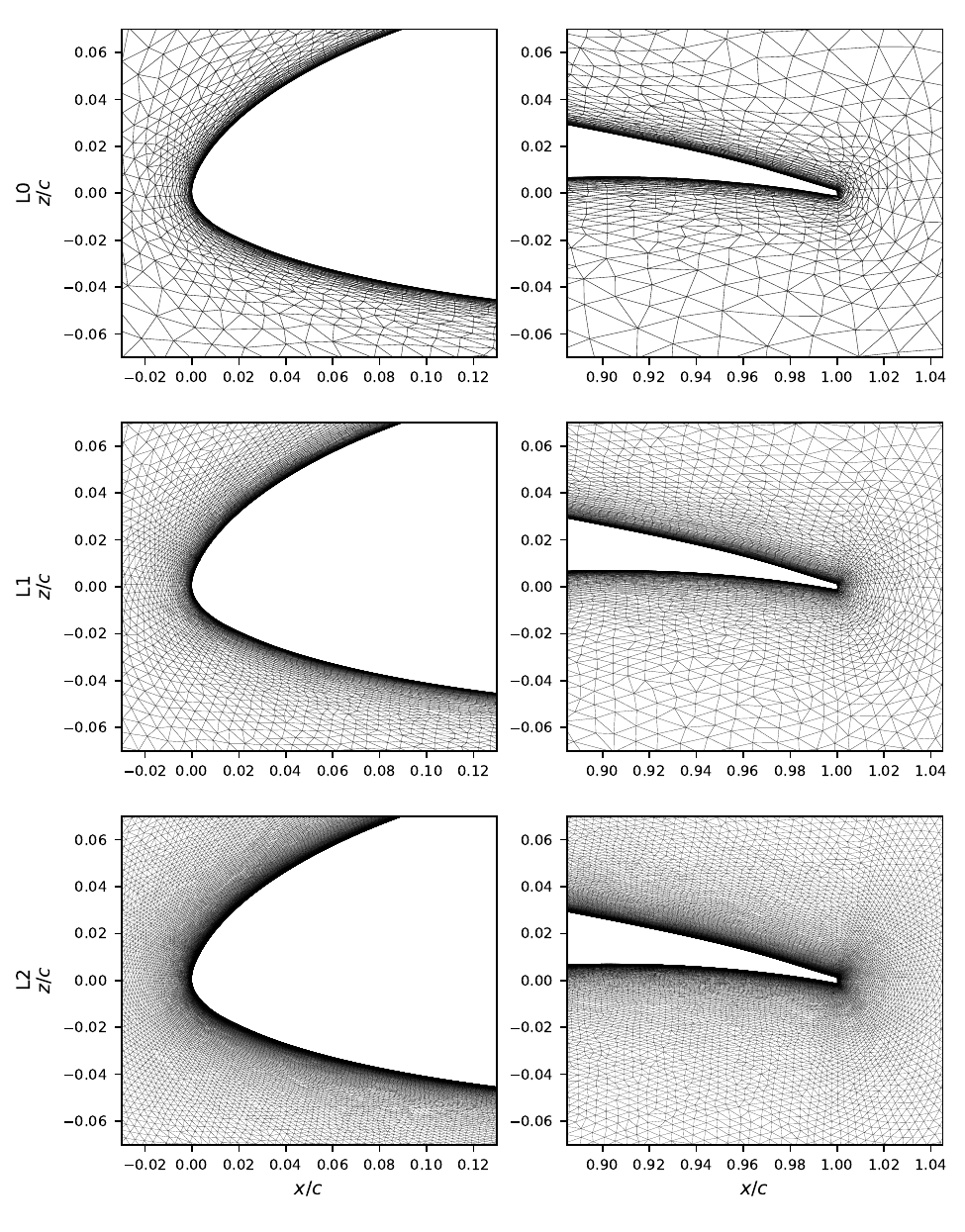}
\caption{NLF(1)-0416 unstructured cavity mesh: same edges and
levels.}\label{f:nlfmeshcav}
\end{figure}
\clearpage
\begin{figure}[H]\centering
\includegraphics[width=\textwidth,height=0.94\textheight,keepaspectratio]{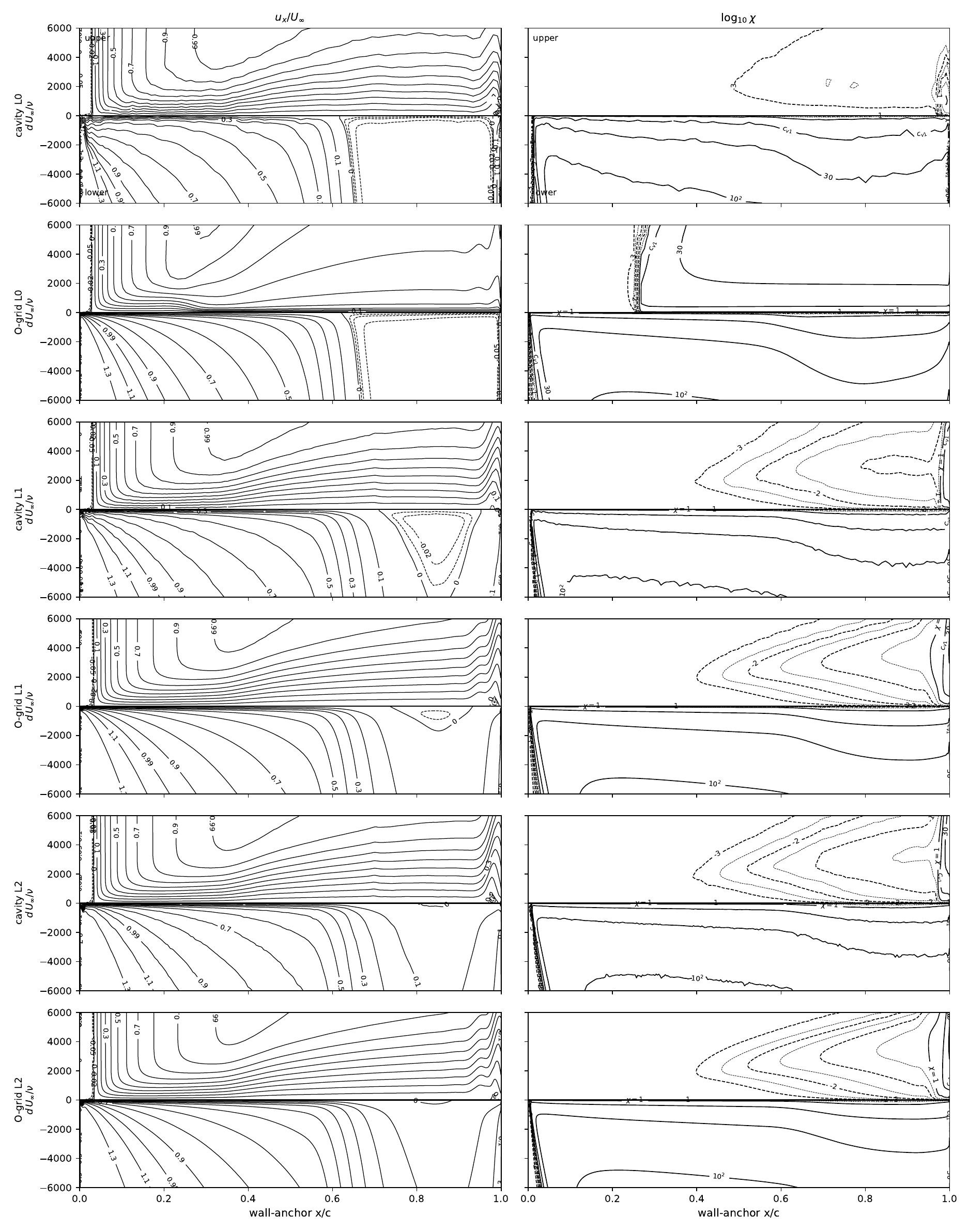}
\caption{NLF(1)-0416 wall-normal sheets, $Re=4\times10^6$, all six grids:
$u_x/U_\infty$ left, $\log_{10}\chi$ right. Panes mirror about the wall---
upper above $y=0$, lower reflected below, same probe depth.
$\alpha=-12^\circ$.}\label{f:nlfsheets}
\end{figure}
\begin{figure}[H]\centering
\includegraphics[width=\textwidth,height=0.94\textheight,keepaspectratio]{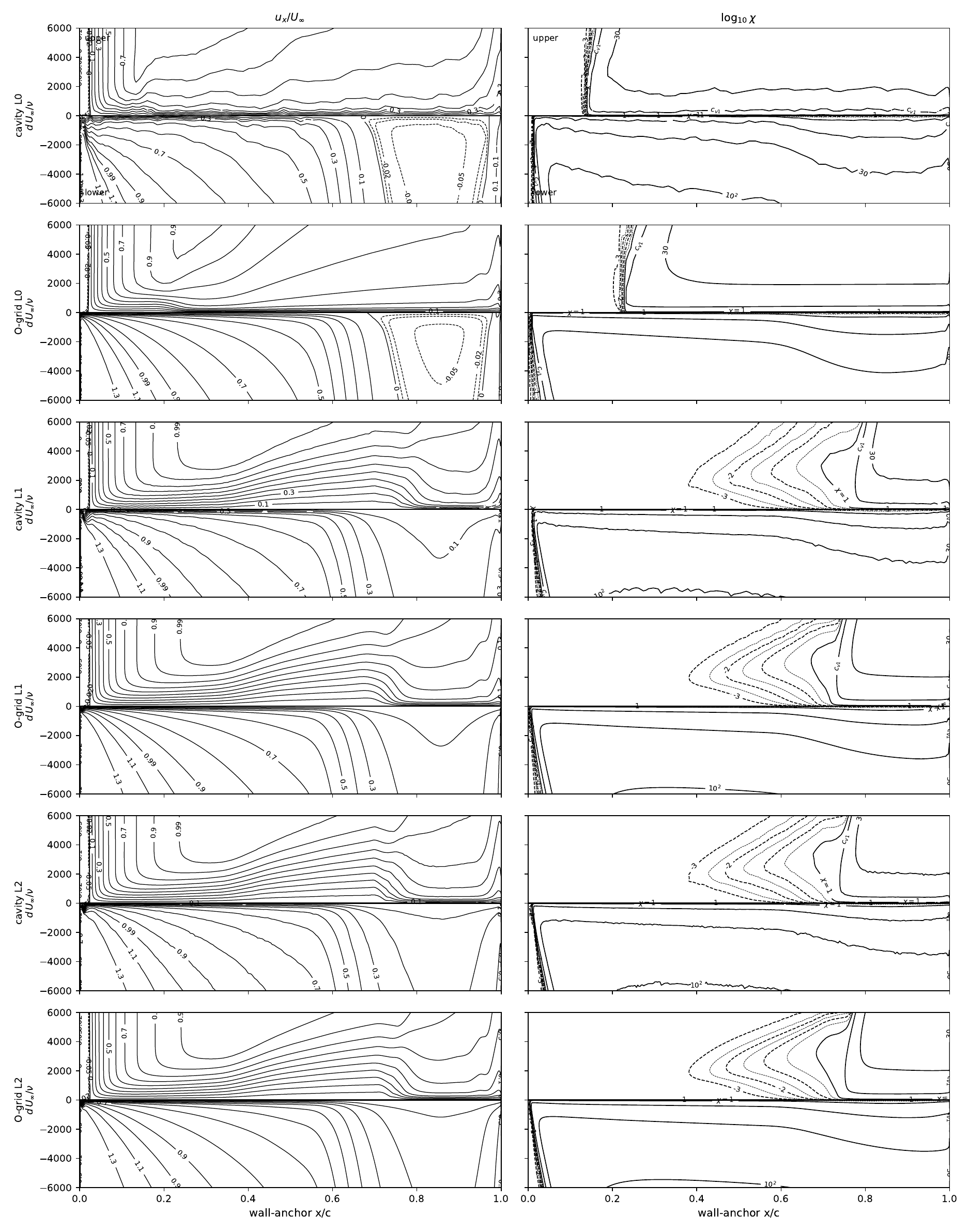}
\figcont{$\alpha=-10^\circ$.}
\end{figure}
\begin{figure}[H]\centering
\includegraphics[width=\textwidth,height=0.94\textheight,keepaspectratio]{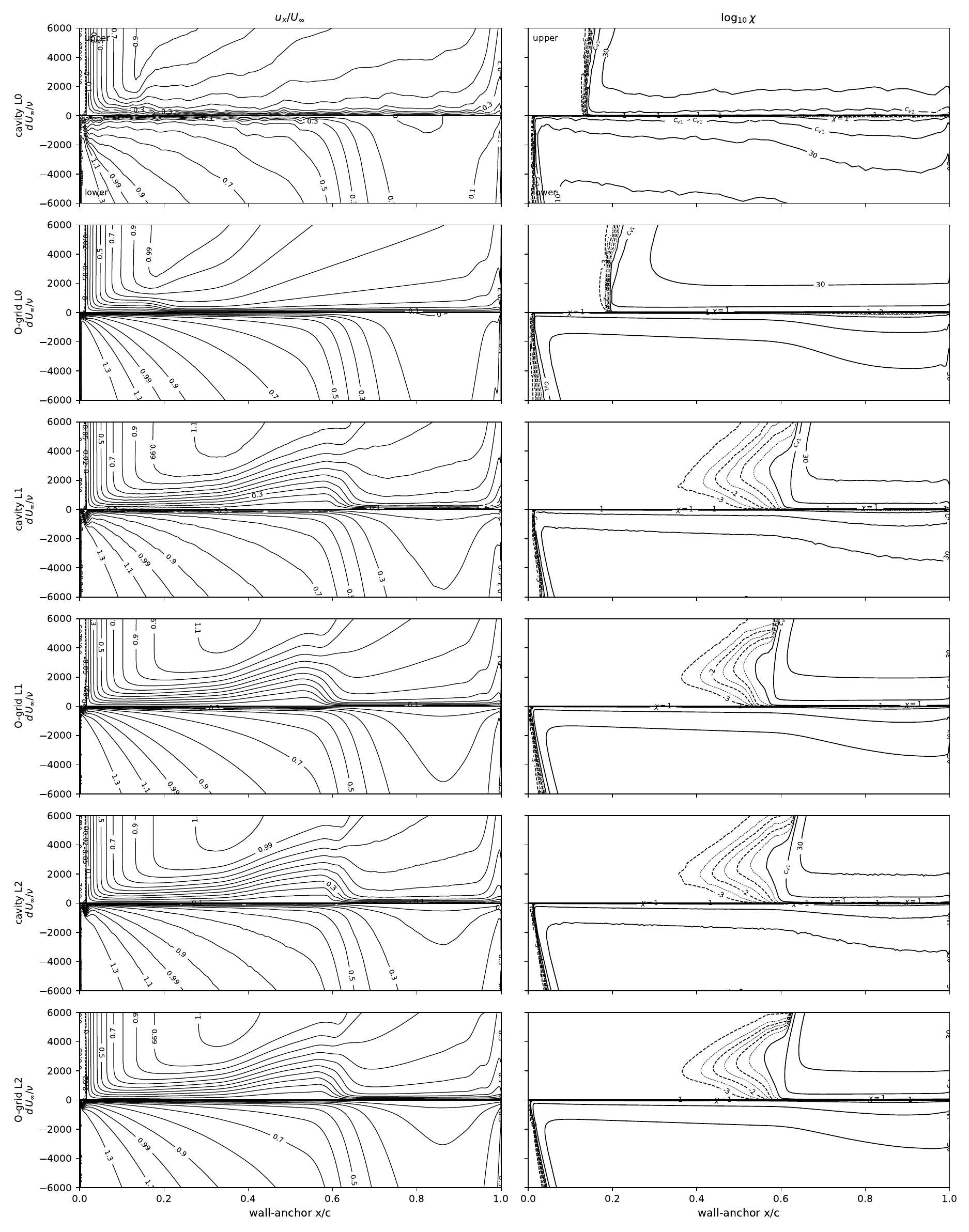}
\figcont{$\alpha=-8^\circ$.}
\end{figure}
\begin{figure}[H]\centering
\includegraphics[width=\textwidth,height=0.94\textheight,keepaspectratio]{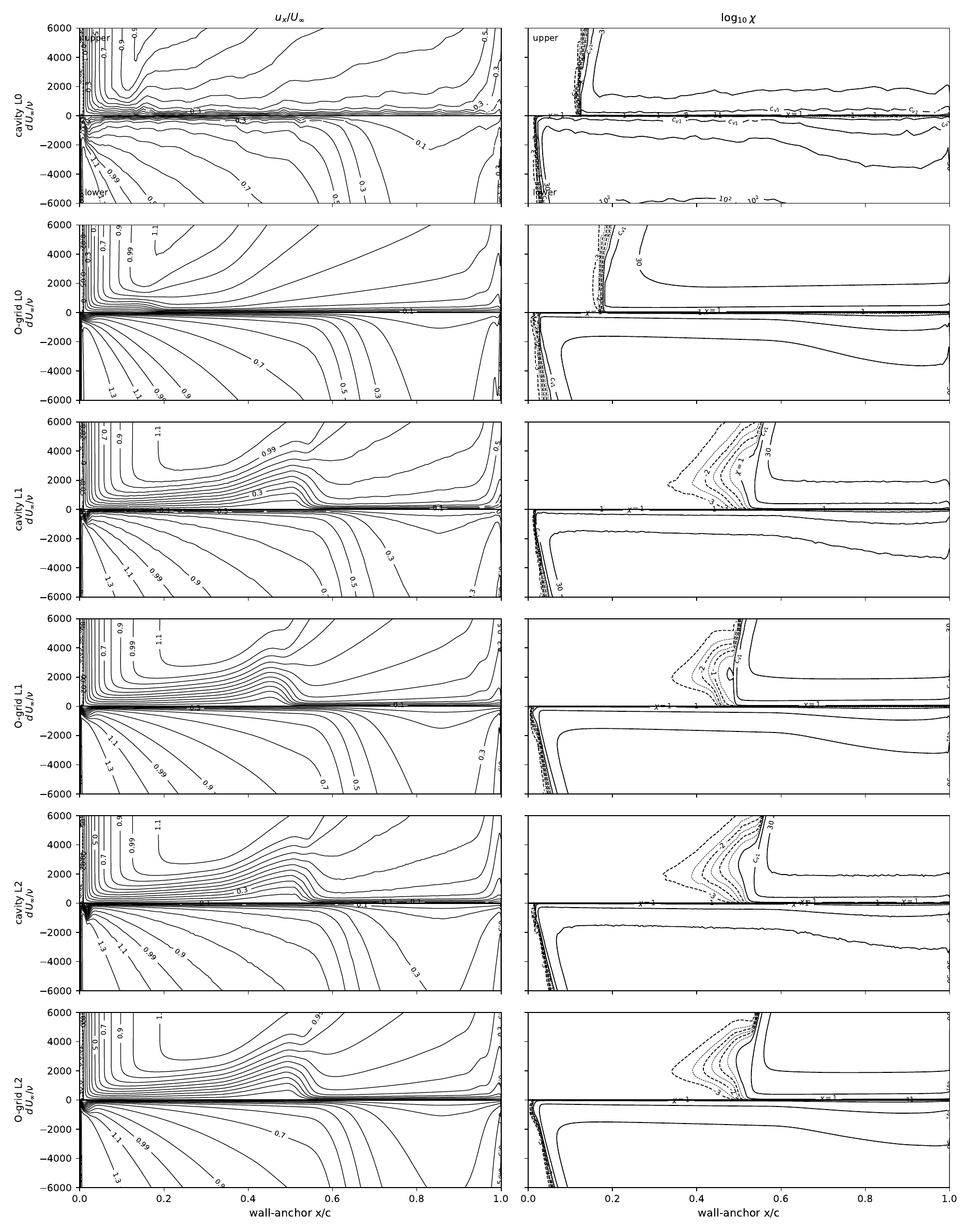}
\figcont{$\alpha=-6^\circ$.}
\end{figure}
\begin{figure}[H]\centering
\includegraphics[width=\textwidth,height=0.94\textheight,keepaspectratio]{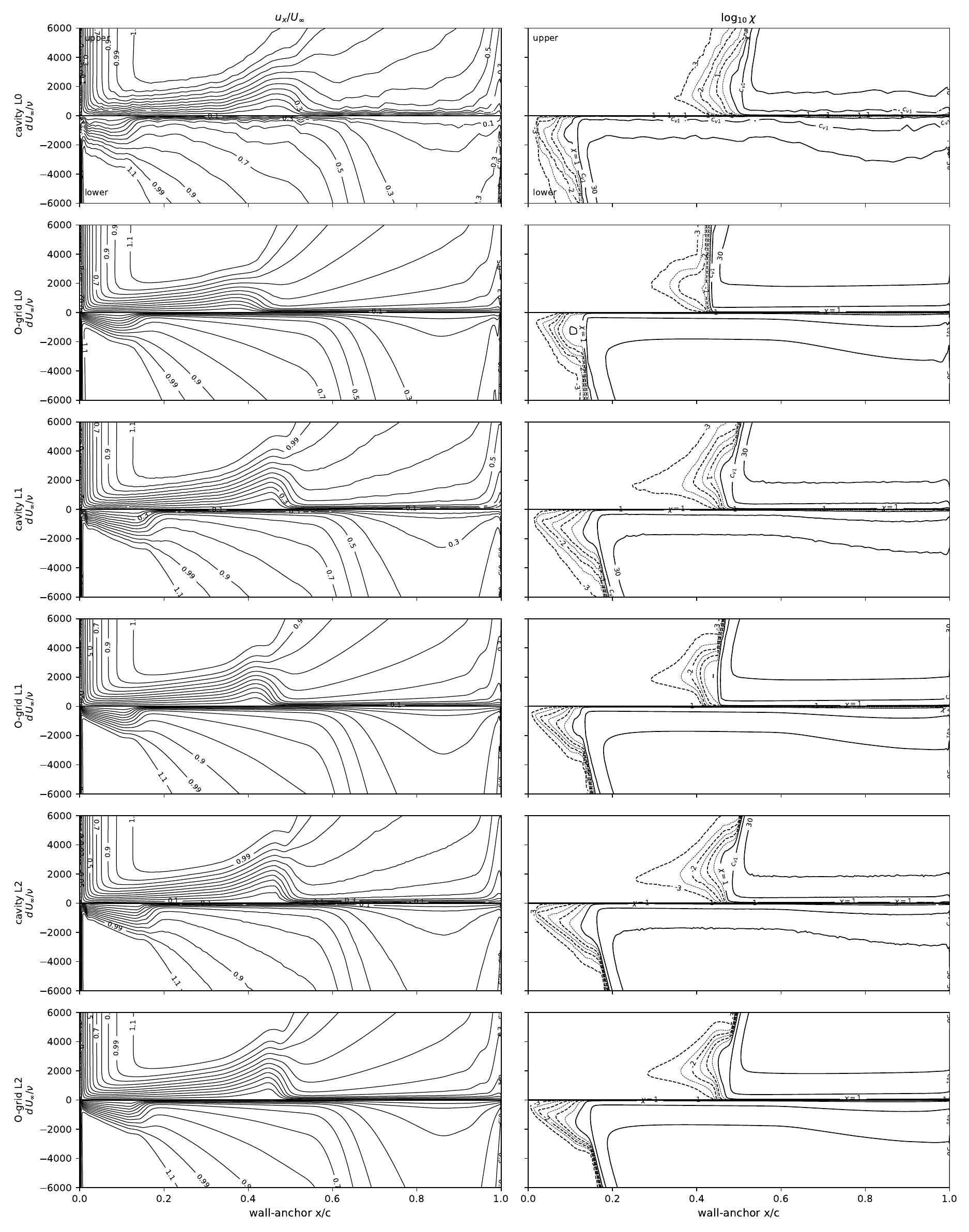}
\figcont{$\alpha=-4^\circ$.}
\end{figure}
\begin{figure}[H]\centering
\includegraphics[width=\textwidth,height=0.94\textheight,keepaspectratio]{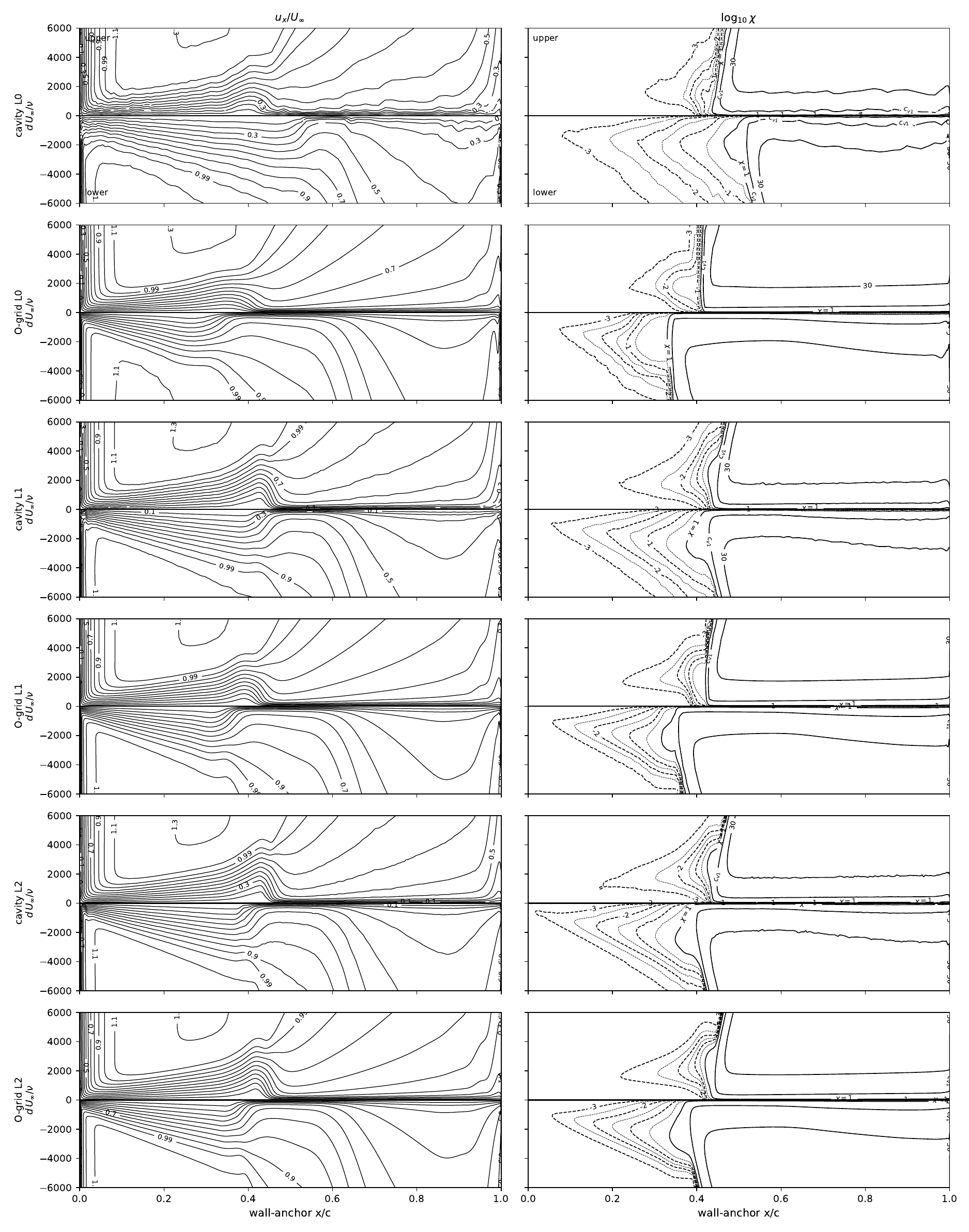}
\figcont{$\alpha=-2^\circ$.}
\end{figure}
\begin{figure}[H]\centering
\includegraphics[width=\textwidth,height=0.94\textheight,keepaspectratio]{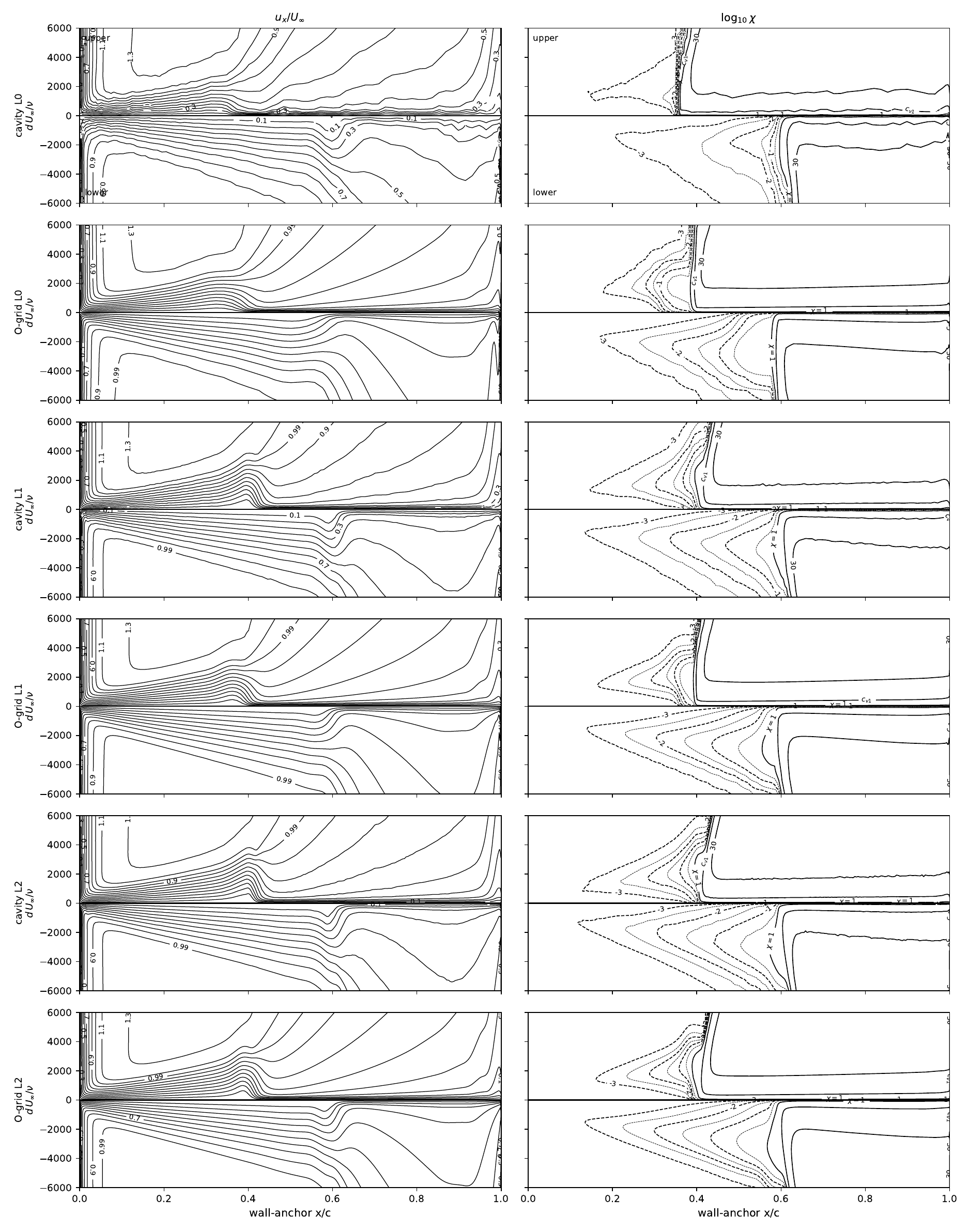}
\figcont{$\alpha=0^\circ$.}
\end{figure}
\begin{figure}[H]\centering
\includegraphics[width=\textwidth,height=0.94\textheight,keepaspectratio]{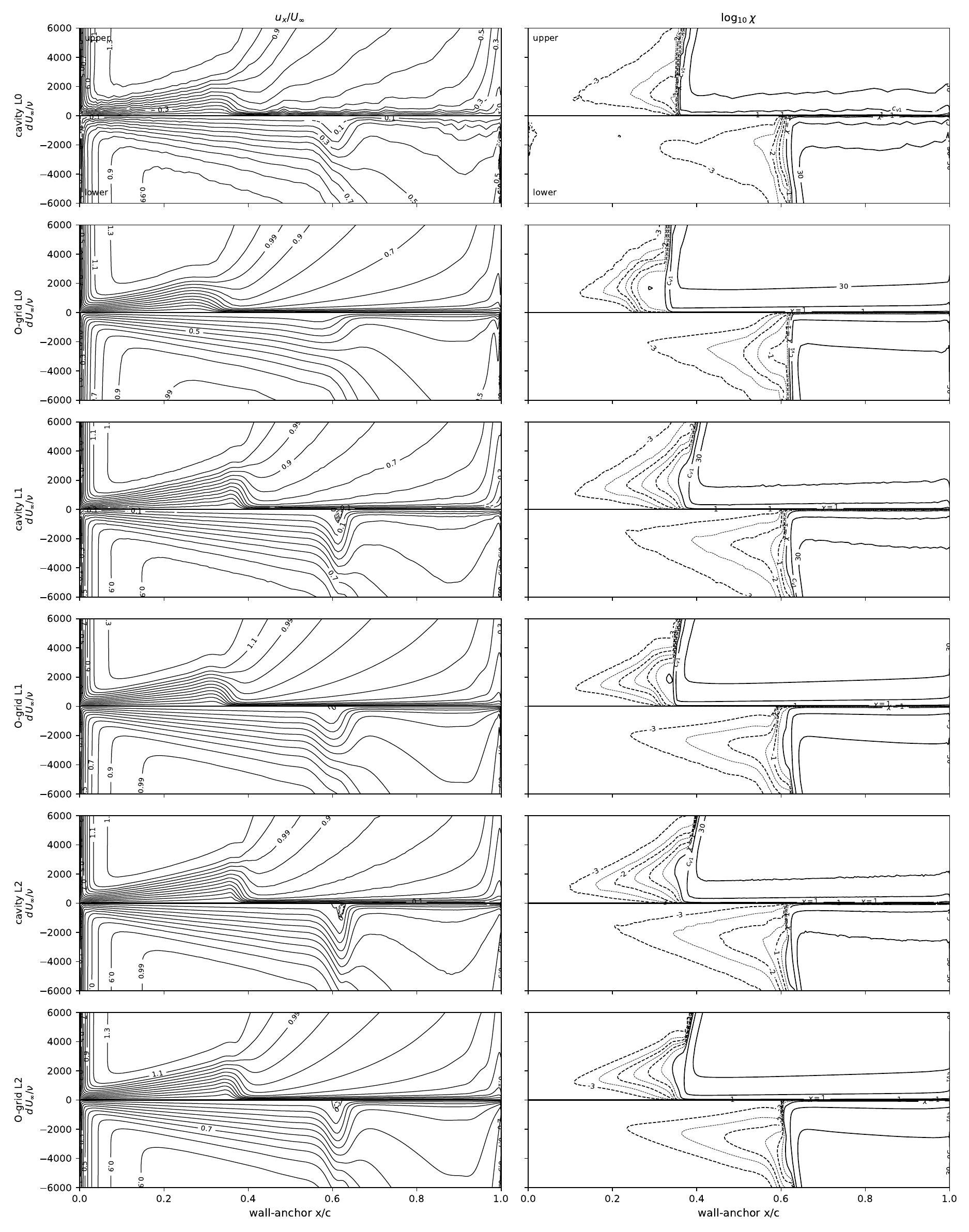}
\figcont{$\alpha=2^\circ$.}
\end{figure}
\begin{figure}[H]\centering
\includegraphics[width=\textwidth,height=0.94\textheight,keepaspectratio]{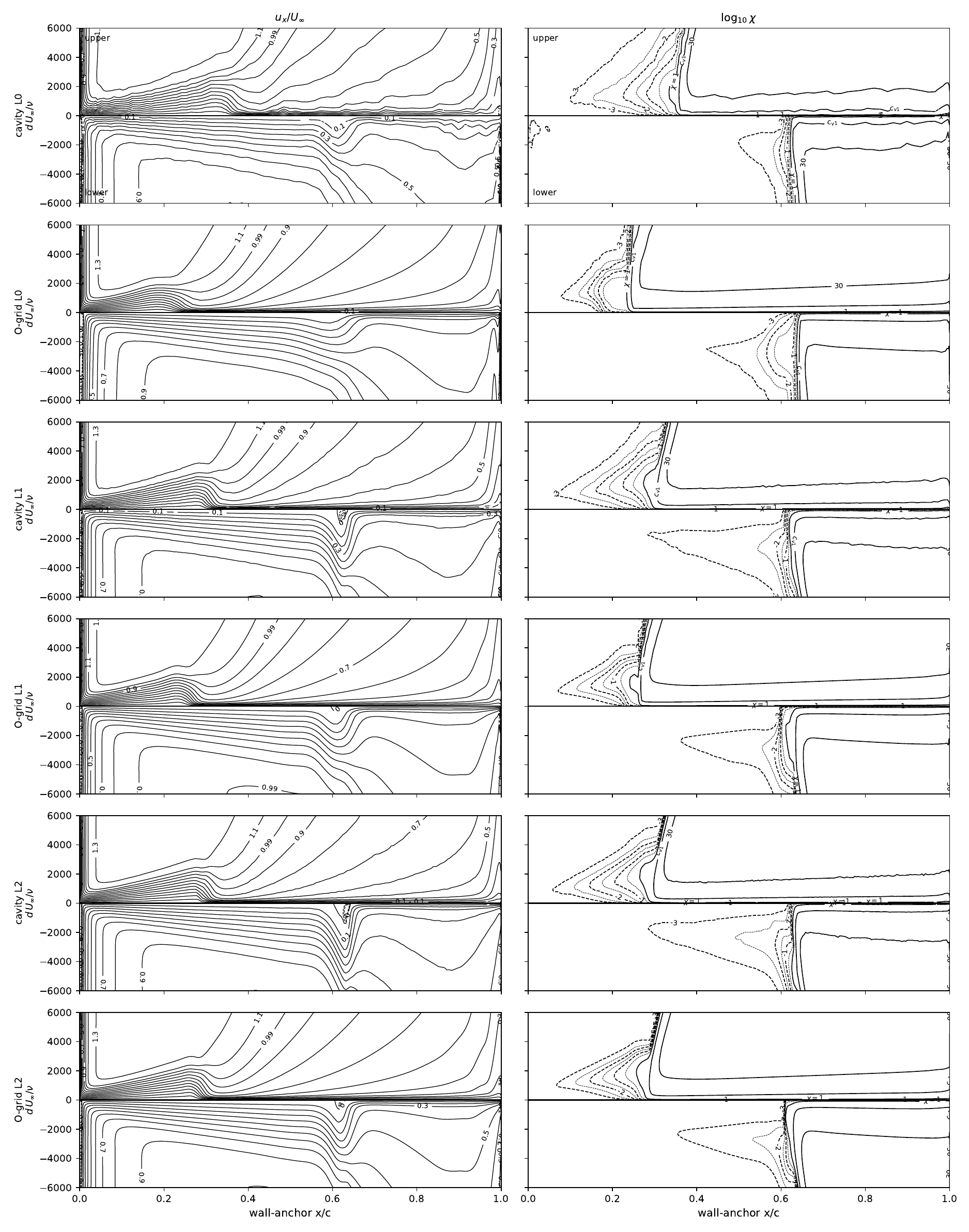}
\figcont{$\alpha=4^\circ$.}
\end{figure}
\begin{figure}[H]\centering
\includegraphics[width=\textwidth,height=0.94\textheight,keepaspectratio]{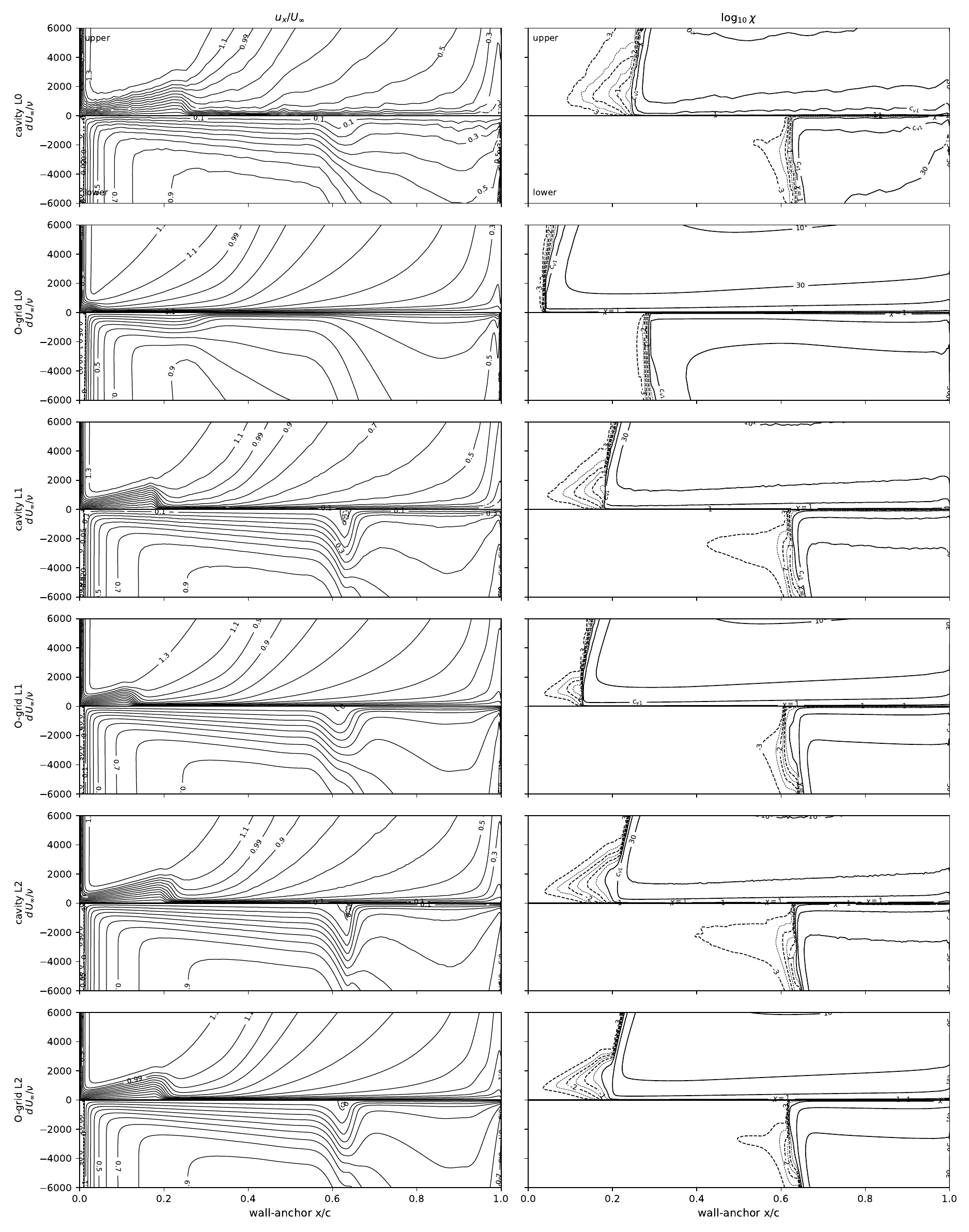}
\figcont{$\alpha=6^\circ$.}
\end{figure}
\begin{figure}[H]\centering
\includegraphics[width=\textwidth,height=0.94\textheight,keepaspectratio]{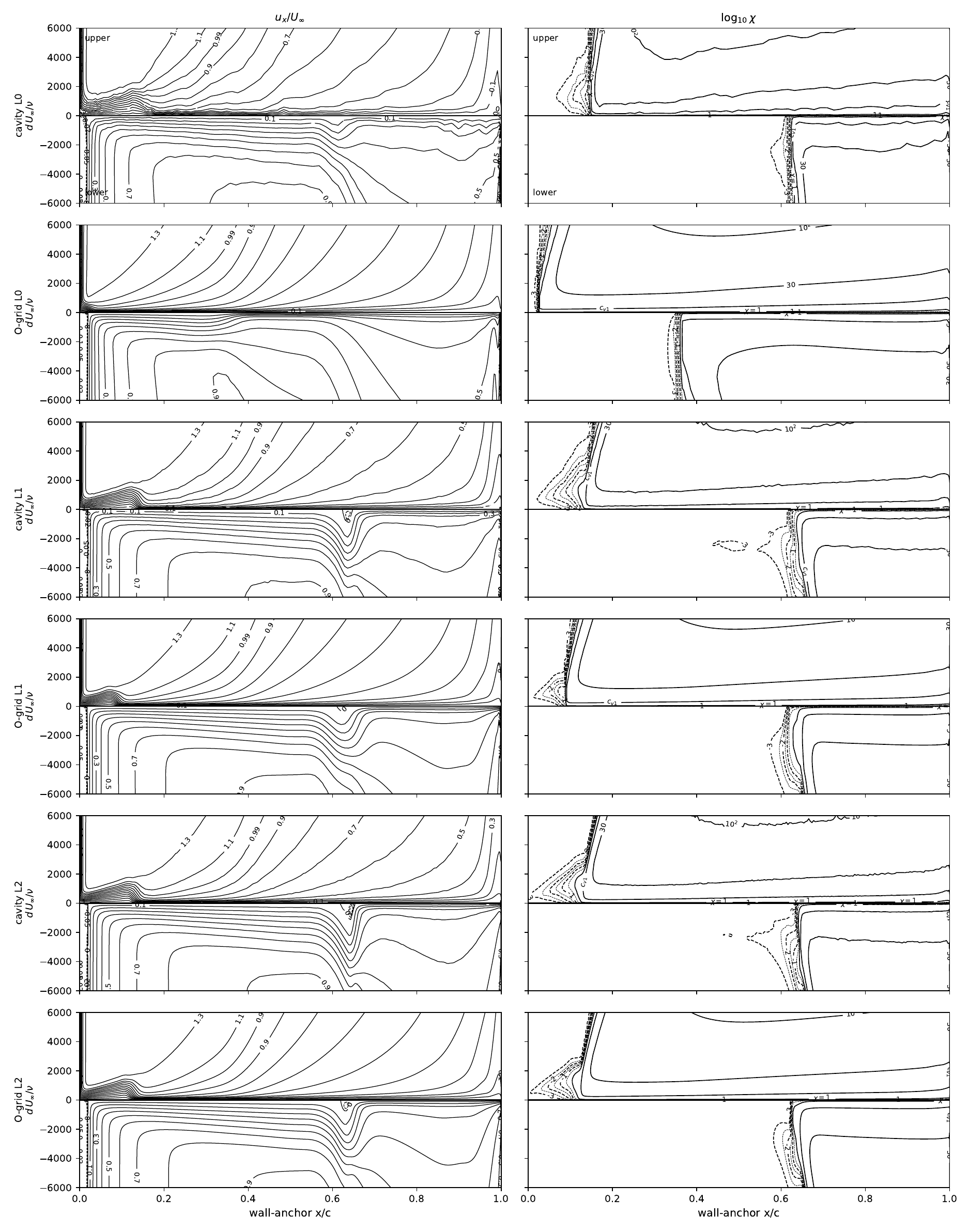}
\figcont{$\alpha=8^\circ$.}
\end{figure}
\begin{figure}[H]\centering
\includegraphics[width=\textwidth,height=0.94\textheight,keepaspectratio]{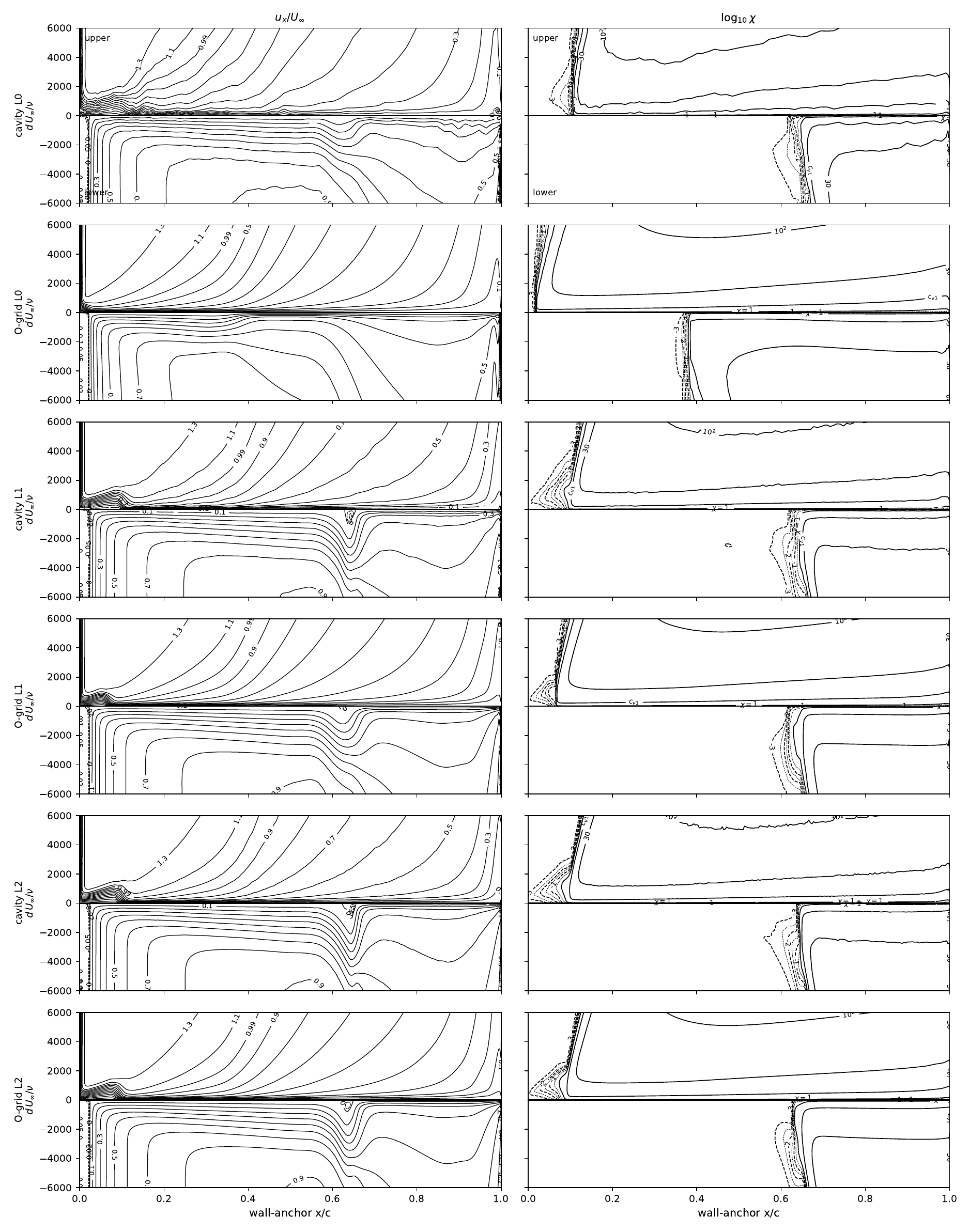}
\figcont{$\alpha=9^\circ$.}
\end{figure}
\begin{figure}[H]\centering
\includegraphics[width=\textwidth,height=0.94\textheight,keepaspectratio]{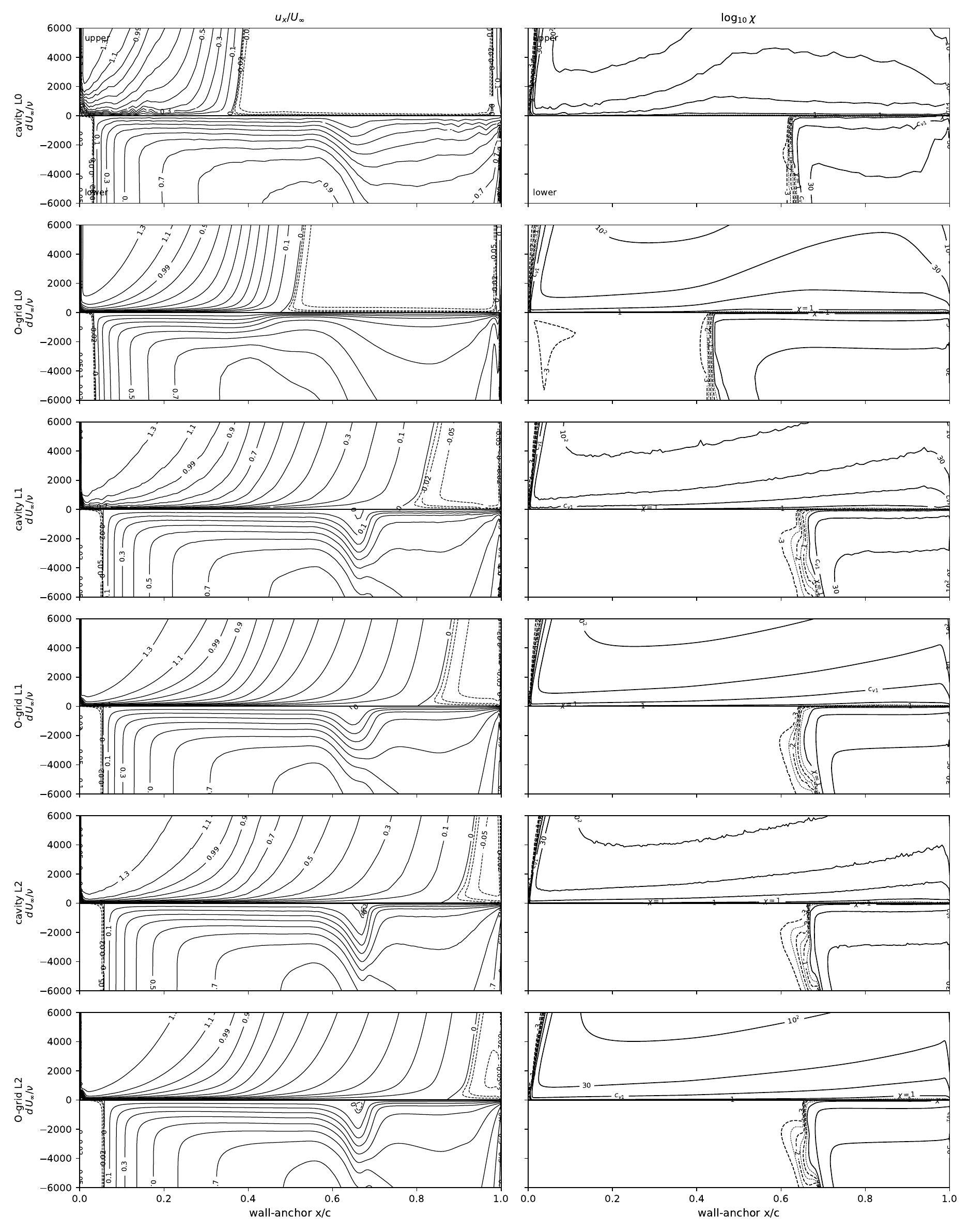}
\figcont{$\alpha=15^\circ$.}
\end{figure}

\subsection{The Eppler 387 airfoil}

The $-1^\circ$ front sits at $0.696$/$0.693\,c$. At $8^\circ$ and $9^\circ$ the
computed upper surface carries no closed bubble and the front collapses
to $0.119$/$0.016\,c$, matching SA-BC's $0.099$/$0.022\,c$ and the
experiment's report of attached transition without a bubble above
$7^\circ$; $\gamma$--$Re_\theta$ holds fronts at $0.451$/$0.150\,c$.

The drag polar sits $-6.4\%$ to $+3.7\%$ against the digitized
$\gamma$--$Re_\theta$ and SA-BC curves over $\alpha=0^\circ$--$7^\circ$.
Reattachment runs $0.02$--$0.04\,c$ late at $\alpha\le5^\circ$, and reaches the
edge of bubble collapse at $7^\circ$. At L2 the lift sits within
$0.5\%$, separation $0.030$--$0.037\,c$ ahead of and reattachment
$0.033$--$0.045\,c$ behind the SA-AI stations.

\begin{figure}[H]\centering
\includegraphics[width=\textwidth]{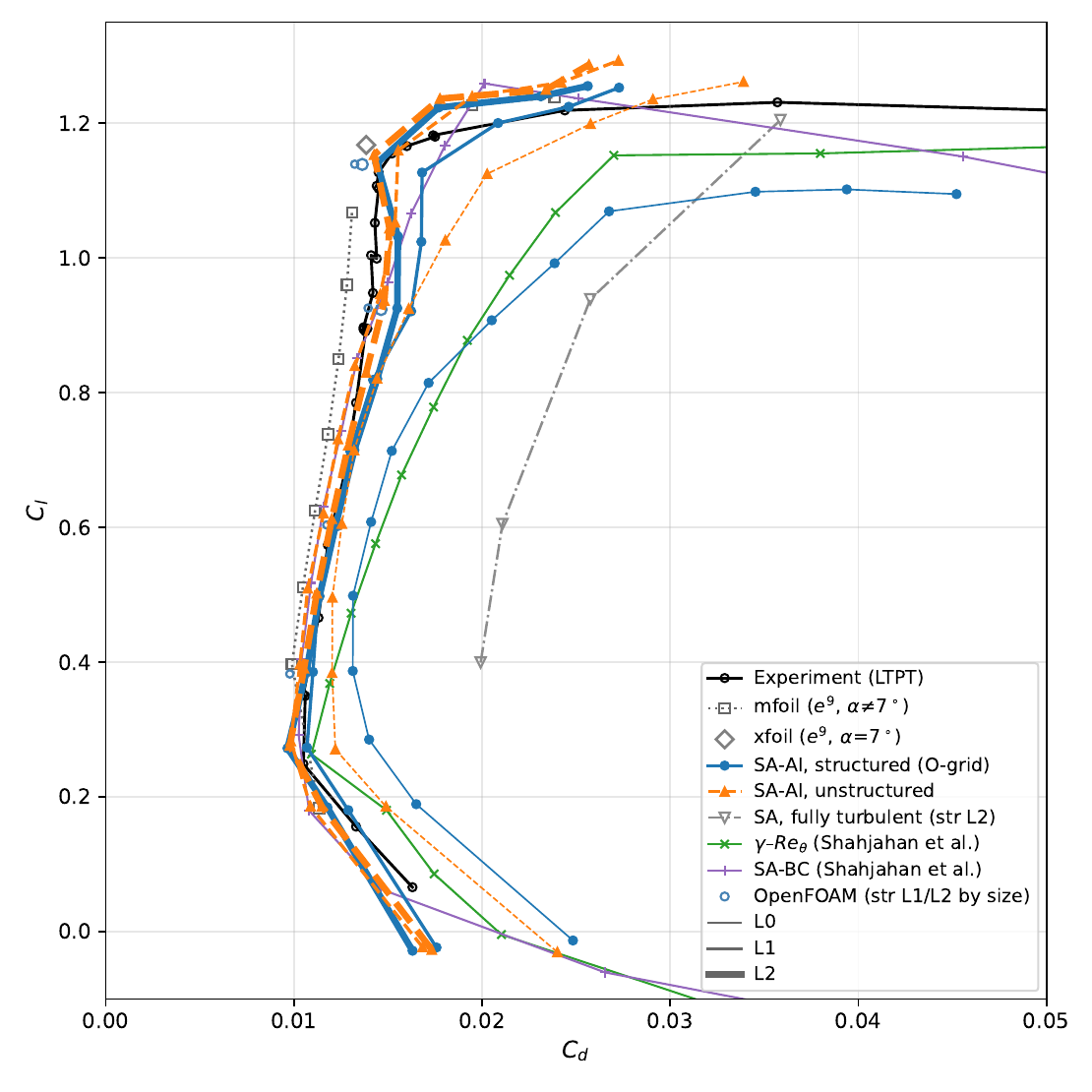}
\caption{Eppler 387 drag polar, $Re=2\times10^5$; all six grids span
$\alpha=-4^\circ$--$8.5^\circ$, the L2 pair to $9^\circ$. Fully turbulent SA
($\chi_\infty=3$) $66$--$87\%$ above the SA-AI drag in the bucket,
$+149\%$ at $\alpha=7^\circ$; at L2 mesh families agree to $\Delta C_d\le
6.9\times10^{-4}$ over $\alpha=-2^\circ$--$7^\circ$ and
$2.7\times10^{-4}$ at $8.5^\circ$; OpenFOAM port omits the bypass,
worth $-6$ to $-10$ counts at $\alpha=0$--$5^\circ$.}\label{f:epppolar}
\end{figure}

\begin{figure}[H]\centering
\includegraphics[width=\textwidth]{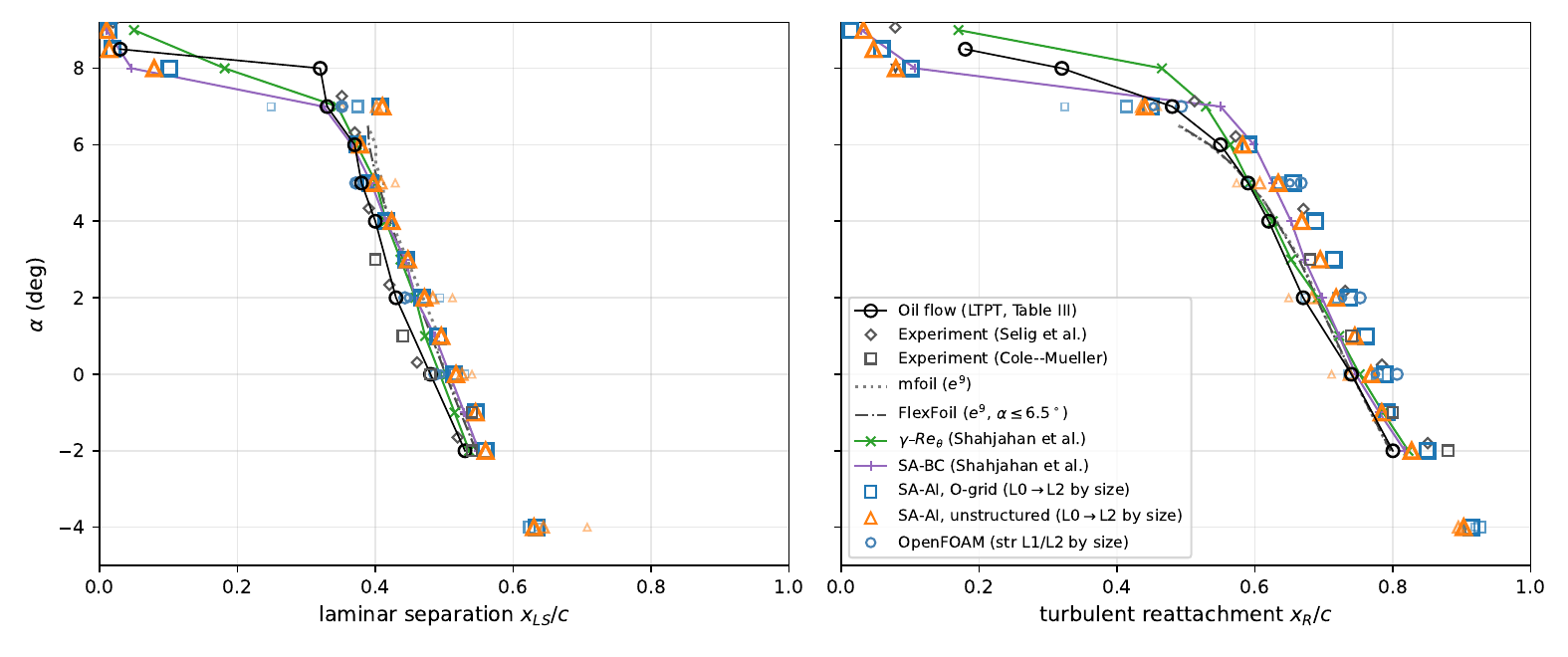}
\caption{Eppler 387 bubble stations, $Re=2\times10^5$; signed-$C_f$
zero crossings against oil flow (TM-4062 Table~III), Selig,
Cole--Mueller, $e^9$. At $\alpha=7^\circ$ XFOIL $C_l=1.17$, mfoil
$C_l=0.928$.}\label{f:eppbubble}
\end{figure}

\begin{figure}[H]\centering
\includegraphics[width=0.99\textwidth]{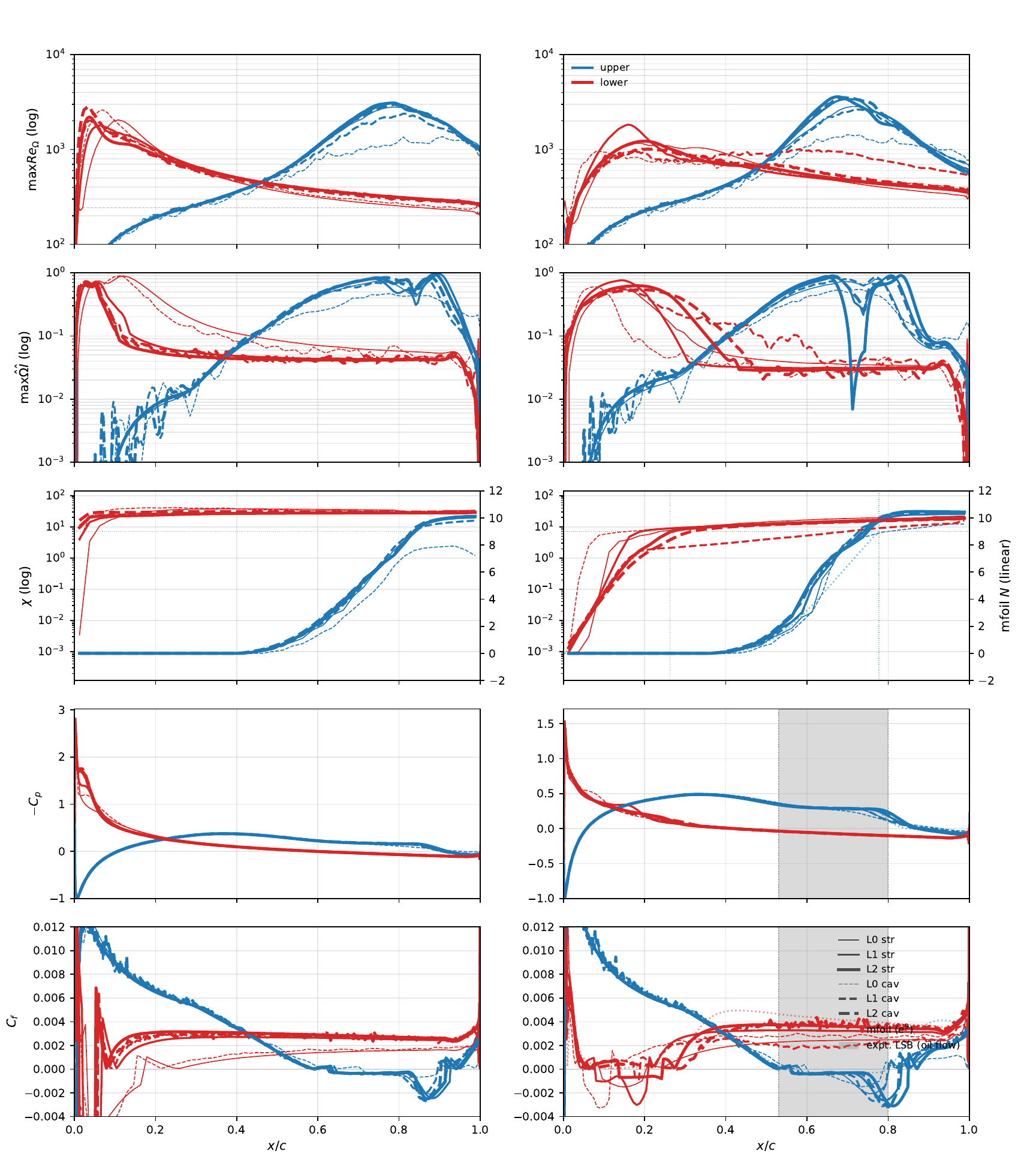}
\caption{Eppler~387 surface distributions, $Re=2\times10^5$, all six grids; pages run in ascending incidence. $\alpha=-4^\circ$, $\alpha=-2^\circ$.}\label{f:eppcflow}
\end{figure}

\begin{figure}[H]\centering
\includegraphics[width=0.99\textwidth]{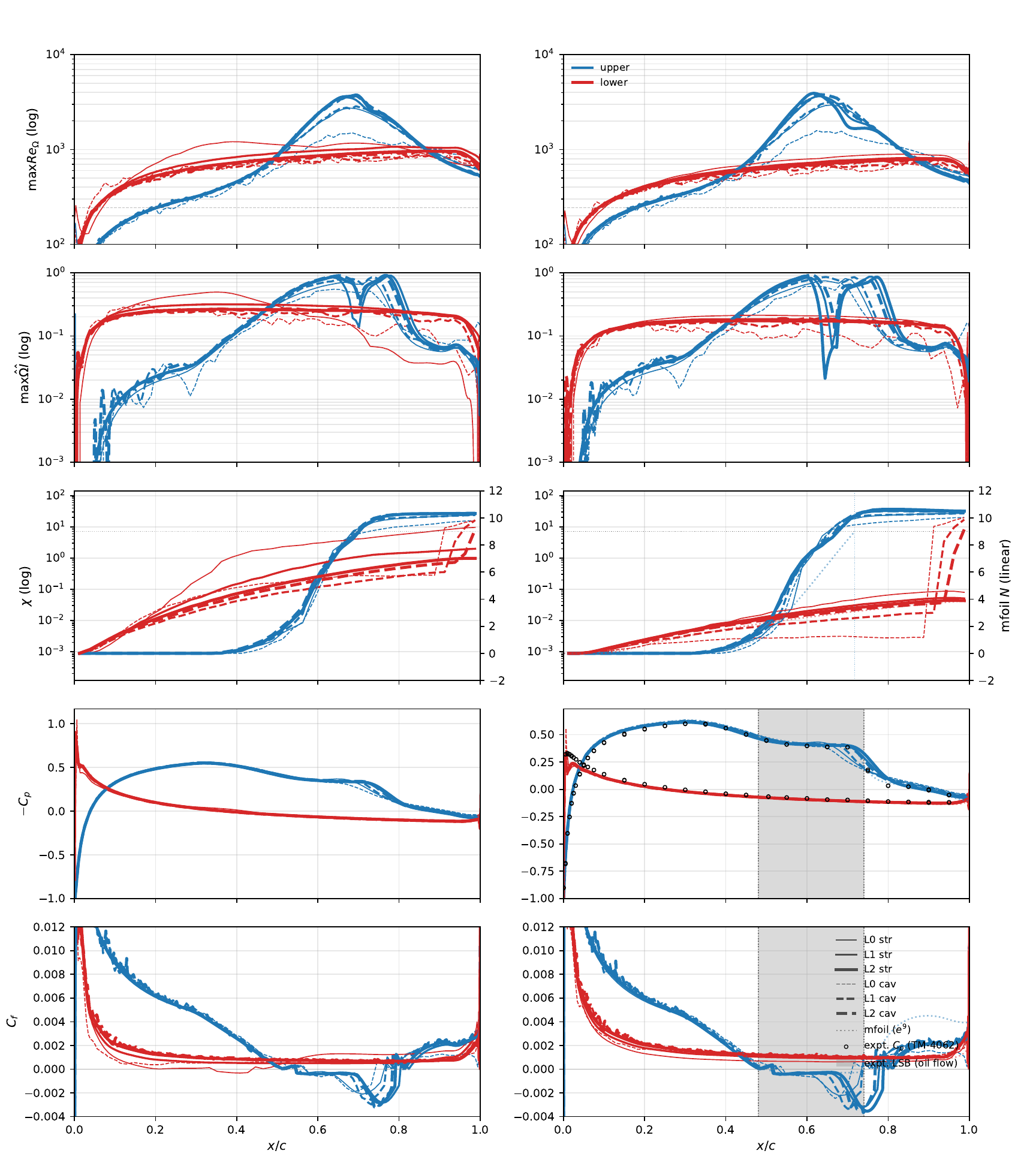}
\figcont{$\alpha=-1^\circ$, $\alpha=0^\circ$.}
\end{figure}

\begin{figure}[H]\centering
\includegraphics[width=0.99\textwidth]{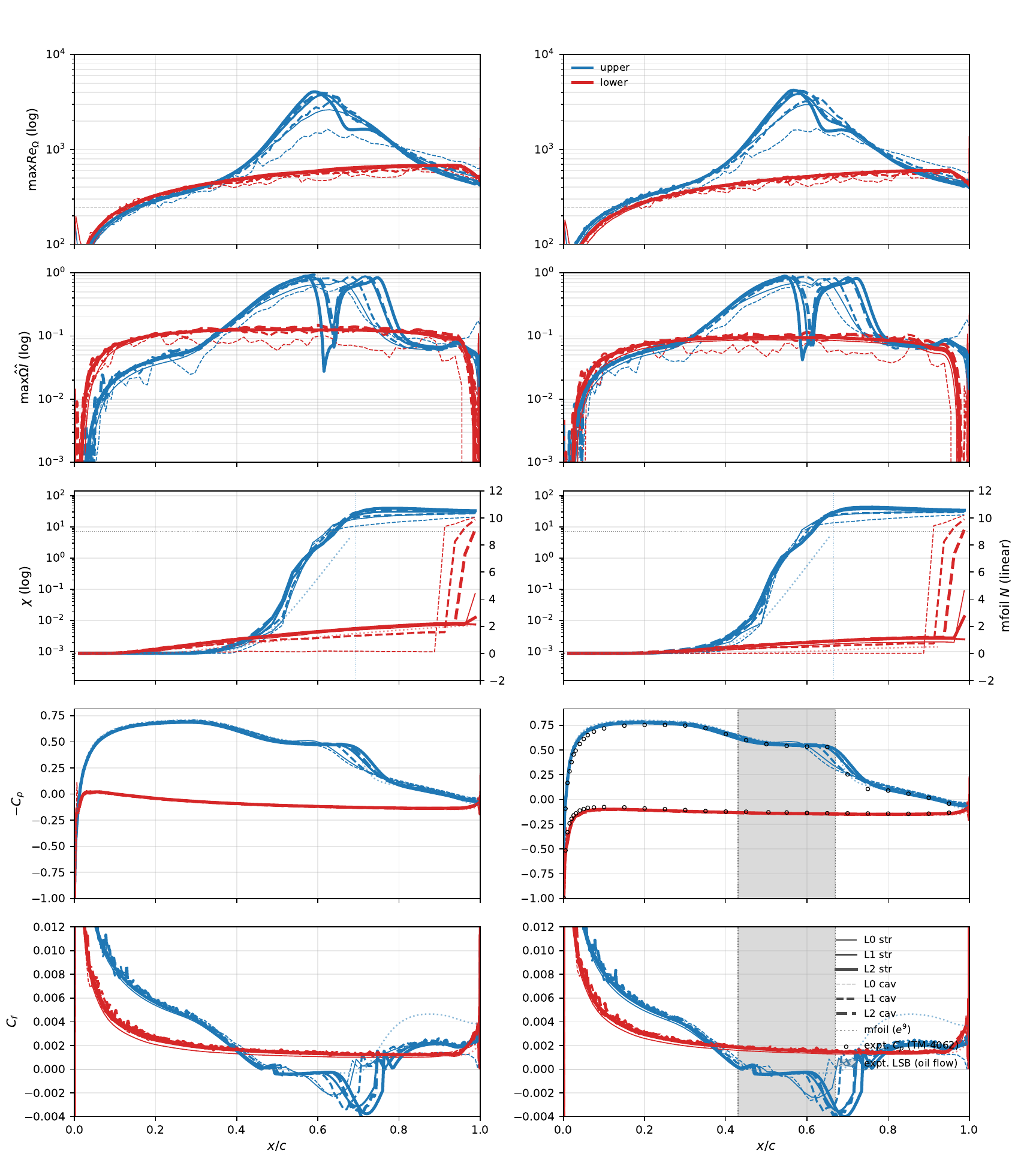}
\figcont{$\alpha=1^\circ$, $\alpha=2^\circ$.}
\end{figure}

\begin{figure}[H]\centering
\includegraphics[width=0.99\textwidth]{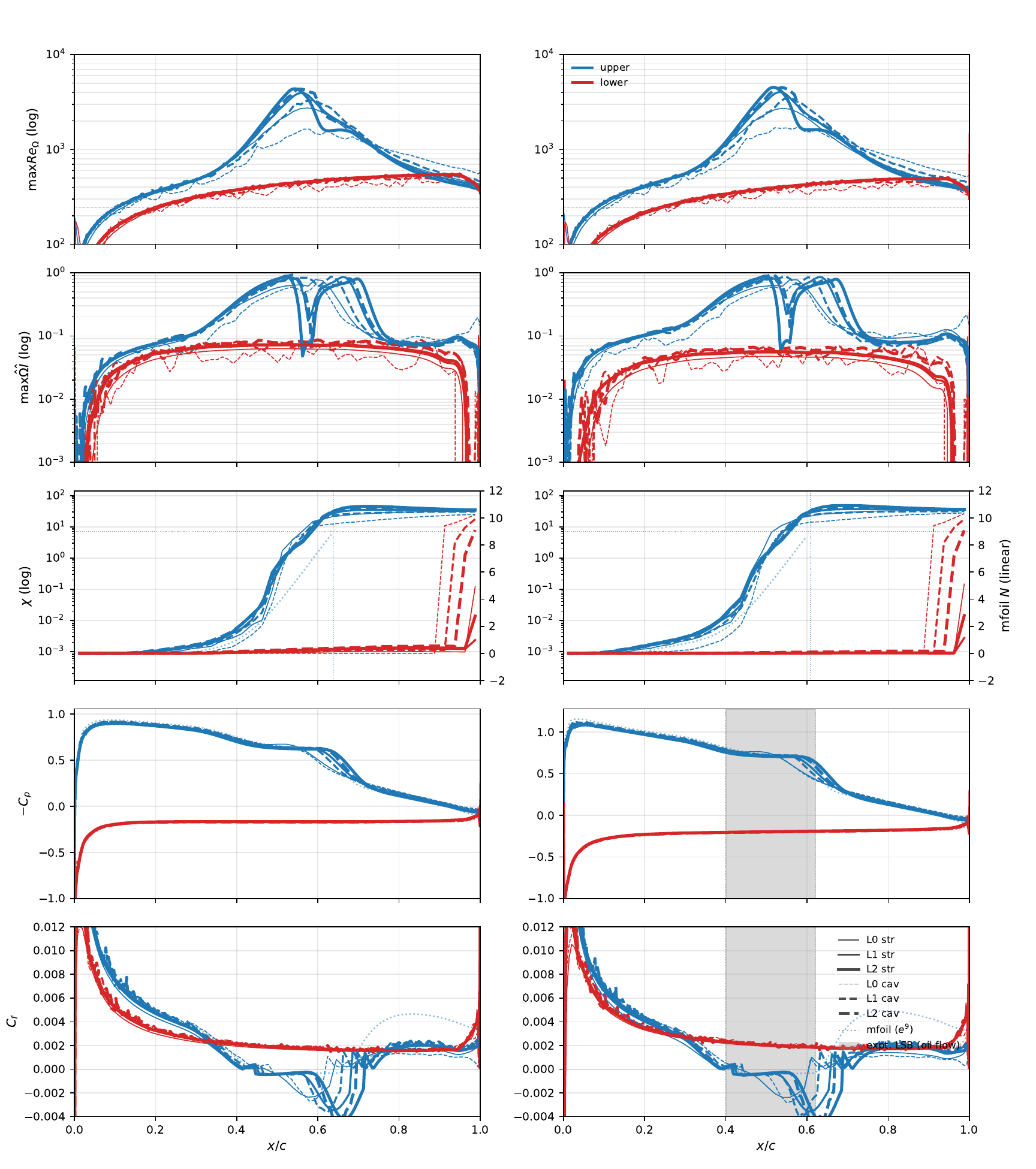}
\figcont{$\alpha=3^\circ$, $\alpha=4^\circ$.}
\end{figure}

\begin{figure}[H]\centering
\includegraphics[width=0.99\textwidth]{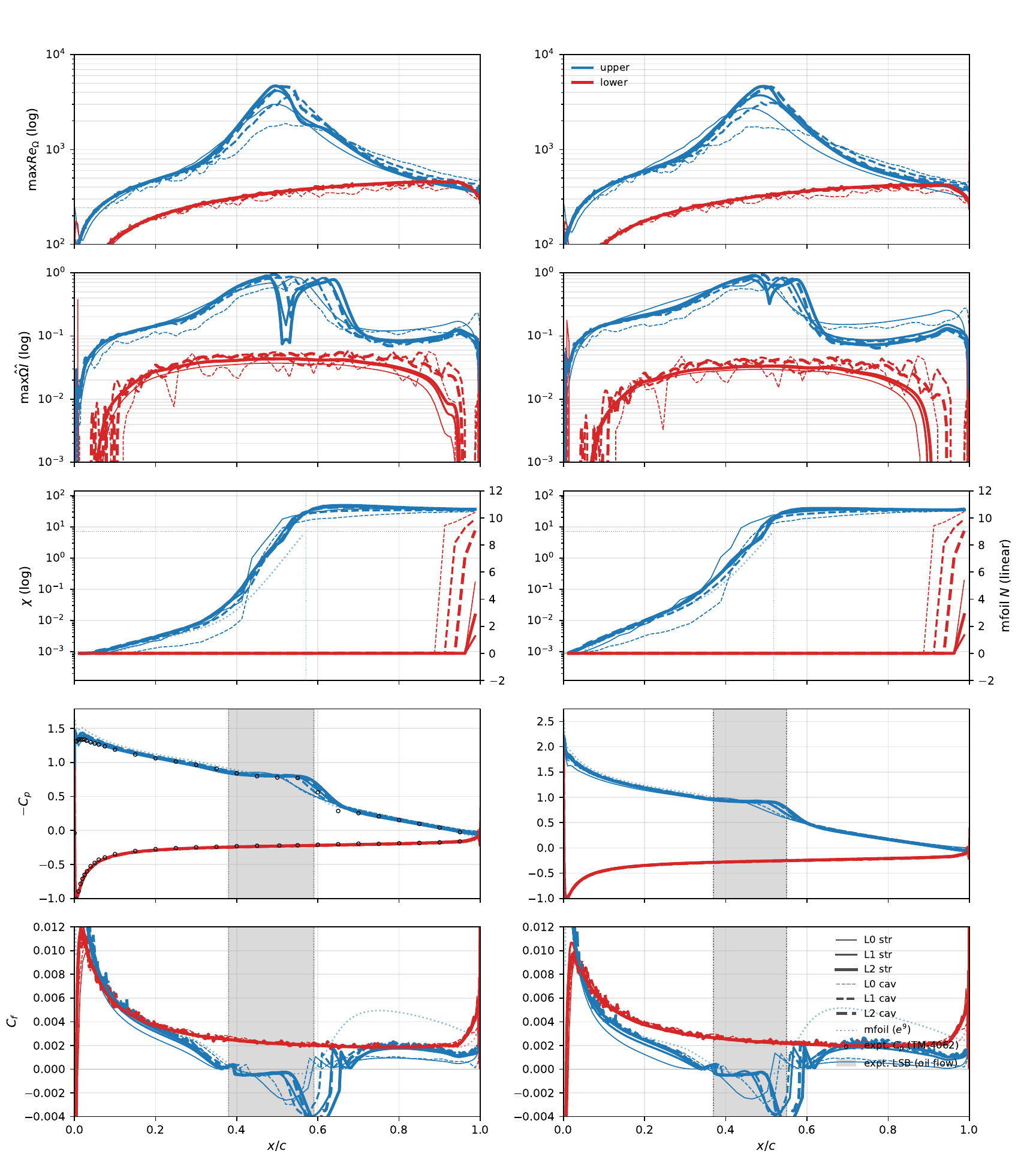}
\figcont{$\alpha=5^\circ$, $\alpha=6^\circ$.}
\end{figure}

\begin{figure}[H]\centering
\includegraphics[width=0.99\textwidth]{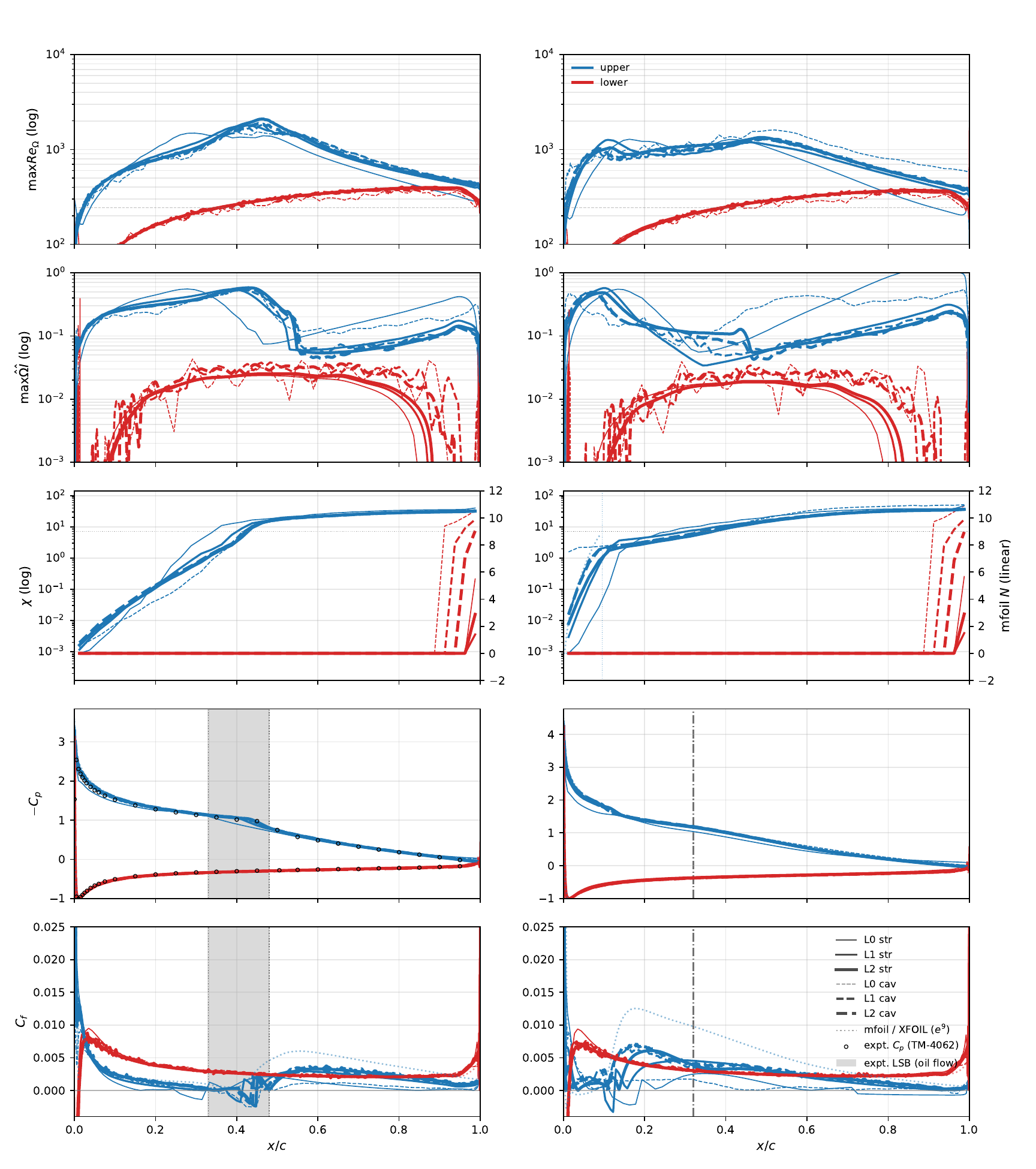}
\figcont{$\alpha=7^\circ$, $\alpha=8^\circ$.}
\end{figure}

\begin{figure}[H]\centering
\includegraphics[width=0.99\textwidth]{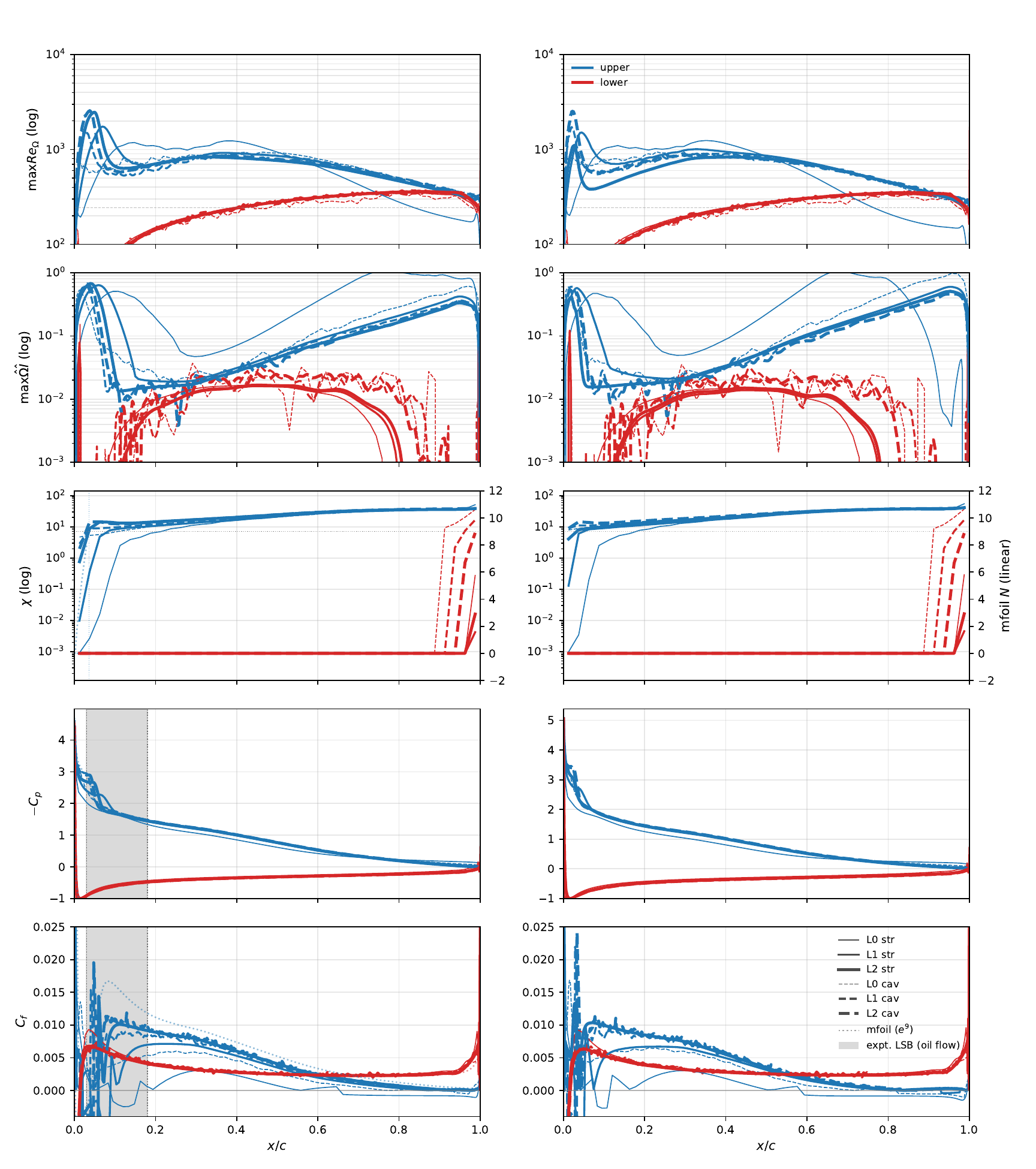}
\figcont{$\alpha=8.5^\circ$, $\alpha=9^\circ$.}
\end{figure}

\begin{table}[H]
  \centering\scriptsize\setlength{\tabcolsep}{2pt}
  \caption{Eppler 387, $Re\!=\!2\!\times\!10^5$: forces and
  upper-surface laminar-separation/turbulent-reattachment stations
  (signed-$C_f$ zero crossings over the whole chord; where the surface
  carries no bubble both stations are the transition front, so the pair
  is equal, and \texttt{--} in $x_R$ alone means separation without
  reattachment ahead of the trailing edge). The $-2^\circ$--$8.5^\circ$ extension incidences exist
  on the finest pair only; $-4^\circ$ runs on all three. Experimental
  forces are TM-4062 Table~B1 ($R\!=\!200{,}000$); experimental stations
  are its Table~III oil flow. At $8^\circ$ the measurement records
  natural transition at $0.32\,c$ and no bubble. The OpenFOAM port is independent, omits the
  $f_{v1}$ bypass, and additionally reports the $\chi\!=\!1$ front
  $x_\mathrm{tr}^\mathrm{up}$.}
  \label{t:epp}
  \begin{tabular}{ll cccc cccc ccccc cccc}
    \toprule
    & & \multicolumn{4}{c}{Flow360, structured O-grid} & \multicolumn{4}{c}{Flow360, unstructured cavity} & \multicolumn{5}{c}{OpenFOAM, structured} & \multicolumn{4}{c}{Experiment (LTPT)} \\
    \cmidrule(lr){3-6}\cmidrule(lr){7-10}\cmidrule(lr){11-15}\cmidrule(lr){16-19}
    $\alpha$ & grid & $C_L$ & $C_D$ & $x_{LS}$ & $x_R$ & $C_L$ & $C_D$ & $x_{LS}$ & $x_R$ & $C_L$ & $C_D$ & $x_\mathrm{tr}^\mathrm{up}$ & $x_{LS}$ & $x_R$ & $C_L$ & $C_D$ & $x_{LS}$ & $x_R$ \\
    \midrule
    $-4$ & L0 & $-0.0130$ & 0.02483 & 0.636 & 0.919 & $-0.0303$ & 0.02400 & 0.707 & 0.900 & -- & -- & -- & -- & -- & -- & -- & -- & -- \\
    $-4$ & L1 & $-0.0234$ & 0.01758 & 0.623 & 0.927 & $-0.0229$ & 0.01684 & 0.644 & 0.895 & -- & -- & -- & -- & -- &  &  &  &  \\
    $-4$ & L2 & $-0.0282$ & 0.01630 & 0.634 & 0.915 & $-0.0269$ & 0.01733 & 0.630 & 0.903 & -- & -- & -- & -- & -- &  &  &  &  \\
    \addlinespace
    $-2$ & L2 & 0.1845 & 0.01178 & 0.560 & 0.850 & 0.1857 & 0.01148 & 0.560 & 0.827 & -- & -- & -- & -- & -- & 0.156 & 0.0133 & 0.53 & 0.80 \\
    \addlinespace
    $-1$ & L2 & 0.2723 & 0.00966 & 0.546 & 0.791 & 0.2763 & 0.00978 & 0.545 & 0.784 & -- & -- & -- & -- & -- & 0.249 & 0.0105 & -- & -- \\
    \addlinespace
    $0$ & L0 & 0.3869 & 0.01313 & 0.530 & 0.735 & 0.3841 & 0.01203 & 0.540 & 0.711 & 0.3887 & 0.01167 & 0.638 & 0.493 & 0.784 & 0.350 & 0.0106 & 0.48 & 0.74 \\
    $0$ & L1 & 0.3855 & 0.01101 & 0.516 & 0.778 & 0.3972 & 0.01032 & 0.526 & 0.737 & 0.3826 & 0.00978 & 0.605 & 0.495 & 0.775 &  &  &  &  \\
    $0$ & L2 & 0.3880 & 0.01068 & 0.514 & 0.788 & 0.3908 & 0.01053 & 0.517 & 0.768 & 0.3857 & 0.01058 & 0.621 & 0.486 & 0.806 &  &  &  &  \\
    \addlinespace
    $1$ & L2 & 0.4979 & 0.01136 & 0.491 & 0.761 & 0.5022 & 0.01121 & 0.496 & 0.745 & -- & -- & -- & -- & -- & 0.466 & 0.0113 & -- & -- \\
    \addlinespace
    $2$ & L0 & 0.6082 & 0.01410 & 0.493 & 0.672 & 0.6055 & 0.01255 & 0.512 & 0.649 & 0.6133 & 0.01406 & 0.568 & 0.447 & 0.739 & 0.574 & 0.0118 & 0.43 & 0.67 \\
    $2$ & L1 & 0.6054 & 0.01228 & 0.470 & 0.721 & 0.6207 & 0.01157 & 0.483 & 0.682 & 0.6037 & 0.01174 & 0.543 & 0.448 & 0.727 &  &  &  &  \\
    $2$ & L2 & 0.6070 & 0.01225 & 0.468 & 0.737 & 0.6126 & 0.01201 & 0.471 & 0.718 & 0.6039 & 0.01229 & 0.564 & 0.443 & 0.752 &  &  &  &  \\
    \addlinespace
    $3$ & L2 & 0.7147 & 0.01321 & 0.444 & 0.714 & 0.7219 & 0.01289 & 0.447 & 0.695 & -- & -- & -- & -- & -- & 0.680 & 0.0127 & -- & -- \\
    \addlinespace
    $4$ & L2 & 0.8206 & 0.01443 & 0.416 & 0.687 & 0.8298 & 0.01384 & 0.424 & 0.668 & -- & -- & -- & -- & -- & 0.785 & 0.0133 & 0.40 & 0.62 \\
    \addlinespace
    $5$ & L0 & 0.9075 & 0.02051 & 0.398 & 0.587 & 0.9246 & 0.01610 & 0.429 & 0.573 & 0.9292 & 0.01665 & 0.457 & 0.344 & 0.654 & 0.891 & 0.0138 & 0.38 & 0.59 \\
    $5$ & L1 & 0.9209 & 0.01623 & 0.394 & 0.633 & 0.9462 & 0.01459 & 0.408 & 0.607 & 0.9256 & 0.01394 & 0.468 & 0.371 & 0.651 &  &  &  &  \\
    $5$ & L2 & 0.9254 & 0.01549 & 0.393 & 0.655 & 0.9365 & 0.01481 & 0.397 & 0.634 & 0.9234 & 0.01463 & 0.487 & 0.372 & 0.667 &  &  &  &  \\
    \addlinespace
    $6$ & L2 & 1.0318 & 0.01553 & 0.374 & 0.591 & 1.0440 & 0.01508 & 0.378 & 0.582 & -- & -- & -- & -- & -- & 1.004 & 0.0141 & 0.37 & 0.55 \\
    \addlinespace
    $7$ & L0 & 1.0691 & 0.02675 & 0.249 & 0.324 & 1.1248 & 0.02027 & 0.351 & 0.459 & 1.1173 & 0.01817 & 0.139 & -- & 0.201 & 1.103 & 0.0145 & 0.33 & 0.48 \\
    $7$ & L1 & 1.1268 & 0.01681 & 0.374 & 0.414 & 1.1596 & 0.01554 & 0.402 & 0.435 & 1.1391 & 0.01323 & 0.329 & 0.353 & 0.453 &  &  &  &  \\
    $7$ & L2 & 1.1411 & 0.01442 & 0.407 & 0.449 & 1.1531 & 0.01426 & 0.410 & 0.440 & 1.1388 & 0.01363 & 0.379 & 0.351 & 0.493 &  &  &  &  \\
    \addlinespace
    $8$ & L2 & 1.2232 & 0.01768 & 0.101 & 0.101 & 1.2364 & 0.01776 & 0.079 & 0.079 & -- & -- & -- & -- & -- & 1.180 & 0.0175 & 0.32 & 0.32 \\
    \addlinespace
    $8.5$ & L2 & 1.2399 & 0.02312 & 0.019 & 0.059 & 1.2496 & 0.02339 & 0.014 & 0.047 & -- & -- & -- & -- & -- & -- & -- & 0.03 & 0.18 \\
    \addlinespace
    $9$ & L2 & 1.2552 & 0.02562 & 0.014 & 0.014 & 1.2869 & 0.02568 & 0.010 & 0.032 & -- & -- & -- & -- & -- & 1.219 & 0.0244 & -- & -- \\
    \bottomrule
  \end{tabular}
\end{table}

\clearpage
\begin{figure}[H]\centering
\includegraphics[width=0.99\textwidth,height=0.94\textheight,keepaspectratio]{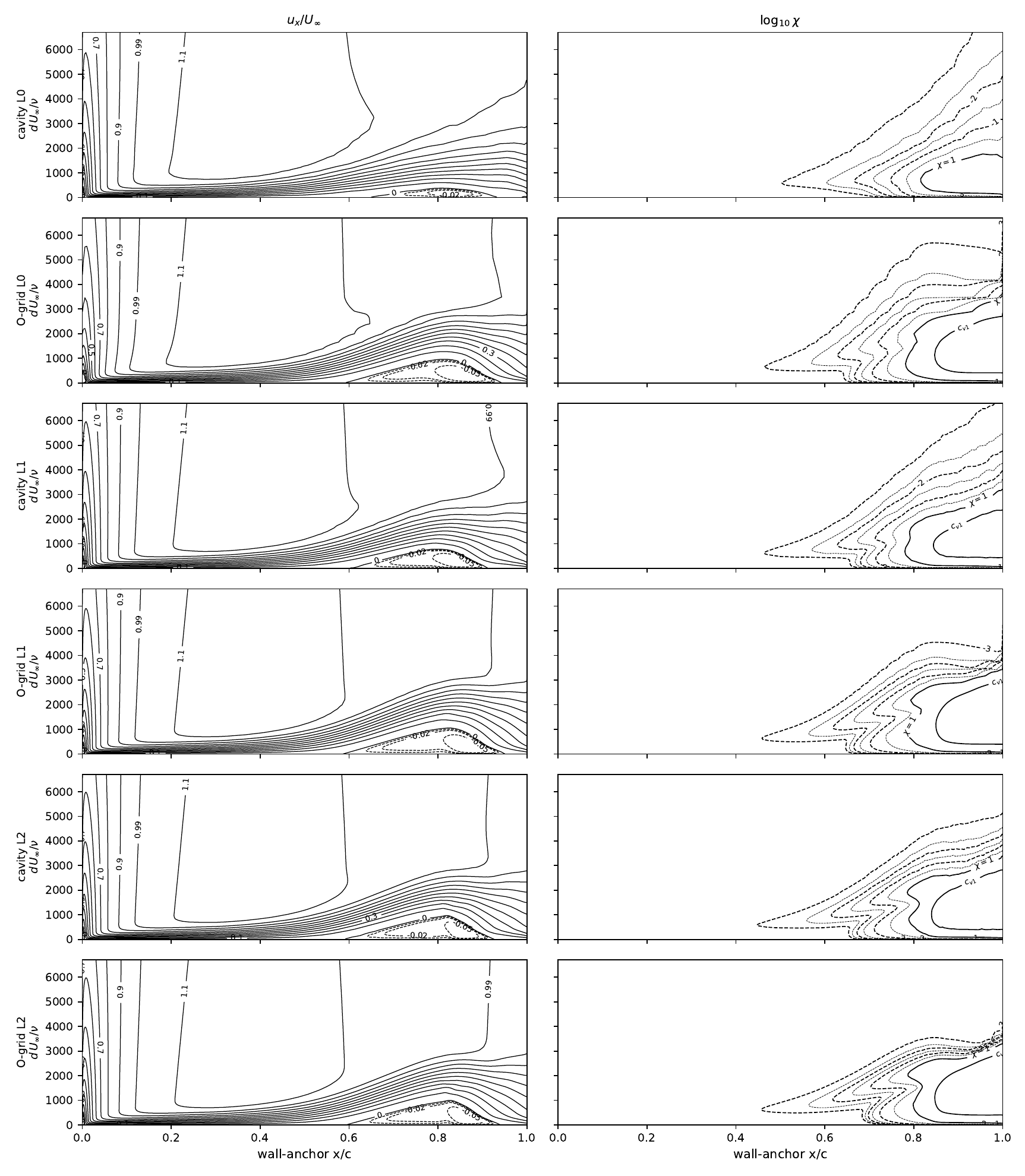}
\caption{Eppler~387 wall-normal sheets, $Re=2\times10^5$, upper surface:
$u_x/U_\infty$ (left) and $\log_{10}\chi$ (right) on all six grids.
$\alpha=-4^\circ$.}\label{f:eppsheets}
\end{figure}
\begin{figure}[H]\centering
\includegraphics[width=0.99\textwidth,height=0.94\textheight,keepaspectratio]{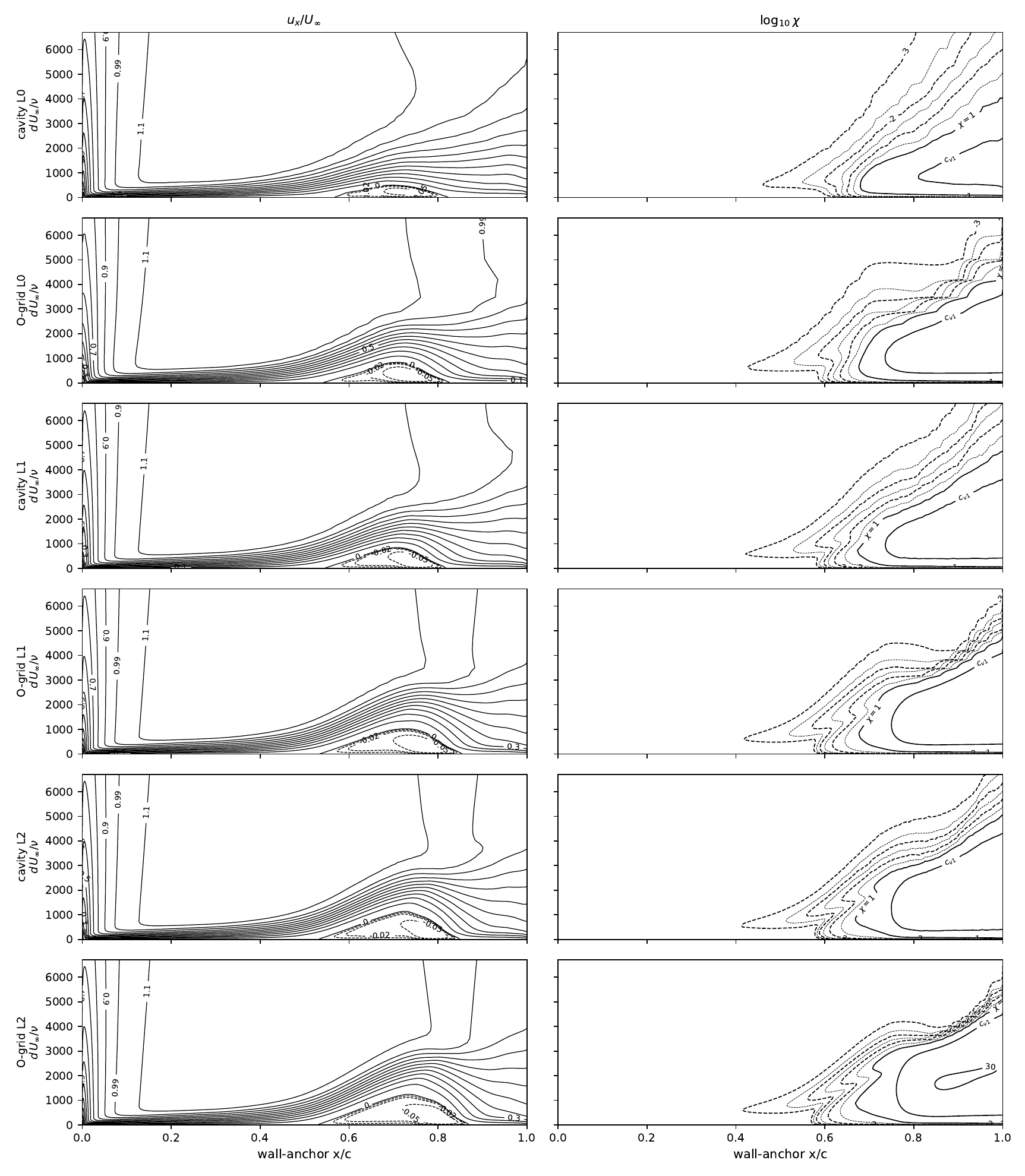}
\figcont{$\alpha=-2^\circ$.}
\end{figure}
\begin{figure}[H]\centering
\includegraphics[width=0.99\textwidth,height=0.94\textheight,keepaspectratio]{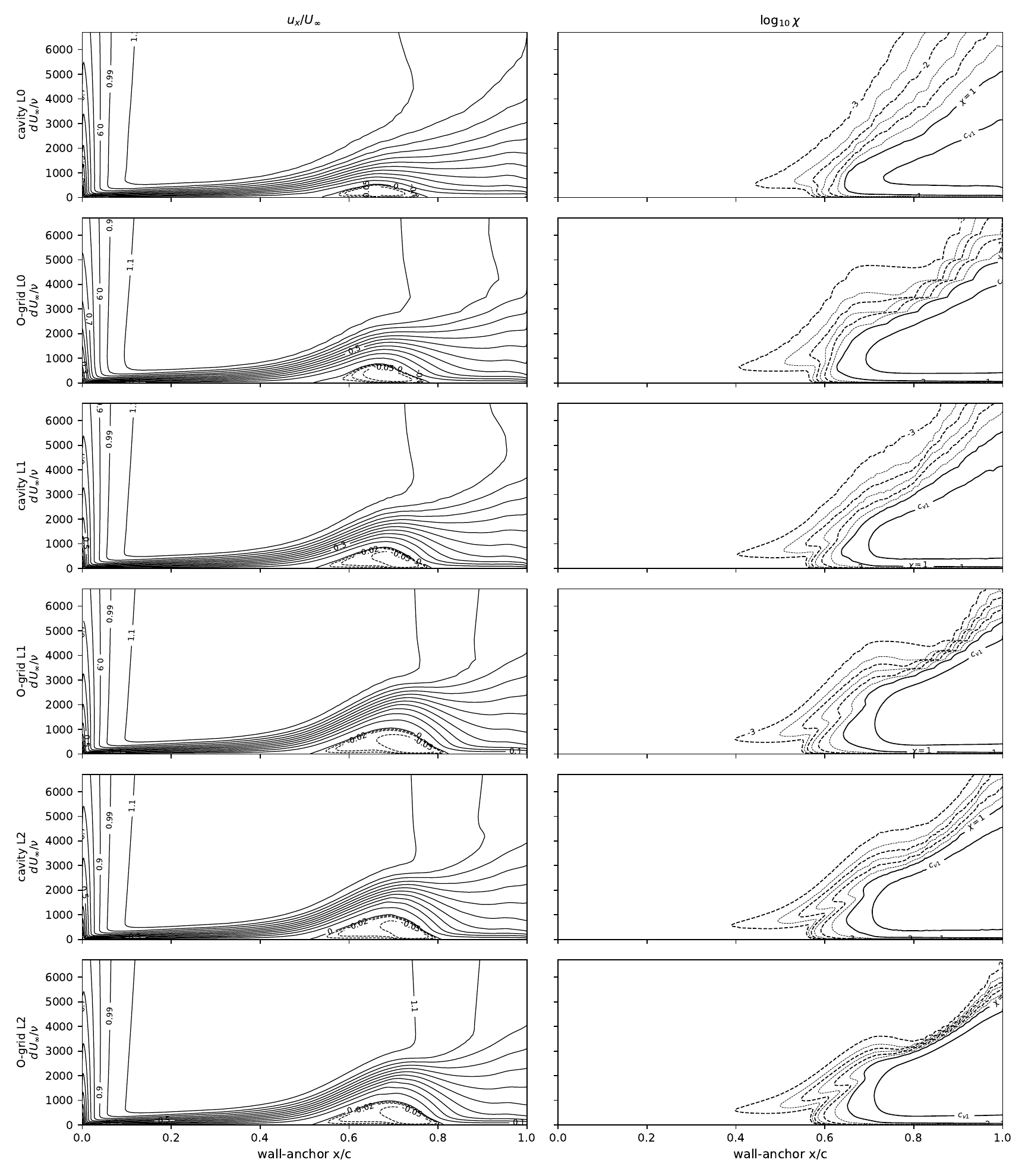}
\figcont{$\alpha=-1^\circ$.}
\end{figure}
\begin{figure}[H]\centering
\includegraphics[width=0.99\textwidth,height=0.94\textheight,keepaspectratio]{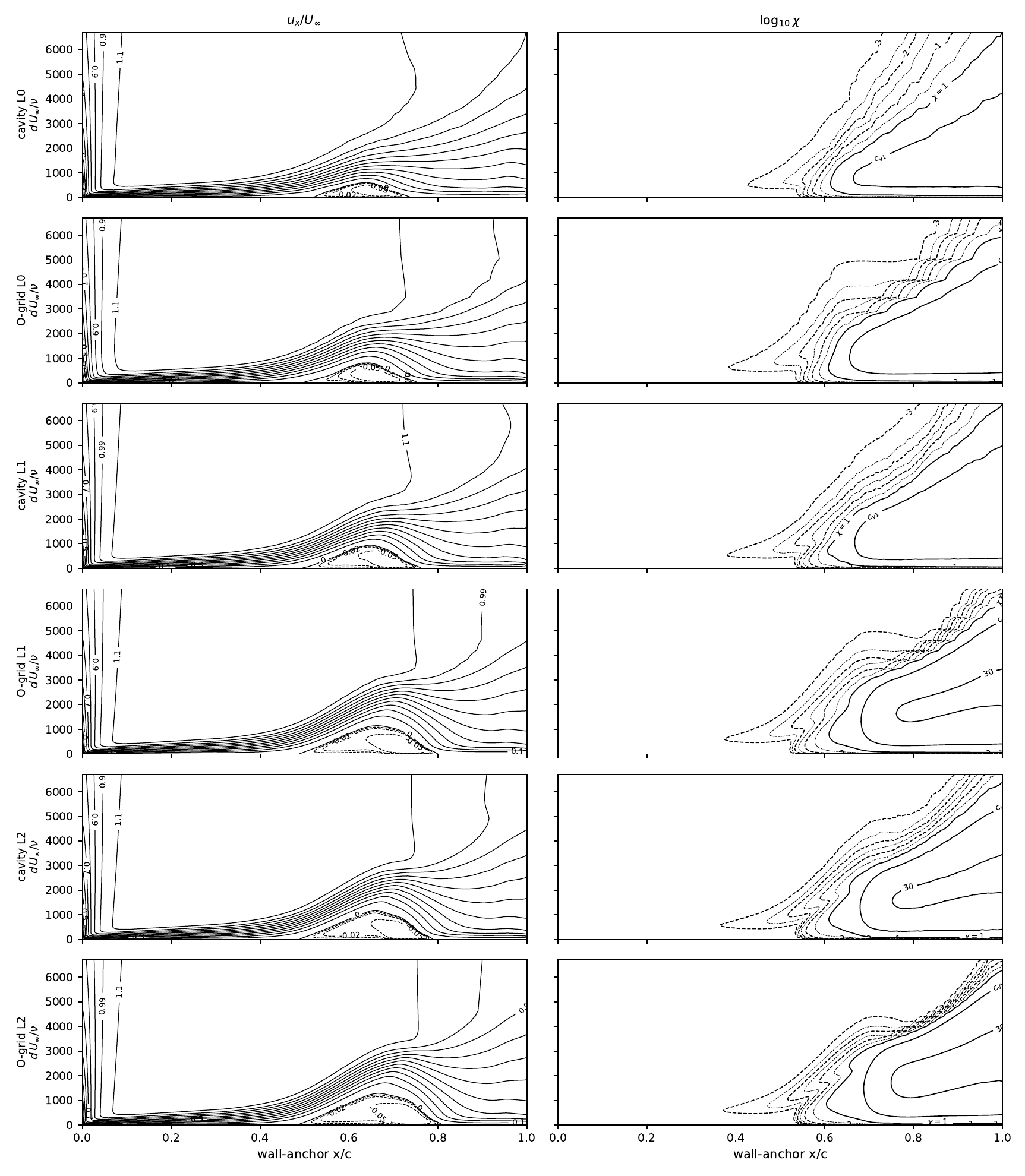}
\figcont{$\alpha=0^\circ$.}
\end{figure}
\begin{figure}[H]\centering
\includegraphics[width=0.99\textwidth,height=0.94\textheight,keepaspectratio]{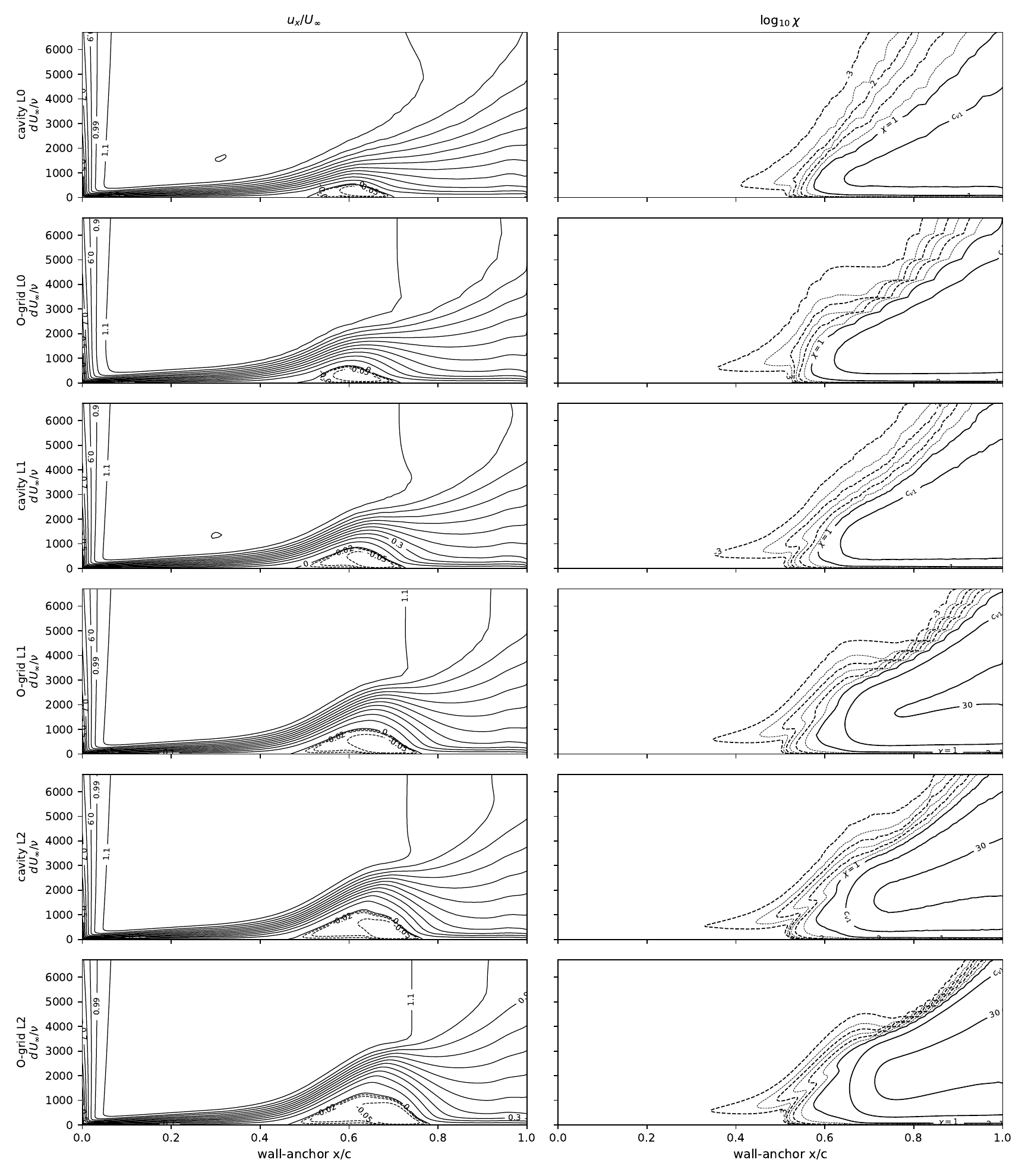}
\figcont{$\alpha=1^\circ$.}
\end{figure}
\begin{figure}[H]\centering
\includegraphics[width=0.99\textwidth,height=0.94\textheight,keepaspectratio]{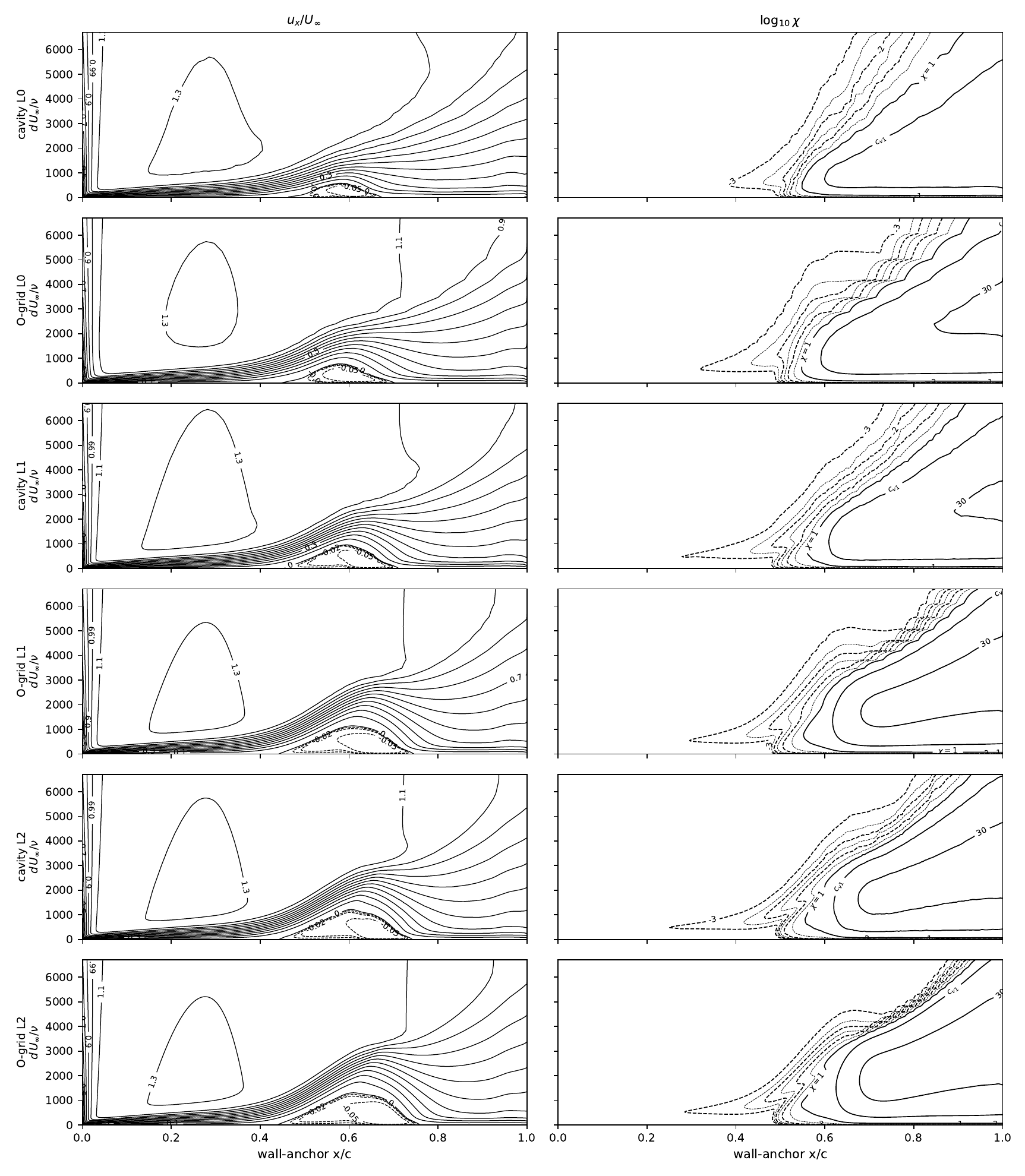}
\figcont{$\alpha=2^\circ$.}
\end{figure}
\begin{figure}[H]\centering
\includegraphics[width=0.99\textwidth,height=0.94\textheight,keepaspectratio]{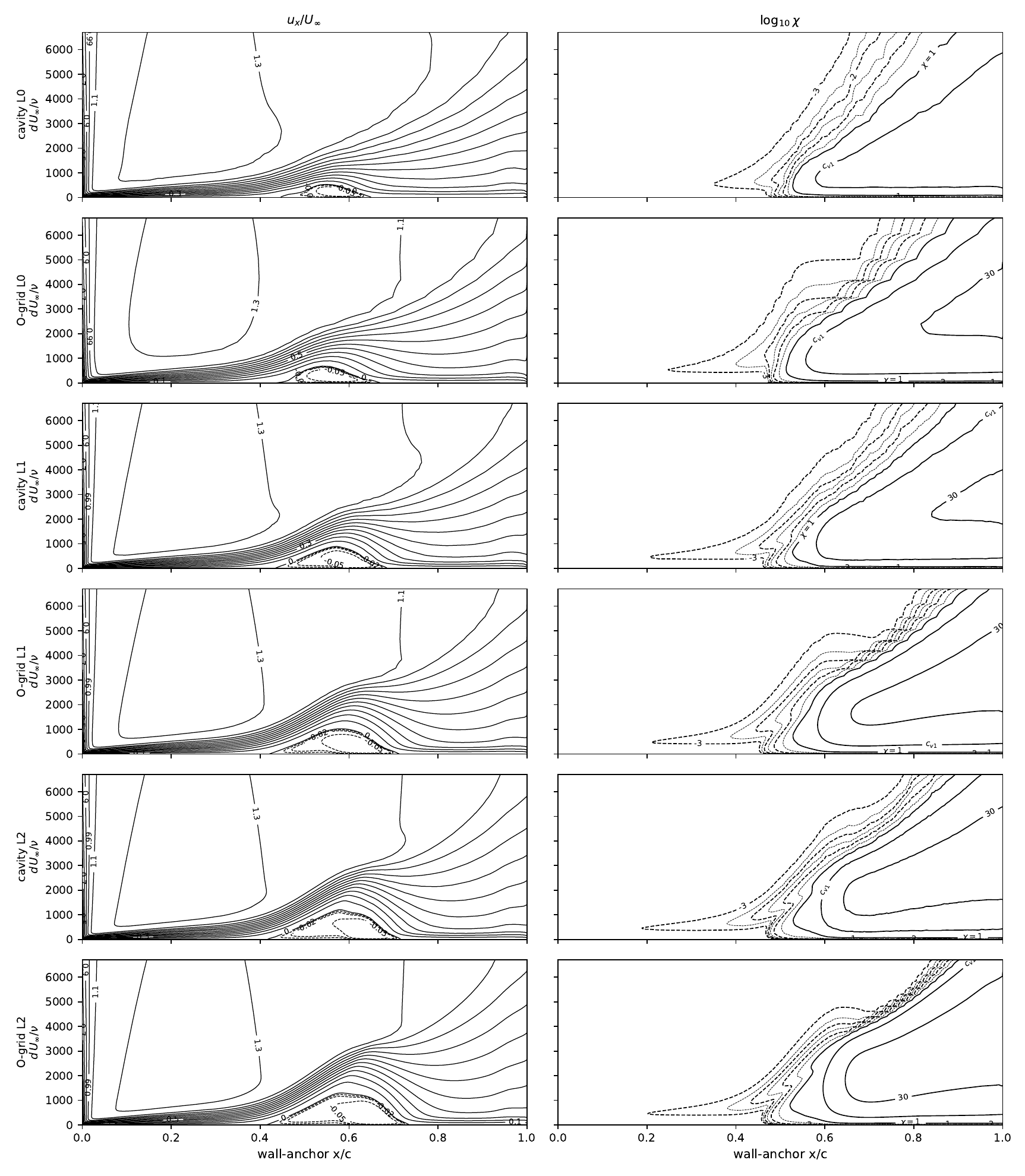}
\figcont{$\alpha=3^\circ$.}
\end{figure}
\begin{figure}[H]\centering
\includegraphics[width=0.99\textwidth,height=0.94\textheight,keepaspectratio]{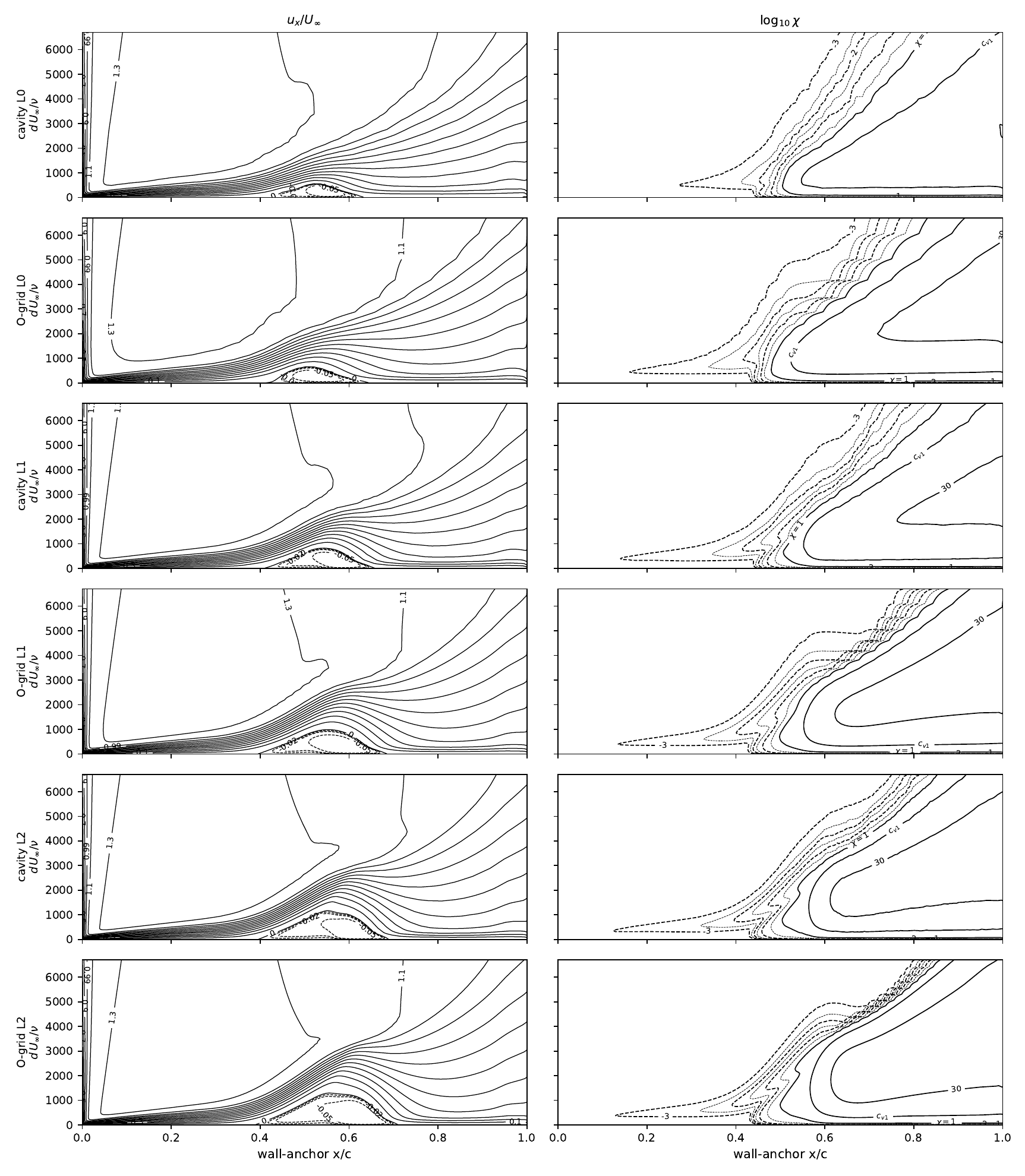}
\figcont{$\alpha=4^\circ$.}
\end{figure}
\begin{figure}[H]\centering
\includegraphics[width=0.99\textwidth,height=0.94\textheight,keepaspectratio]{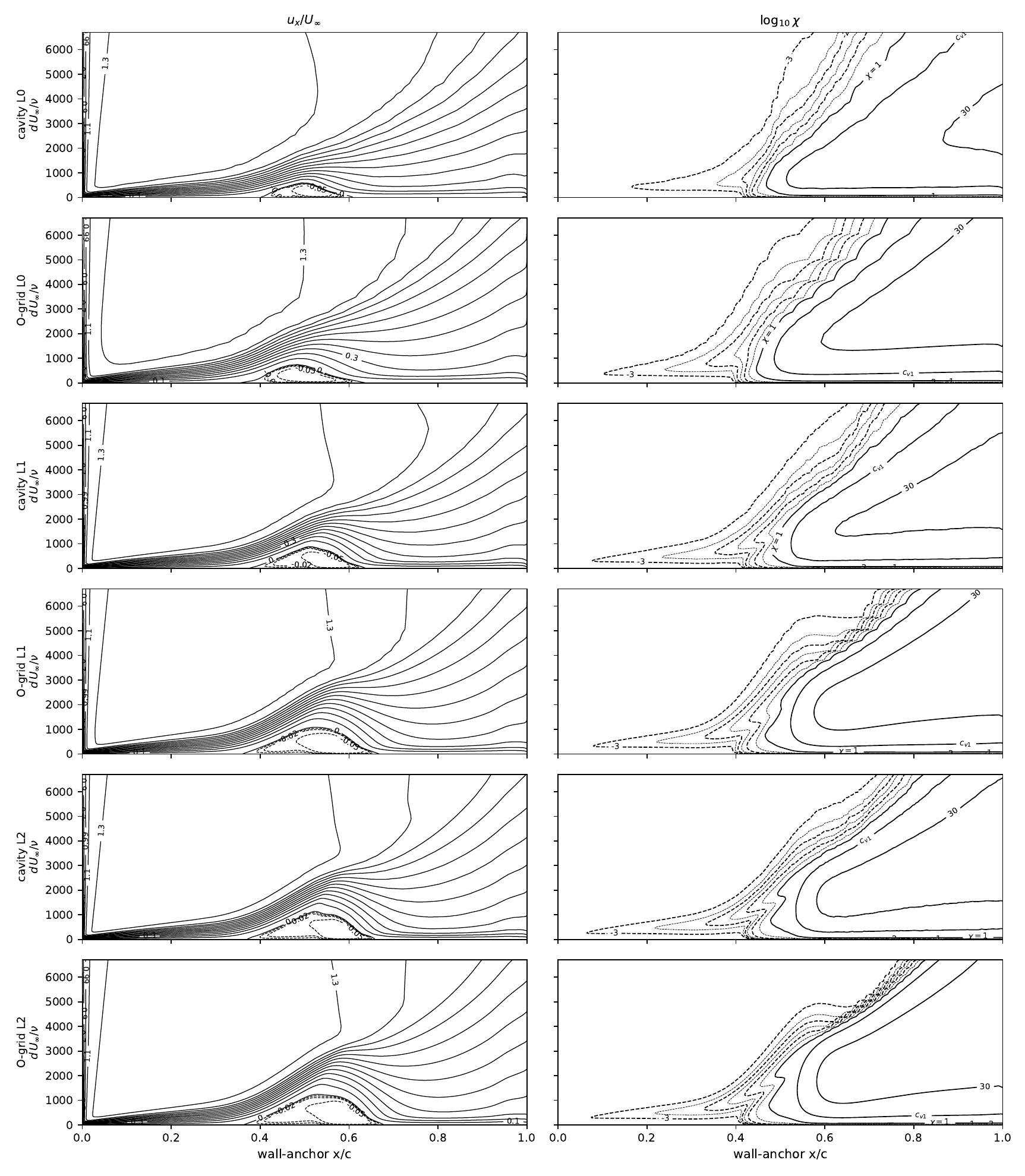}
\figcont{$\alpha=5^\circ$.}
\end{figure}
\begin{figure}[H]\centering
\includegraphics[width=0.99\textwidth,height=0.94\textheight,keepaspectratio]{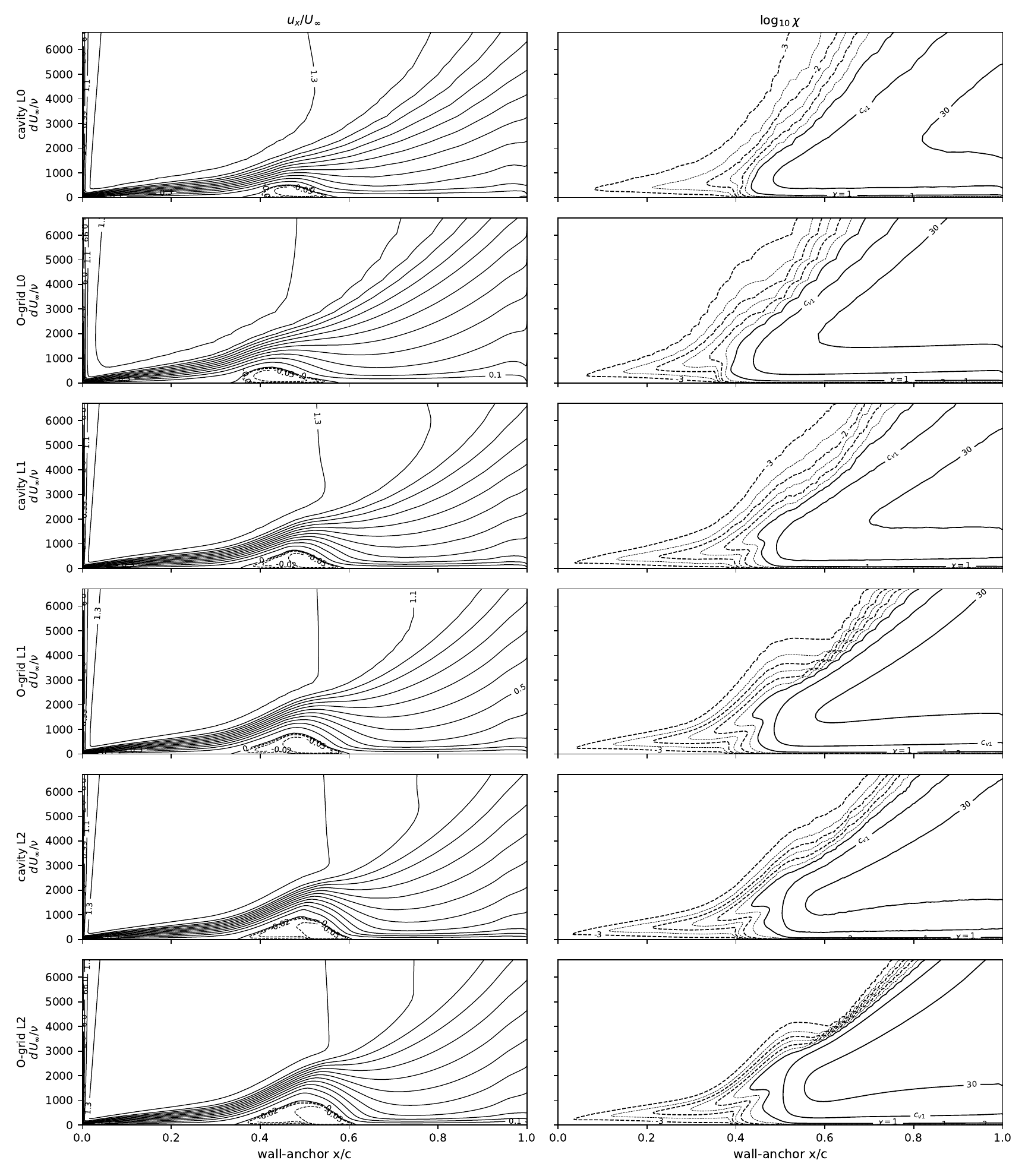}
\figcont{$\alpha=6^\circ$.}
\end{figure}
\begin{figure}[H]\centering
\includegraphics[width=0.99\textwidth,height=0.94\textheight,keepaspectratio]{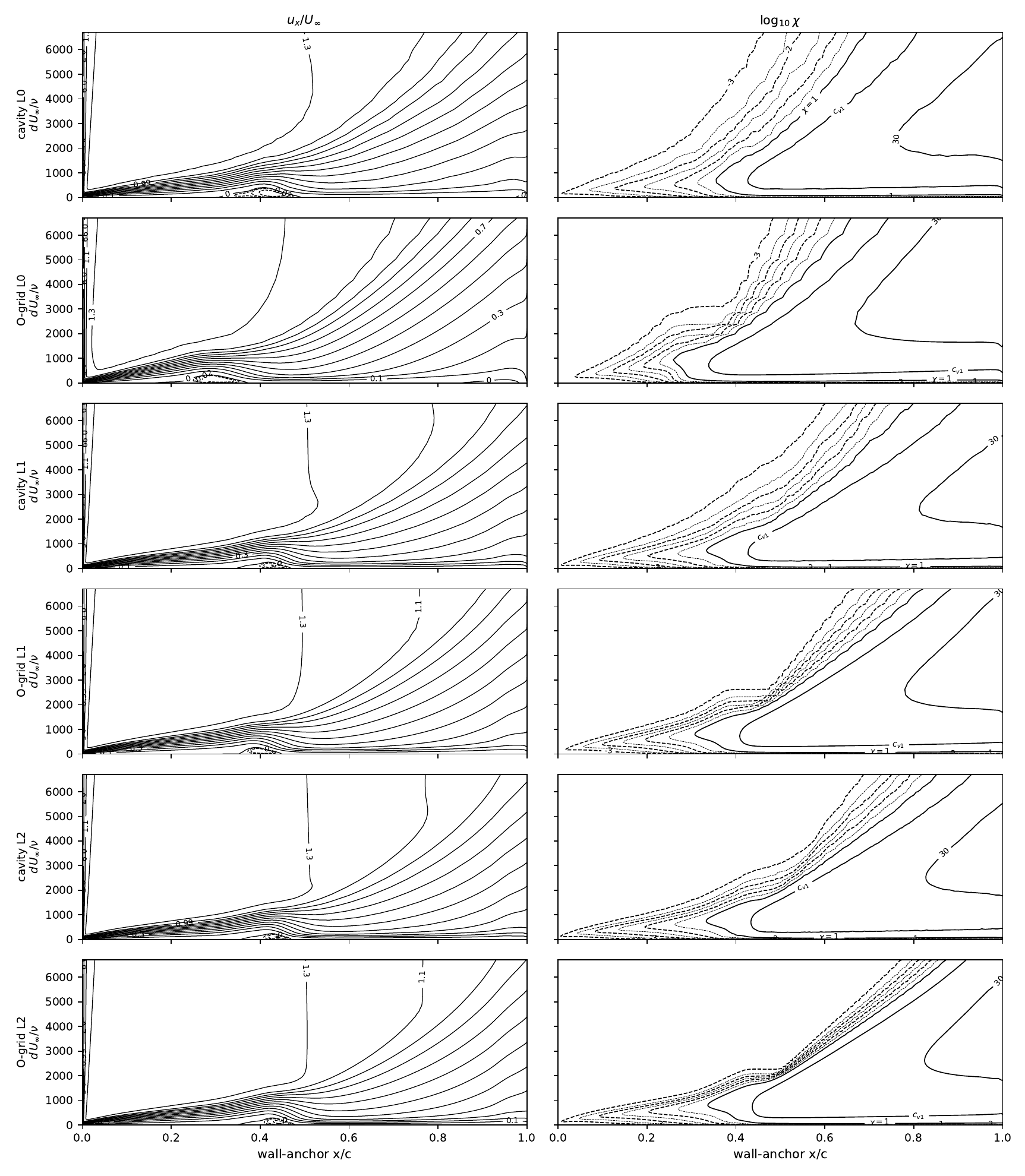}
\figcont{$\alpha=7^\circ$.}
\end{figure}
\begin{figure}[H]\centering
\includegraphics[width=0.99\textwidth,height=0.94\textheight,keepaspectratio]{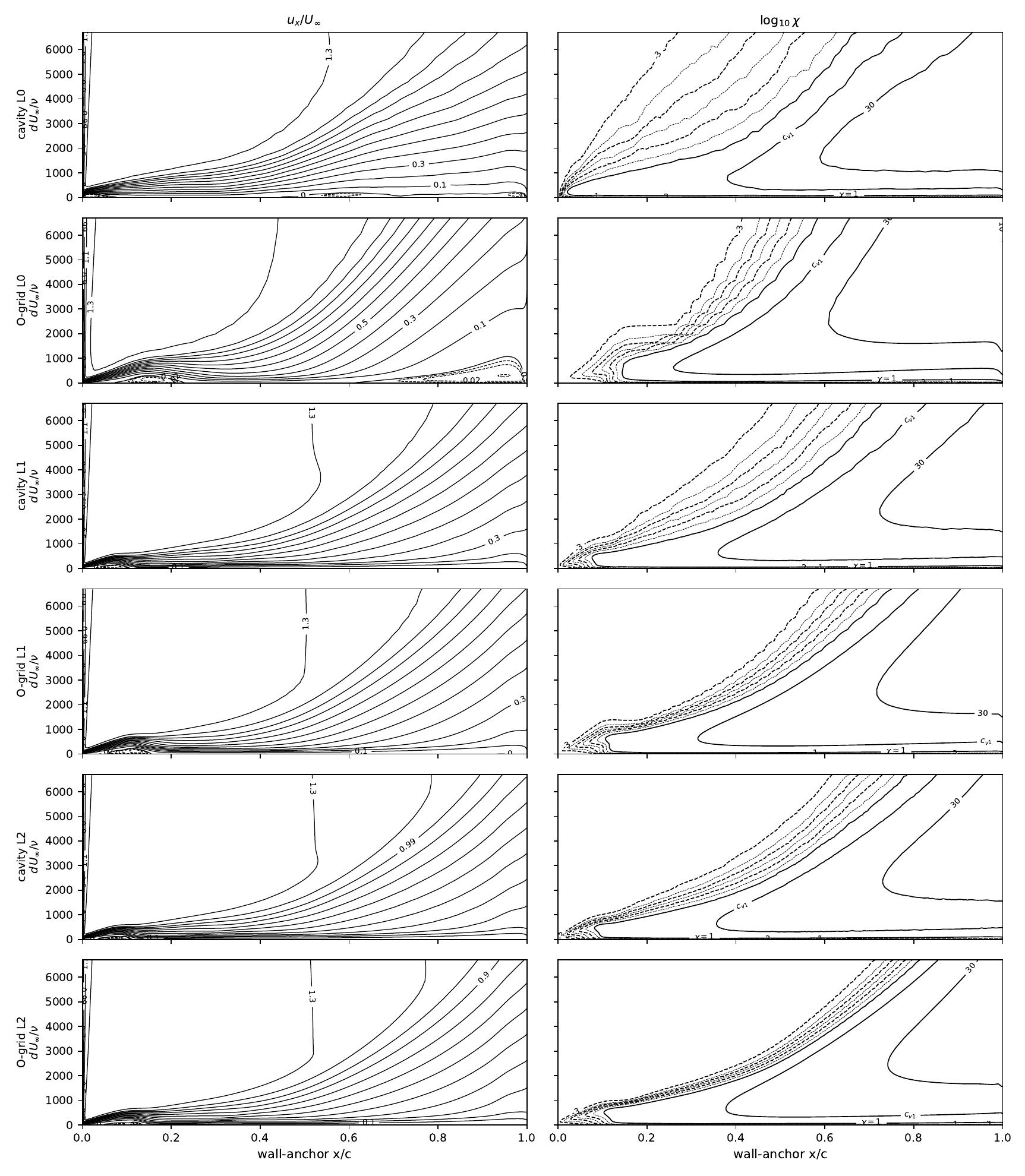}
\figcont{$\alpha=8^\circ$.}
\end{figure}
\begin{figure}[H]\centering
\includegraphics[width=0.99\textwidth,height=0.94\textheight,keepaspectratio]{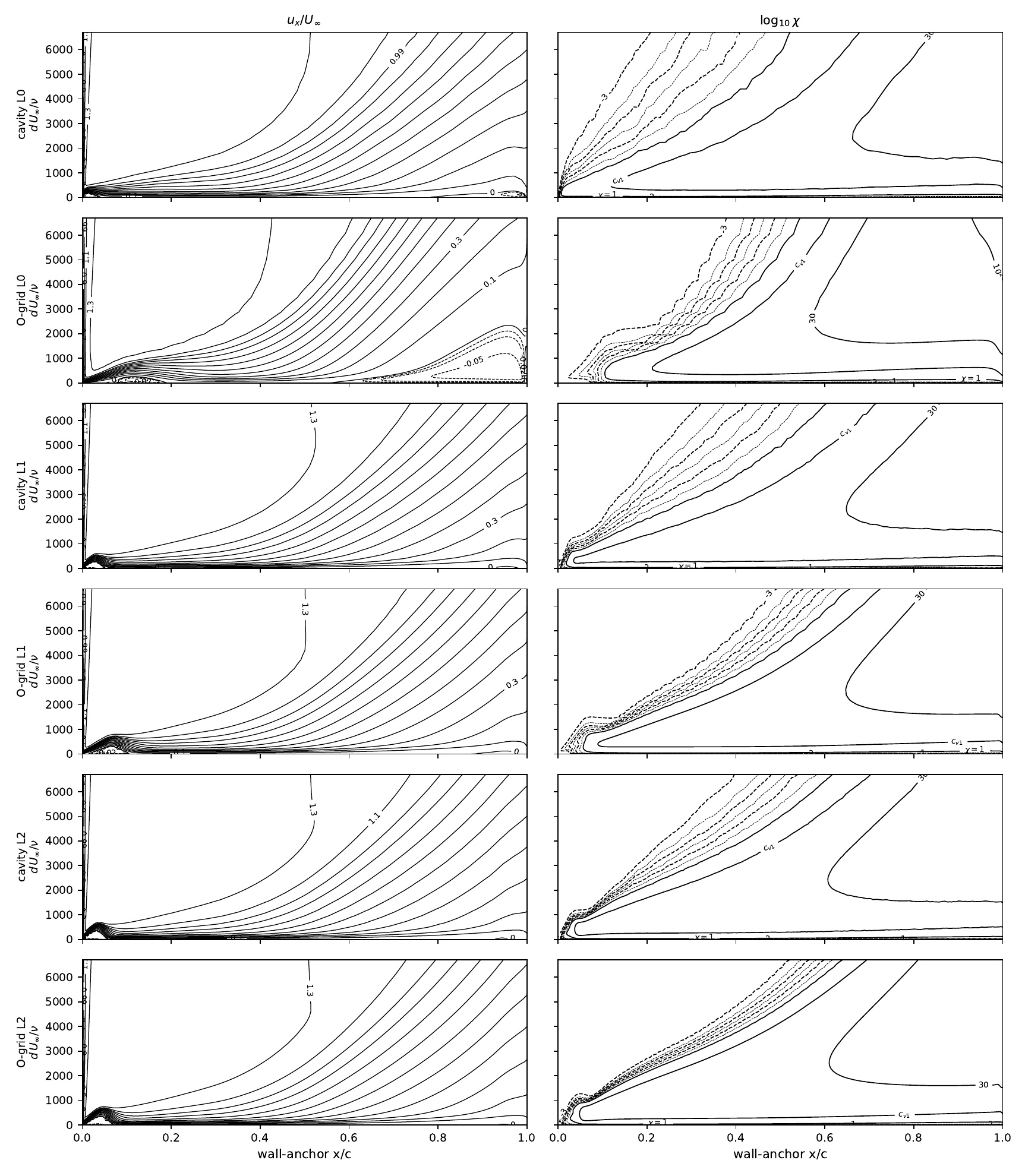}
\figcont{$\alpha=8.5^\circ$.}
\end{figure}
\begin{figure}[H]\centering
\includegraphics[width=0.99\textwidth,height=0.94\textheight,keepaspectratio]{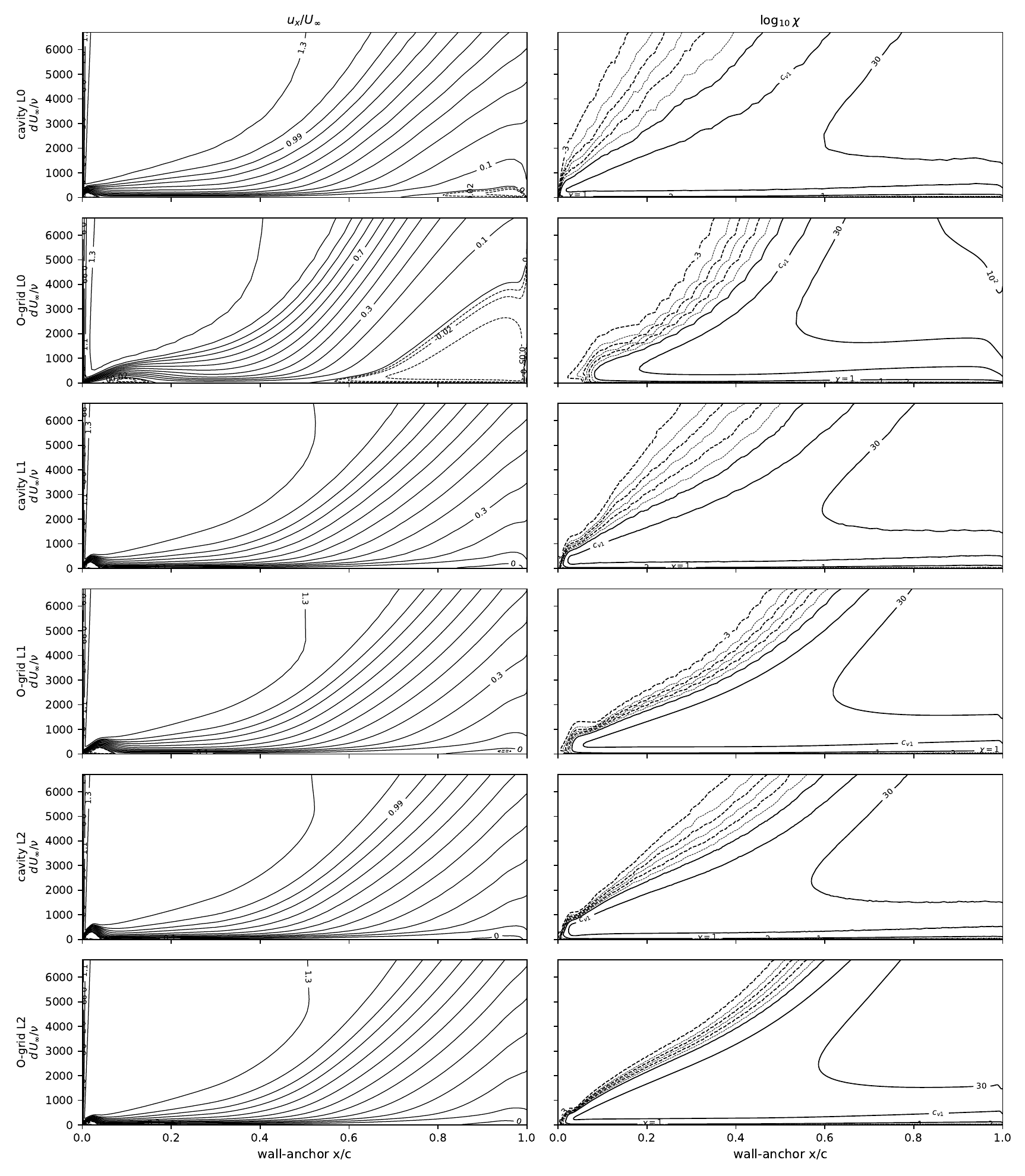}
\figcont{$\alpha=9^\circ$.}
\end{figure}

\subsection{Reynolds-number study at $\alpha=5^\circ$}

$Re\in\{6\times10^4,\,10^5,\,3\times10^5,\,4.6\times10^5\}$ on the
same grids (24 further solutions).
Figure~\ref{f:eppresweepforces}: the refinement fan closes onto the
measurement at $2$--$4.6\times10^5$ and opens at $10^5$, the model's
bursting boundary, one Reynolds step above the measurement's
($6\times10^4$).
Figure~\ref{f:eppresweeplow}: surface
suites. Table~\ref{t:eppresweep}: L2 coefficients.

\begin{figure}[H]\centering
\includegraphics[width=0.82\textwidth]{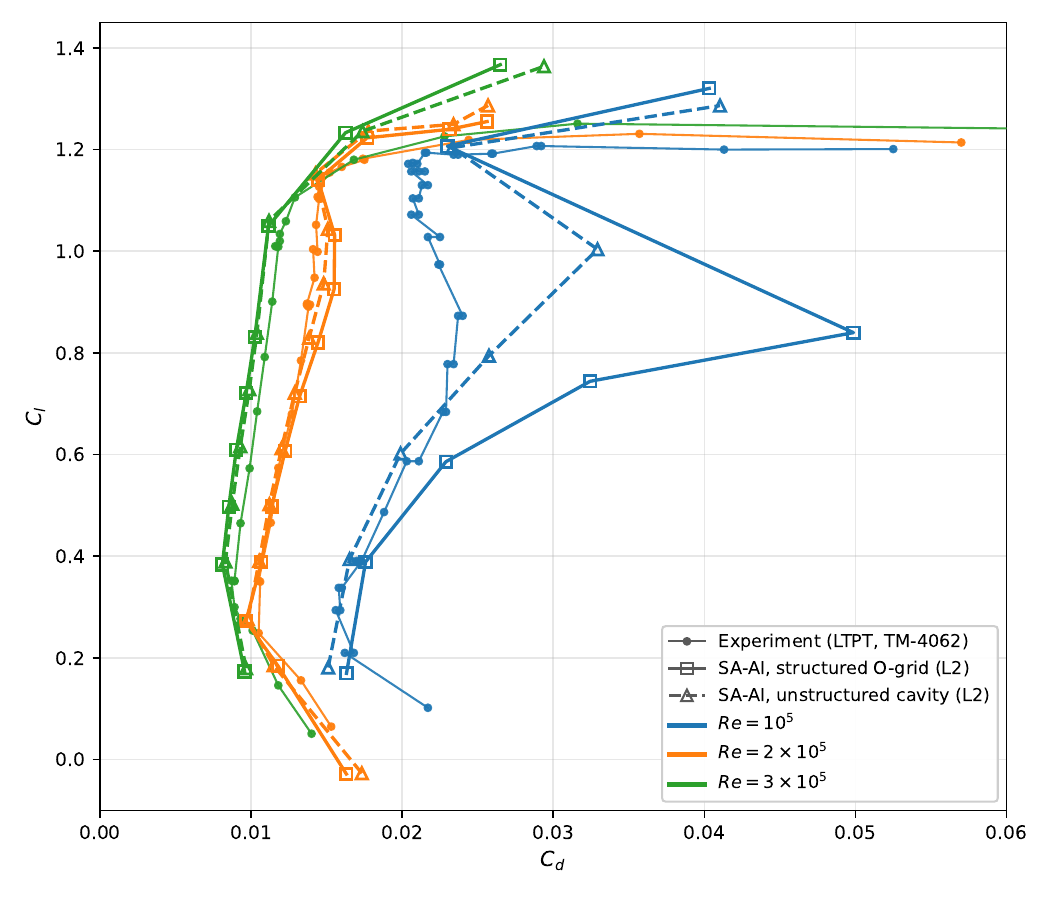}
\caption{Eppler 387 drag polar at $10^5$, $2\times10^5$ and $3\times10^5$,
L2 pair, against the measurement. Colour is Reynolds number. At $10^5$
the computed solution stalls by $\alpha=10^\circ$.}\label{f:epprepolar}
\end{figure}

\begin{figure}[H]\centering
\includegraphics[width=\textwidth]{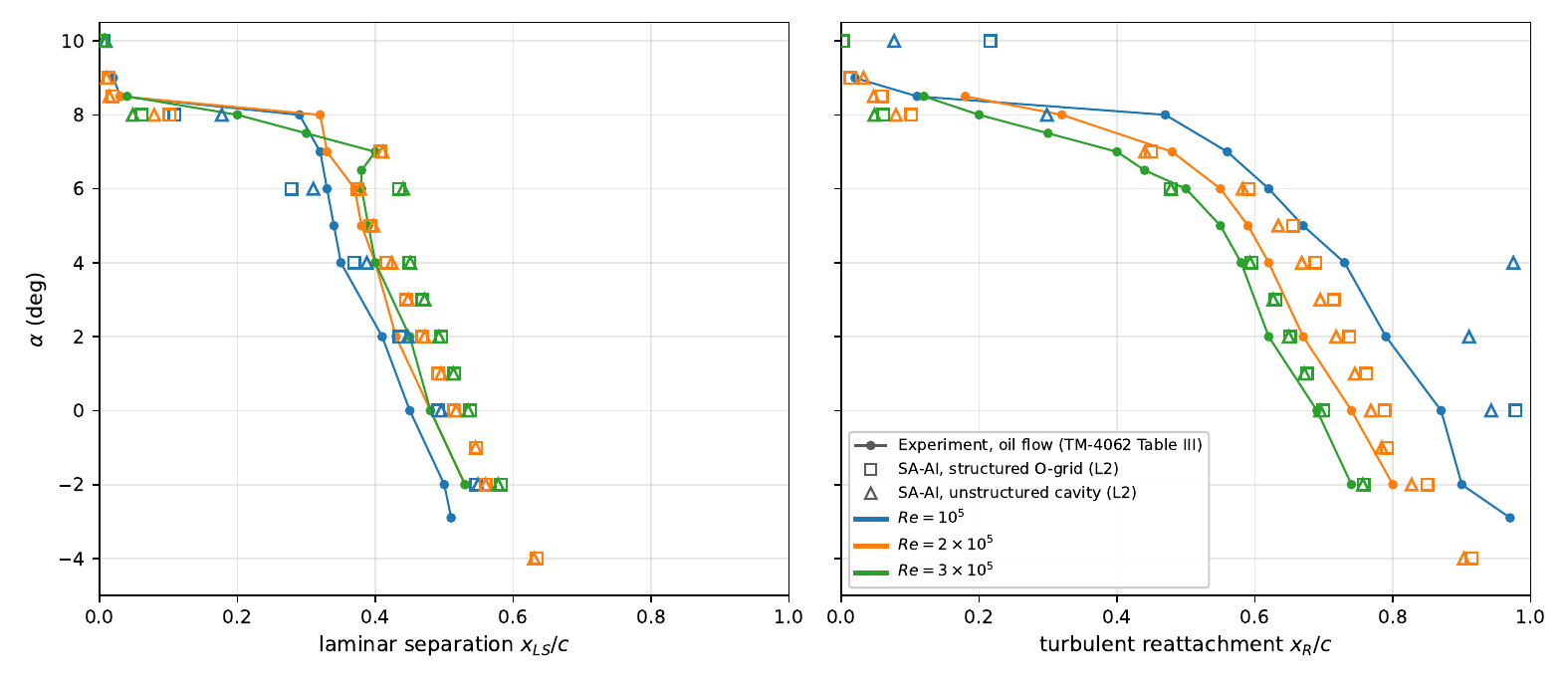}
\caption{Eppler 387 bubble stations at $10^5$, $2\times10^5$ and
$3\times10^5$, L2 pair, against the oil flow. Colour is Reynolds
number. Where there is no bubble---the oil flow reports natural
transition, or the computed surface stays attached---both stations lie
at the transition front, so the point sits at the same $x/c$ in each
panel.}\label{f:epprebubble}
\end{figure}

\begin{figure}[H]\centering
\includegraphics[width=\textwidth]{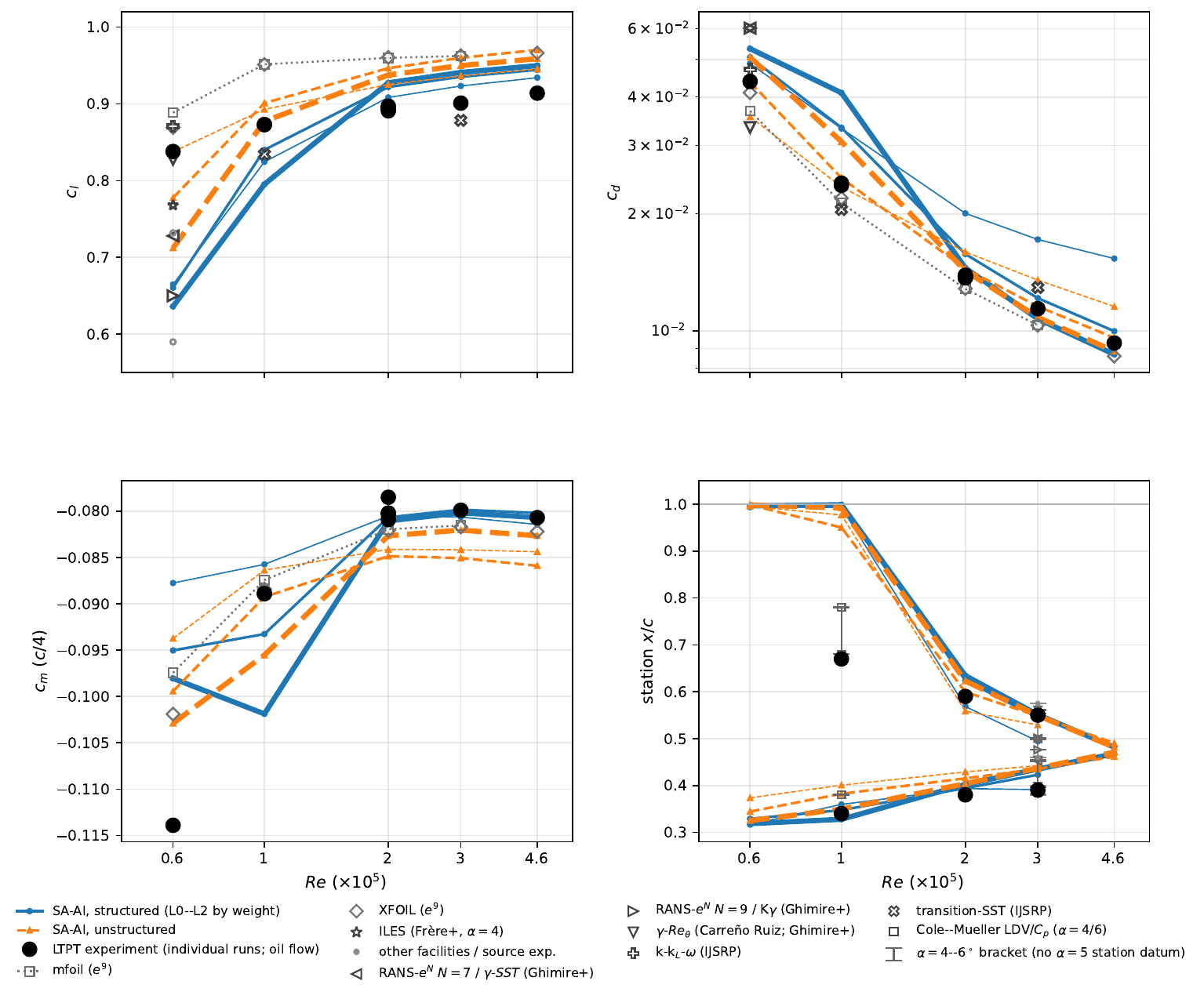}
\caption{Reynolds-sweep forces and stations, $\alpha=5^\circ$,
$Re=6\times10^4$--$4.6\times10^5$. LTPT runs $1$/$2$/$4$/$1$/$1$ at
$0.6$/$1$/$2$/$3$/$4.6\times10^5$.}\label{f:eppresweepforces}
\end{figure}

\begin{figure}[H]\centering
\includegraphics[width=\textwidth]{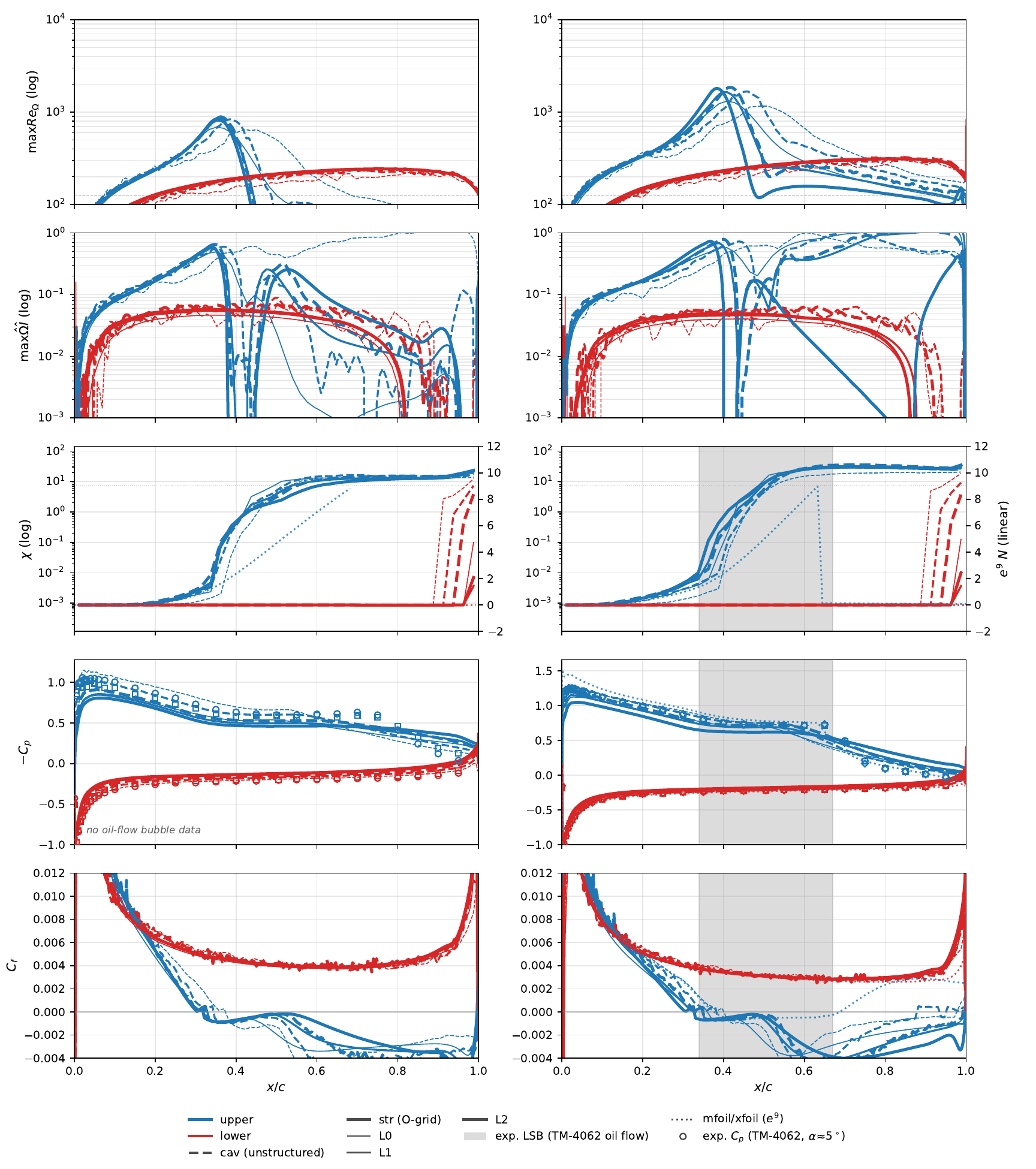}
\caption{Eppler~387 Reynolds-sweep surface distributions, $\alpha=5^\circ$:
$Re=6\times10^4$, $10^5$.}\label{f:eppresweeplow}
\end{figure}

Incidence sweeps at the two Reynolds numbers the literature samples
densely, on the L2 pair, same seed and constants (32 further
solutions): $\alpha=-2^\circ$--$10^\circ$ at $10^5$
(Table~\ref{t:extepp100k}; Cole and Mueller's eleven measured bubble
stations, Shahjahan's $\gamma$--$Re_\theta$ and SA-BC fronts) and
$\alpha=-2^\circ$--$10^\circ$ at $3\times10^5$
(Table~\ref{t:extepp300k}; Ghimire's three OpenFOAM transitional models
at $2^\circ$, $4^\circ$, $6^\circ$, Cole and Mueller at $-2^\circ$,
$0^\circ$, $1^\circ$, $3^\circ$). At $3\times10^5$ the transition
station is $0.579$/$0.532$/$0.443\,c$ at $\alpha=2^\circ$/$4^\circ$/$6^\circ$
against a measured $0.66$/$0.55$/$0.46\,c$ and a three-model spread of
$0.592$--$0.646$/$0.490$--$0.542$/$0.470$--$0.479\,c$; separation lands
within $0.06\,c$ and reattachment within $0.07\,c$ of Cole and Mueller. At $10^5$ the
front is $0.07$--$0.12\,c$ ahead of the measurement at every incidence
and the bubble closes late or not at all
($0.978\,c$ at $0^\circ$ against $0.90\,c$; no closure at $6^\circ$
against $0.68\,c$), with $c_d$ reaching $0.0499$ at $6^\circ$.
$6\times10^4$ and $4.6\times10^5$ keep their single $\alpha=5^\circ$
stations.
Figures~\ref{f:epprepolar} and~\ref{f:epprebubble}: polar and
bubble stations at the three Reynolds numbers against the
measurement alone. Figures~\ref{f:eppcfRe100k}
and~\ref{f:eppcfRe300k}: surface suites, L2 pair.
Figures~\ref{f:eppsheetRe100k} and~\ref{f:eppsheetRe300k}:
wall-normal sheets.

\begin{table}[H]
  \centering\small
  \caption{Eppler 387 at $Re=3\times10^5$, incidence sweep on the L2 pair. $x_\mathrm{tr}$ is the near-wall $\chi\!=\!1$ front; $x_\mathrm{sep}$ and $x_\mathrm{reatt}$ are the signed-$C_f$ crossings, with reattachment required to lie downstream of separation; ``--'' means the upper surface carries no closed bubble ahead of the trailing edge.}
  \label{t:extepp300k}
  \begin{tabular}{c ccccc ccccc}
    \toprule
    & \multicolumn{5}{c}{structured} & \multicolumn{5}{c}{cavity} \\
    \cmidrule(lr){2-6}\cmidrule(lr){7-11}
    $\alpha$ & $C_L$ & $C_D$ & $x_\mathrm{tr}$ & $x_\mathrm{sep}$ & $x_\mathrm{reatt}$
             & $C_L$ & $C_D$ & $x_\mathrm{tr}$ & $x_\mathrm{sep}$ & $x_\mathrm{reatt}$ \\
    \midrule
    $-2$ & 0.1733 & 0.00957 & 0.698 & 0.582 & 0.758 & 0.1797 & 0.00966 & 0.699 & 0.578 & 0.756 \\
    $0$ & 0.3840 & 0.00810 & 0.632 & 0.537 & 0.699 & 0.3896 & 0.00830 & 0.629 & 0.535 & 0.695 \\
    $1$ & 0.4974 & 0.00852 & 0.607 & 0.514 & 0.675 & 0.5035 & 0.00876 & 0.602 & 0.513 & 0.672 \\
    $2$ & 0.6098 & 0.00902 & 0.579 & 0.496 & 0.651 & 0.6164 & 0.00929 & 0.579 & 0.492 & 0.649 \\
    $3$ & 0.7212 & 0.00967 & 0.561 & 0.468 & 0.630 & 0.7284 & 0.00988 & 0.558 & 0.471 & 0.627 \\
    $4$ & 0.8315 & 0.01024 & 0.532 & 0.449 & 0.596 & 0.8394 & 0.01039 & 0.531 & 0.451 & 0.593 \\
    $6$ & 1.0496 & 0.01115 & 0.443 & 0.435 & 0.477 & 1.0596 & 0.01118 & 0.444 & 0.440 & 0.478 \\
    $8$ & 1.2333 & 0.01625 & 0.070 & -- & -- & 1.2359 & 0.01737 & 0.055 & -- & -- \\
    $10$ & 1.3670 & 0.02649 & 0.005 & -- & -- & 1.3638 & 0.02938 & 0.003 & 0.006 & 0.987 \\
    \bottomrule
  \end{tabular}
\end{table}

\begin{table}[H]
  \centering\small
  \caption{Eppler 387 at $Re=10^5$, incidence sweep on the L2 pair, columns as in Table~\ref{t:extepp300k}. The structured $\alpha\!=\!8^\circ$ entry is the one unsteady solution of the matrix: its front cycles over $0.24$--$0.57\,c$ without settling through $8\!\times\!10^4$ pseudo-steps and the final value is tabulated; its cavity counterpart is steady at $0.279\,c$.}
  \label{t:extepp100k}
  \begin{tabular}{c ccccc ccccc}
    \toprule
    & \multicolumn{5}{c}{structured} & \multicolumn{5}{c}{cavity} \\
    \cmidrule(lr){2-6}\cmidrule(lr){7-11}
    $\alpha$ & $C_L$ & $C_D$ & $x_\mathrm{tr}$ & $x_\mathrm{sep}$ & $x_\mathrm{reatt}$
             & $C_L$ & $C_D$ & $x_\mathrm{tr}$ & $x_\mathrm{sep}$ & $x_\mathrm{reatt}$ \\
    \midrule
    $-2$ & 0.1698 & 0.01631 & 0.830 & 0.546 & -- & 0.1813 & 0.01510 & 0.821 & 0.549 & 0.987 \\
    $0$ & 0.3888 & 0.01758 & 0.750 & 0.491 & 0.978 & 0.3944 & 0.01652 & 0.731 & 0.496 & 0.943 \\
    $2$ & 0.5860 & 0.02289 & 0.677 & 0.435 & -- & 0.6021 & 0.01989 & 0.661 & 0.447 & 0.910 \\
    $4$ & 0.7442 & 0.03244 & 0.590 & 0.369 & -- & 0.7944 & 0.02574 & 0.587 & 0.388 & 0.975 \\
    $6$ & 0.8399 & 0.04986 & 0.482 & 0.278 & -- & 1.0039 & 0.03293 & 0.523 & 0.310 & -- \\
    $8$ & 1.2067 & 0.02299 & 0.571 & 0.109 & -- & 1.2039 & 0.02332 & 0.279 & 0.177 & 0.990 \\
    $10$ & 1.3204 & 0.04032 & 0.018 & 0.006 & 0.216 & 1.2865 & 0.04103 & 0.025 & 0.009 & 0.077 \\
    \bottomrule
  \end{tabular}
\end{table}

\begin{figure}[H]\centering
\includegraphics[width=\textwidth]{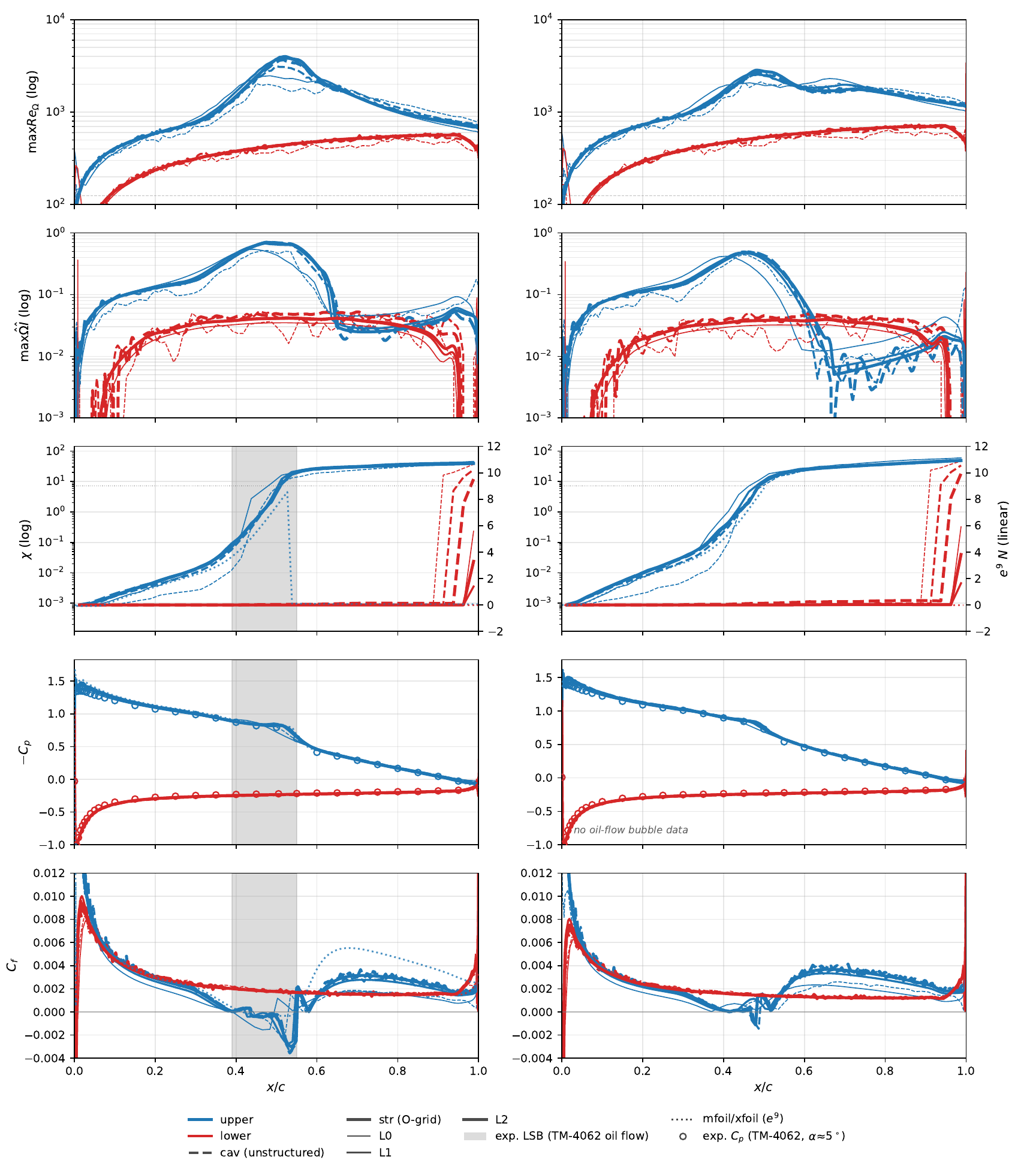}
\figcont{$Re=3$, $4.6\times10^5$.}
\end{figure}

\begin{table}[H]
  \centering\footnotesize
  \setlength{\tabcolsep}{3.5pt}
  \caption{Eppler 387 at $\alpha\!=\!5^\circ$ across the Reynolds
  sweep: L2 section coefficients against the $e^9$ panel reference and
  the LTPT measurement (read at the tabulated angle nearest $5^\circ$).
  Computed moments transferred to the quarter chord.}
  \label{t:eppresweep}
  \begin{tabular}{l ccc ccc ccc ccc}
    \toprule
    & \multicolumn{3}{c}{SA-AI, structured L2} & \multicolumn{3}{c}{SA-AI, unstructured L2}
    & \multicolumn{3}{c}{$e^9$ panel} & \multicolumn{3}{c}{Exp (LTPT)}\\
    \cmidrule(lr){2-4}\cmidrule(lr){5-7}\cmidrule(lr){8-10}\cmidrule(lr){11-13}
    $Re$ & $c_l$ & $c_d$ & $c_m$ & $c_l$ & $c_d$ & $c_m$ & $c_l$ & $c_d$ & $c_m$ & $c_l$ & $c_d$ & $c_m$ \\
    \midrule
    $6\!\times\!10^4$   & 0.636 & 0.0533 & $-$0.098 & 0.712 & 0.0506 & $-$0.103 & 0.869$^\ddagger$ & 0.0411$^\ddagger$ & $-$0.102$^\ddagger$ & 0.838 & 0.0439 & $-$0.114 \\
    $1\!\times\!10^5$   & 0.795 & 0.0410 & $-$0.102 & 0.877 & 0.0307 & $-$0.096 & 0.952 & 0.0213 & $-$0.087 & 0.873 & 0.0237 & $-$0.089 \\
    $2\!\times\!10^5$   & 0.927 & 0.0145 & $-$0.081 & 0.938 & 0.0143 & $-$0.083 & 0.960 & 0.0128 & $-$0.082 & 0.891 & 0.0138 & $-$0.081 \\
    $3\!\times\!10^5$   & 0.941 & 0.0107 & $-$0.080 & 0.950 & 0.0108 & $-$0.082 & 0.962 & 0.0103 & $-$0.082 & 0.901 & 0.0114 & $-$0.080 \\
    $4.6\!\times\!10^5$ & 0.950 & 0.0087 & $-$0.081 & 0.959 & 0.0089 & $-$0.083 & 0.966 & 0.0086 & $-$0.082 & 0.914 & 0.0093 & $-$0.081 \\
    \bottomrule
  \end{tabular}
  \\[2pt]{\footnotesize $^\ddagger$XFOIL, whose forces sit closest to
  the measurement, is the reference at this bistable condition; mfoil
  gives $c_l\!=\!0.889$, $c_d\!=\!0.0368$, $c_m\!=\!-0.097$, and
  FlexFoil sides with XFOIL ($c_l\!=\!0.868$, $c_d\!=\!0.0406$,
  $c_m\!=\!-0.102$).}
\end{table}

\clearpage
\begin{figure}[H]\centering
\includegraphics[width=0.99\textwidth,height=0.94\textheight,keepaspectratio]{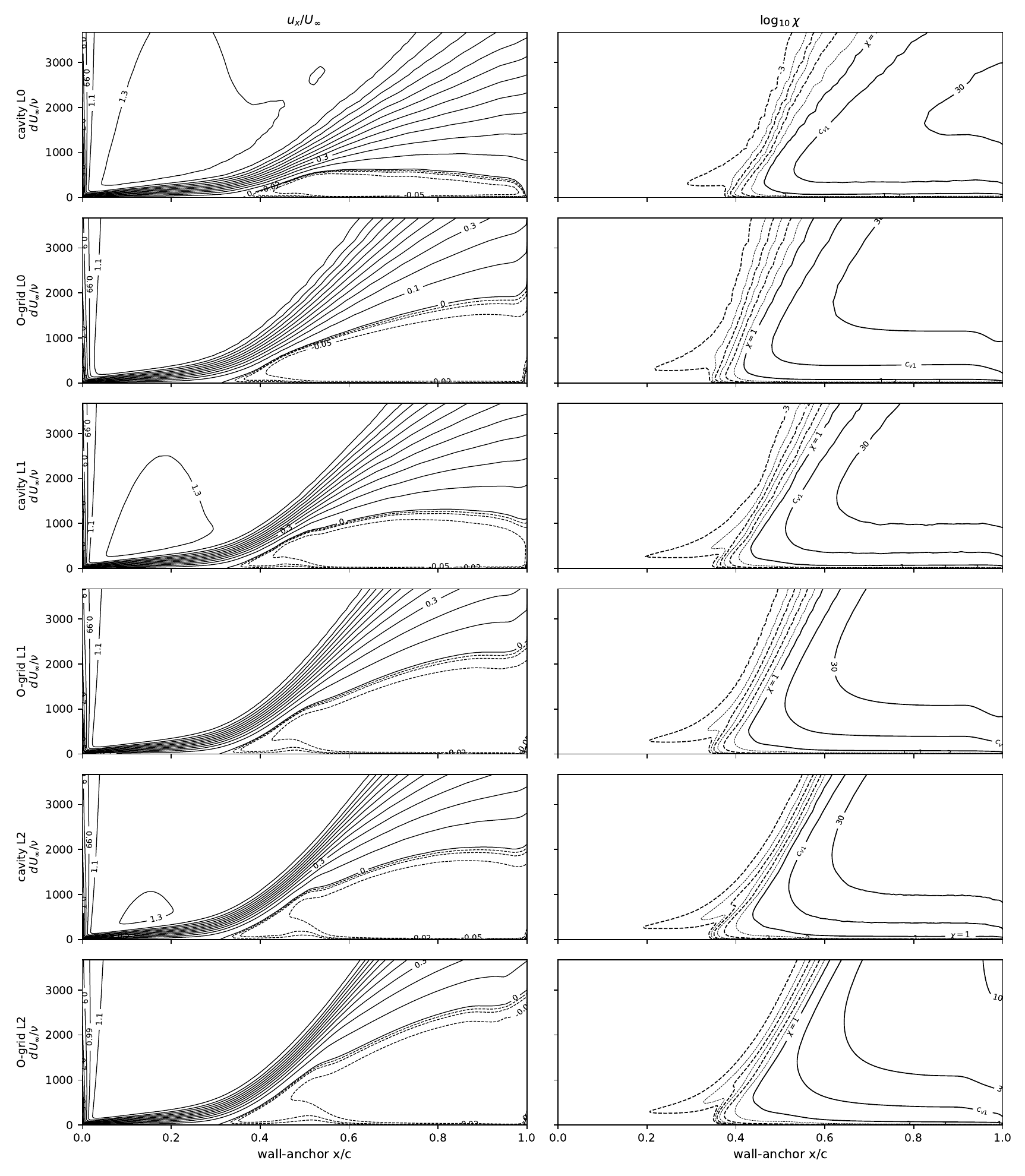}
\caption{Eppler~387 wall-normal sheets at $\alpha=5^\circ$, upper surface:
$u_x/U_\infty$ (left) and $\log_{10}\chi$ (right) on all six grids.
$Re=6\times10^4$.}\label{f:eppresweepsheets}
\end{figure}
\begin{figure}[H]\centering
\includegraphics[width=0.99\textwidth,height=0.94\textheight,keepaspectratio]{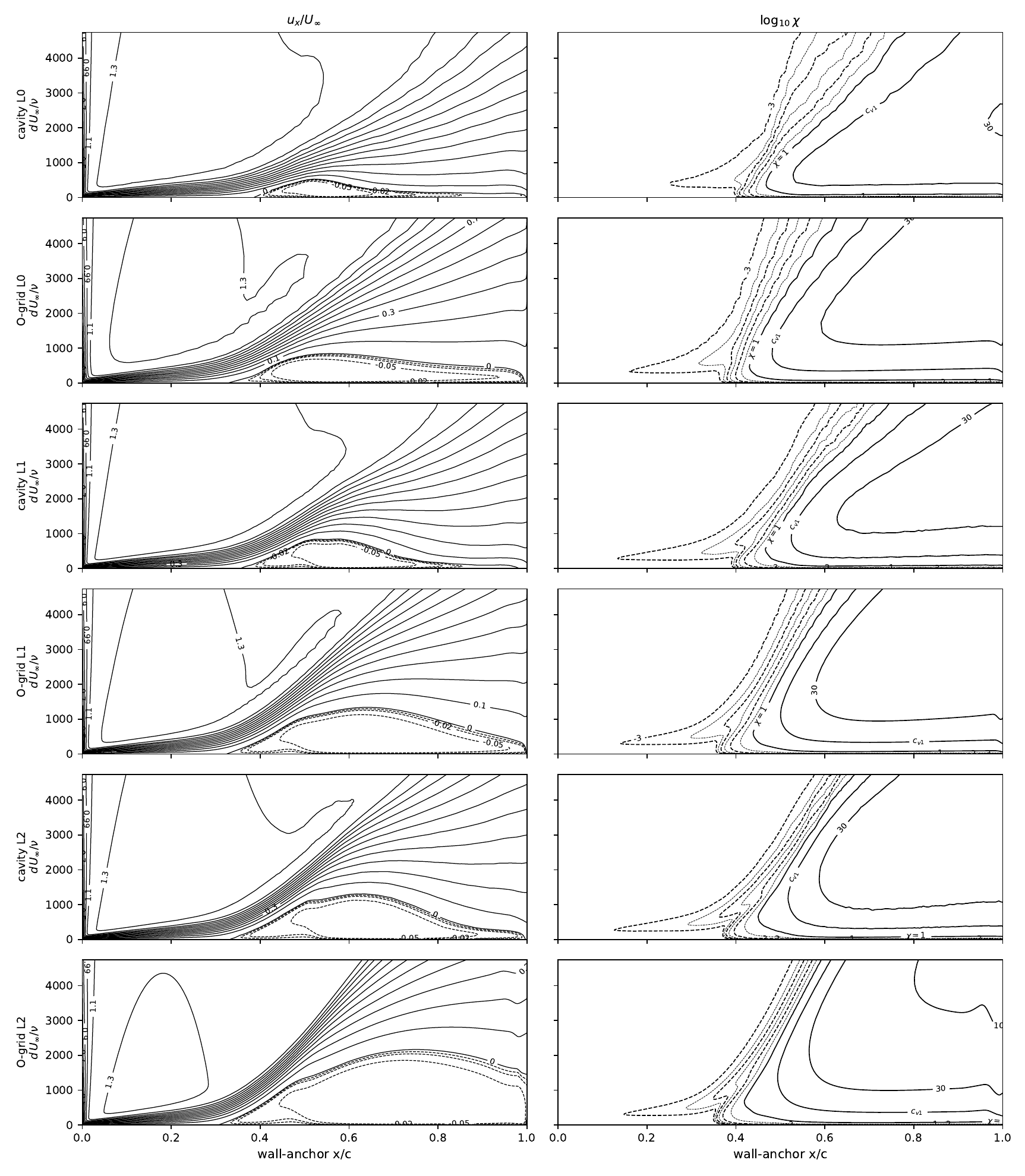}
\figcont{$Re=10^5$.}
\end{figure}
\begin{figure}[H]\centering
\includegraphics[width=0.99\textwidth,height=0.94\textheight,keepaspectratio]{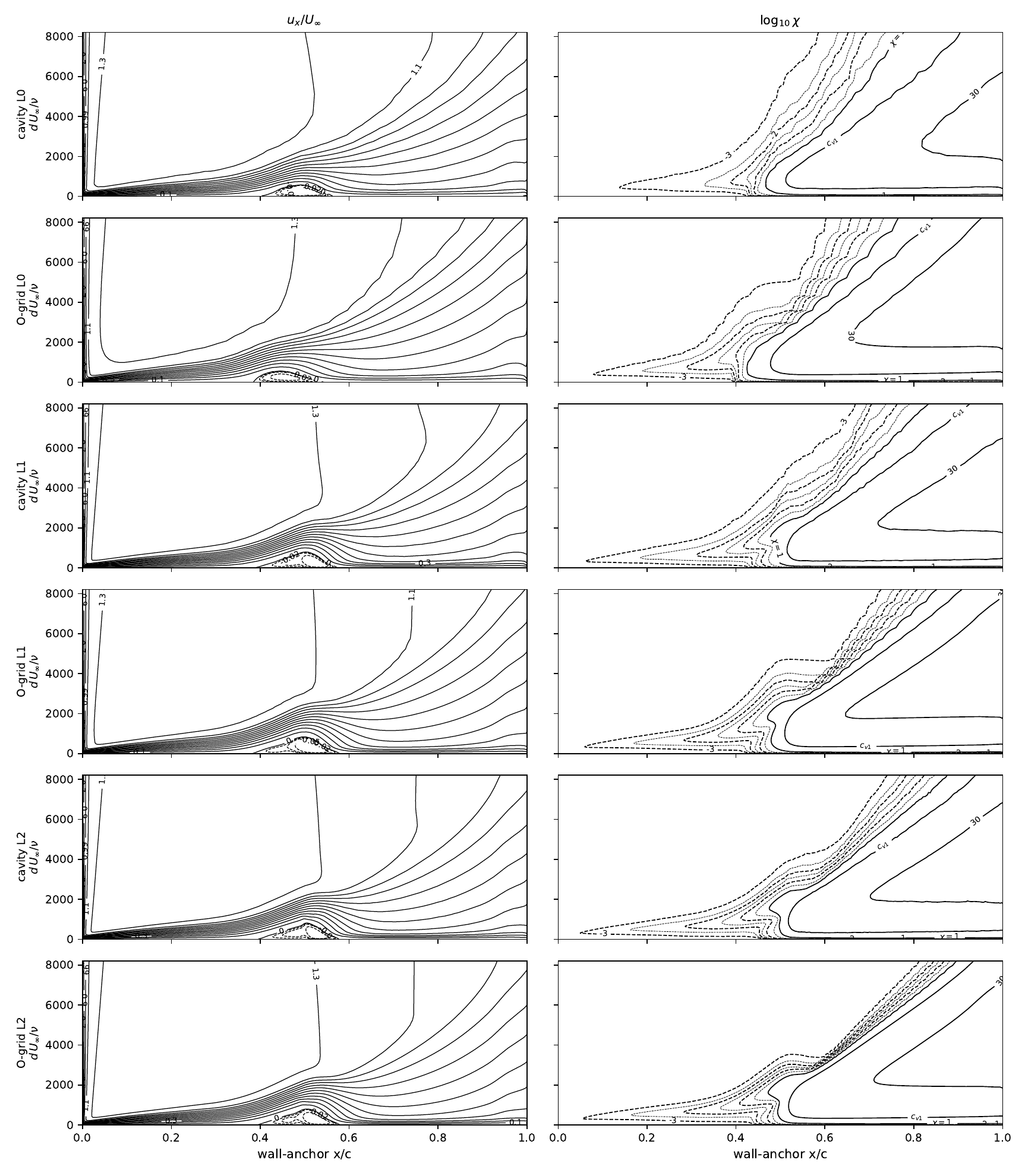}
\figcont{$Re=3\times10^5$.}
\end{figure}
\begin{figure}[H]\centering
\includegraphics[width=0.99\textwidth,height=0.94\textheight,keepaspectratio]{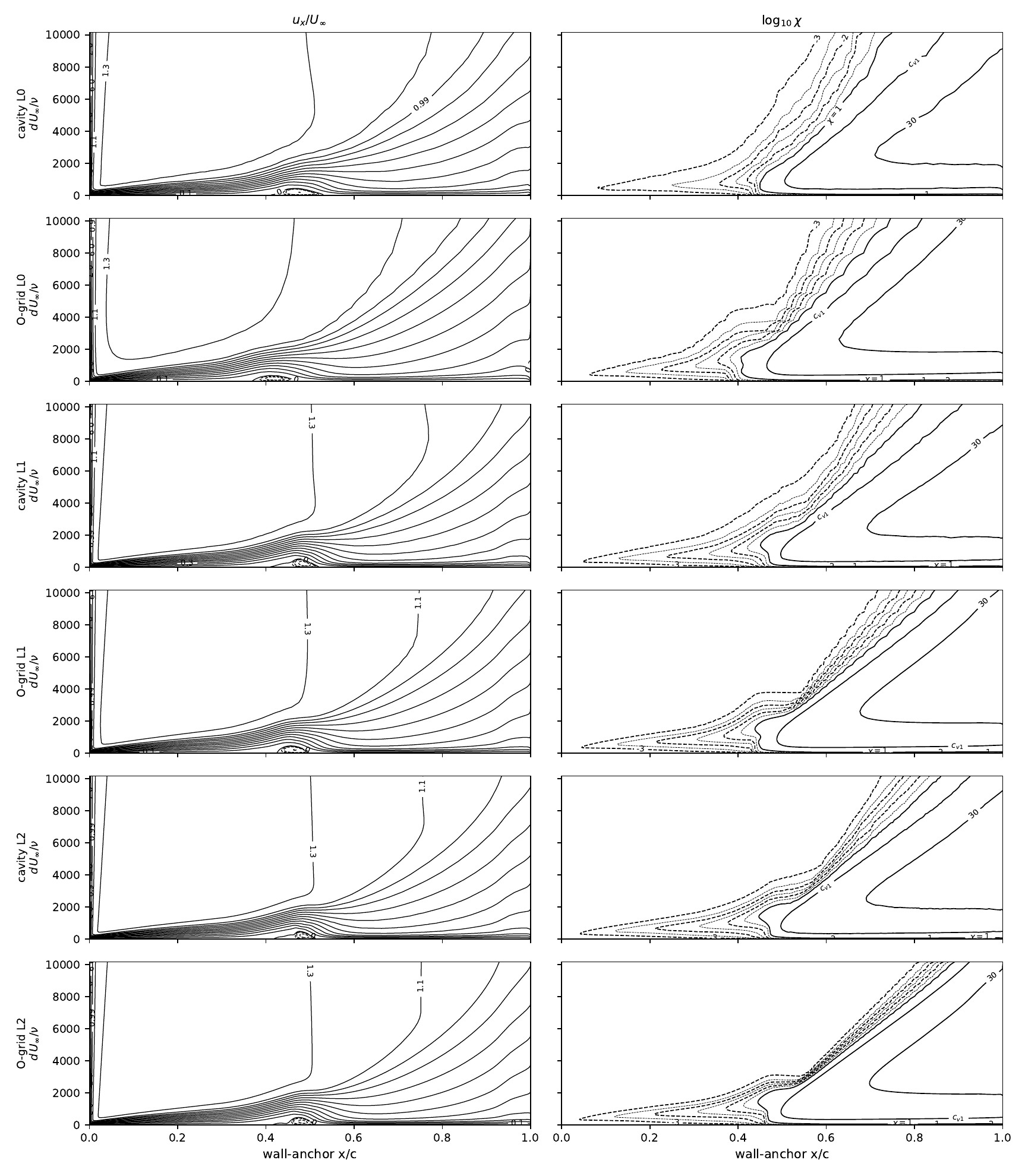}
\figcont{$Re=4.6\times10^5$.}
\end{figure}

\clearpage
\begin{figure}[H]\centering
\includegraphics[width=0.99\textwidth,height=0.94\textheight,keepaspectratio]{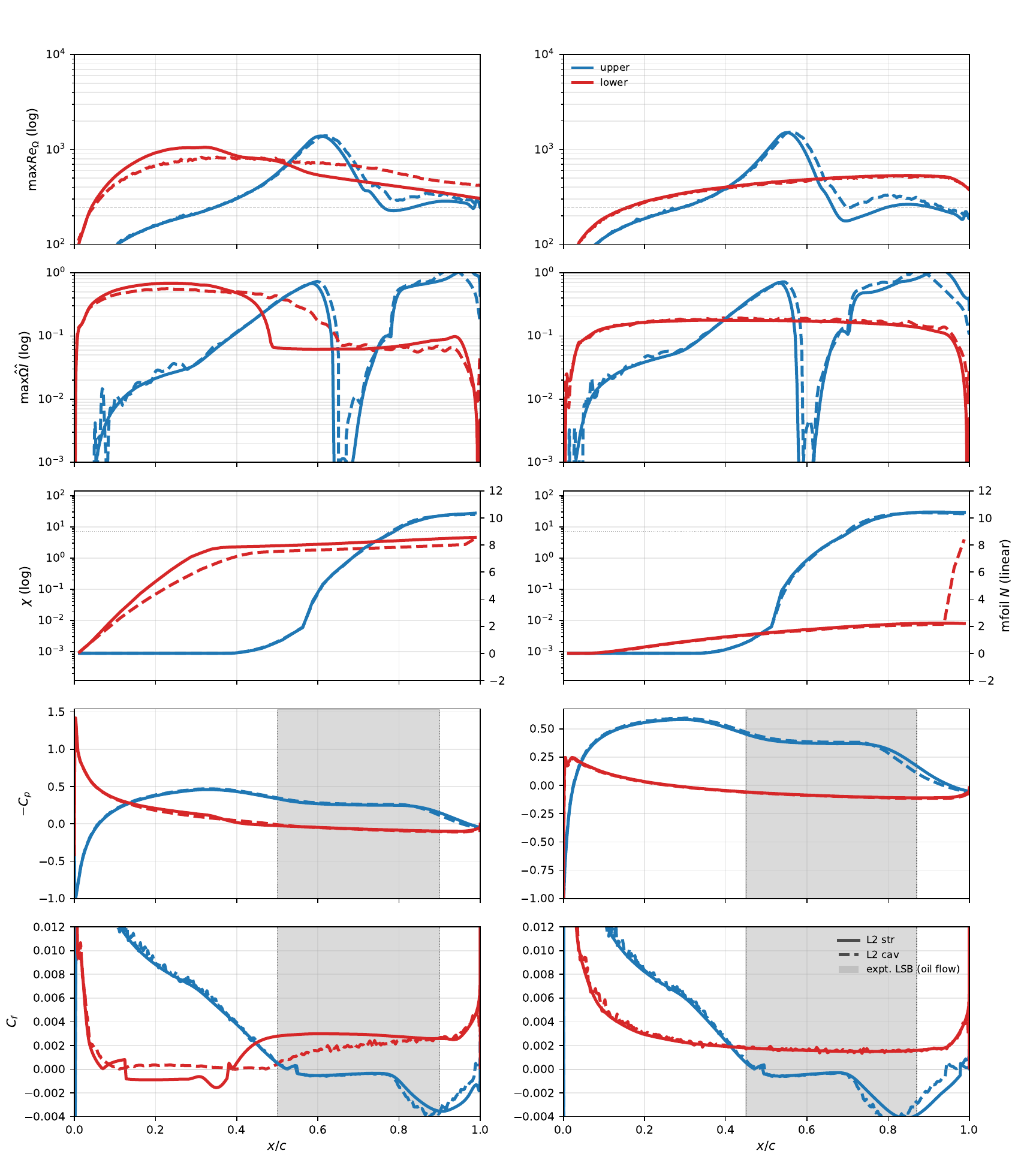}
\caption{Eppler~387 surface distributions, $Re=10^5$, L2 pair; pages run in ascending incidence. $\alpha=-2^\circ$, $\alpha=0^\circ$.}\label{f:eppcfRe100k}
\end{figure}
\begin{figure}[H]\centering
\includegraphics[width=0.99\textwidth,height=0.94\textheight,keepaspectratio]{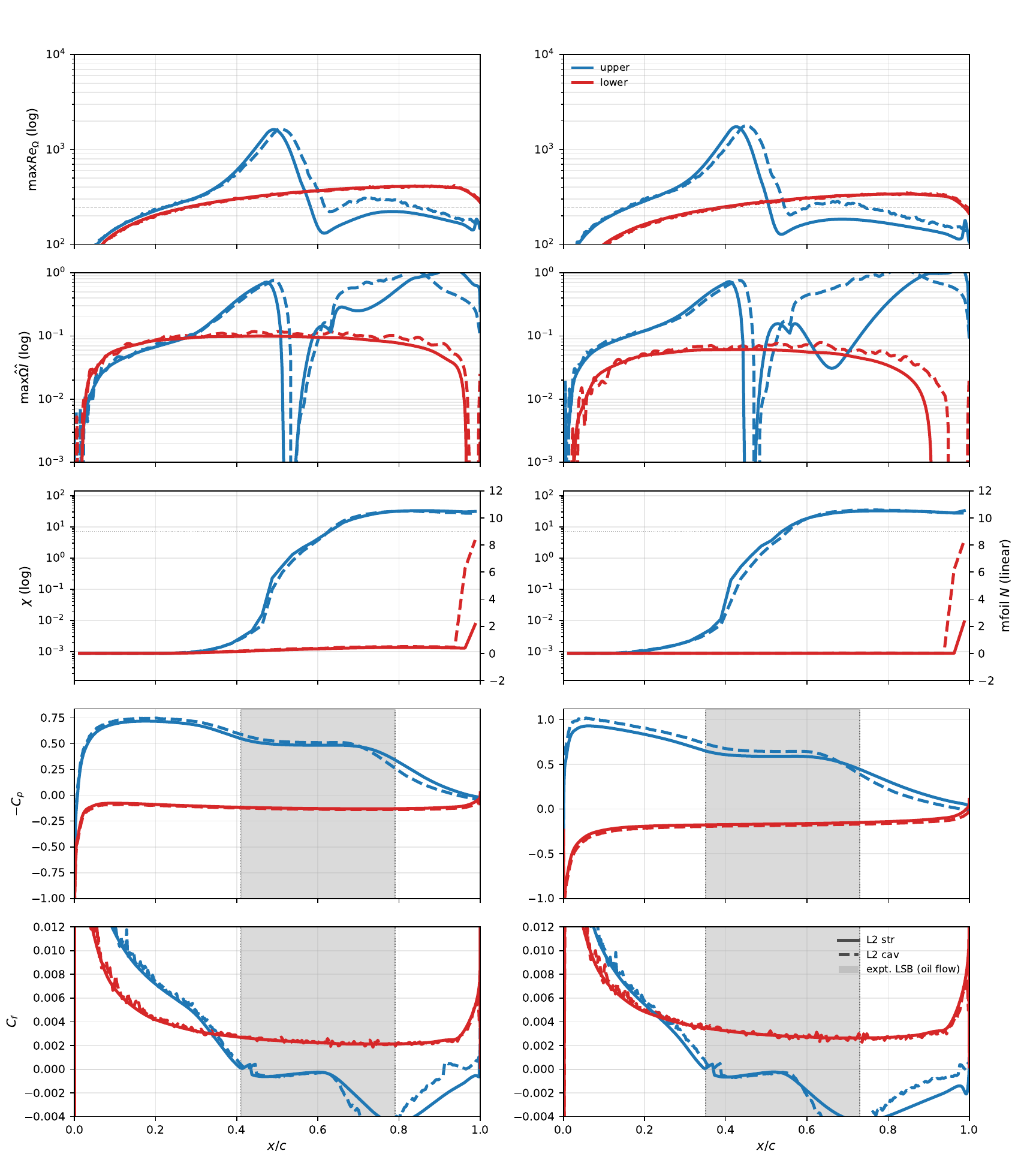}
\figcont{$\alpha=2^\circ$, $\alpha=4^\circ$.}
\end{figure}
\begin{figure}[H]\centering
\includegraphics[width=0.99\textwidth,height=0.94\textheight,keepaspectratio]{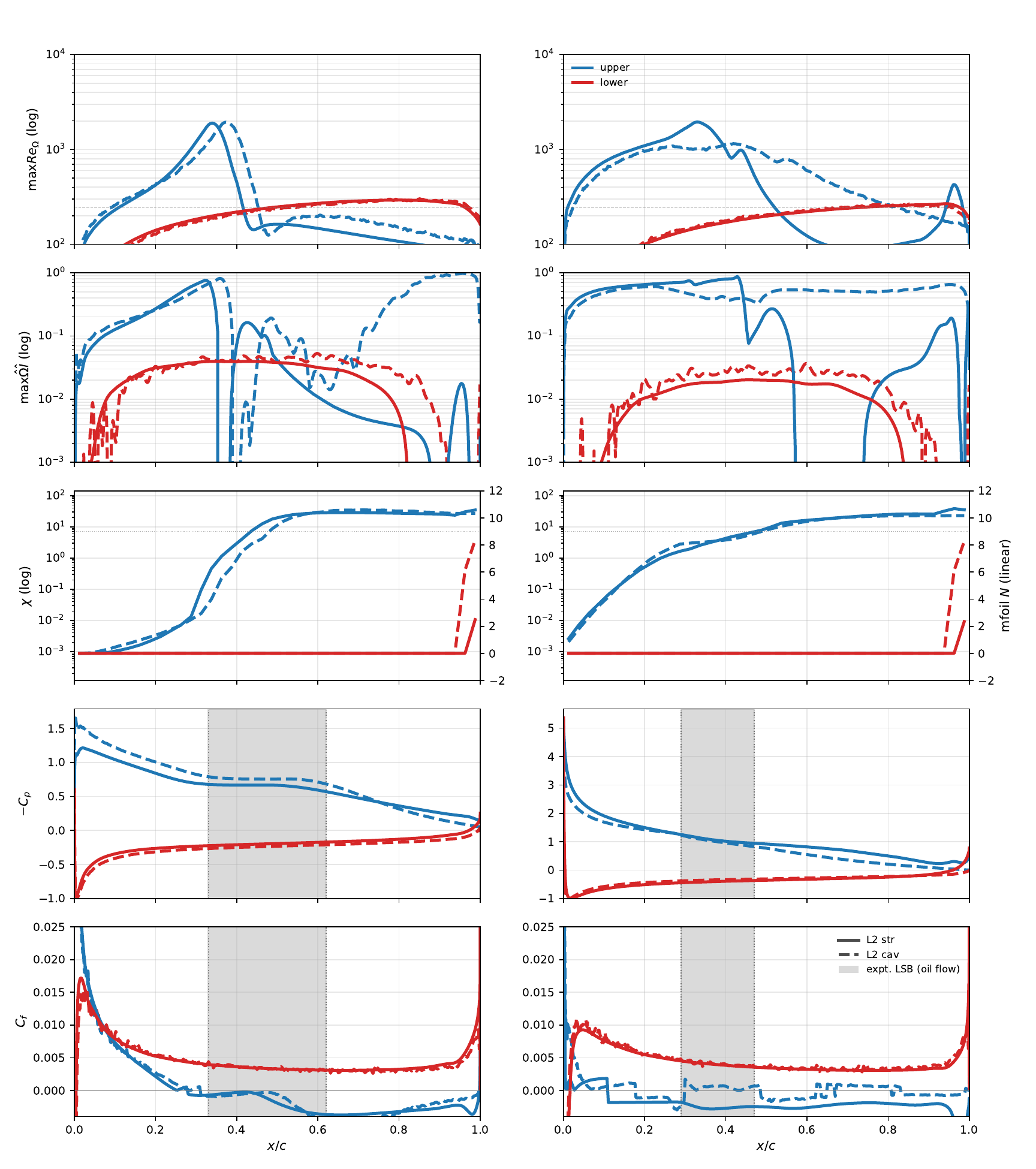}
\figcont{$\alpha=6^\circ$, $\alpha=8^\circ$.}
\end{figure}
\begin{figure}[H]\centering
\includegraphics[width=0.99\textwidth,height=0.94\textheight,keepaspectratio]{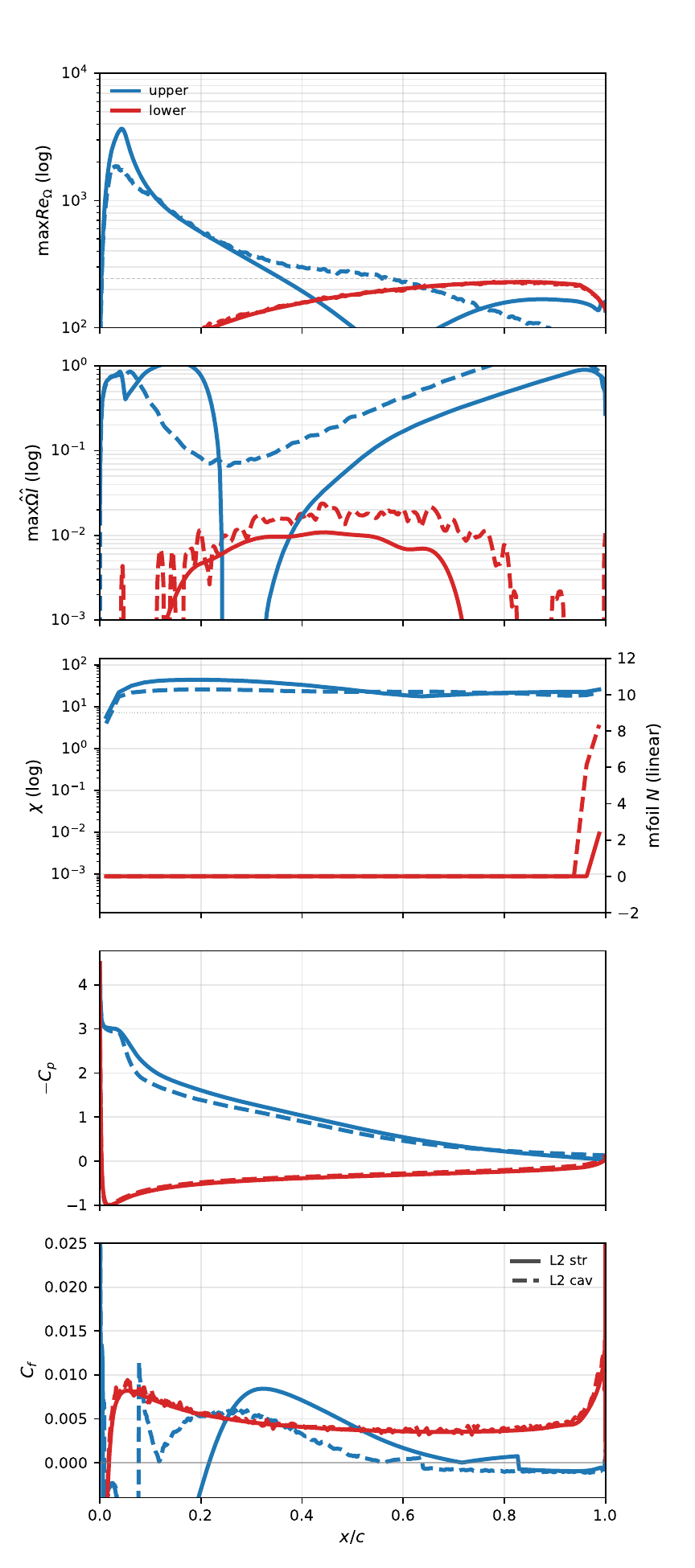}
\figcont{$\alpha=10^\circ$.}
\end{figure}
\begin{figure}[H]\centering
\includegraphics[width=0.99\textwidth,height=0.94\textheight,keepaspectratio]{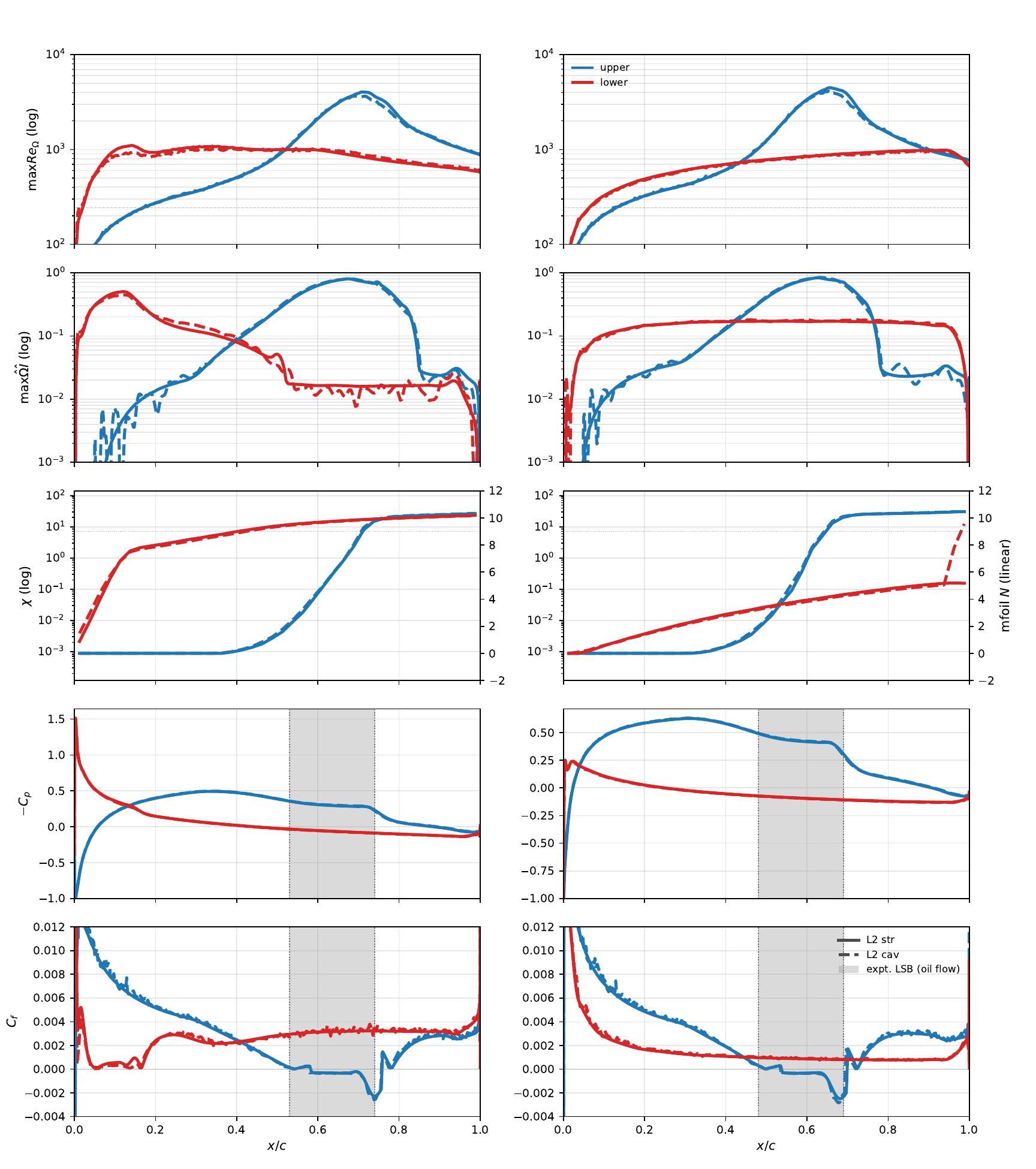}
\caption{Eppler~387 surface distributions, $Re=3\times10^5$, L2 pair; pages run in ascending incidence. $\alpha=-2^\circ$, $\alpha=0^\circ$.}\label{f:eppcfRe300k}
\end{figure}
\begin{figure}[H]\centering
\includegraphics[width=0.99\textwidth,height=0.94\textheight,keepaspectratio]{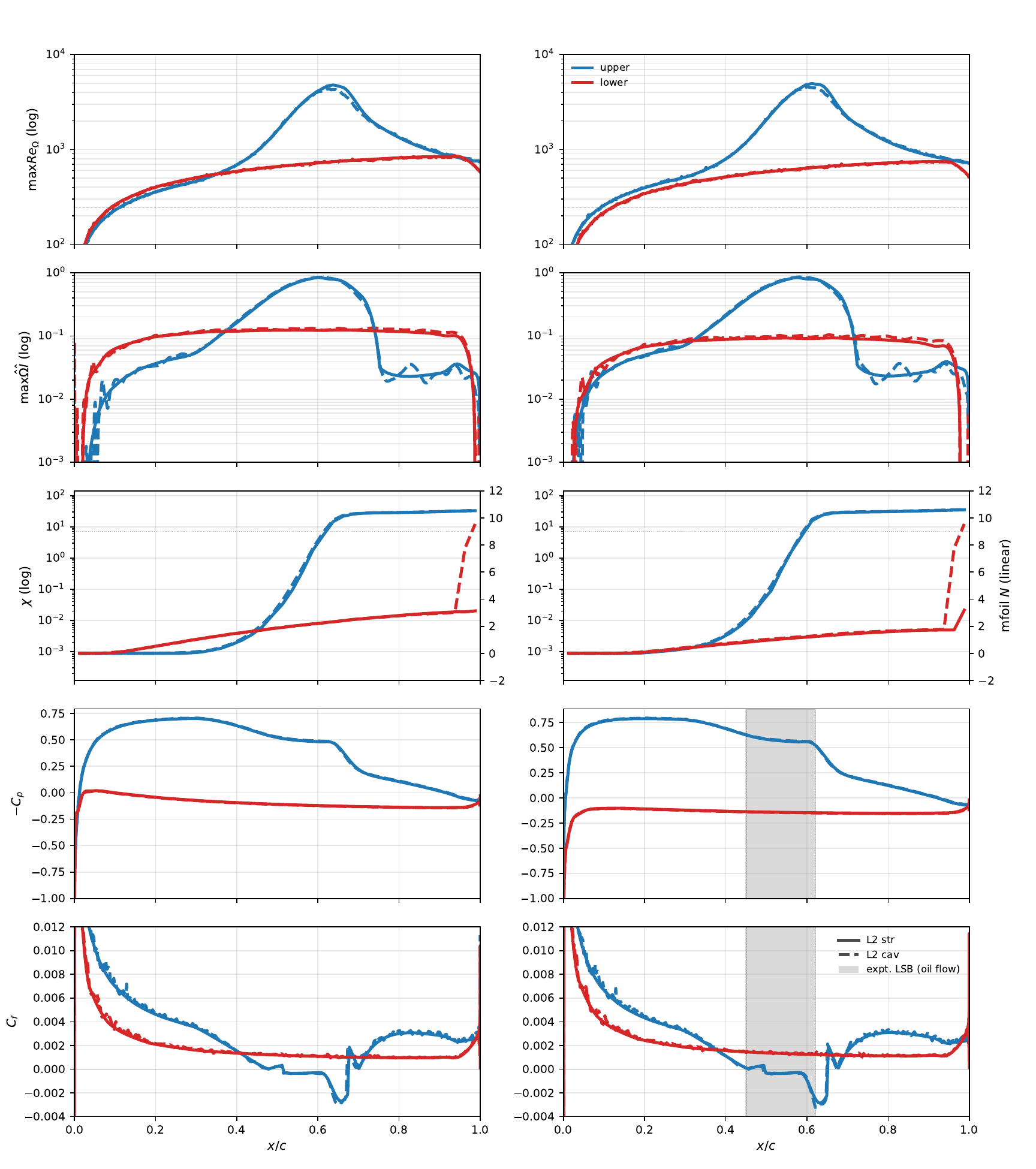}
\figcont{$\alpha=1^\circ$, $\alpha=2^\circ$.}
\end{figure}
\begin{figure}[H]\centering
\includegraphics[width=0.99\textwidth,height=0.94\textheight,keepaspectratio]{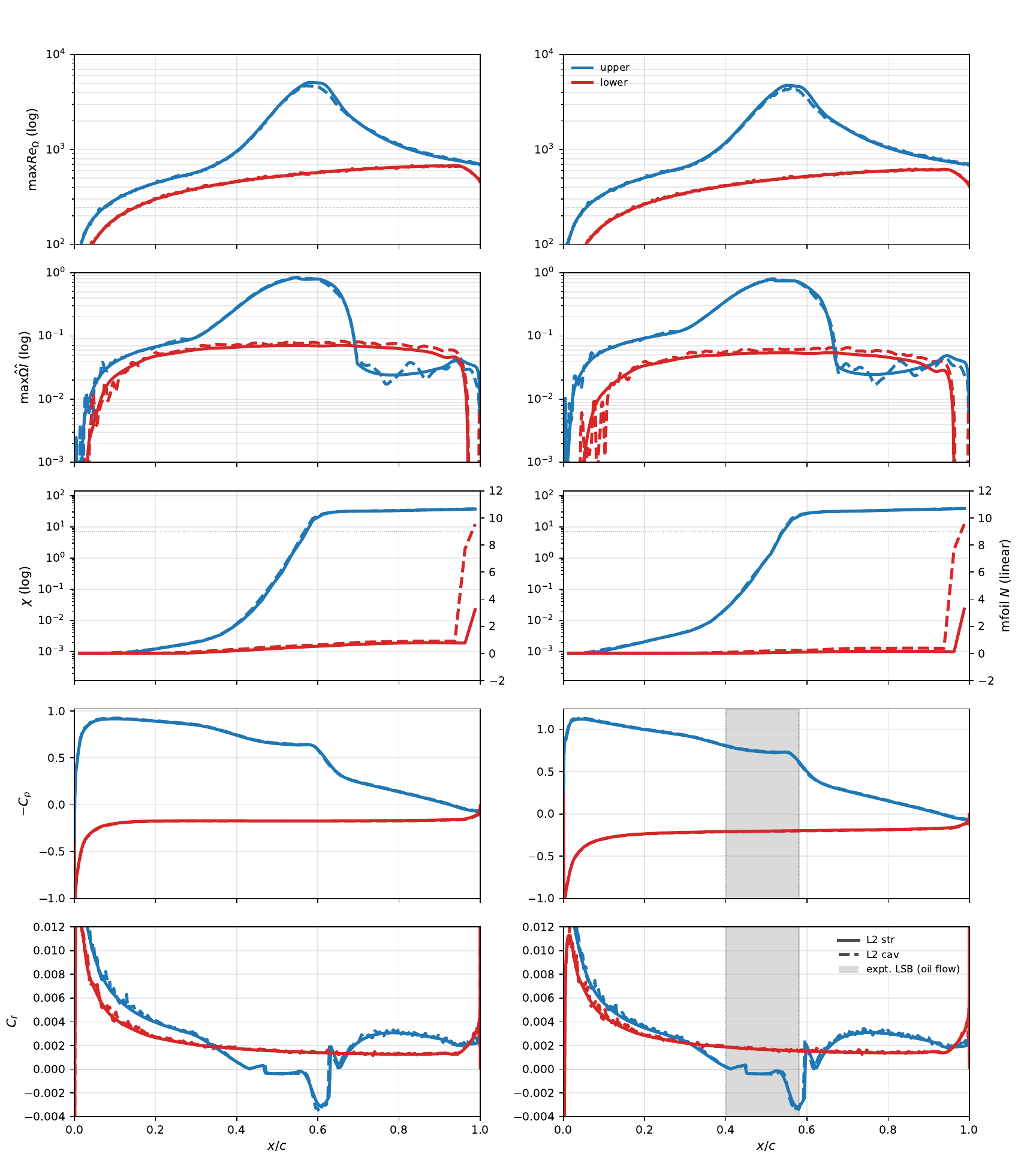}
\figcont{$\alpha=3^\circ$, $\alpha=4^\circ$.}
\end{figure}
\begin{figure}[H]\centering
\includegraphics[width=0.99\textwidth,height=0.94\textheight,keepaspectratio]{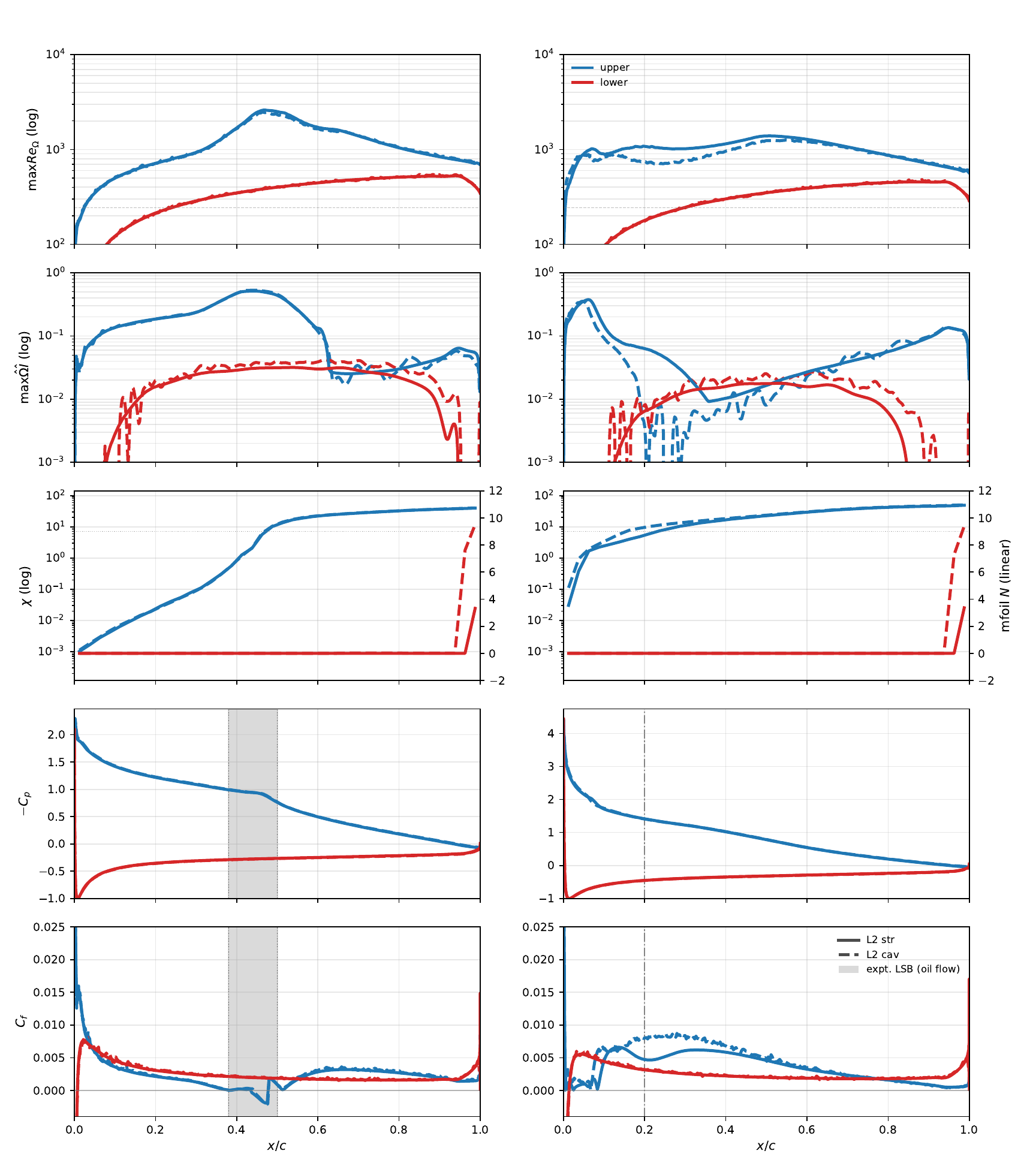}
\figcont{$\alpha=6^\circ$, $\alpha=8^\circ$.}
\end{figure}
\begin{figure}[H]\centering
\includegraphics[width=0.99\textwidth,height=0.94\textheight,keepaspectratio]{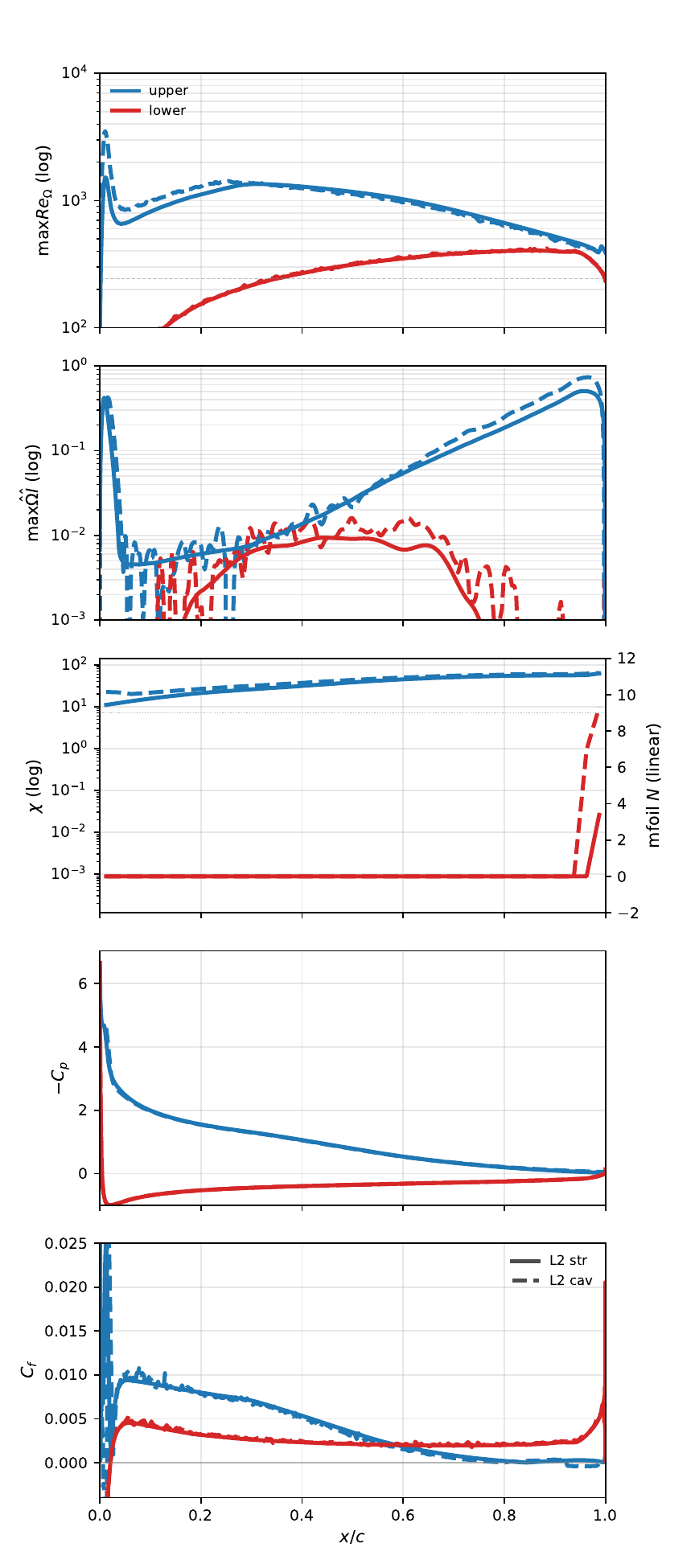}
\figcont{$\alpha=10^\circ$.}
\end{figure}
\begin{figure}[H]\centering
\includegraphics[width=0.99\textwidth,height=0.94\textheight,keepaspectratio]{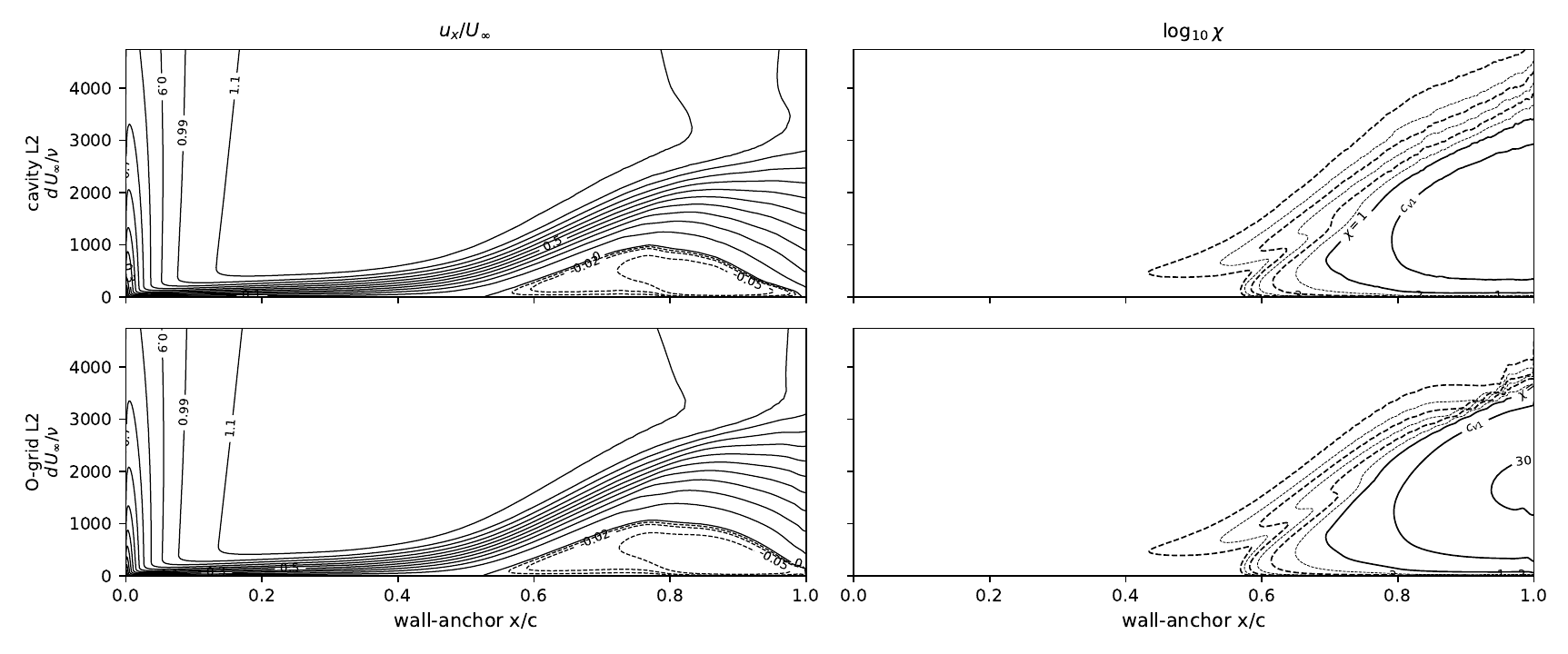}
\caption{Eppler~387 wall-normal sheets, $Re=10^5$, upper surface: $u_x/U_\infty$ (left) and $\log_{10}\chi$ (right) on the L2 pair. $\alpha=-2^\circ$.}\label{f:eppsheetRe100k}
\end{figure}
\begin{figure}[H]\centering
\includegraphics[width=0.99\textwidth,height=0.94\textheight,keepaspectratio]{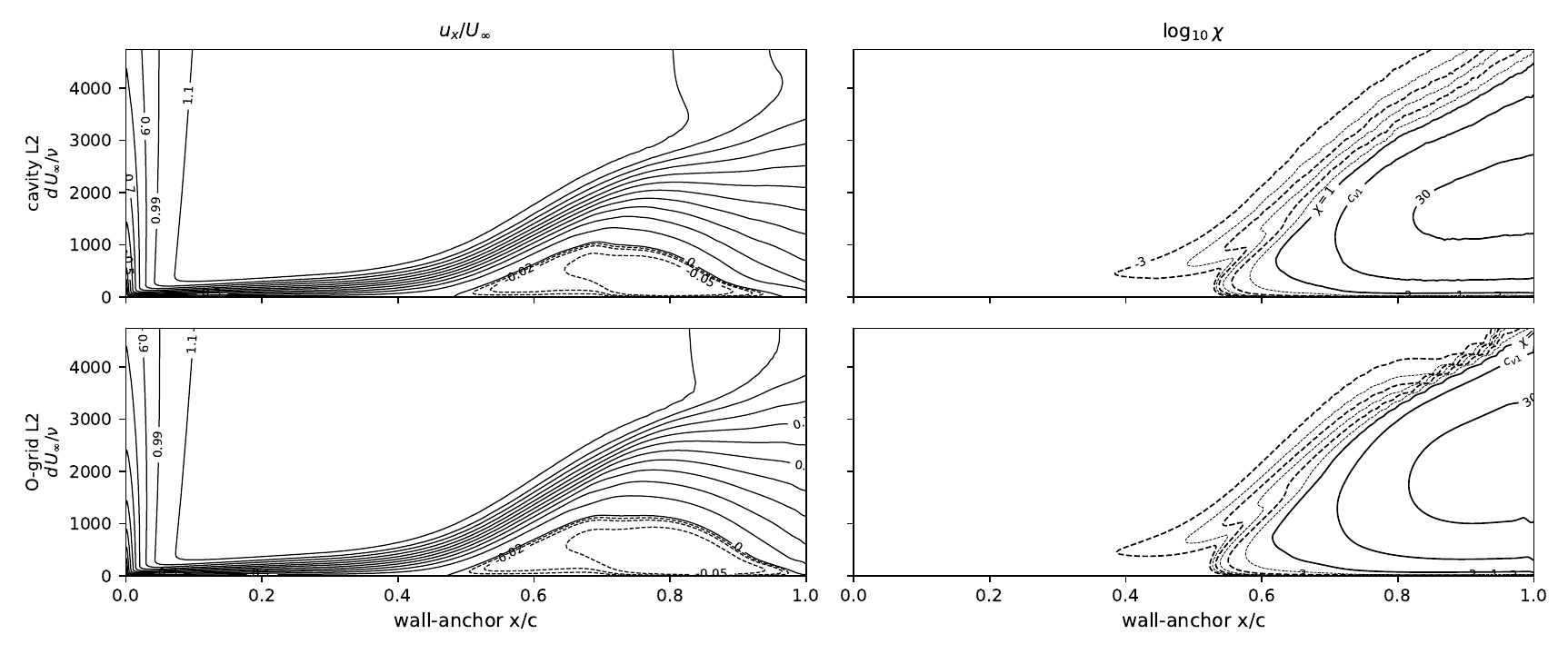}
\figcont{$\alpha=0^\circ$.}
\end{figure}
\begin{figure}[H]\centering
\includegraphics[width=0.99\textwidth,height=0.94\textheight,keepaspectratio]{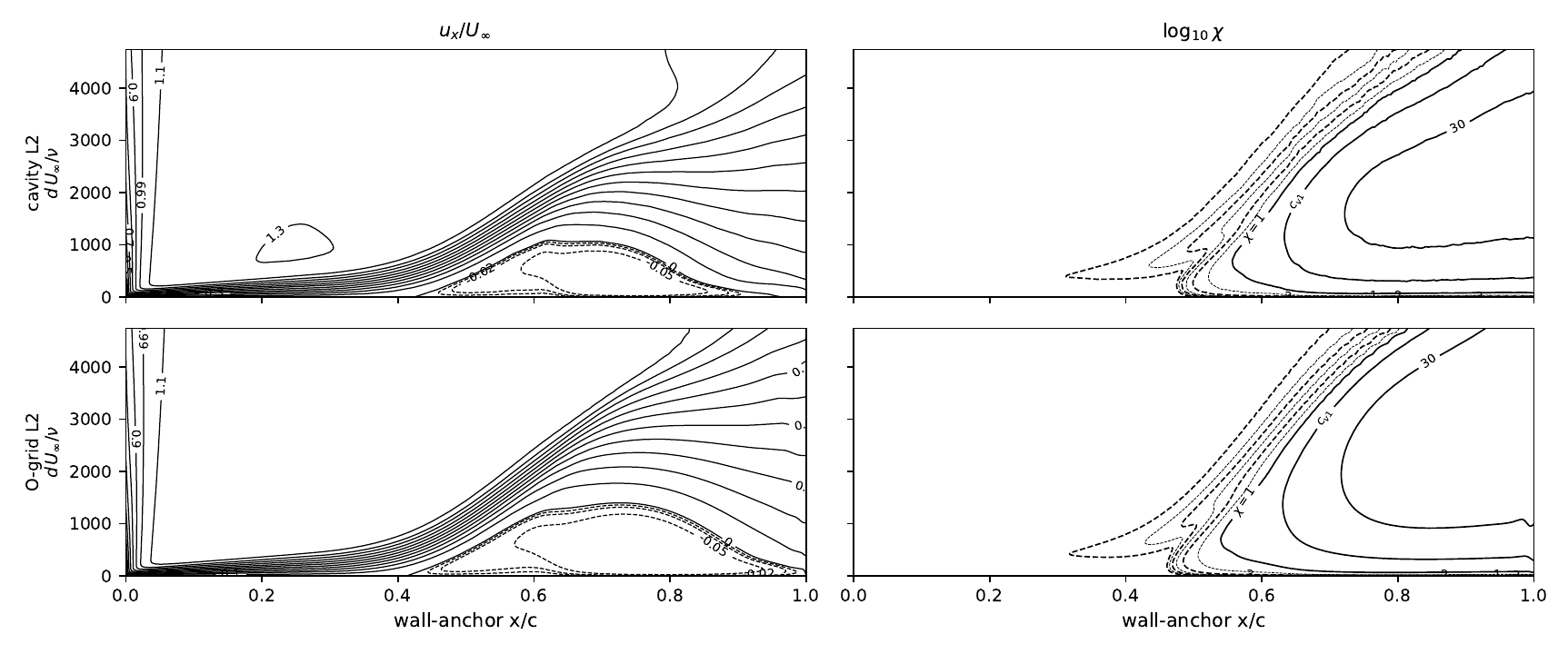}
\figcont{$\alpha=2^\circ$.}
\end{figure}
\begin{figure}[H]\centering
\includegraphics[width=0.99\textwidth,height=0.94\textheight,keepaspectratio]{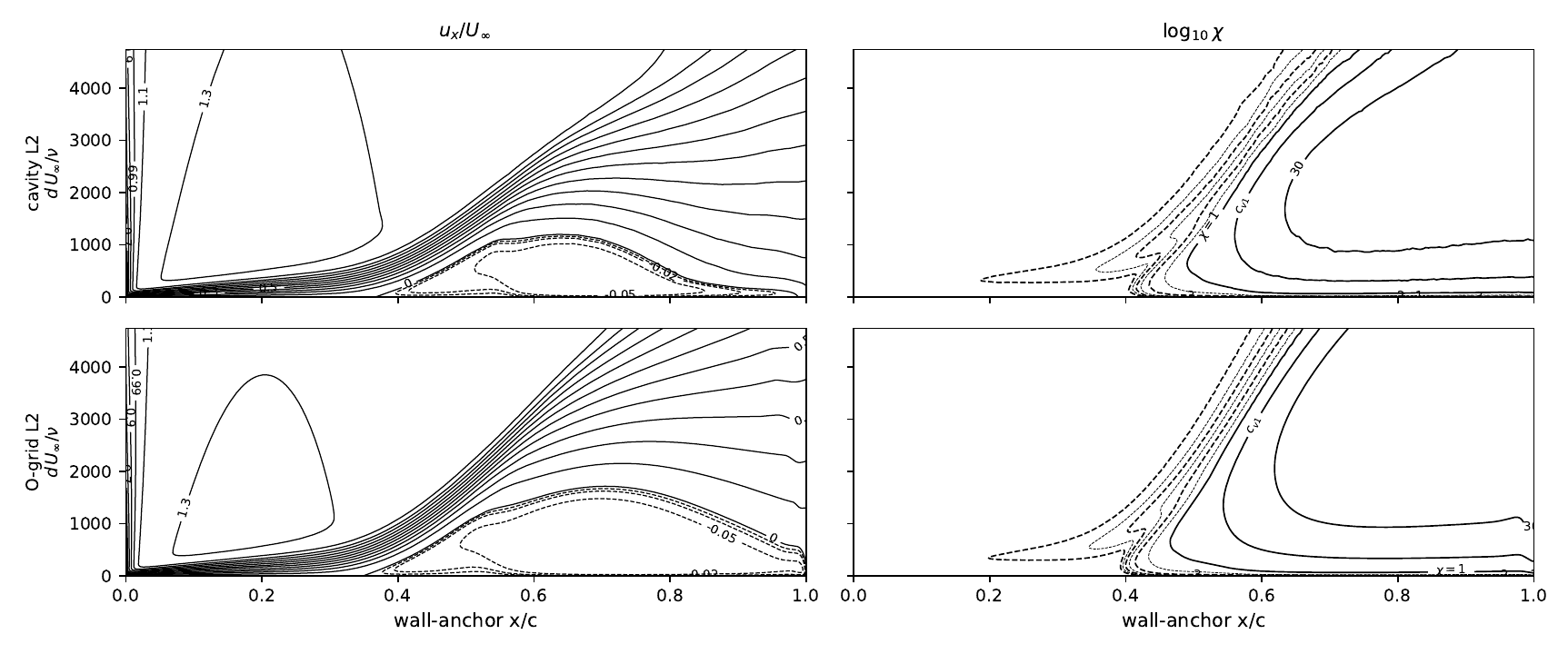}
\figcont{$\alpha=4^\circ$.}
\end{figure}
\begin{figure}[H]\centering
\includegraphics[width=0.99\textwidth,height=0.94\textheight,keepaspectratio]{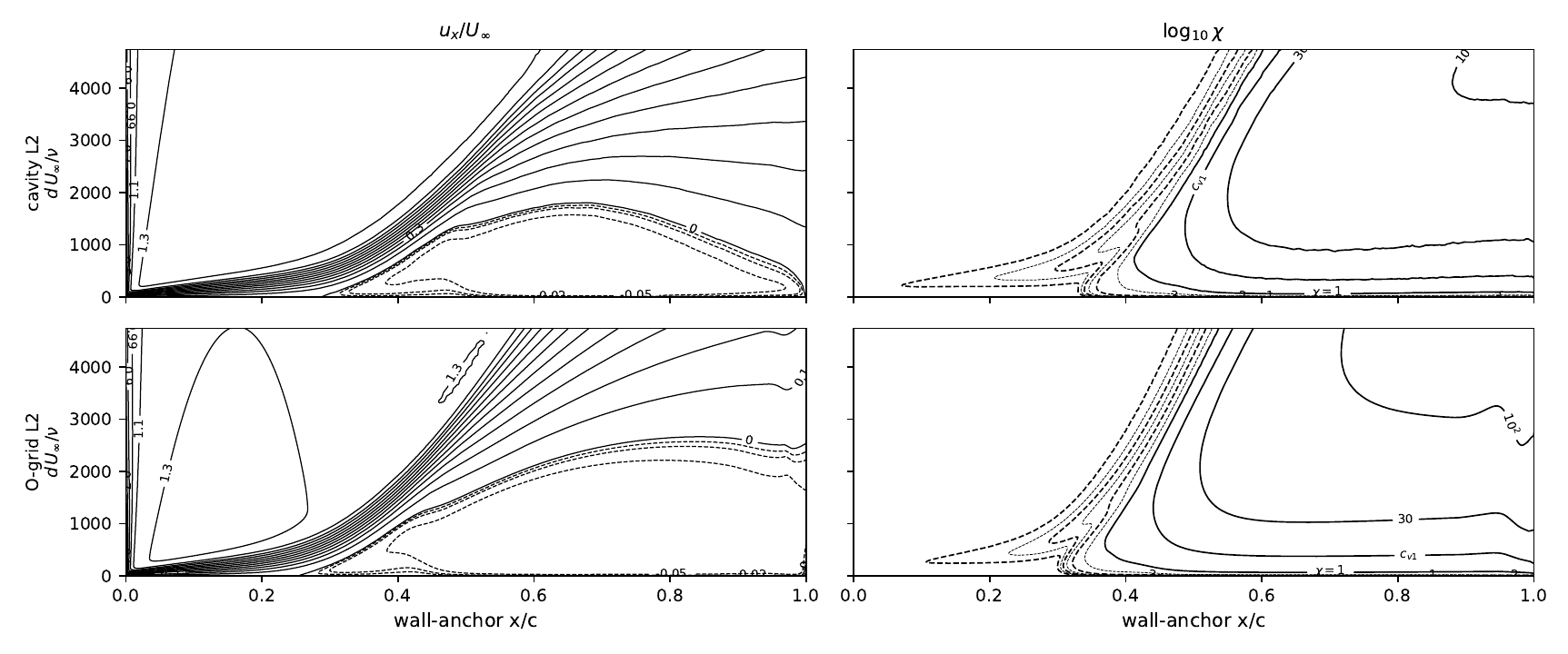}
\figcont{$\alpha=6^\circ$.}
\end{figure}
\begin{figure}[H]\centering
\includegraphics[width=0.99\textwidth,height=0.94\textheight,keepaspectratio]{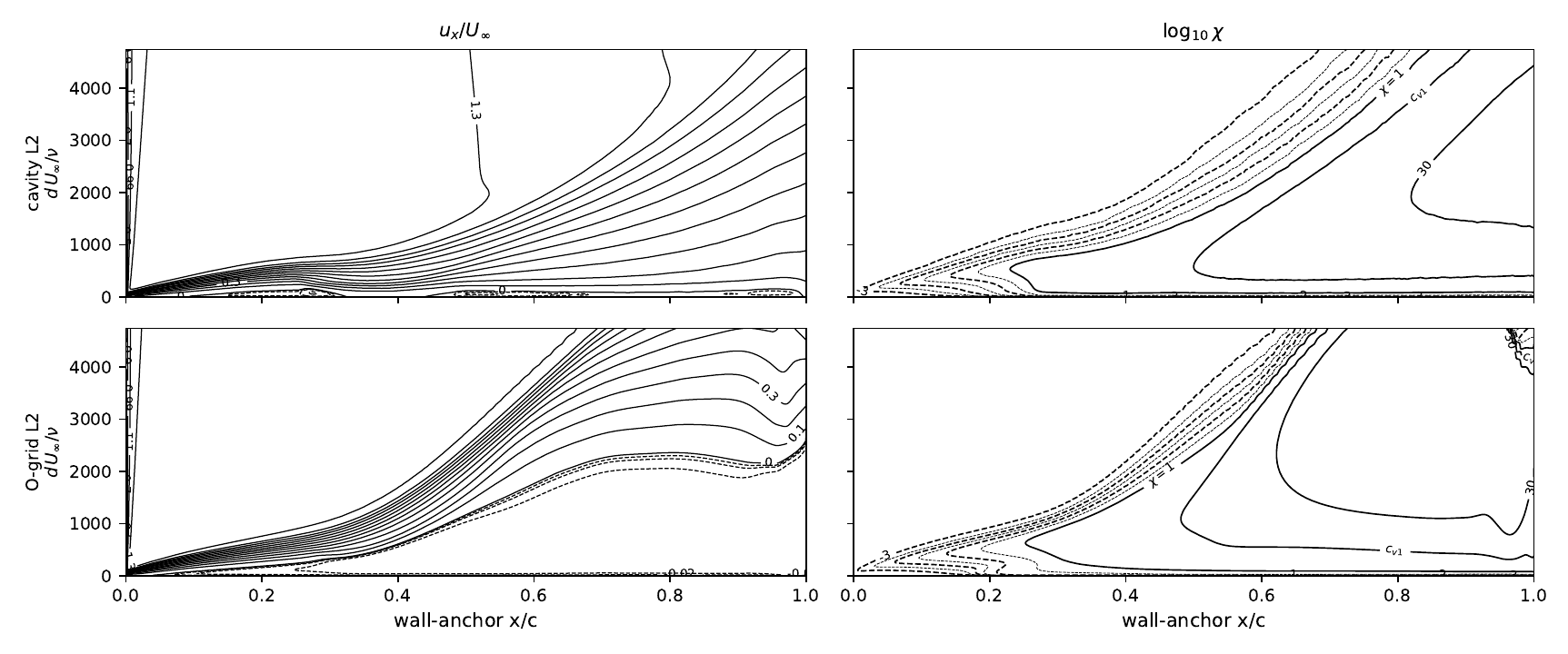}
\figcont{$\alpha=8^\circ$.}
\end{figure}
\begin{figure}[H]\centering
\includegraphics[width=0.99\textwidth,height=0.94\textheight,keepaspectratio]{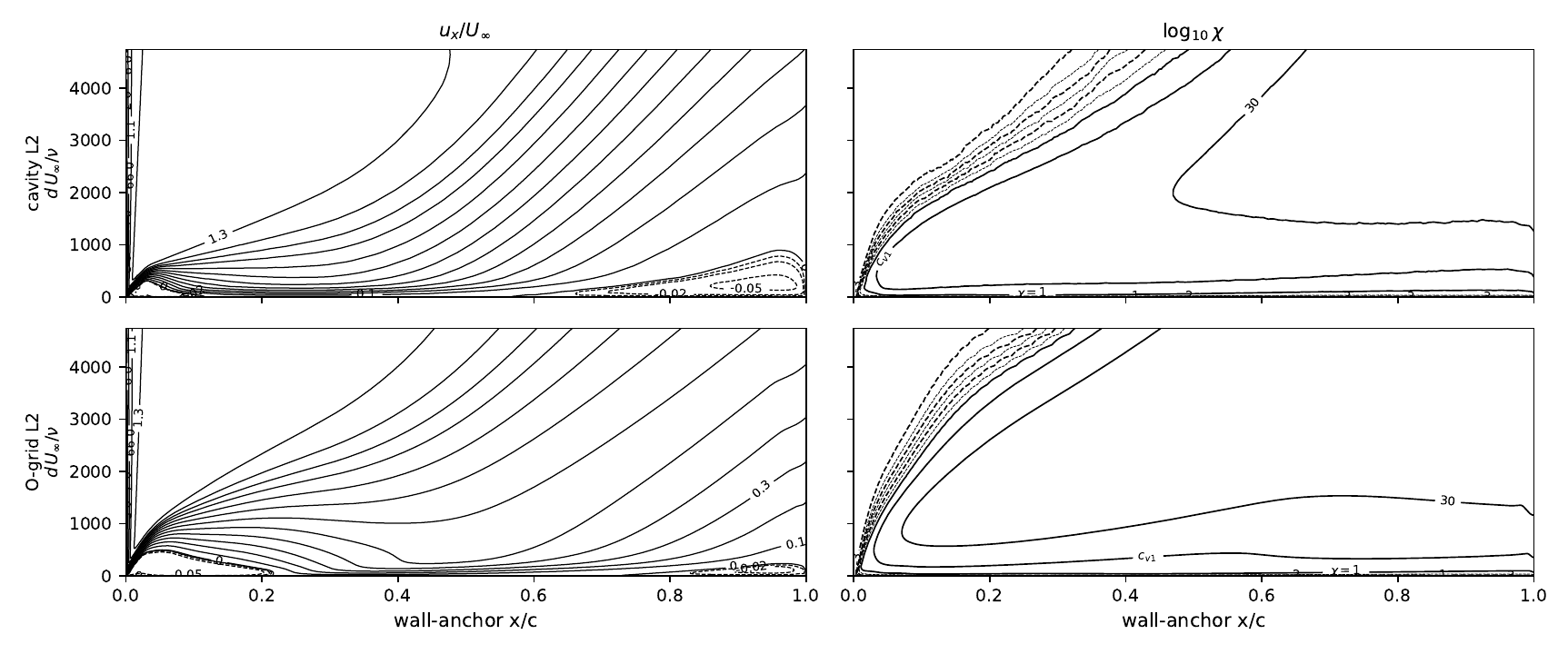}
\figcont{$\alpha=10^\circ$.}
\end{figure}
\begin{figure}[H]\centering
\includegraphics[width=0.99\textwidth,height=0.94\textheight,keepaspectratio]{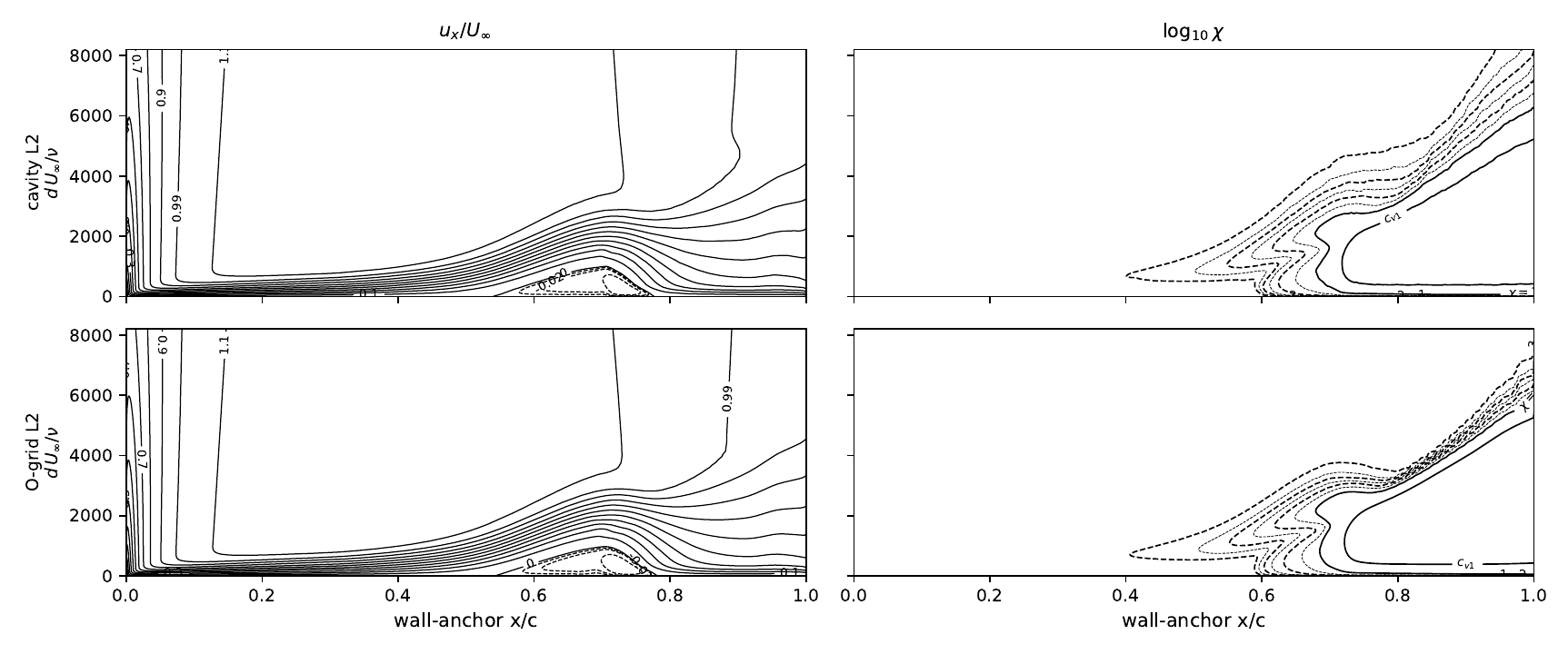}
\caption{Eppler~387 wall-normal sheets, $Re=3\times10^5$, upper surface: $u_x/U_\infty$ (left) and $\log_{10}\chi$ (right) on the L2 pair. $\alpha=-2^\circ$.}\label{f:eppsheetRe300k}
\end{figure}
\begin{figure}[H]\centering
\includegraphics[width=0.99\textwidth,height=0.94\textheight,keepaspectratio]{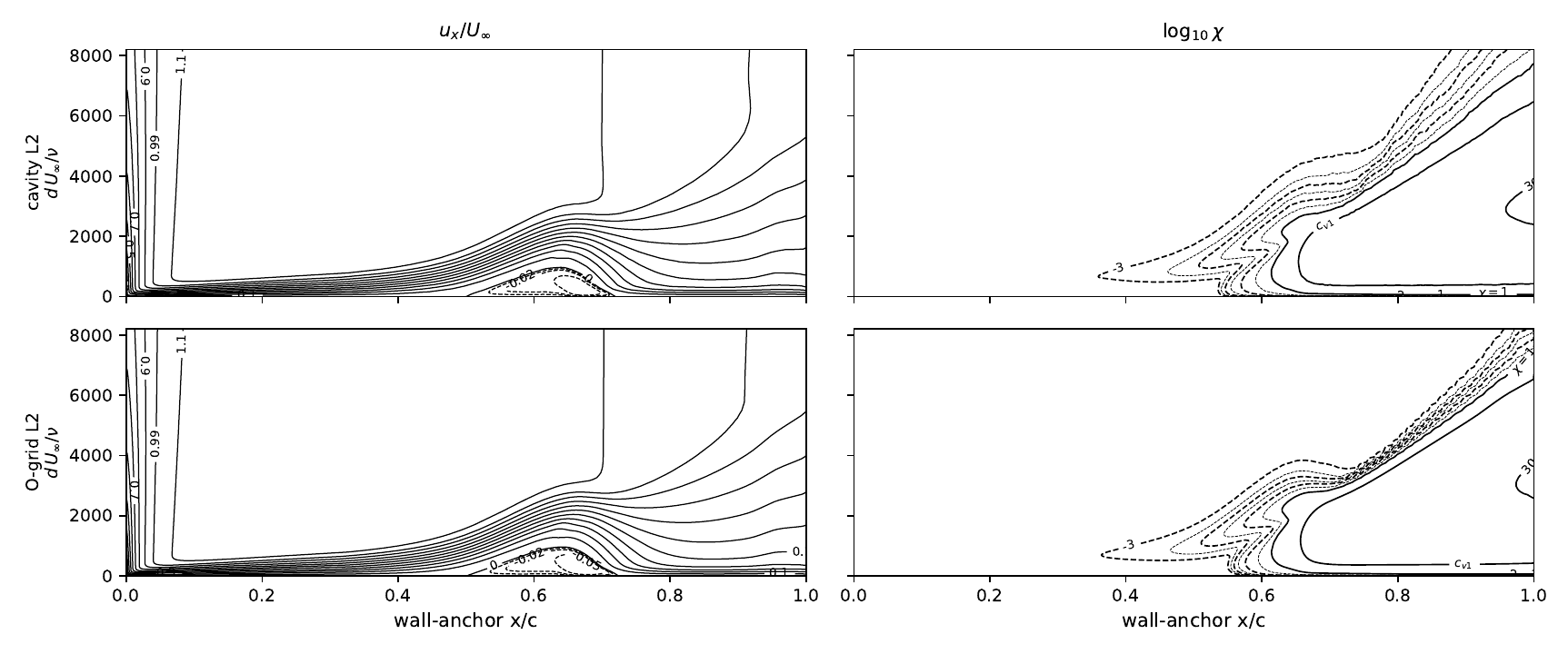}
\figcont{$\alpha=0^\circ$.}
\end{figure}
\begin{figure}[H]\centering
\includegraphics[width=0.99\textwidth,height=0.94\textheight,keepaspectratio]{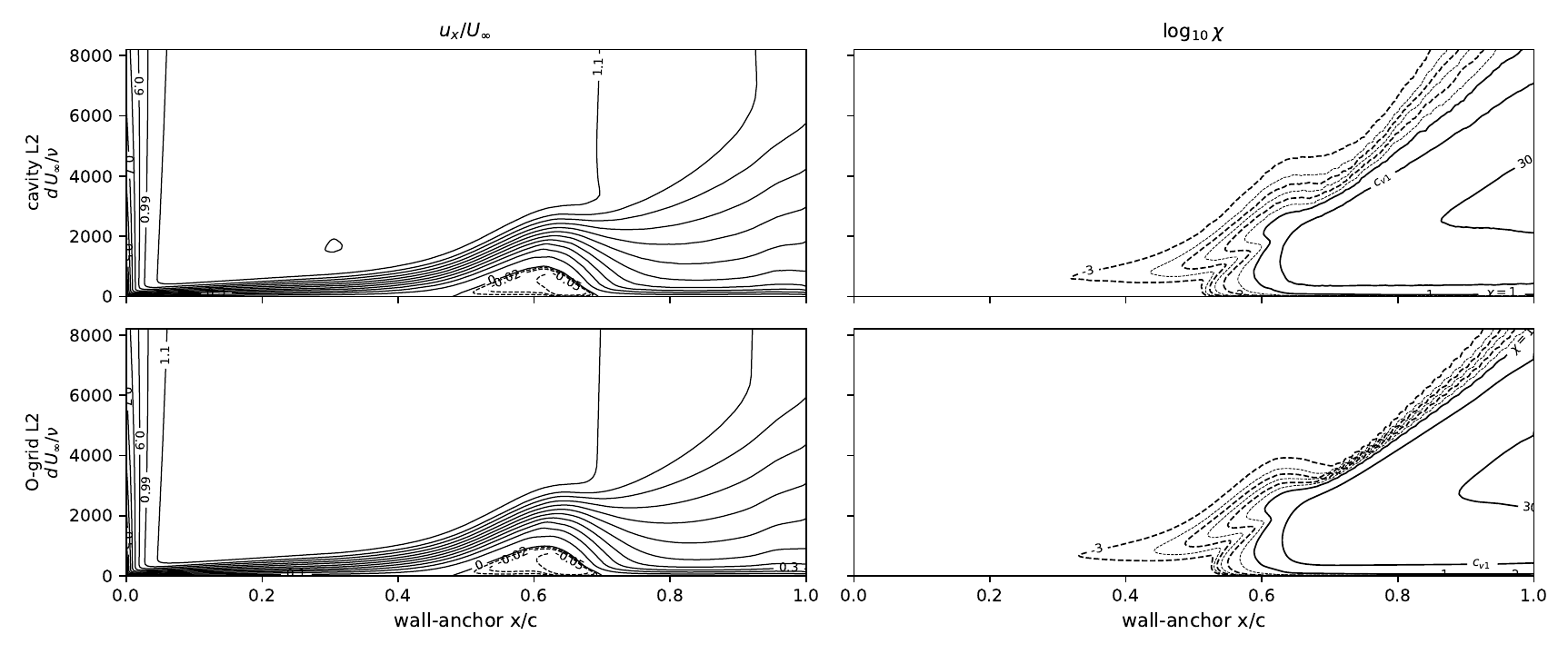}
\figcont{$\alpha=1^\circ$.}
\end{figure}
\begin{figure}[H]\centering
\includegraphics[width=0.99\textwidth,height=0.94\textheight,keepaspectratio]{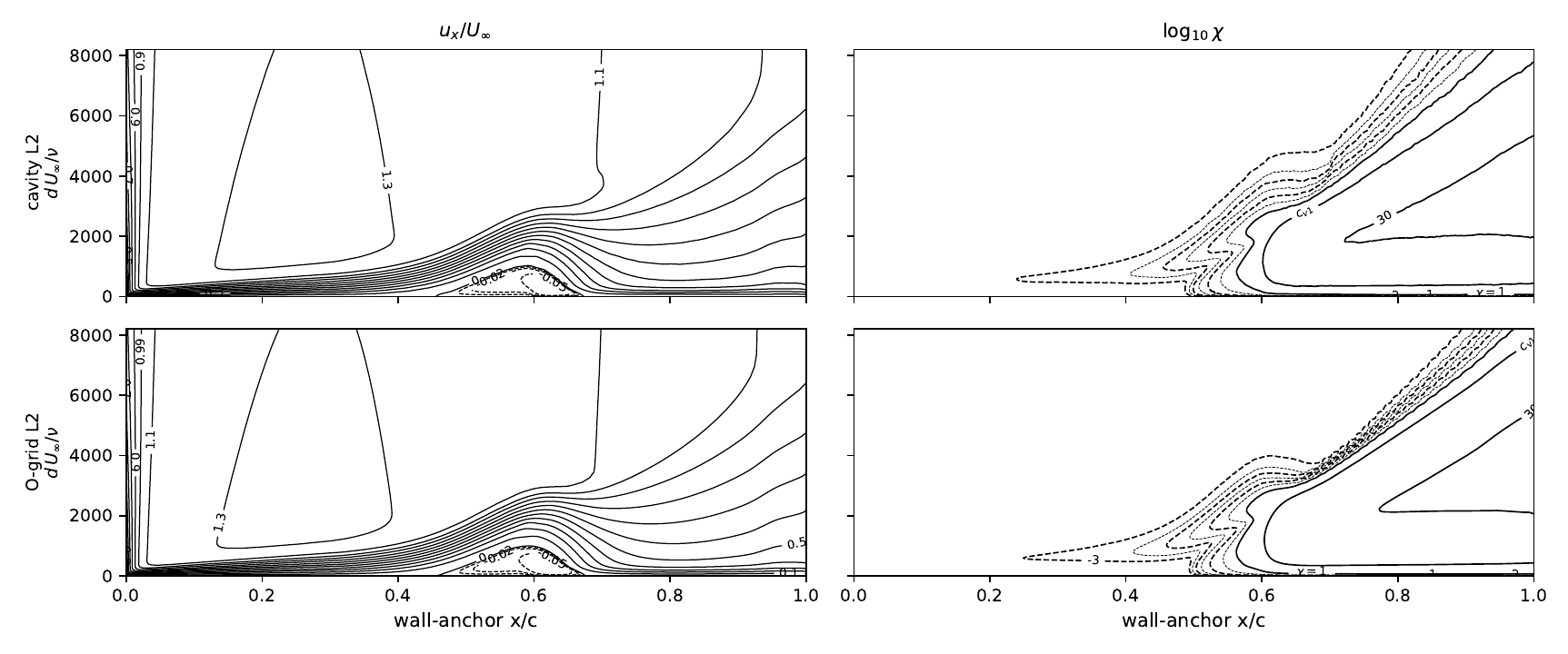}
\figcont{$\alpha=2^\circ$.}
\end{figure}
\begin{figure}[H]\centering
\includegraphics[width=0.99\textwidth,height=0.94\textheight,keepaspectratio]{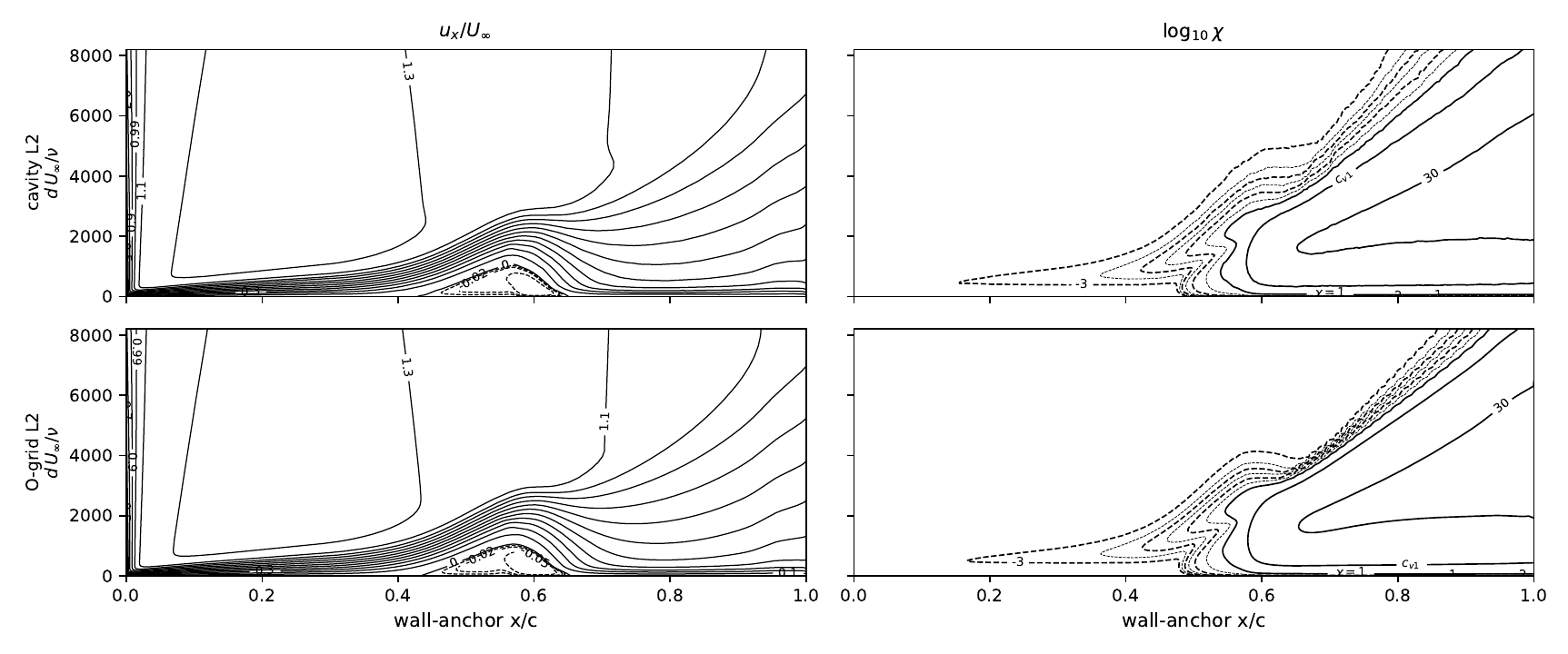}
\figcont{$\alpha=3^\circ$.}
\end{figure}
\begin{figure}[H]\centering
\includegraphics[width=0.99\textwidth,height=0.94\textheight,keepaspectratio]{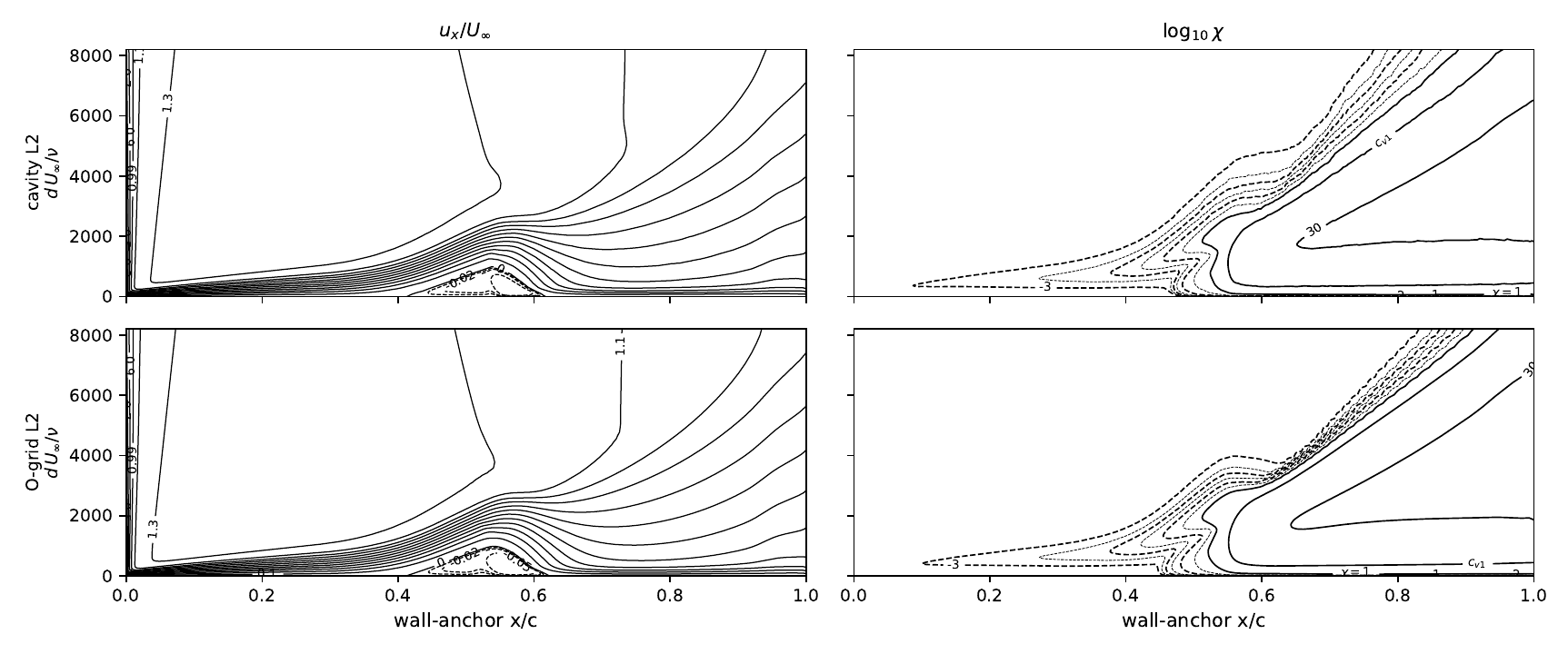}
\figcont{$\alpha=4^\circ$.}
\end{figure}
\begin{figure}[H]\centering
\includegraphics[width=0.99\textwidth,height=0.94\textheight,keepaspectratio]{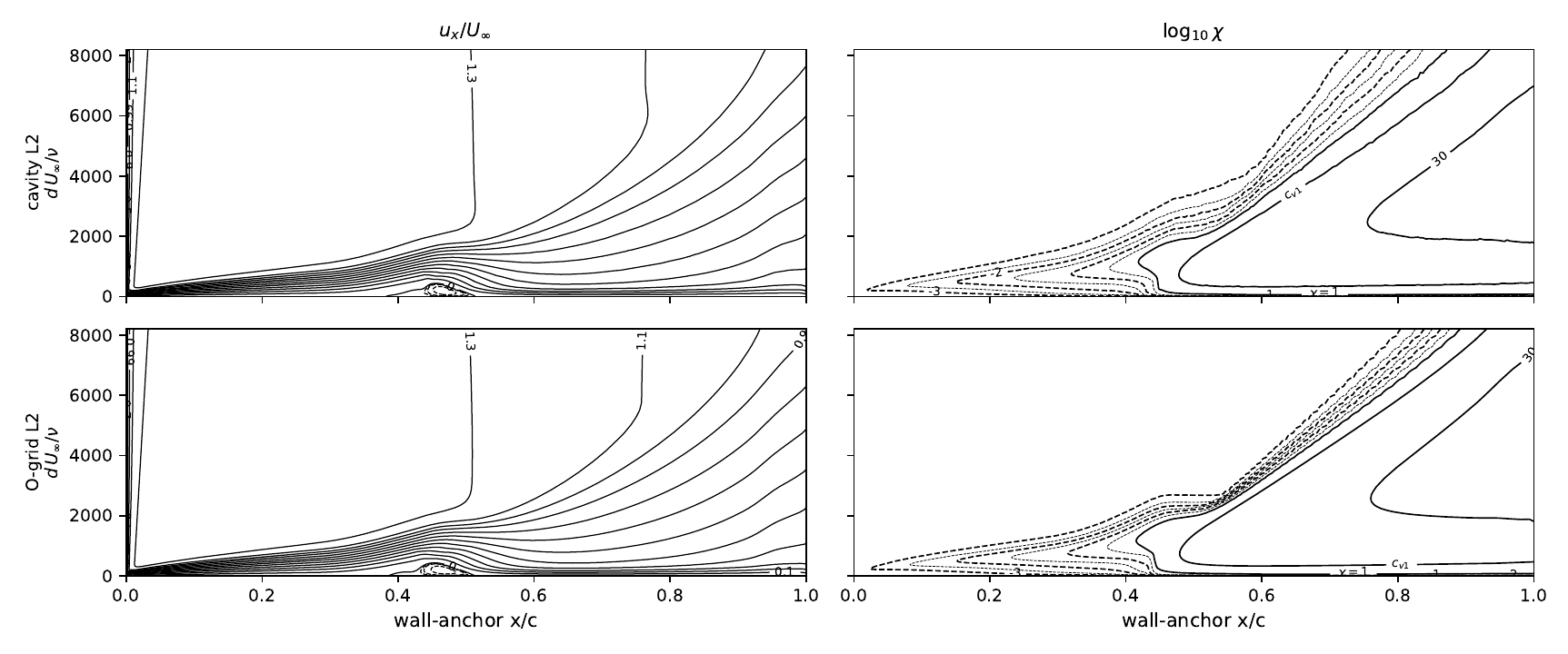}
\figcont{$\alpha=6^\circ$.}
\end{figure}
\begin{figure}[H]\centering
\includegraphics[width=0.99\textwidth,height=0.94\textheight,keepaspectratio]{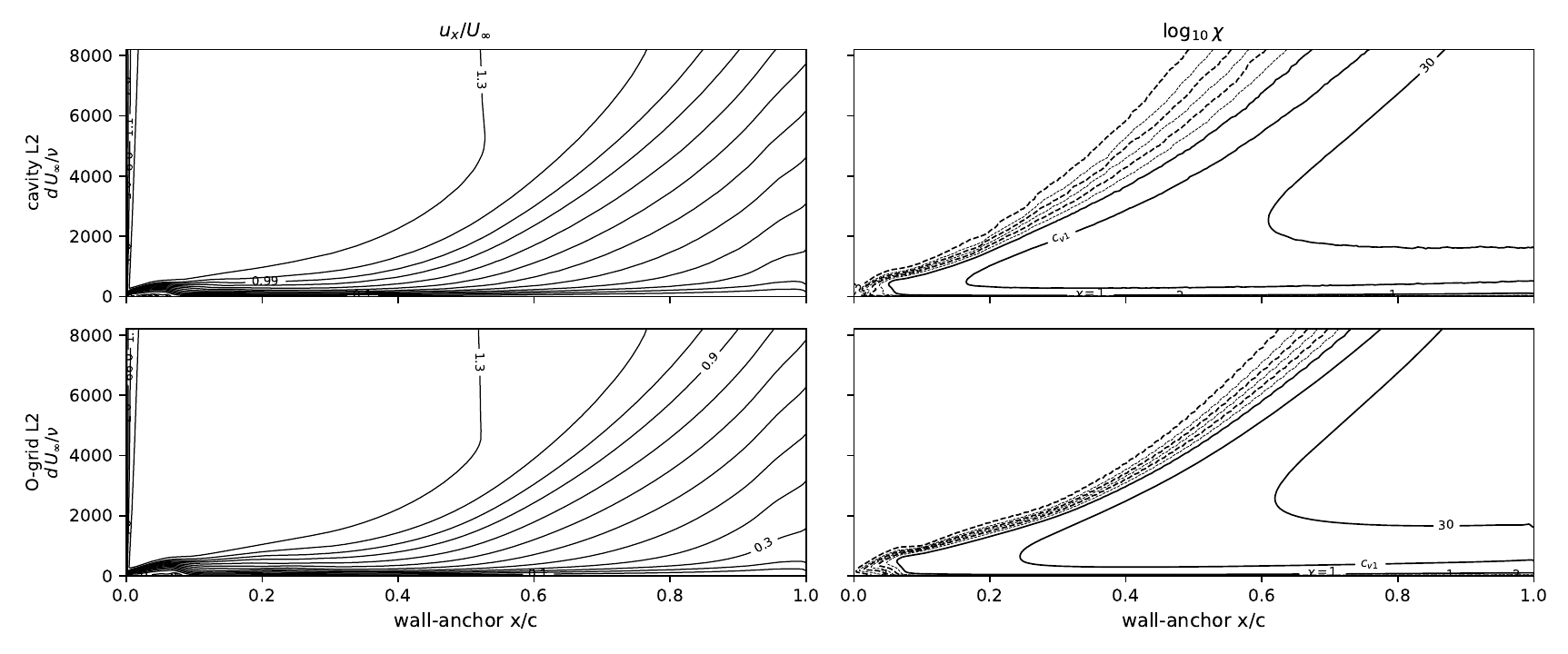}
\figcont{$\alpha=8^\circ$.}
\end{figure}
\begin{figure}[H]\centering
\includegraphics[width=0.99\textwidth,height=0.94\textheight,keepaspectratio]{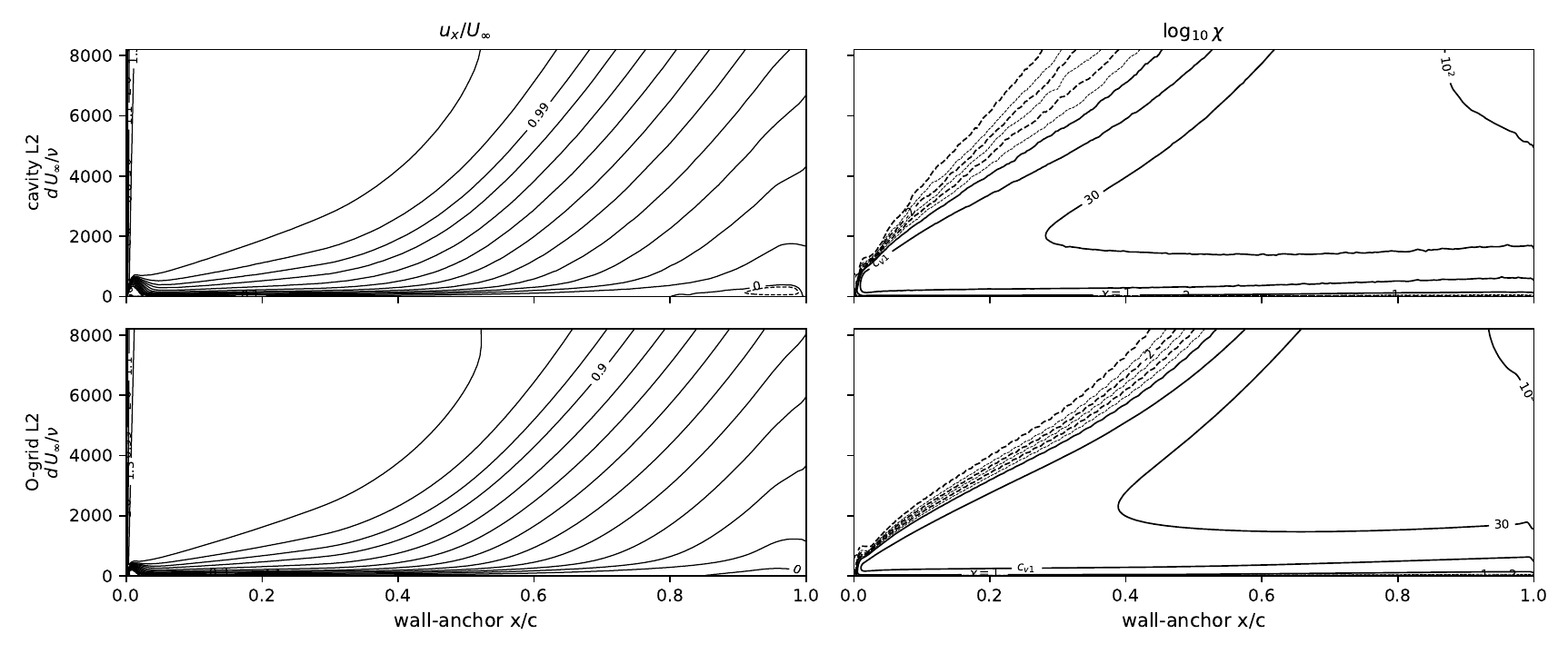}
\figcont{$\alpha=10^\circ$.}
\end{figure}

\subsection{A two-element section}

\begin{figure}[H]\centering
\includegraphics[width=0.58\textwidth]{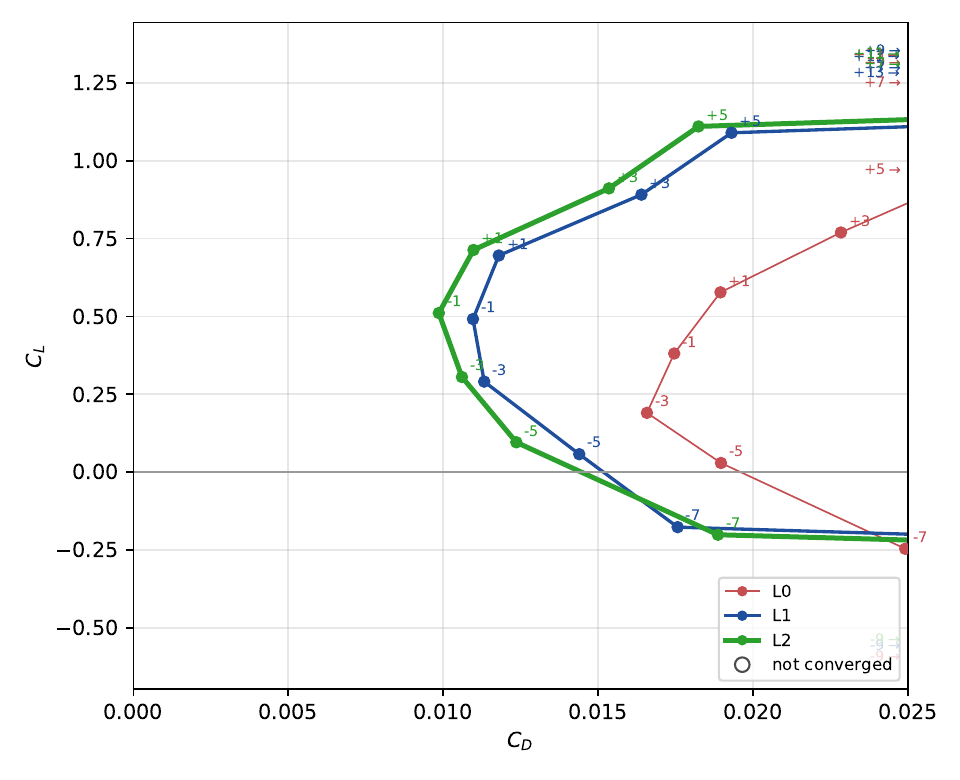}
\caption{Incidence sweep, $\alpha=-9$--$+13^\circ$; L2 to $+11^\circ$.
Separated points lie beyond the clip. The open marker, L2 at
$\alpha=-9^\circ$, stopped at its step ceiling.}
\label{f:twoelempolar}
\end{figure}

\begin{figure}[H]\centering
\includegraphics[width=\textwidth]{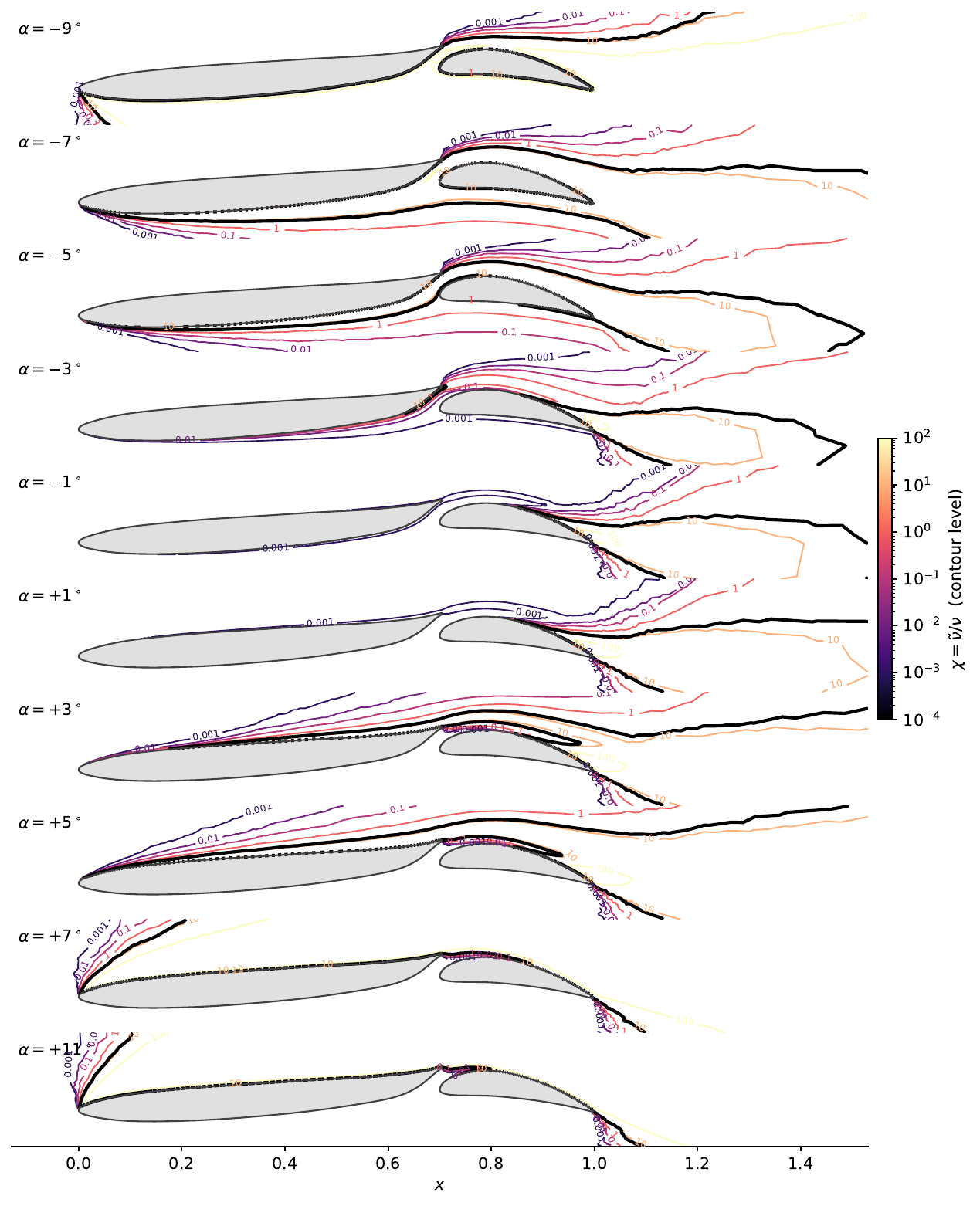}
\caption{Two-element mid-plane $\chi$ contours, one row per incidence;
decades from $10^{-3}$ to $10^{2}$ coloured by value, heavy black
$\chi=c_{v1}$. Pages run coarse to fine. L0.}
\label{f:twoelemchiL1}
\end{figure}

Purpose-built section: laminar main element ahead of a laminar
separation bubble on the flap, so the upstream wake and the downstream
bubble can be varied against one another. Definition in
Table~\ref{t:twoelementgeom}. $Re=10^6$ and $M=0.10$ on the total
chord, seed $\chi_\infty=c_{v1}e^{-9}$, L0/L1/L2 as elsewhere; every
solution converged to the $10^{-9}$ momentum tolerance.
Figure~\ref{f:twoelempolar}: polar.
Figure~\ref{f:twoelemchiL1}: $\chi$ contours by
incidence, one page per grid level -- the section itself, before the traces.
Figure~\ref{f:twoelemsuite1}: surface
suites, two incidences per page. Figure~\ref{f:twoelemflapchi}: flap upper surface unrolled,
wall distance to $0.06c$.

\begin{table}[H]
  \centering\footnotesize
  \setlength{\tabcolsep}{4pt}
  \caption{Two-element section definition, in total-chord units. Fore
  element: CST mean line
  $z_c=x_e[\hat x(1-\hat x)\sum_{i=0}^{10}A^{c}_iB^{10}_i(\hat x)]
  +\hat x\,\Delta z_{te}$ and half-thickness
  $t=x_e[\hat x^{1/2}(1-\hat x)\sum_{i=0}^{9}A^{t}_iB^{9}_i(\hat x)
  +\hat x\,t_{te}/(2x_e)]$, with $\hat x=x/x_e$, $B^{n}_i$ the Bernstein
  basis, surfaces offset $\pm t$ normal to the mean line, and the leading
  edge at the origin. These are the only independent numbers: the
  trailing-edge camber slope and included angle, the leading-edge radius,
  the maximum thickness and the slot gap all follow from them.}
  \label{t:twoelementgeom}
  \begin{tabular}{ll}
    \toprule
    \multicolumn{2}{l}{\emph{Fore element}}\\
    $x_e$ & 0.7039 \\
    $A^{c}_0\ldots A^{c}_{10}$ &
      $-0.05320$, $-0.16218$, $+0.17632$, $-0.65937$, $+0.84104$, $-1.32895$, \\
    & $+1.13471$, $-1.25791$, $+0.67079$, $-0.83361$, $-0.25106$ \\
    $\Delta z_{te}$ & $+0.08272$ \\
    $A^{t}_0\ldots A^{t}_{9}$ &
      $+0.10668$, $+0.07402$, $+0.29533$, $-0.36313$, $+0.99921$, $-1.10675$, \\
    & $+1.44742$, $-0.91005$, $+0.83270$, $+0.08962$ \\
    \midrule
    \multicolumn{2}{l}{\emph{Flap}}\\
    section & NACA 9416 \\
    chord & 0.30 \\
    incidence & $-8.0^\circ$ \\
    leading edge & $(0.70,\,0.04)$ \\
    \midrule
    \multicolumn{2}{l}{\emph{Both}}\\
    trailing-edge base $t_{te}$ & 0.003 \\
    \bottomrule
  \end{tabular}
\end{table}

\begin{figure}[H]\centering
\includegraphics[width=\textwidth]{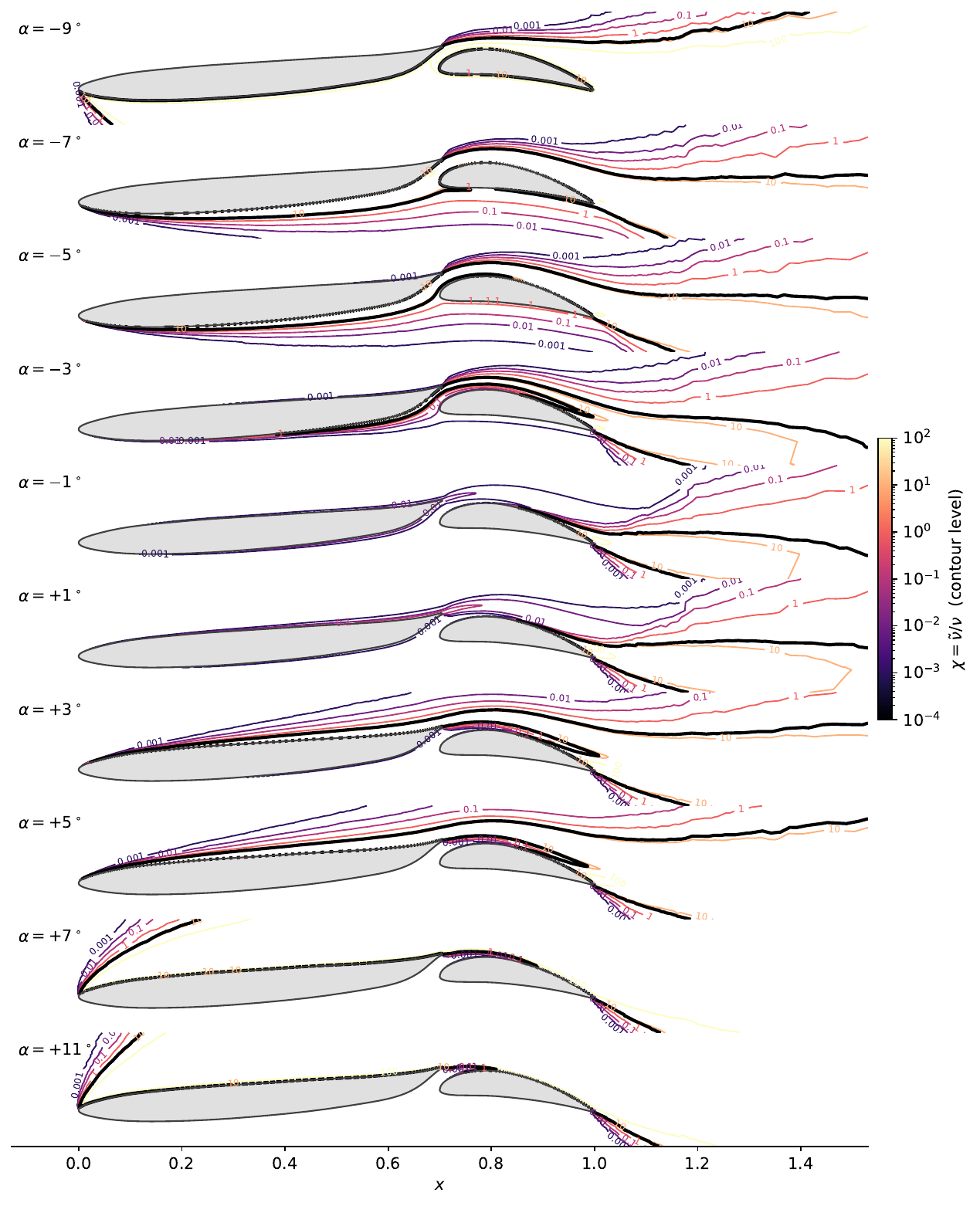}
\figcont{L1.}
\end{figure}

\begin{figure}[H]\centering
\includegraphics[width=\textwidth]{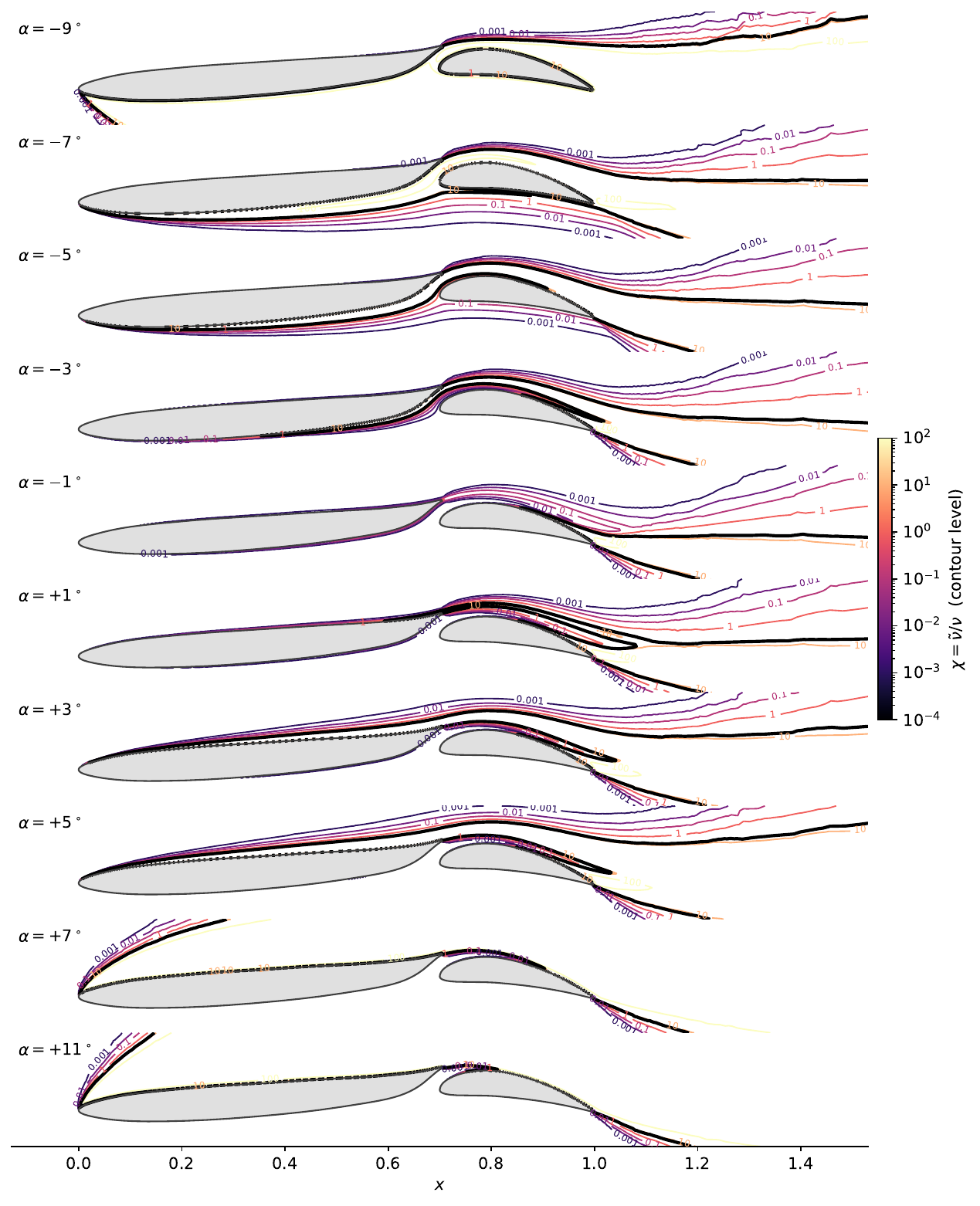}
\figcont{L2.}
\end{figure}

\begin{figure}[H]\centering
\includegraphics[width=\textwidth,page=1]{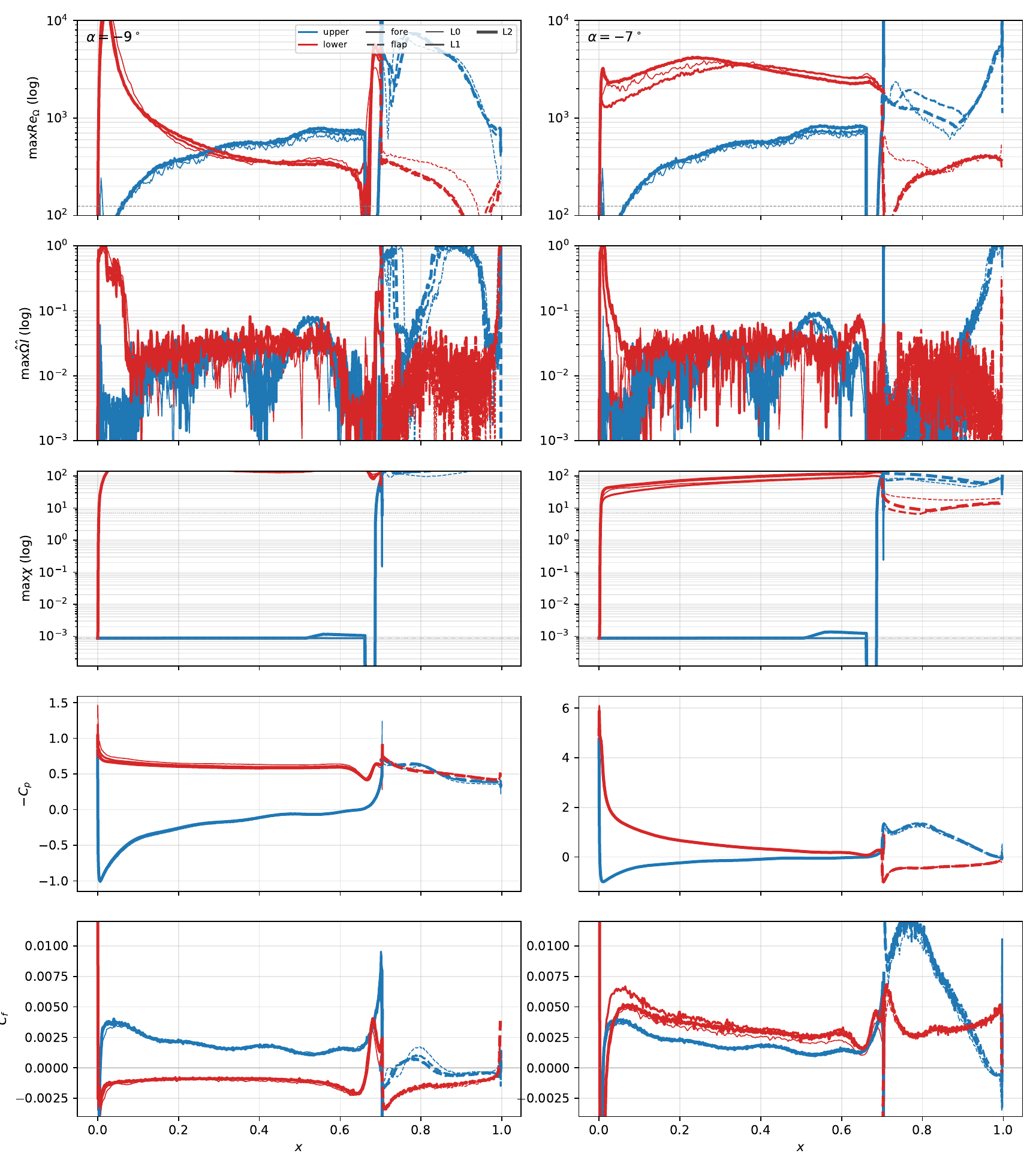}
\caption{Two-element surface suites, two incidences per page. Rows: $\max Re_\Omega$; $\max\hat\Omega\hat I$; $\max\chi$; $-C_p$; signed $C_f$. Upper blue, lower red; fore solid, flap dashed; L0--L2 by line thickness. $\alpha=-9$, $-7^\circ$.}\label{f:twoelemsuite1}
\end{figure}

\begin{figure}[H]\centering
\includegraphics[width=\textwidth,page=2]{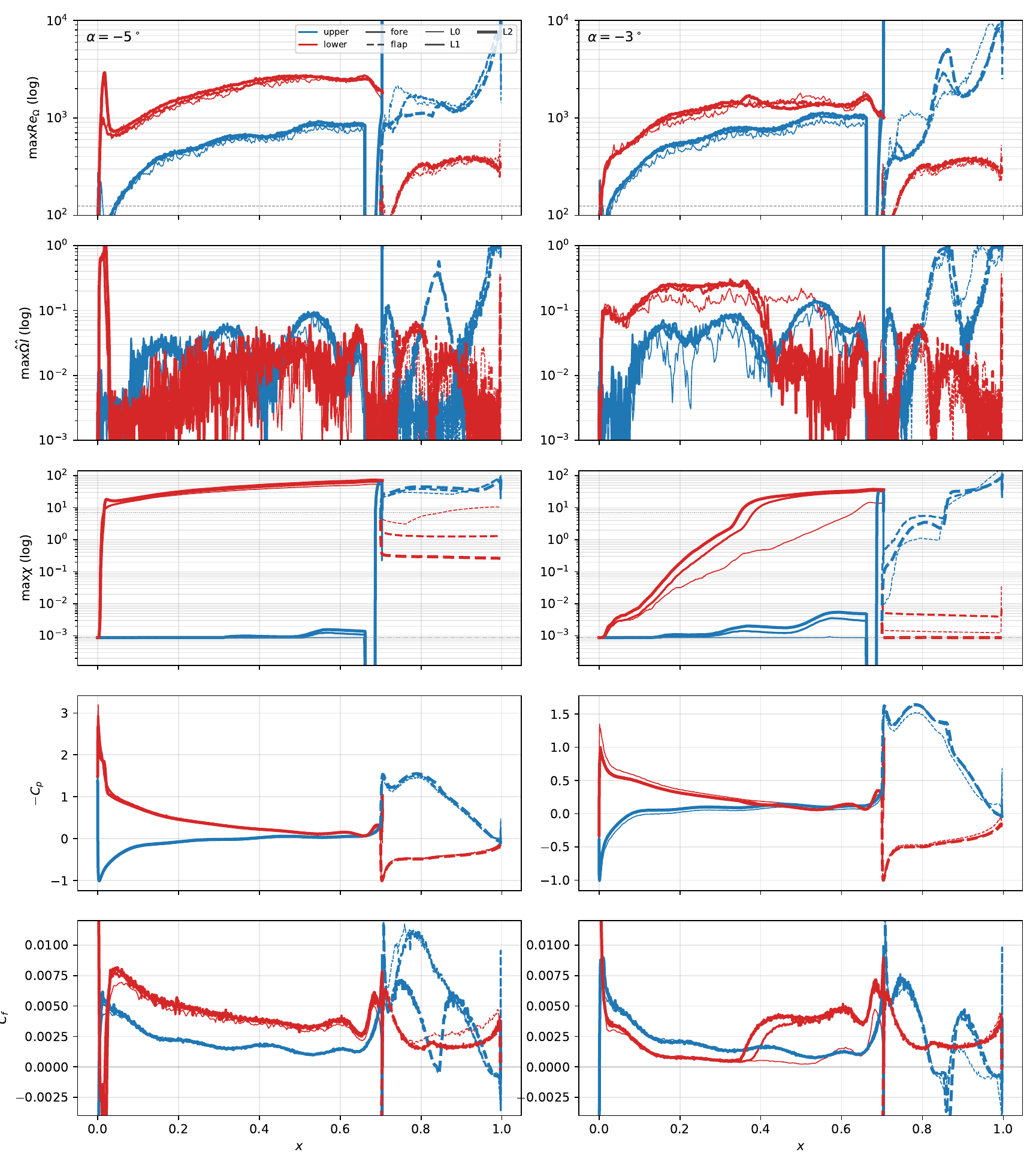}
\figcont{$\alpha=-5$, $-3^\circ$.}
\end{figure}

\begin{figure}[H]\centering
\includegraphics[width=\textwidth,page=3]{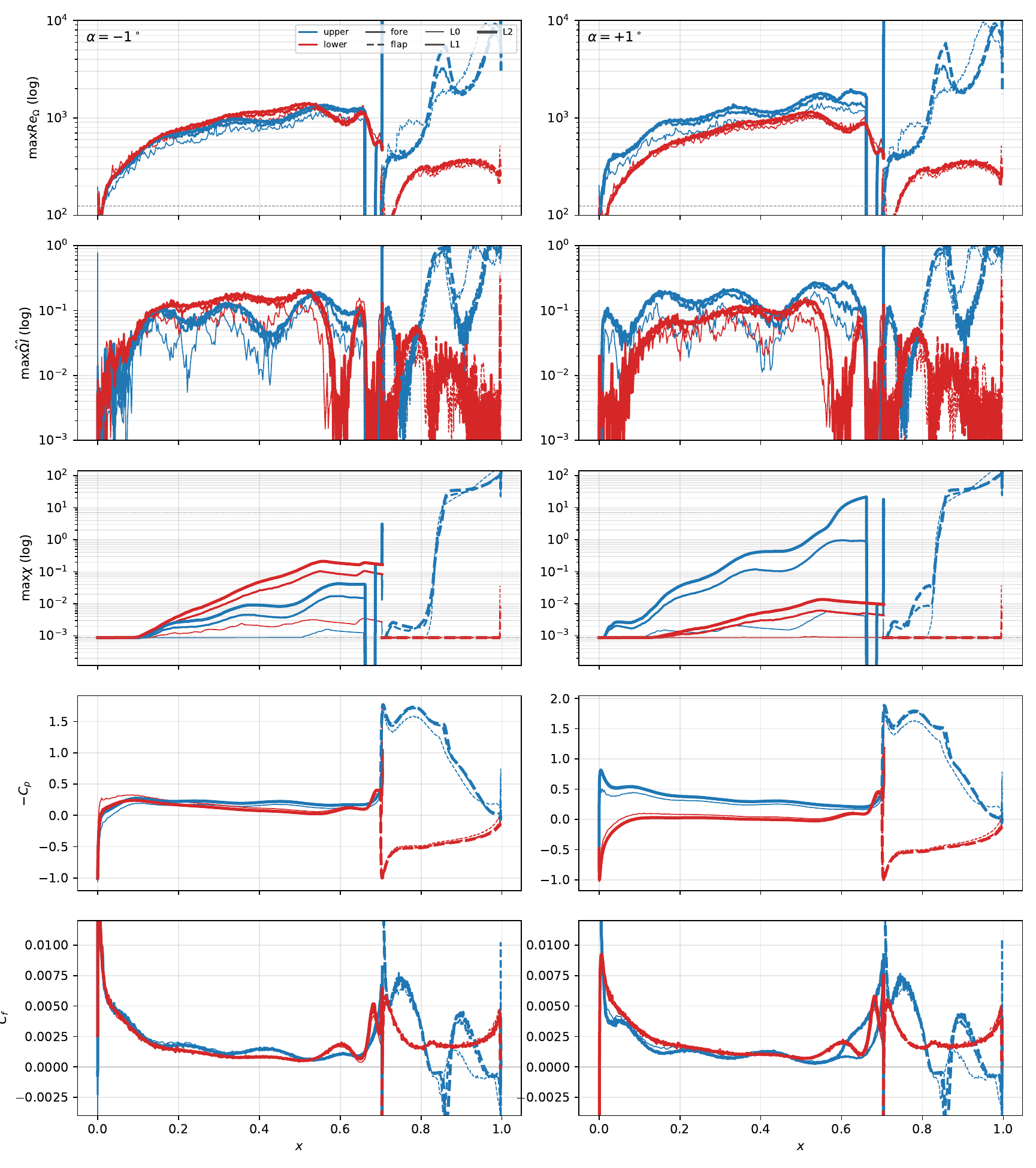}
\figcont{$\alpha=-1$, $+1^\circ$.}
\end{figure}

\begin{figure}[H]\centering
\includegraphics[width=\textwidth,page=4]{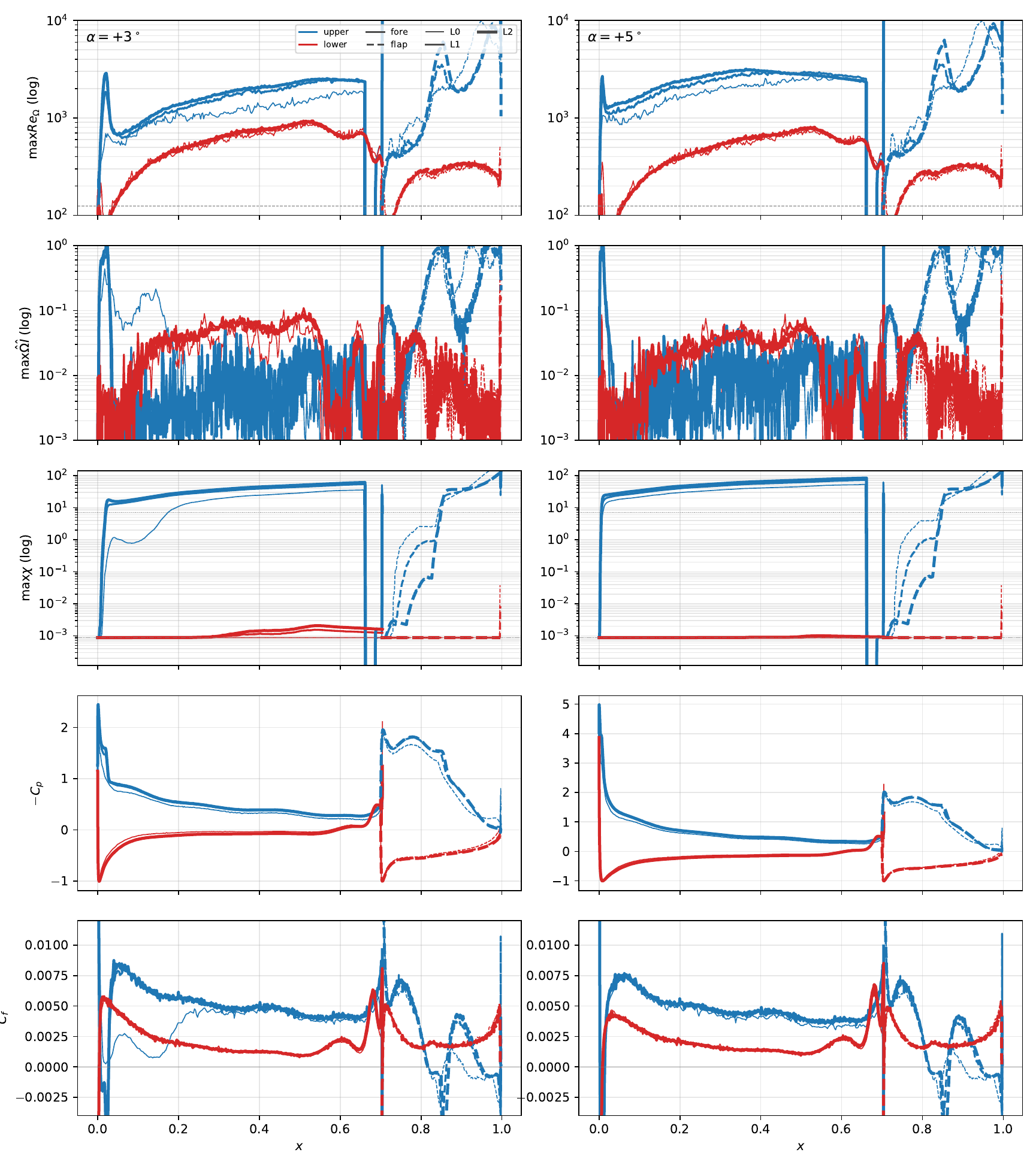}
\figcont{$\alpha=+3$, $+5^\circ$.}
\end{figure}

\begin{figure}[H]\centering
\includegraphics[width=\textwidth,page=5]{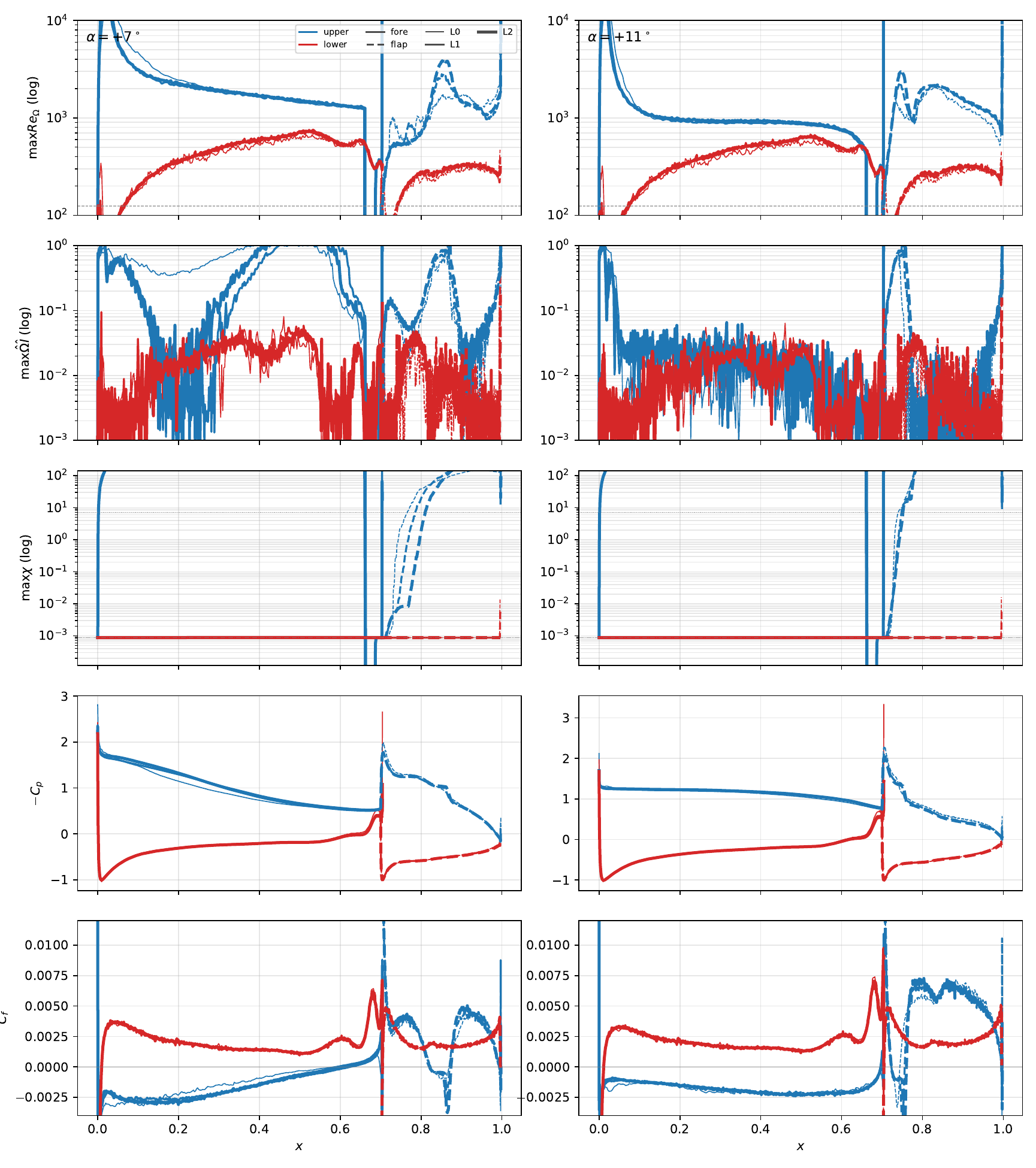}
\figcont{$\alpha=+7$, $+11^\circ$.}
\end{figure}

\begin{figure}[H]\centering
\includegraphics[width=\textwidth]{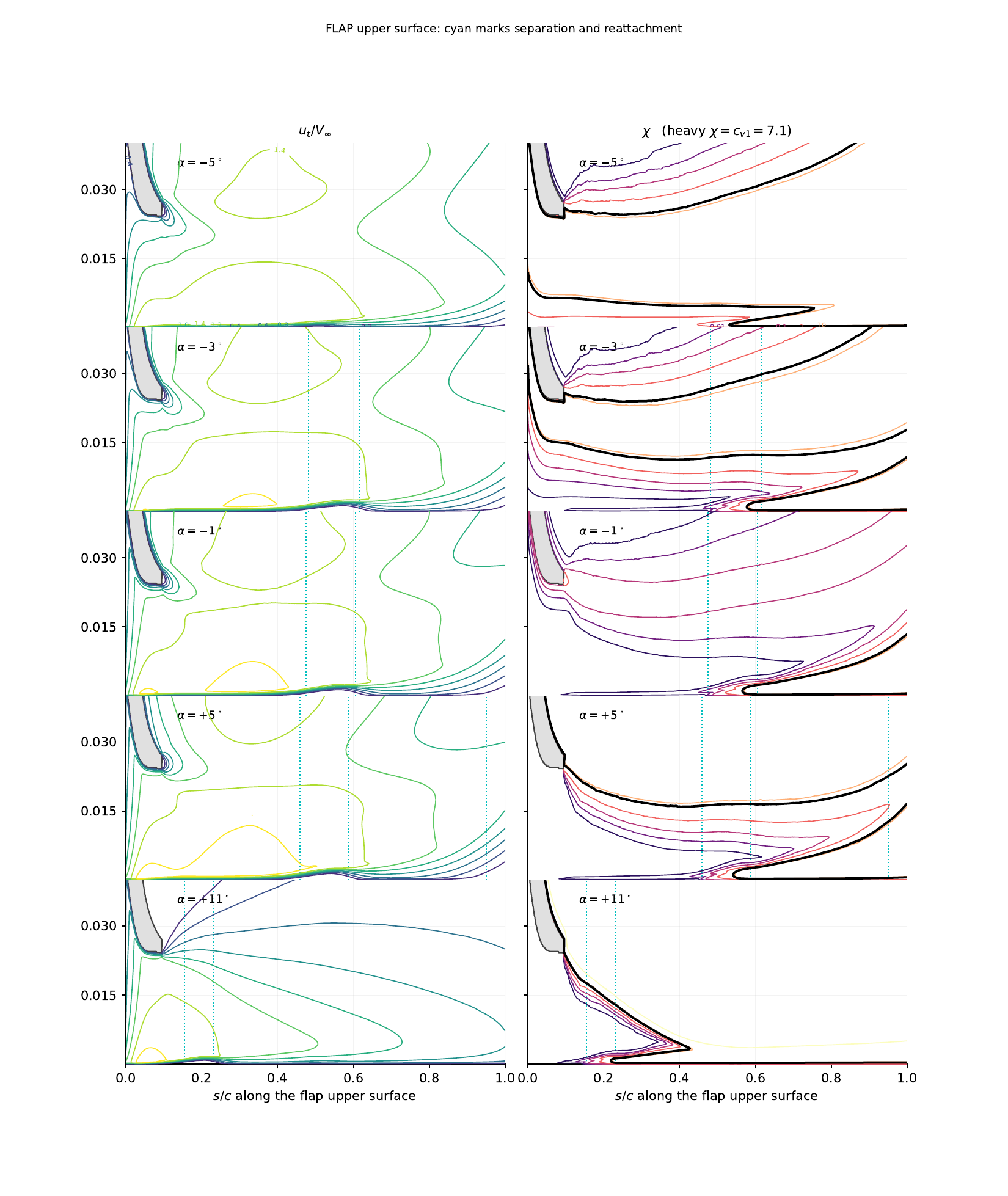}
\caption{Flap upper surface, L2, one row per incidence:
$u_t/V_\infty$ left, $\chi$ right (heavy $\chi=c_{v1}$), against arc
length and wall distance. Grey: the fore element's trailing edge, which the
flap's wall normals run into over the first $0.09$ of flap arc length. Cyan
marks separation and reattachment. Rows
$\alpha=-5$, $-3$, $-1$, $+5$, $+11^\circ$.}
\label{f:twoelemflapchi}
\end{figure}

\subsection{The cylinder drag crisis}

A steady-branch traverse over ten decades at a single seed
($\chi_\infty=1.1\times10^{-2}$; Mack's map at $Tu=0.2\%$), swept as
one upward continuation ladder on matched quasi-two-dimensional
O-grids ($y^+\lesssim1$), identified throughout by first-layer thickness
$y_1/R$: $2\times10^{-3}$ ($600\times121$) for $Re_D\le10^3$,
$1.6\times10^{-5}$ ($1200\times148$), then $2\times10^{-6}$ and
$2\times10^{-9}$ ($1600\times170$); twenty-five points
$Re_D=1$--$10^{10}$. Every case lands on the symmetric branch,
$|C_L|<2\times10^{-6}$.

\begin{figure}[H]\centering
\includegraphics[width=\textwidth,height=0.94\textheight,keepaspectratio]{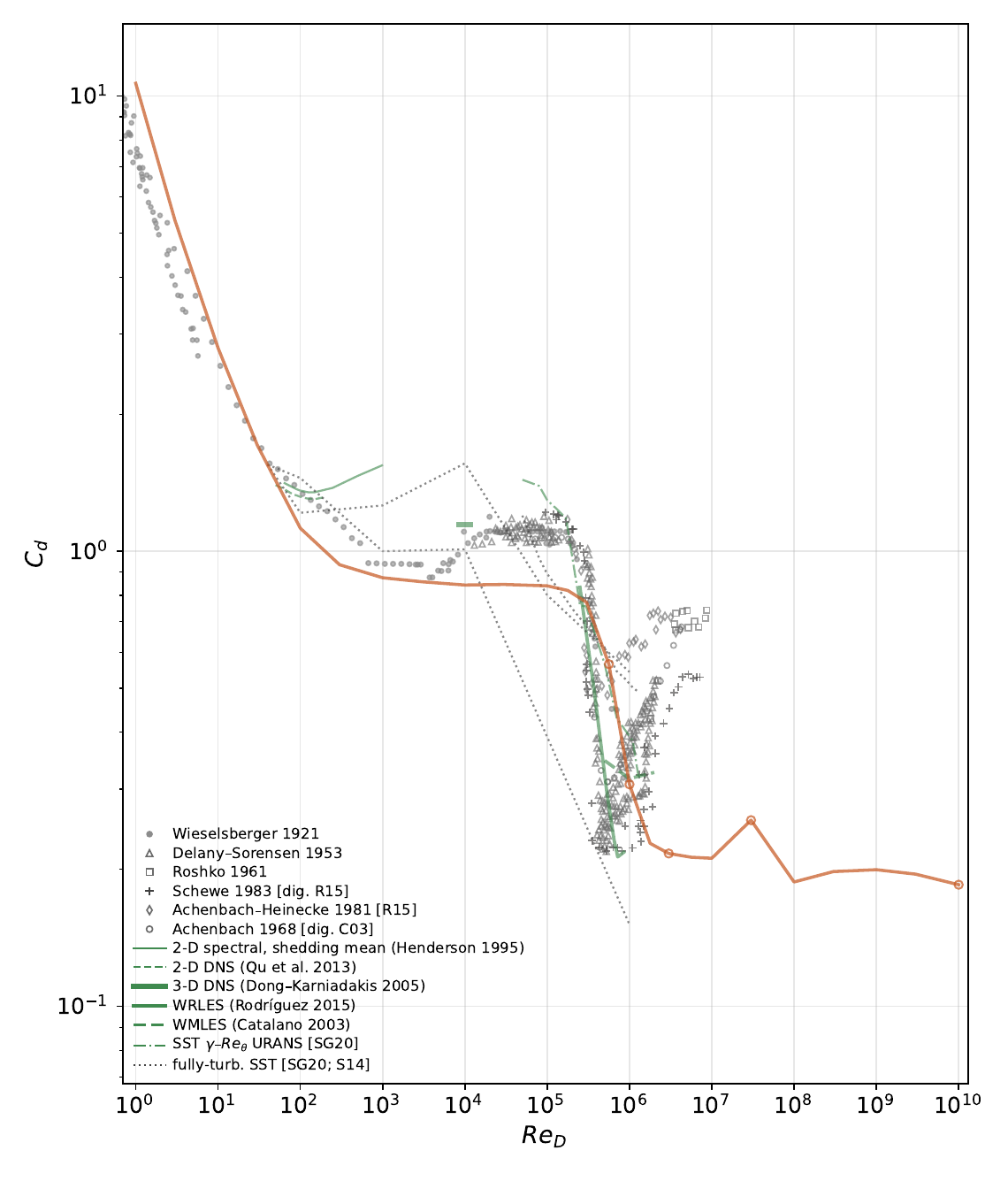}
\caption{Cylinder drag crisis, $Re_D=1$--$10^{10}$; single seed
$Tu=0.2\%$, up-ladder (clean systematic sweep, matched $Re_D$ grid).
Symbols experiments, lines computations; open rings mark the cases
the steady monitor did not accept. The SA-AI curve is drawn unbroken
across the four grids the ladder crosses---$y_1/R=2\times10^{-3}$,
$Re_D=1$--$10^3$; $1.6\times10^{-5}$, $3\times10^3$--$1.78\times10^6$;
$2\times10^{-6}$, $3\times10^6$--$10^7$; $2\times10^{-9}$,
$3\times10^7$--$10^{10}$---because it is one
continuation path: each grid change restarts from the previous grid's
converged field, interpolated onto the new grid.}\label{f:dragcrisiscd}
\end{figure}

\begin{figure}[H]\centering
\includegraphics[width=\textwidth]{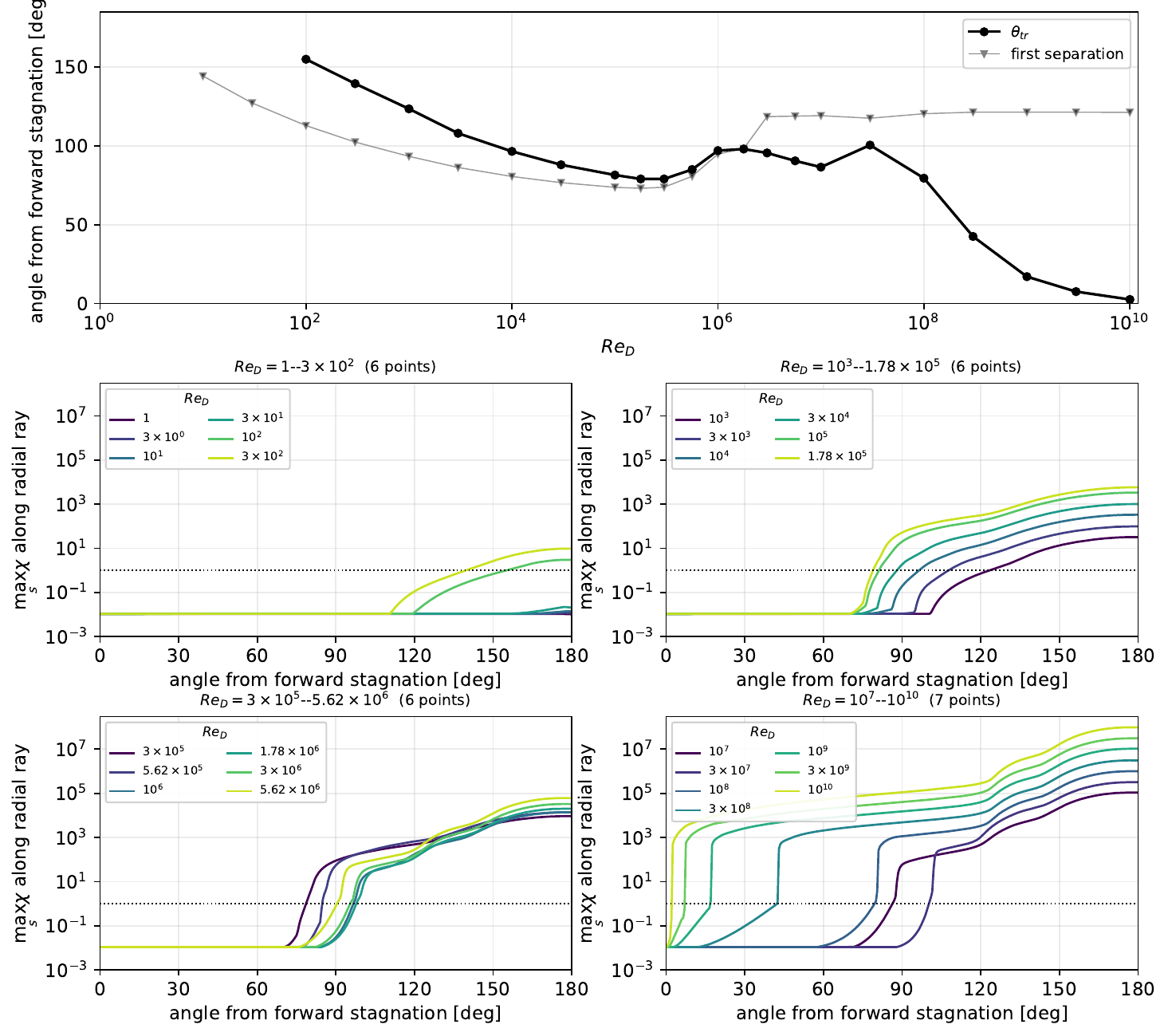}
\caption{$Tu=0.2\%$, $Re_D=1$--$10^{10}$, up-ladder. (a) transition angle
$\theta_{tr}$ ($\chi=1$) and first wall
separation vs $Re_D$ (below the crisis $\theta_{tr}$ is the wake $\chi=1$
front); the front marches from ${\sim}100^\circ$ to ${\sim}2^\circ$ by
$10^{10}$. (b)--(e) $\max_s\chi(\theta)$ for every Reynolds number of the ladder,
split into panes of six or seven over contiguous
ranges.}\label{f:dragcrisisangles}
\end{figure}

\begin{figure}[H]\centering
\includegraphics[width=\textwidth,height=0.90\textheight,keepaspectratio,page=1]{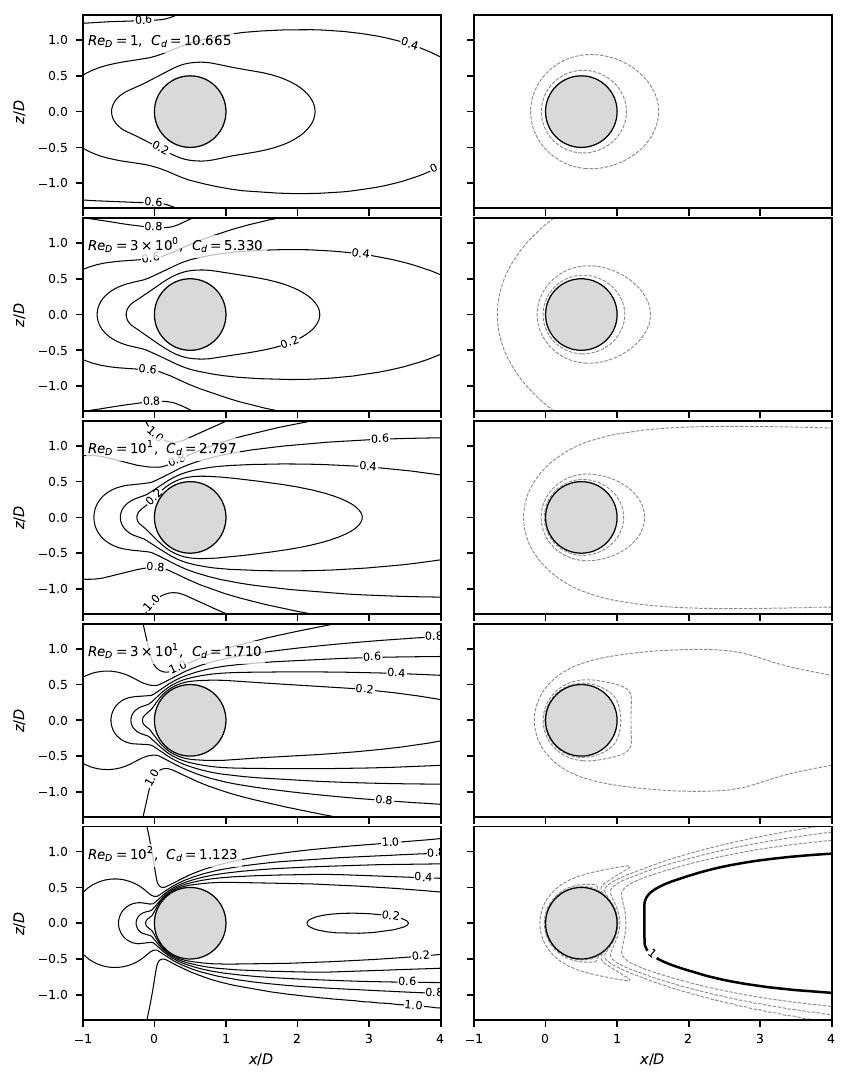}
\caption{Drag-crisis fields, up-ladder $Tu=0.2\%$: every Reynolds
number of Fig.~\ref{f:dragcrisiscd}, five per page in ascending
order. Columns $|\mathbf{u}|/U_\infty$ and $\chi$ (dashed
$\chi<1$, solid $\chi=1$, $c_{v1}$, $30$, $10^2$--$10^4$); each row
carries its own $Re_D$ and $C_d$. $Re_D=1$--$10^2$.}\label{f:dragcrisisfields}
\end{figure}

Figure~\ref{f:dragcrisiscd}: the crisis falls between
$Re_D=3\times10^5$ ($C_d=0.773$) and $1.78\times10^6$ ($0.228$); the
supercritical floor settles near $C_d=0.21$--$0.23$
($1.78\times10^6$--$10^7$) and stays flat to $10^{10}$.
Figure~\ref{f:dragcrisisangles}: transition and separation angles.
Figure~\ref{f:dragcrisisfields}: fields across the traverse.
Table~\ref{t:dragcrisis}: the up-ladder $C_d$. Below shedding onset
($Re_D\approx46$) the steady solution is physical: $C_d=10.67$ at
$Re_D=1$, $2.797$ at $10$ ($-1.7\%$ vs Dennis \& Chang).

\begin{figure}[H]\centering
\includegraphics[width=\textwidth,height=0.90\textheight,keepaspectratio,page=2]{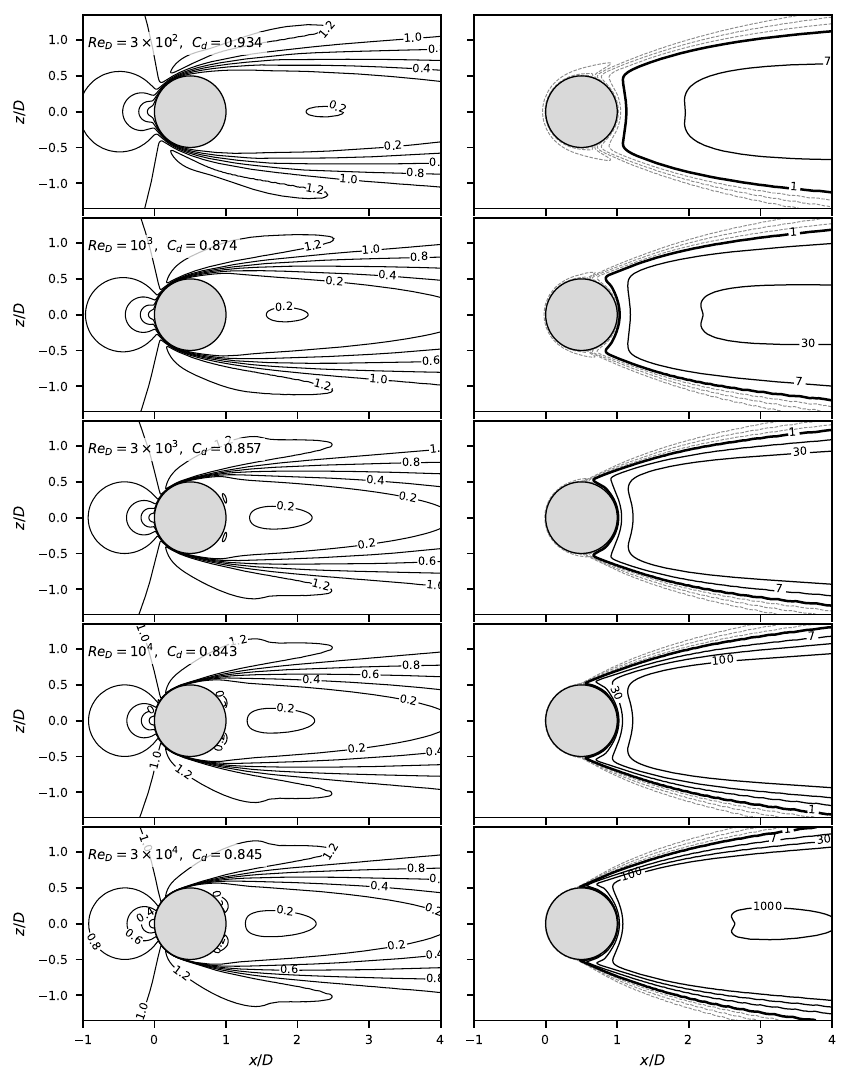}
\figcont{$Re_D=3\times10^2$--$3\times10^4$.}
\end{figure}

\begin{figure}[H]\centering
\includegraphics[width=\textwidth,height=0.90\textheight,keepaspectratio,page=3]{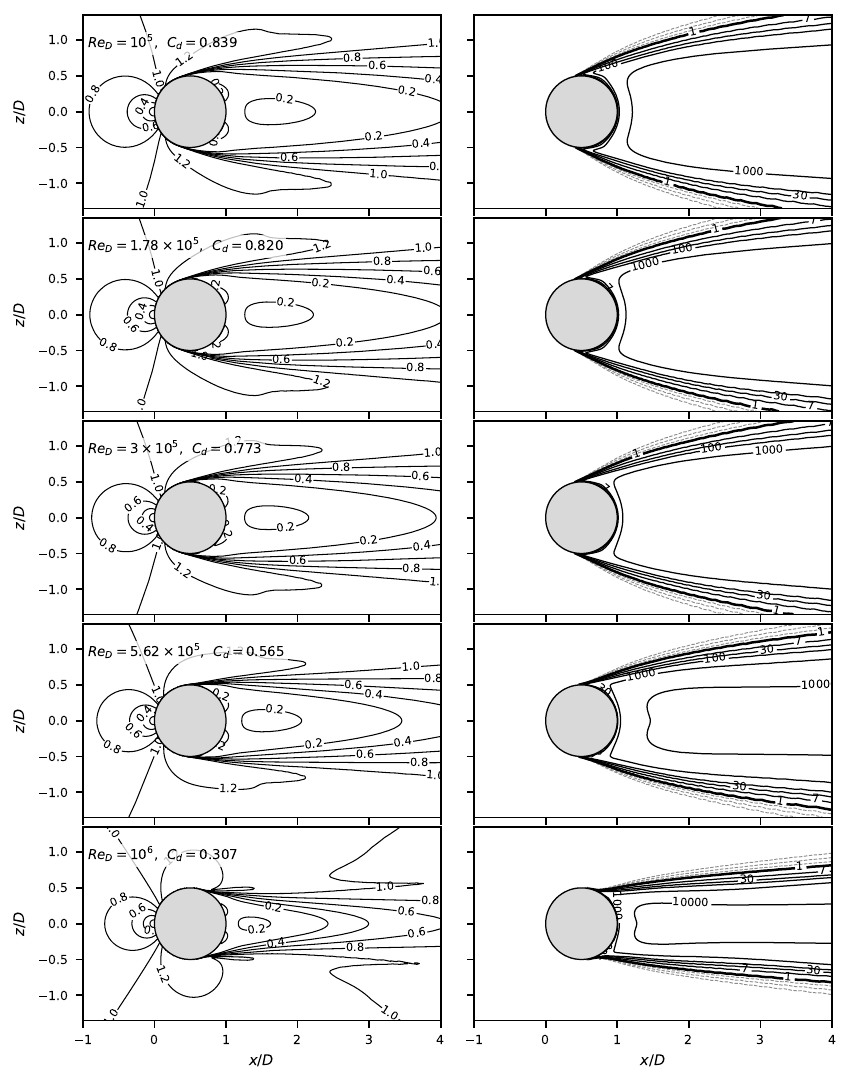}
\figcont{$Re_D=10^5$--$10^6$.}
\end{figure}

\begin{figure}[H]\centering
\includegraphics[width=\textwidth,height=0.90\textheight,keepaspectratio,page=4]{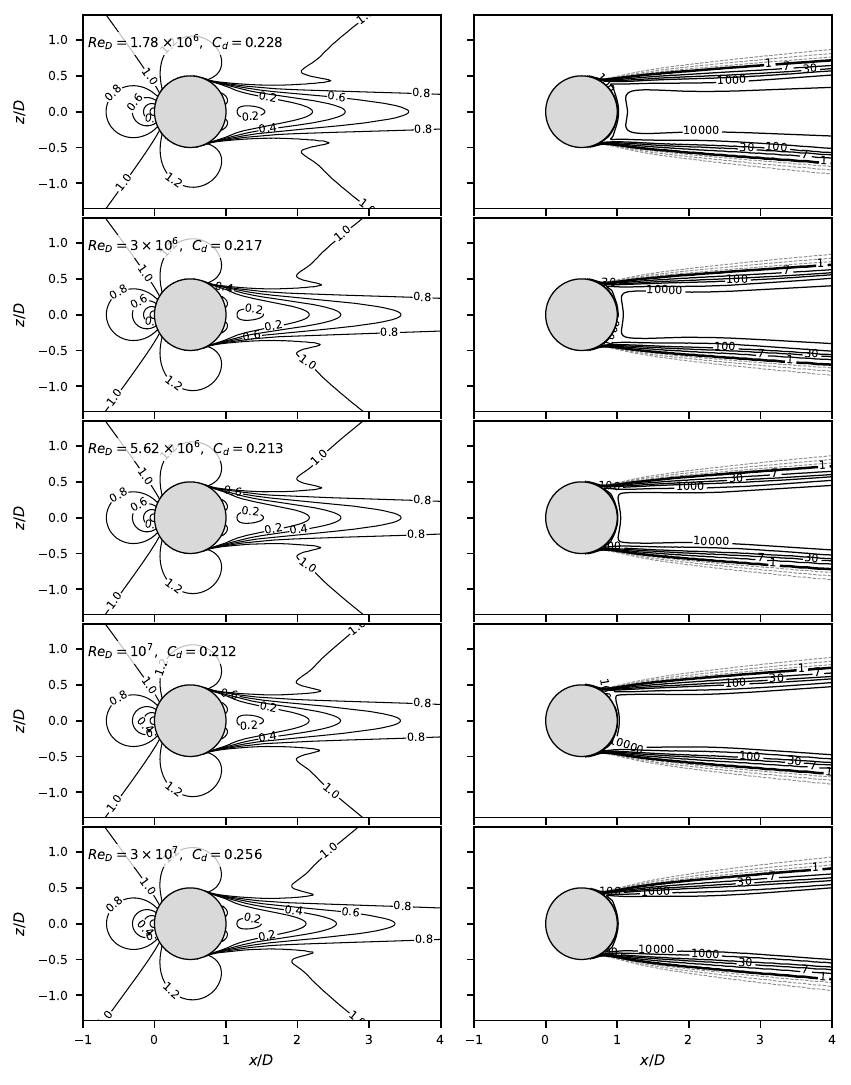}
\figcont{$Re_D=1.78\times10^6$--$3\times10^7$.}
\end{figure}

\begin{figure}[H]\centering
\includegraphics[width=\textwidth,height=0.90\textheight,keepaspectratio,page=5]{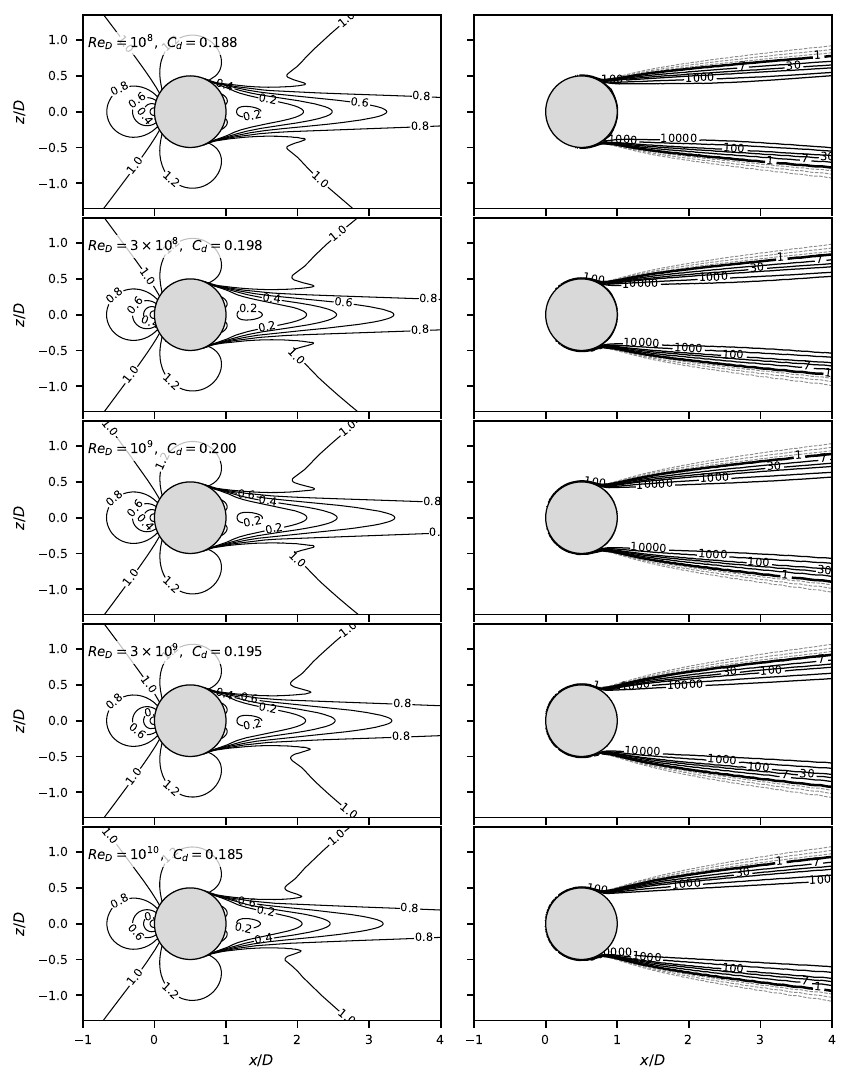}
\figcont{$Re_D=10^8$--$10^{10}$.}
\end{figure}

\begin{figure}[H]\centering
\includegraphics[width=\textwidth,height=0.92\textheight,keepaspectratio]{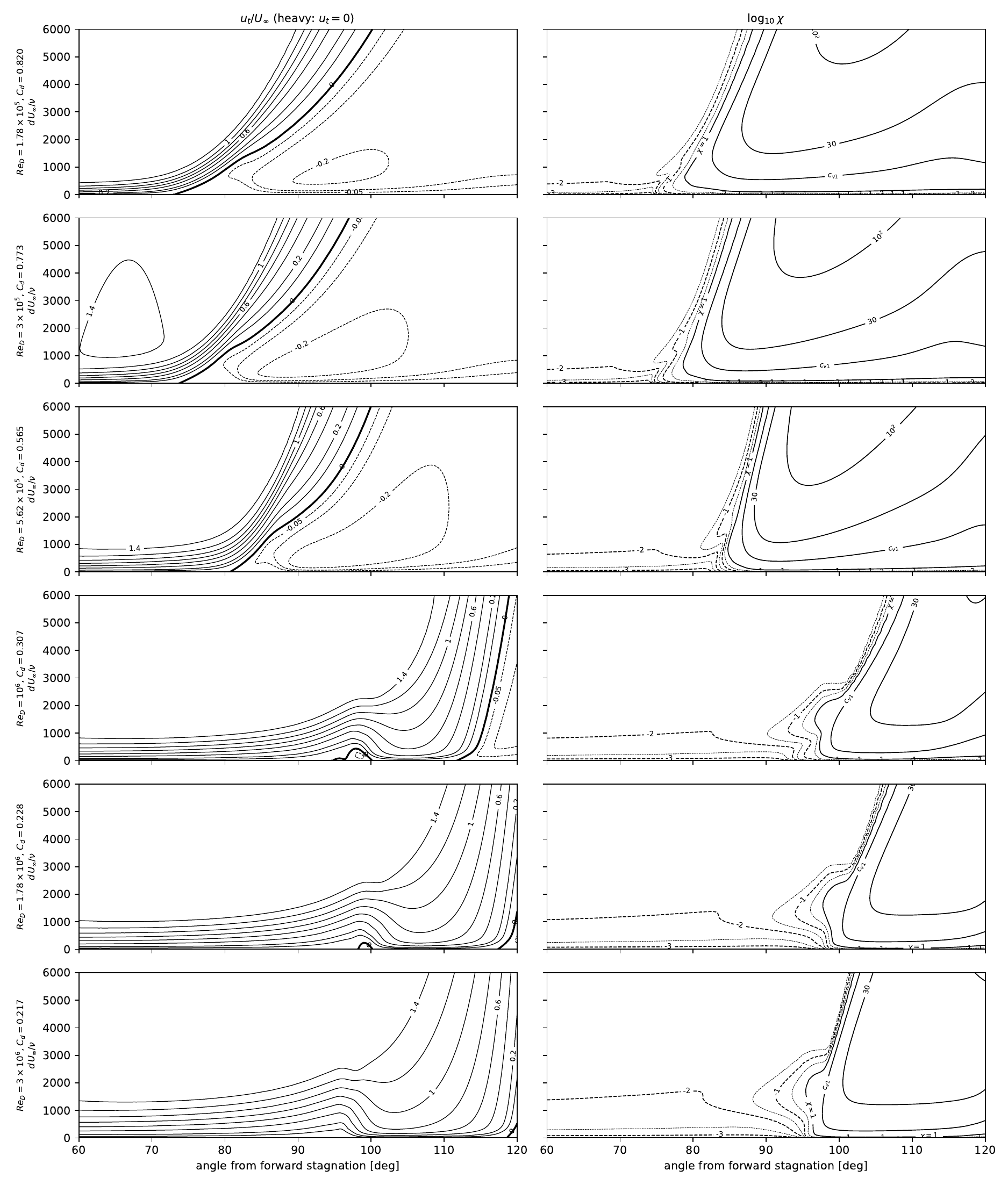}
\caption{The crisis at boundary-layer scale, $Re_D=1.78\times10^5$ to
$3\times10^6$ (rows): $u_t/U_\infty$ left (heavy $u_t=0$),
$\log_{10}\chi$ right, against angle from forward stagnation and wall
distance in viscous units.}\label{f:dragcrisisblsheet}
\end{figure}

\begin{table}[H]
  \centering\small
  \caption{Cylinder steady drag-crisis traverse, systematic $Tu\!=\!0.2\%$ up-ladder ($C_d$ median over the tail window, [peak-to-peak] bracketed where the steady monitor did not accept; $\theta_{tr}$ the $\chi\!=\!1$ transition angle, $\theta_{sep}$ the first separation; $|C_L|\!<\!2\times10^{-6}$ throughout).}
  \label{t:dragcrisis}
  \begin{tabular}{l l c cc}
    \toprule
    $Re_D$ & $y_1/R$ & $C_d$ & $\theta_{tr}\,(^\circ)$ & $\theta_{sep}\,(^\circ)$ \\
    \midrule
    $1$ & $2\times10^{-3}$ & 10.665 & -- & -- \\
    $3$ & $2\times10^{-3}$ & 5.330 & -- & -- \\
    $10$ & $2\times10^{-3}$ & 2.797 & -- & 144.3 \\
    $30$ & $2\times10^{-3}$ & 1.710 & -- & 127.2 \\
    $100$ & $2\times10^{-3}$ & 1.123 & 155.0 & 112.9 \\
    $300$ & $2\times10^{-3}$ & 0.934 & 139.5 & 102.4 \\
    $1000$ & $2\times10^{-3}$ & 0.874 & 123.5 & 93.4 \\
    $3000$ & $1.6\times10^{-5}$ & 0.857 & 108.0 & 86.3 \\
    $10^{4}$ & $1.6\times10^{-5}$ & 0.843 & 96.5 & 80.6 \\
    $3\!\times\!10^{4}$ & $1.6\times10^{-5}$ & 0.845 & 88.0 & 76.7 \\
    $10^{5}$ & $1.6\times10^{-5}$ & 0.839 & 81.5 & 73.7 \\
    $1.78\!\times\!10^{5}$ & $1.6\times10^{-5}$ & 0.820 & 79.0 & 73.0 \\
    $3\!\times\!10^{5}$ & $1.6\times10^{-5}$ & 0.773 & 79.0 & 73.8 \\
    $5.62\!\times\!10^{5}$ & $1.6\times10^{-5}$ & 0.565\,[0.075] & 85.0 & 80.7 \\
    $10^{6}$ & $1.6\times10^{-5}$ & 0.307\,[0.044] & 97.0 & 94.8 \\
    $1.78\!\times\!10^{6}$ & $1.6\times10^{-5}$ & 0.228 & 98.0 & 98.3 \\
    $3\!\times\!10^{6}$ & $2\times10^{-6}$ & 0.217\,[0.010] & 95.5 & 118.6 \\
    $5.62\!\times\!10^{6}$ & $2\times10^{-6}$ & 0.213 & 90.5 & 118.9 \\
    $10^{7}$ & $2\times10^{-6}$ & 0.212 & 86.5 & 119.2 \\
    $3\!\times\!10^{7}$ & $2\times10^{-9}$ & 0.256\,[0.171] & 100.5 & 117.6 \\
    $10^{8}$ & $2\times10^{-9}$ & 0.188 & 79.5 & 120.4 \\
    $3\!\times\!10^{8}$ & $2\times10^{-9}$ & 0.198 & 42.5 & 121.4 \\
    $10^{9}$ & $2\times10^{-9}$ & 0.200 & 17.0 & 121.4 \\
    $3\!\times\!10^{9}$ & $2\times10^{-9}$ & 0.195 & 7.5 & 121.3 \\
    $10^{10}$ & $2\times10^{-9}$ & 0.185\,[0.009] & 2.5 & 121.3 \\
    \bottomrule
  \end{tabular}
\end{table}

\subsection{The Daedalus wing}

\begin{figure}[H]\centering
\includegraphics[width=0.99\textwidth]{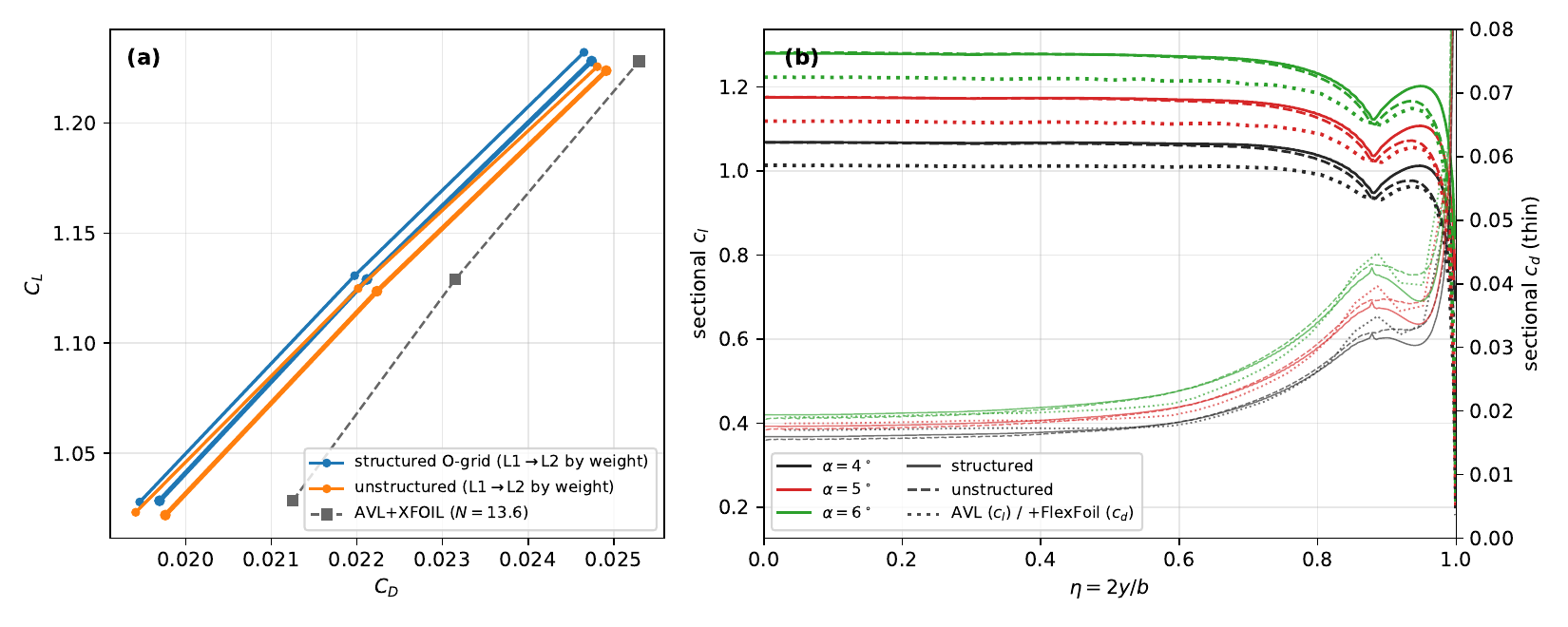}
\caption{Daedalus polar and sections. Families agree at L1 to
$0.46$--$0.53\%$ in $C_L$ and $0.4$--$1.6$ counts in $C_D$, at L0 to
$0.74$--$0.97\%$ and $18$--$21$ counts; taper break at
$\eta=0.88$.}\label{f:daepolar}
\end{figure}

\begin{figure}[H]\centering
\includegraphics[width=0.99\textwidth]{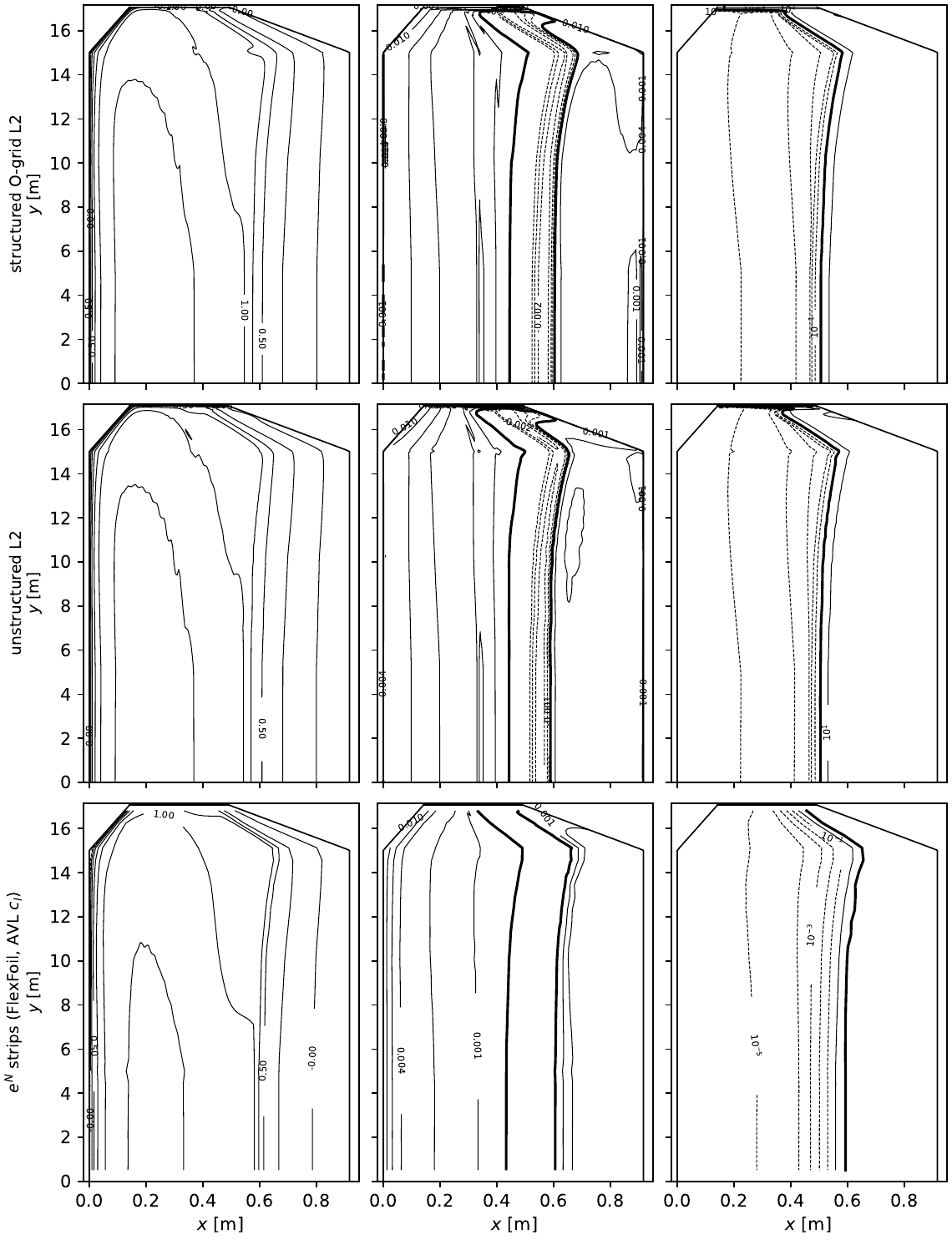}
\caption{Daedalus surface maps, $\alpha=4^\circ$,
$Re_\mathrm{root}=5\times10^5$.}\label{f:daesurf}
\end{figure}

The Daedalus human-powered-aircraft wing: half-span $17.07\,$m, root
chord $0.917\,$m, aspect ratio $37.8$, DAE-11/21/31 sections;
$Re_\mathrm{root}=5\times10^5$,
$\alpha\in\{4^\circ,5^\circ,6^\circ\}$ ($C_L\approx1.02$--$1.23$),
flight-quiet seed $\chi_\infty=8.76\times10^{-6}$
($N_\mathrm{crit}\approx13.6$); two mesh families
(spanwise-stacked structured O-grid, $0.72$--$36.7$M nodes;
unstructured prism/tet/hex, $1.70$--$54.9$M nodes) at three levels,
eighteen solutions; the reference is AVL with strip-wise XFOIL
section polars at matched local $c_l$ and $N_\mathrm{crit}$, trimmed
to the RANS lift. Figure~\ref{f:daepolar}: the families agree at L2
to $0.35$--$0.63\%$ in $C_L$ and
$0.7$--$1.7$ counts in $C_D$, and the finest-grid drag runs
$15.6$/$10.3$/$5.6$ counts ($7.3$/$4.4$/$2.2\%$) below the
matched-lift reference at $\alpha=4^\circ$/$5^\circ$/$6^\circ$.
Figure~\ref{f:daesurf}: surface maps at $\alpha=4^\circ$ against the
$e^N$ strips---at
$\eta=0.31$, separation within $0.017\,c$ and reattachment within
$0.009\,c$ at every incidence, the $\chi=1$ front sits
$\sim0.09\,c$ ahead of the strips' $N_\mathrm{crit}$ contour, and the
tip vortex trips transition to the leading
edge for $\eta\gtrsim0.992$. Table~\ref{t:daetotals}: totals by
level ($\alpha=5^\circ$, $6^\circ$ surface maps and section sheets
below).

\begin{table}[H]
  \centering\small
  \caption{Daedalus wing totals by grid level: SA-AI on both mesh
  families against the AVL$+$XFOIL reference trimmed to the RANS lift
  at each incidence.}
  \label{t:daetotals}
  \begin{tabular}{cl cc cc}
    \toprule
    & & \multicolumn{2}{c}{structured O-grid} & \multicolumn{2}{c}{unstructured}\\
    \cmidrule(lr){3-4}\cmidrule(lr){5-6}
    $\alpha$ & level & $C_L$ & $C_D$ & $C_L$ & $C_D$\\
    \midrule
    $4^\circ$ & L0 & 1.0244 & 0.01958 & 1.0156 & 0.02134\\
              & L1 & 1.0279 & 0.01946 & 1.0232 & 0.01941\\
              & L2 & 1.0284 & 0.01969 & 1.0219 & 0.01976\\
              & AVL$+$XFOIL & 1.0284 & 0.02125 & &\\
    \midrule
    $5^\circ$ & L0 & 1.1254 & 0.02219 & 1.1145 & 0.02427\\
              & L1 & 1.1307 & 0.02197 & 1.1248 & 0.02201\\
              & L2 & 1.1290 & 0.02212 & 1.1236 & 0.02223\\
              & AVL$+$XFOIL & 1.1290 & 0.02315 & &\\
    \midrule
    $6^\circ$ & L0 & 1.2262 & 0.02505 & 1.2171 & 0.02713\\
              & L1 & 1.2321 & 0.02465 & 1.2256 & 0.02481\\
              & L2 & 1.2282 & 0.02474 & 1.2238 & 0.02491\\
              & AVL$+$XFOIL & 1.2282 & 0.02530 & &\\
    \bottomrule
  \end{tabular}
\end{table}

\clearpage
\begin{figure}[H]\centering
\includegraphics[width=0.99\textwidth,height=0.94\textheight,keepaspectratio]{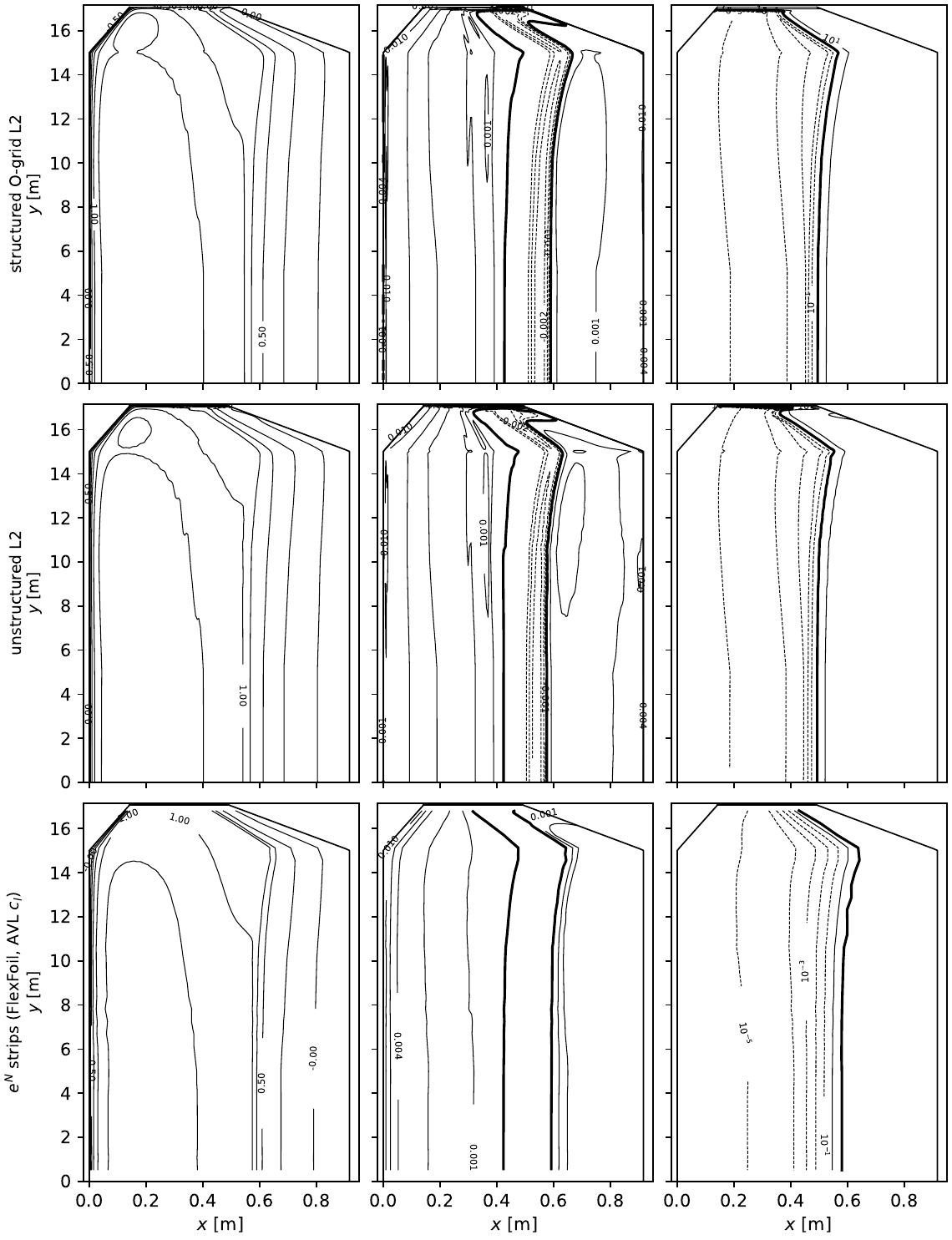}
\figcont{$\alpha=5^\circ$.}
\end{figure}
\begin{figure}[H]\centering
\includegraphics[width=0.99\textwidth,height=0.94\textheight,keepaspectratio]{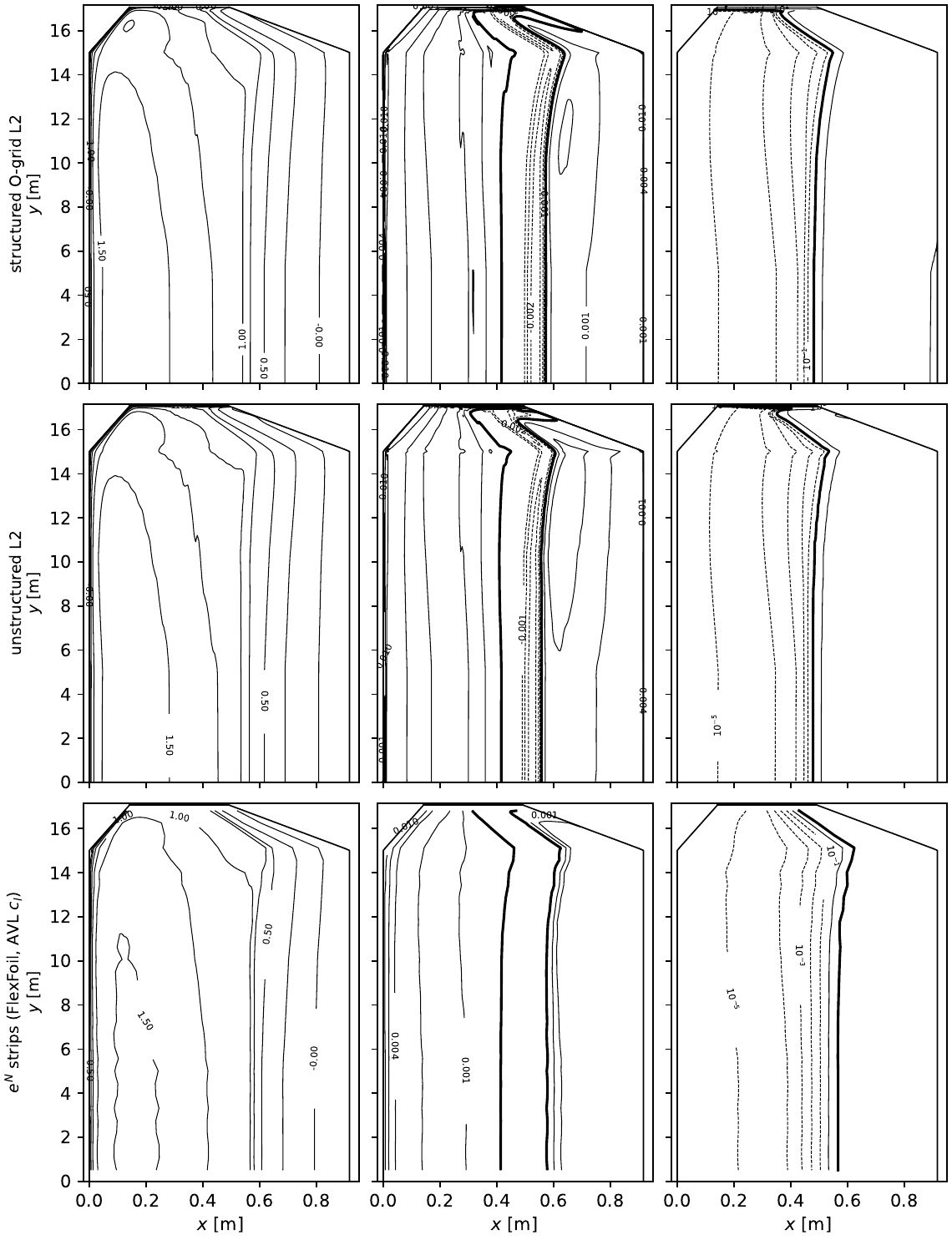}
\figcont{$\alpha=6^\circ$.}
\end{figure}
\begin{figure}[H]\centering
\includegraphics[width=0.99\textwidth,height=0.94\textheight,keepaspectratio]{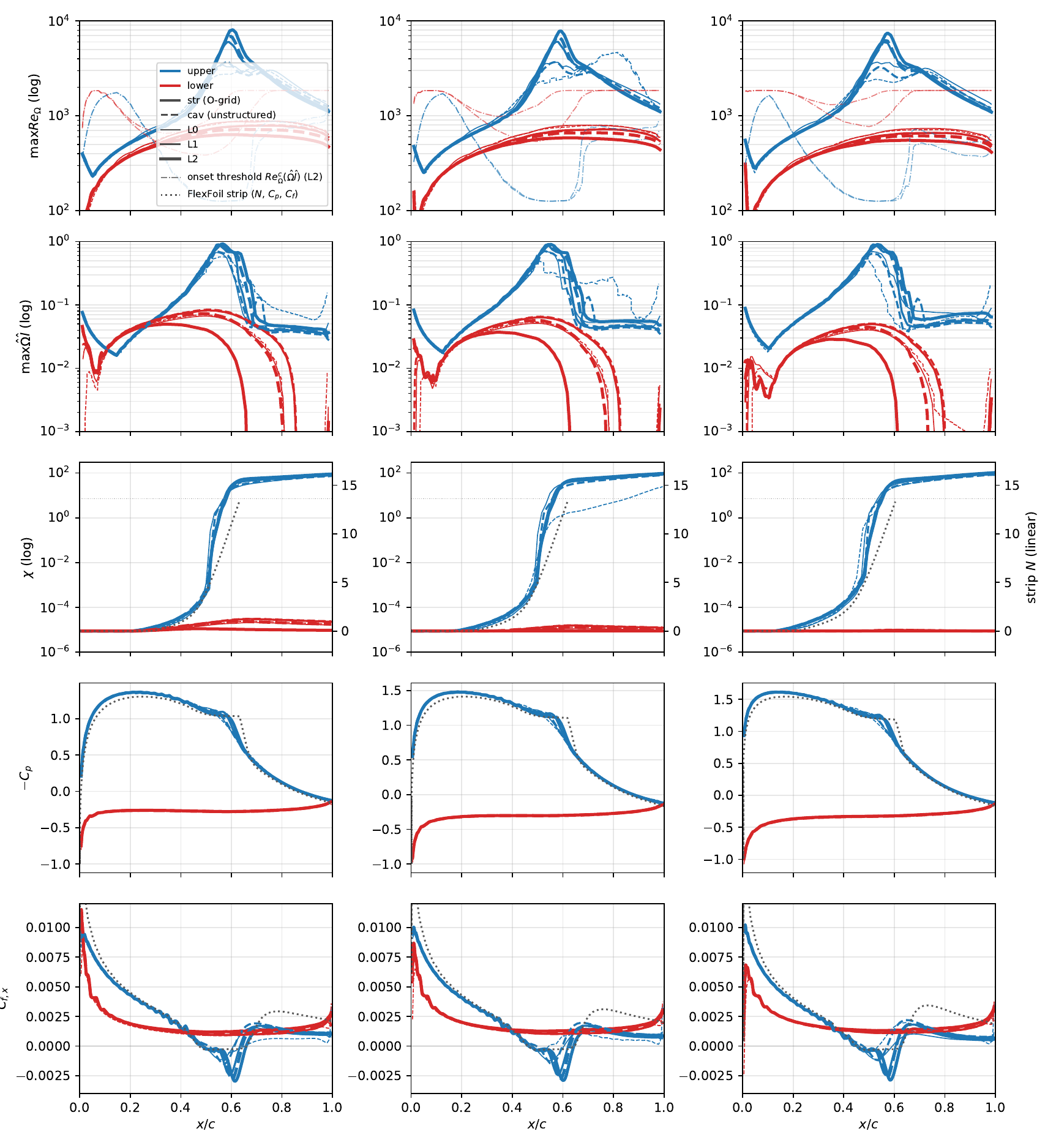}
\caption{Daedalus wing section cuts, all six grids against the AVL-trimmed
XFOIL strip, at $\alpha=4^\circ$, $5^\circ$, $6^\circ$ (columns).
$\eta=0.10$.}\label{f:daesections}
\end{figure}
\begin{figure}[H]\centering
\includegraphics[width=0.99\textwidth,height=0.94\textheight,keepaspectratio]{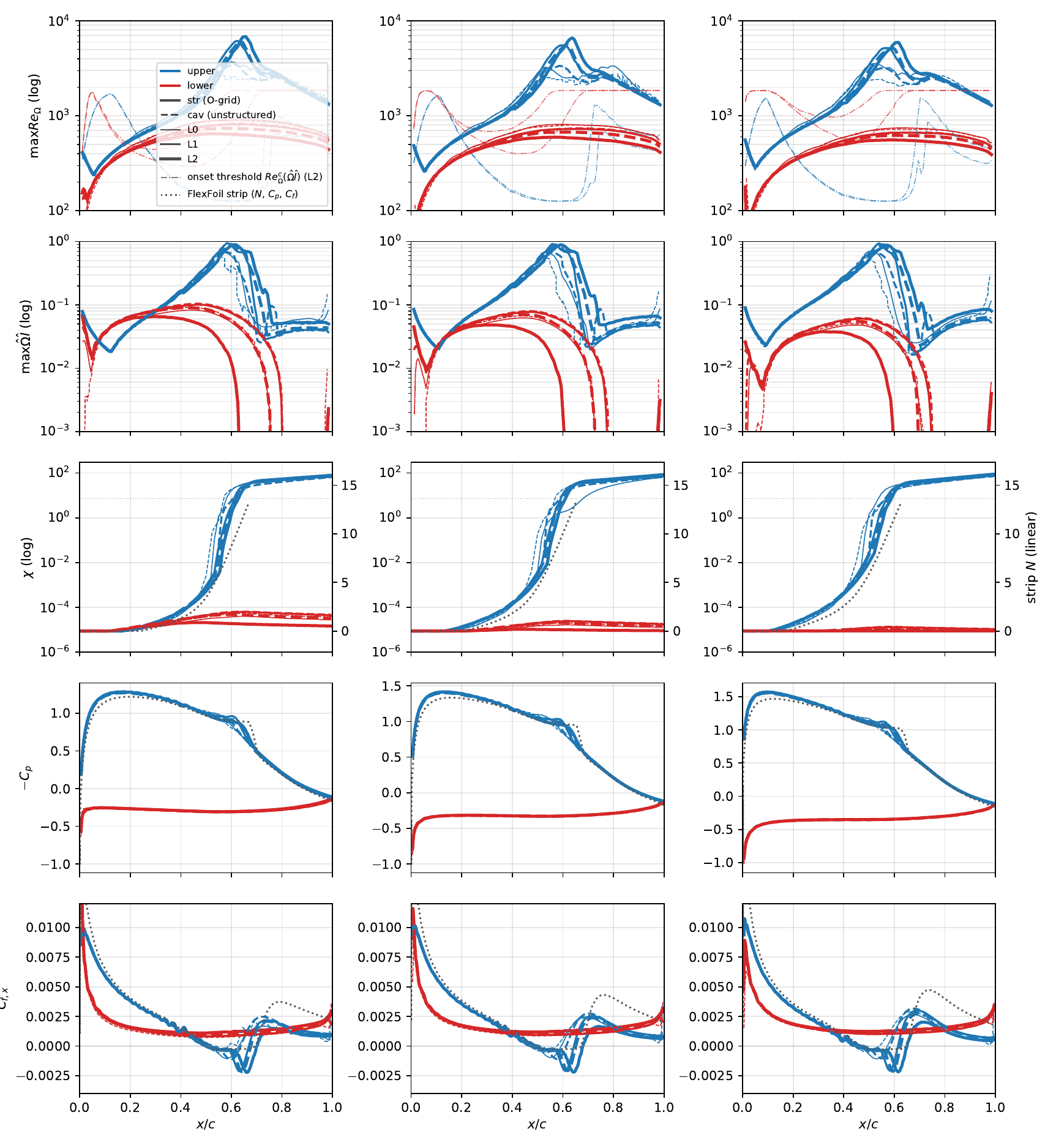}
\figcont{$\eta=0.75$.}
\end{figure}
\begin{figure}[H]\centering
\includegraphics[width=0.99\textwidth,height=0.94\textheight,keepaspectratio]{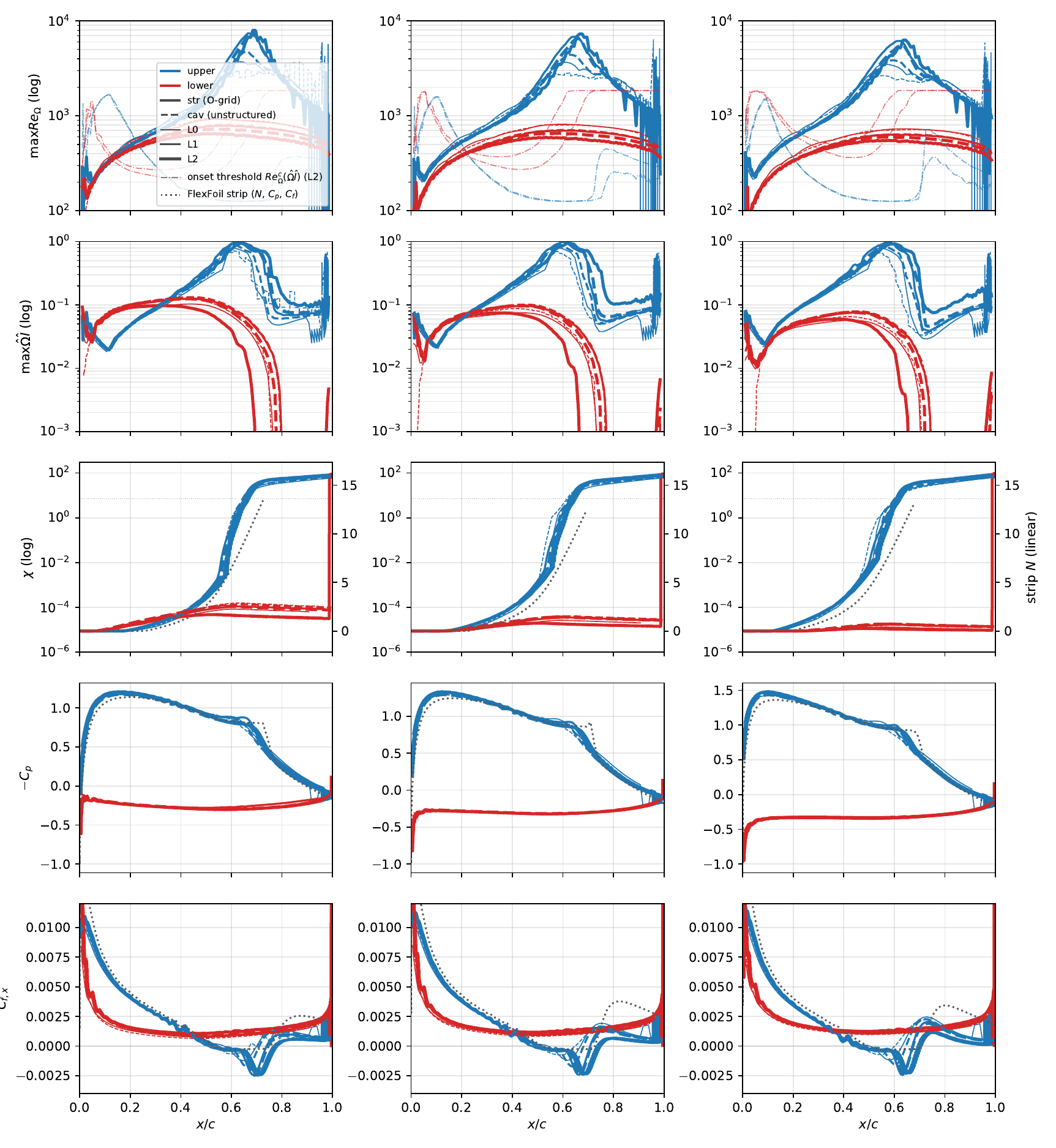}
\figcont{$\eta=0.92$.}
\end{figure}
\begin{figure}[H]\centering
\includegraphics[width=0.99\textwidth,height=0.94\textheight,keepaspectratio]{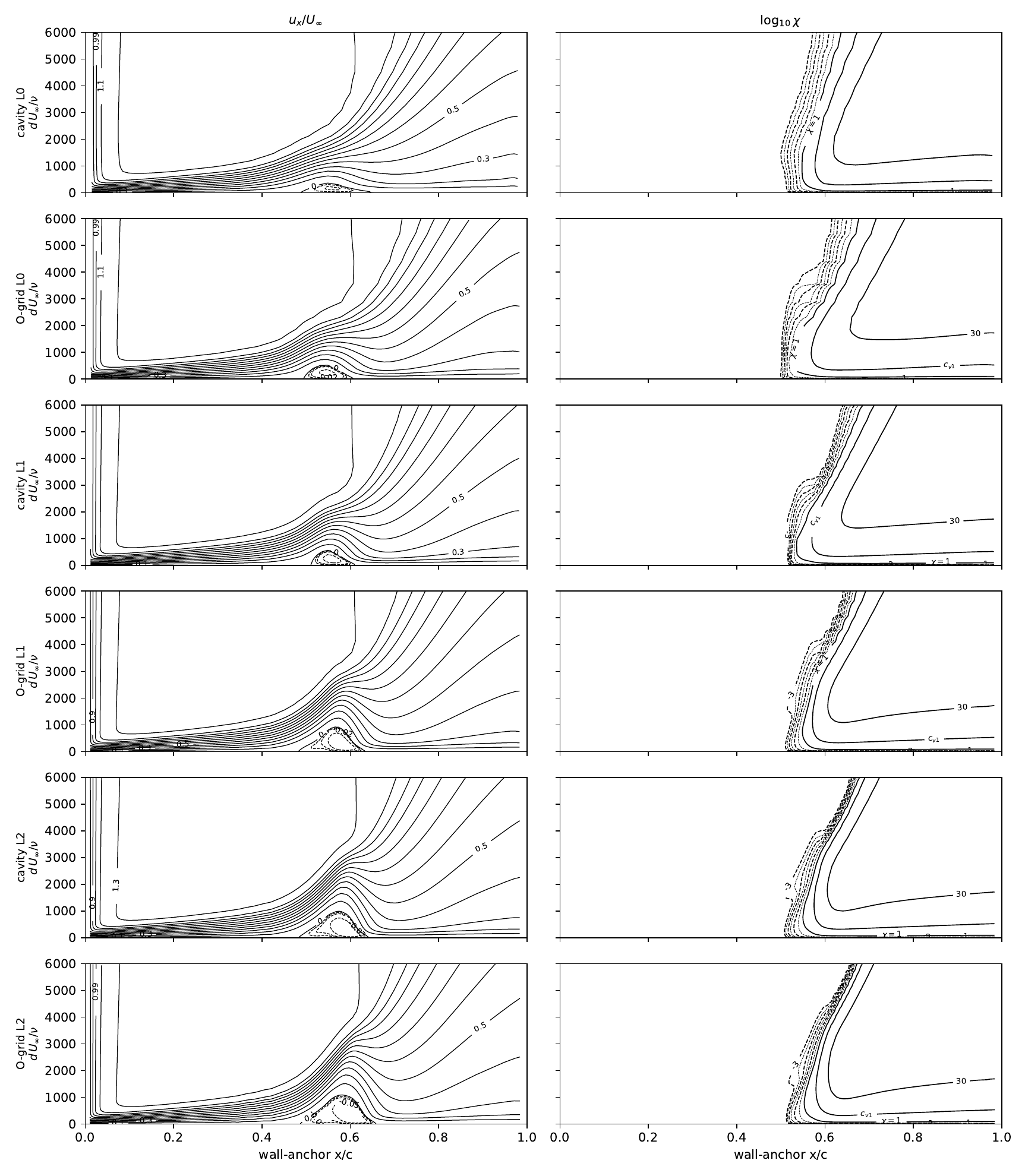}
\caption{Daedalus wing wall-normal sheets, upper surface: $u_x/U_\infty$
(left) and $\log_{10}\chi$ (right) on the archived meshes.
$\alpha=4^\circ$, $\eta=0.10$.}\label{f:daesheets}
\end{figure}
\begin{figure}[H]\centering
\includegraphics[width=0.99\textwidth,height=0.94\textheight,keepaspectratio]{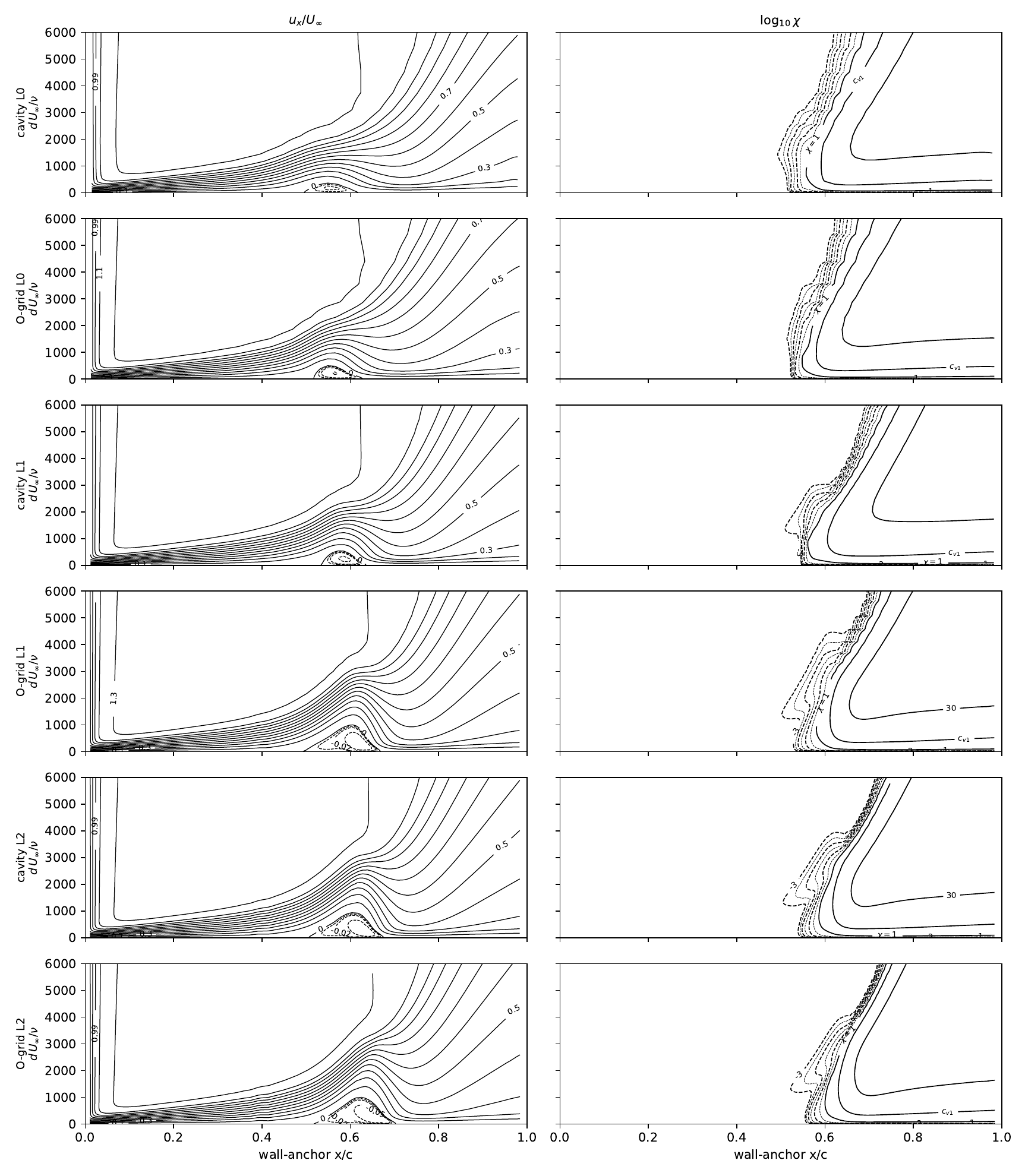}
\figcont{$\alpha=4^\circ$, $\eta=0.75$.}
\end{figure}
\begin{figure}[H]\centering
\includegraphics[width=0.99\textwidth,height=0.94\textheight,keepaspectratio]{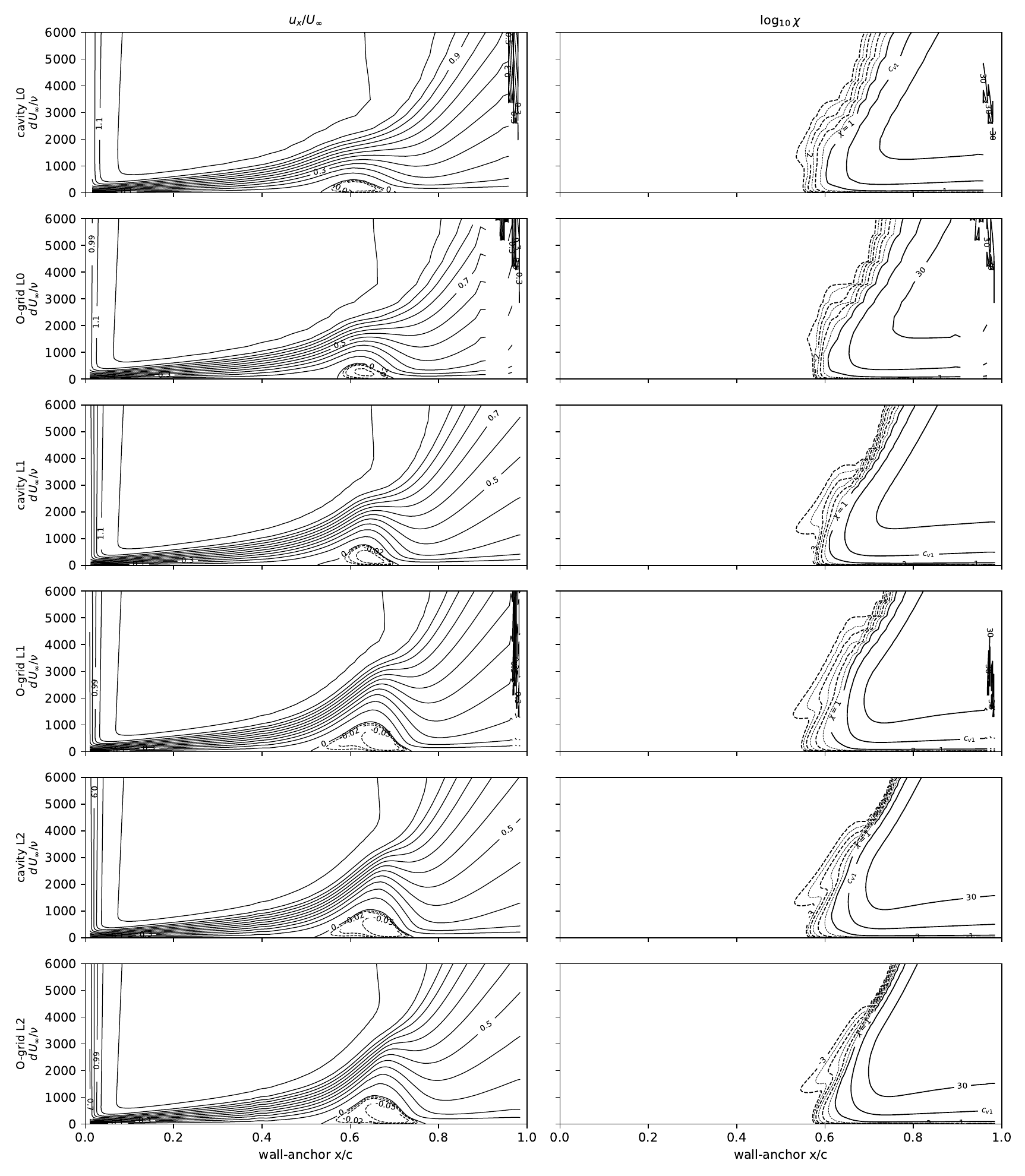}
\figcont{$\alpha=4^\circ$, $\eta=0.92$.}
\end{figure}
\begin{figure}[H]\centering
\includegraphics[width=0.99\textwidth,height=0.94\textheight,keepaspectratio]{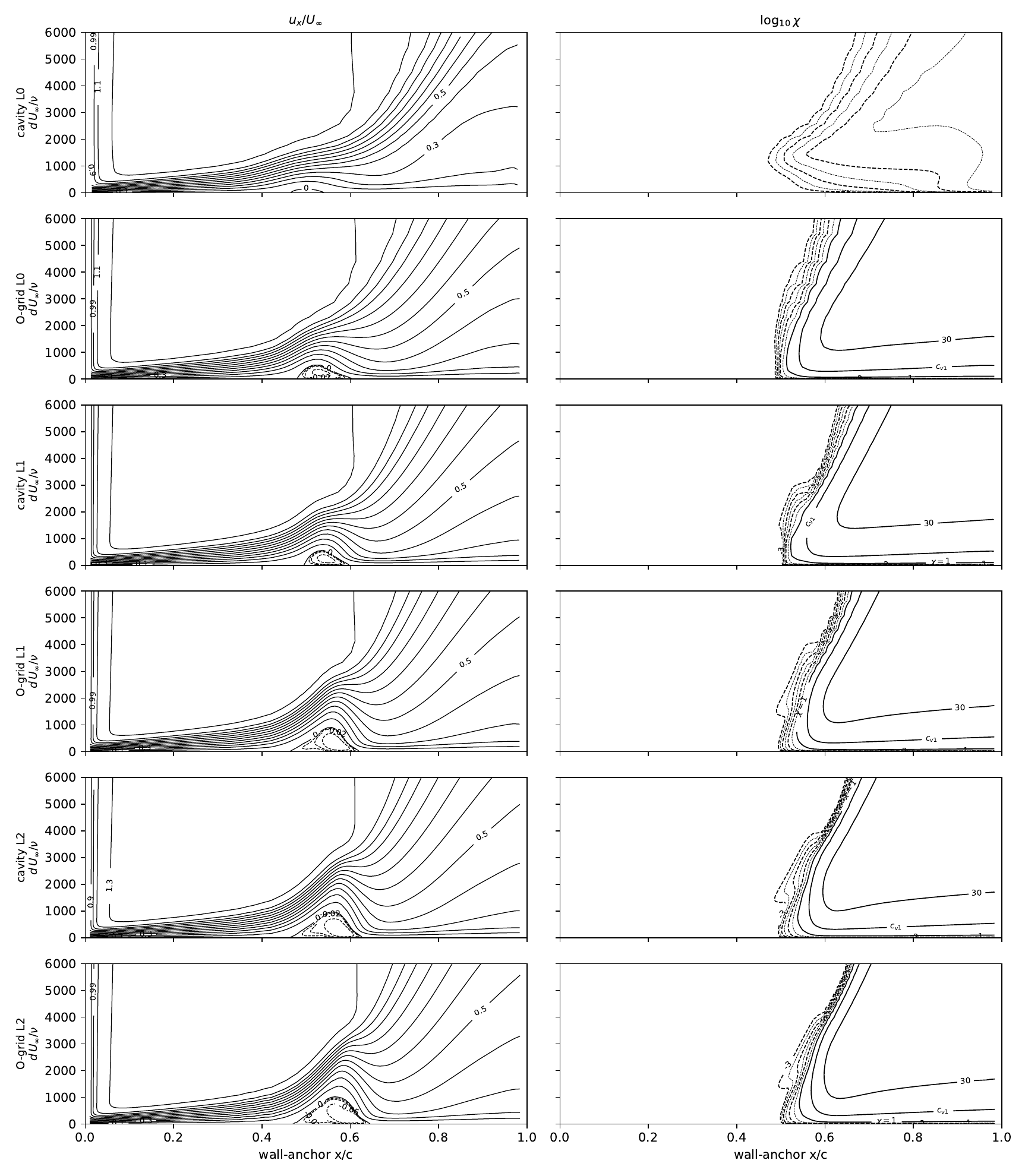}
\figcont{$\alpha=5^\circ$, $\eta=0.10$.}
\end{figure}
\begin{figure}[H]\centering
\includegraphics[width=0.99\textwidth,height=0.94\textheight,keepaspectratio]{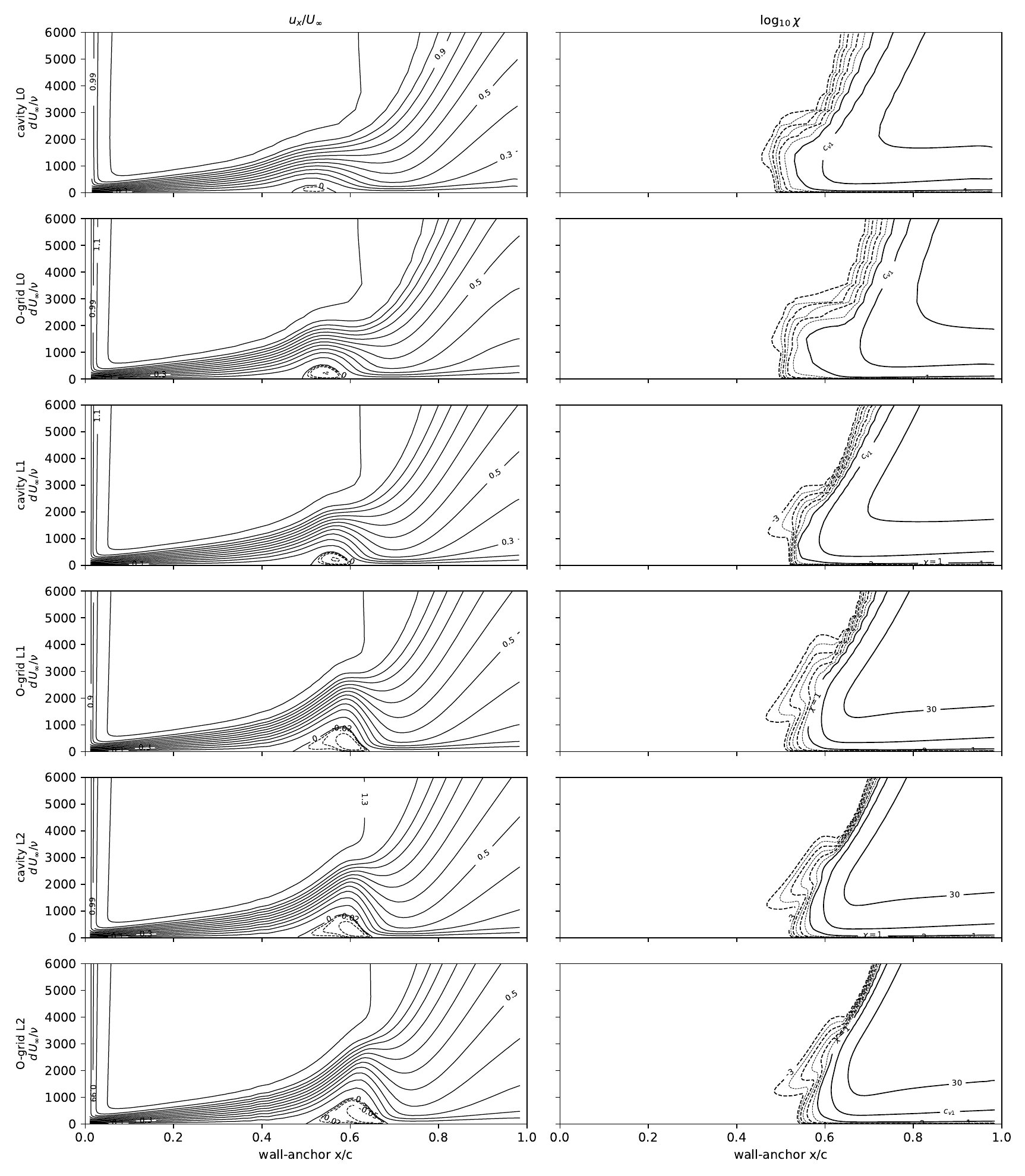}
\figcont{$\alpha=5^\circ$, $\eta=0.75$.}
\end{figure}
\begin{figure}[H]\centering
\includegraphics[width=0.99\textwidth,height=0.94\textheight,keepaspectratio]{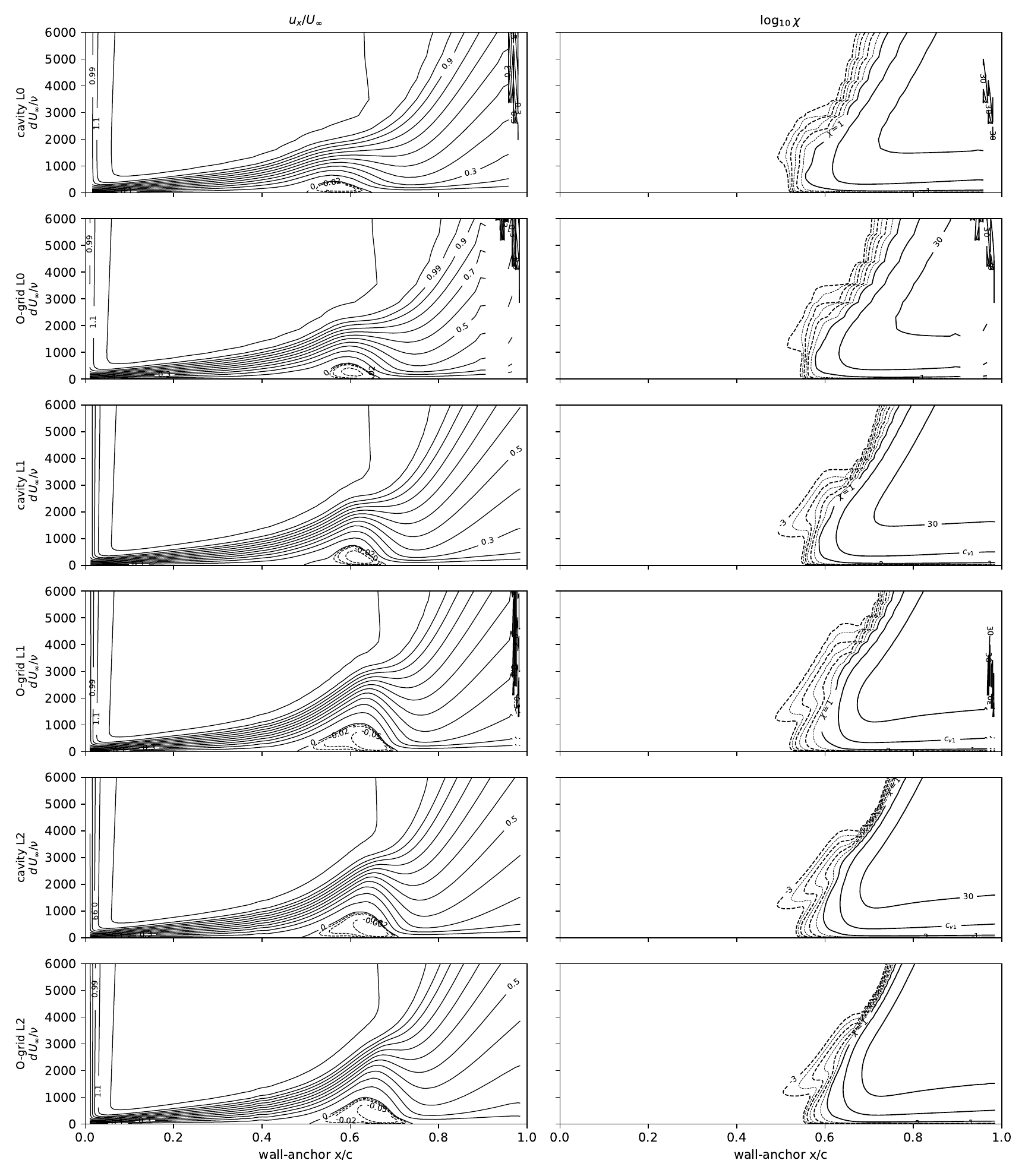}
\figcont{$\alpha=5^\circ$, $\eta=0.92$.}
\end{figure}
\begin{figure}[H]\centering
\includegraphics[width=0.99\textwidth,height=0.94\textheight,keepaspectratio]{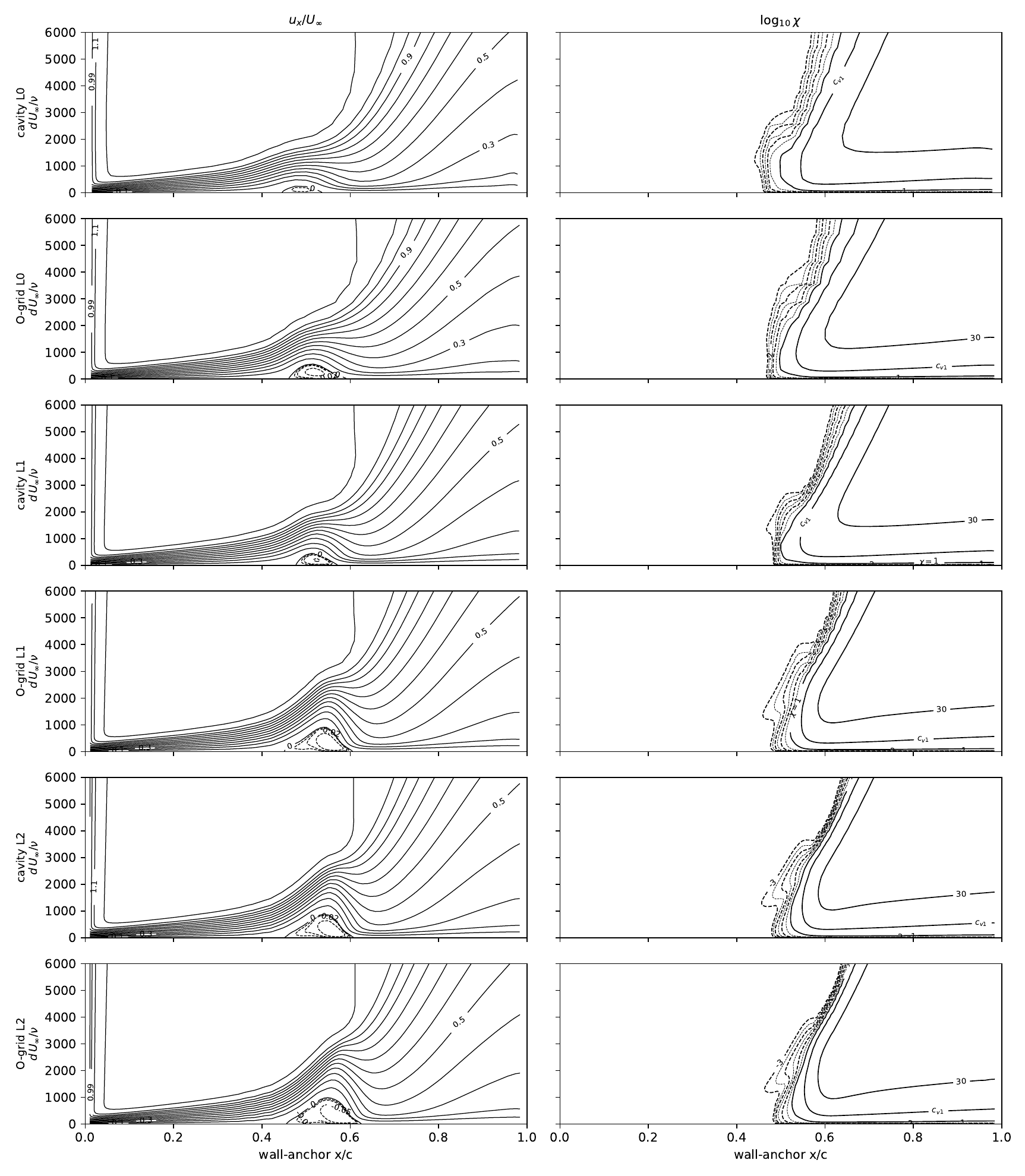}
\figcont{$\alpha=6^\circ$, $\eta=0.10$.}
\end{figure}
\begin{figure}[H]\centering
\includegraphics[width=0.99\textwidth,height=0.94\textheight,keepaspectratio]{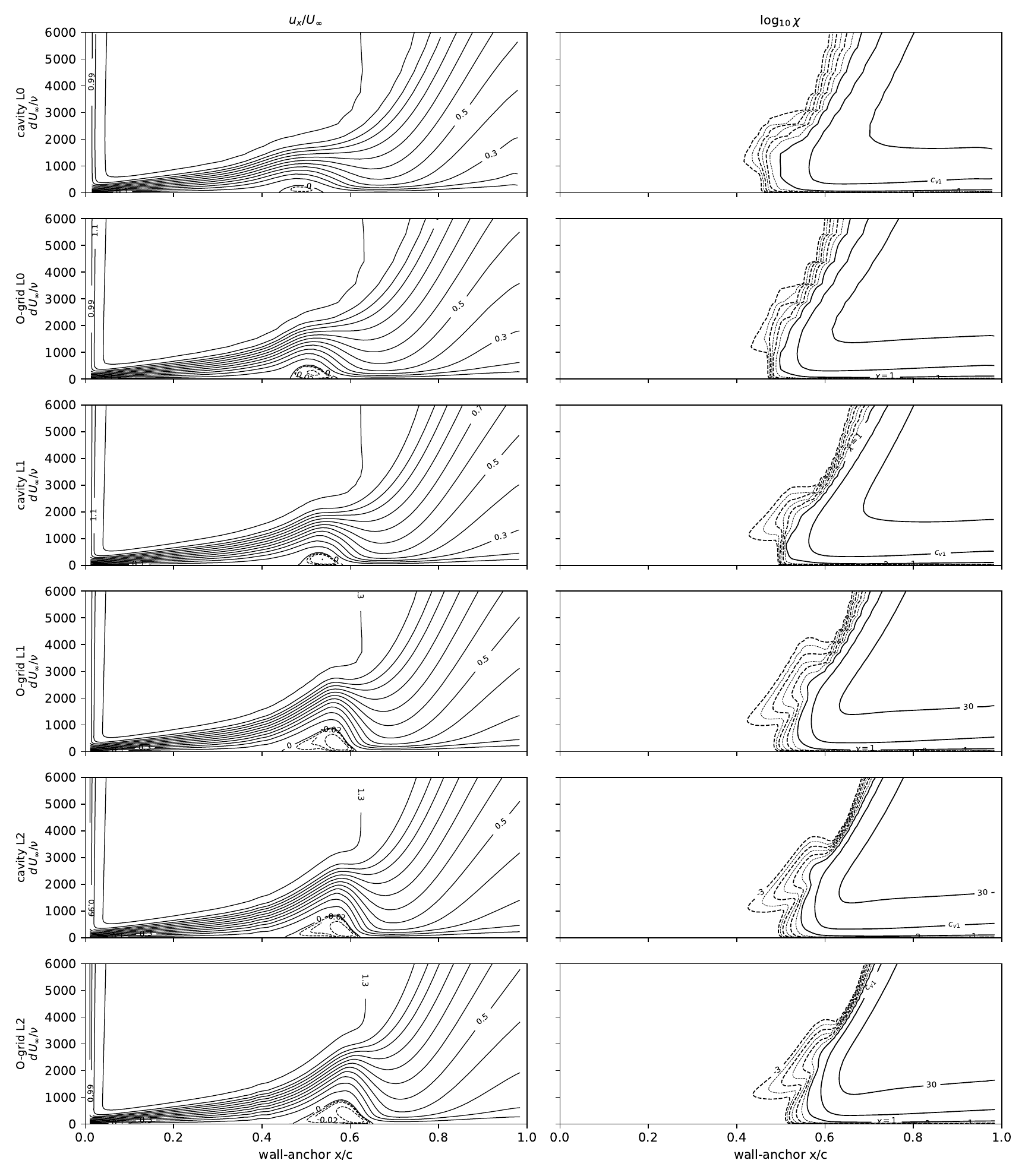}
\figcont{$\alpha=6^\circ$, $\eta=0.75$.}
\end{figure}
\begin{figure}[H]\centering
\includegraphics[width=0.99\textwidth,height=0.94\textheight,keepaspectratio]{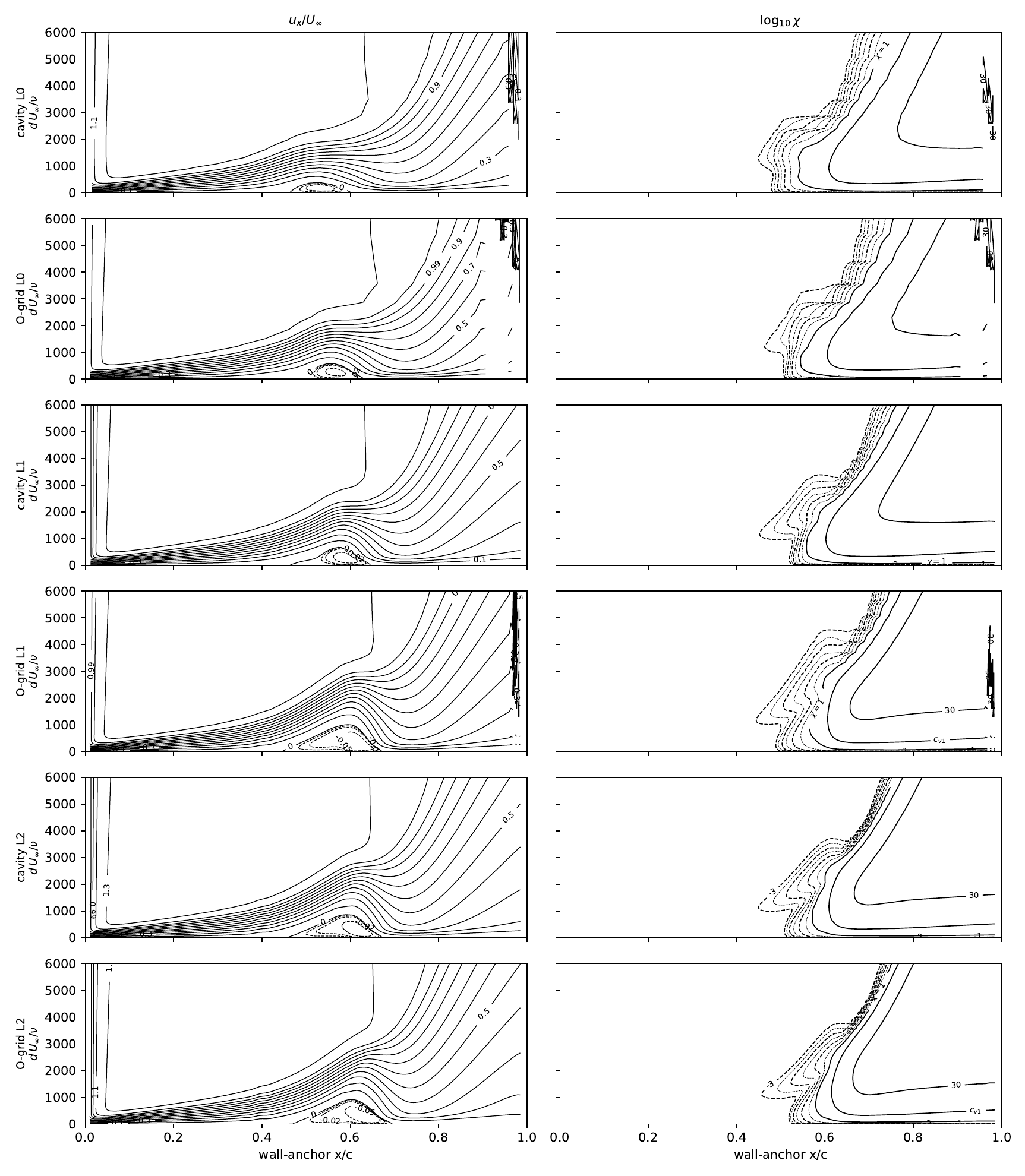}
\figcont{$\alpha=6^\circ$, $\eta=0.92$.}
\end{figure}

\clearpage
\subsection{The 6:1 spheroid}

Analytic spheroid ($x^2/A^2+r^2/B^2=1$, $A=L/2$, $B=L/12$, half-model) on
the L1 structured O-grid ($2.2$M nodes, first wall spacing
$1.5\times10^{-6}L$), campaign constants and kernel unchanged, at the
DFVLR/ONERA measured conditions. Mach is the tunnel's: the Göttingen
$3\times3$~m tunnel is atmospheric, so $M=1.994\times10^{-8}\,Re_L$
(calibrated on its two published $(Re_L,U_\infty)$ pairs, $7.70\times10^6$ at
$55.0$~m/s and $6.54\times10^6$ at $45.0$~m/s, reproduced to $2\%$), i.e.
$M=0.030$ at $Re_L=1.52\times10^6$ to $0.144$ at $7.2\times10^6$; the
pressurised ONERA F1 tunnel does not have $M$ fixed by $Re_L$, so its
condition is run at its Göttingen twin's Mach. $45\,000$ cold-start
pseudo-steps, seed applied in the freestream and the far-field boundary with
a cold-start freestream initial condition. Each condition at the
facility-calibrated seed (limiting $N_\mathrm{TS}=8.0$ Göttingen, $7.0$
ONERA F1) and at the facility-measured hot-wire level (Göttingen $0.33\%$
streamwise, ONERA F1 ${<}0.1\%$).

Table~\ref{tab:sphtunnel}: computed near-wall $\chi=1$ front minus measured,
azimuth by azimuth, low Reynolds number first. At $Re_L=1.52\times10^6$ the
layer runs laminar to the open free-vortex-layer separation at every measured
incidence, so the measured front is that separation line: $\alpha=5^\circ$,
rms $0.028$ at the calibrated seed over the four azimuths that resolve, with
no front at all at the four most leeward stations ($\chi$ never reaches $1$
within $x/L\le0.97$); $\alpha=10^\circ$, rms $0.027$ at the measured seed,
$8/8$ azimuths; $\alpha=29.7^\circ$, $Re_L=1.53\times10^6$ (the largest
incidence measured at this Reynolds number, open free-vortex-layer separation
over most of the leeward half), $12/12$ azimuths, mean $+0.041$ rms $0.095$ at
the calibrated seed and $-0.048$ rms $0.098$ at the measured level, with the
residual smallest leeward ($-0.024$, $-0.045$) and largest windward ($+0.264$,
$+0.185$). At the high Reynolds numbers, $\alpha=0$,
$Re_L=7.2\times10^6$: $+0.077$ and $-0.097$ at the calibrated and measured
seed, uniform to $\pm0.001$ across seventeen azimuths. $\alpha=2.5^\circ$:
$+0.016$ leeward, $+0.132$ windward, against a measured azimuthal spread of
$0.018$. $\alpha=5^\circ$: $-0.088$ to $+0.246$ at $6.49\times10^6$.
$\alpha=10^\circ$, $6.56\times10^6$: the windward residual is $+0.375$ to
$+0.549$. Seed swing in the windward residual between the two facility
levels: $0.174$ at $\alpha=0$, $0.181$ at $2.5^\circ$, $0.239$ at
$5^\circ$ ($6.49\times10^6$), $0.012$ at $10^\circ$ ($6.56\times10^6$),
$0.000$ on the ONERA twin.

\begin{table}[H]
  \centering\small
  \caption{6:1 prolate spheroid at the tunnel conditions: computed near-wall $\chi\!=\!1$ front minus the measured front, at every azimuth carrying a measurement, on the L1 O-grid at the measured Mach number. Each condition is run at the facility-calibrated seed (the tunnel-calibrated limiting $N_{TS}$) and at the facility-measured hot-wire level. $\Delta$ lee / $\Delta$ wind are the residuals at the most leeward and most windward measured azimuth ($\phi\!=\!0$ windward). The $n$ column is the number of measured azimuths at which the computed front was found, over the number measured: where they differ the near-wall $\chi$ never reaches $1$ within $x/L\!\le\!0.97$, i.e. the computed layer is still laminar at the tail and the residual there is a bound, not a value. The last two columns are the same residual for Stocks own two-N-factor $e^N$ computation, digitized from the printed panels at the same azimuths; at $Re_L\!=\!1.52\!\times\!10^6$ his computed transition line is the laminar separation line. Rows are grouped by Reynolds number: the low-$Re$ band first, where the boundary layer runs laminar to the open free-vortex-layer separation at every measured incidence, then the high-$Re$ band where transition is real; within each band the walk is by incidence.}
  \label{tab:sphtunnel}
  \begin{tabular}{cc l l c c cccc cc}
    \toprule
    $\alpha$ & $Re_L$ & tunnel & seed & $Tu$ & $n$ & \multicolumn{4}{c}{SA-AI $-$ measured} & \multicolumn{2}{c}{Stock $-$ measured} \\
    & $[10^6]$ & & & [\%] & & mean & rms & lee & wind & mean & rms \\
    \midrule
    $5^\circ$ & $1.52$ & G\"ottingen & measured & 0.330 & 7/8 & -0.020 & 0.068 & +0.074 & -0.037 & +0.020 & 0.058 \\
    $5^\circ$ & $1.52$ & G\"ottingen & calibrated & 0.106 & 4/8 & -0.006 & 0.028 & +0.026 & -0.040 & -0.064 & 0.064 \\
    $10^\circ$ & $1.52$ & G\"ottingen & measured & 0.330 & 8/8 & -0.013 & 0.027 & +0.002 & -0.041 & -0.091 & 0.100 \\
    $10^\circ$ & $1.52$ & G\"ottingen & calibrated & 0.106 & 8/8 & +0.019 & 0.082 & +0.160 & -0.053 & -0.091 & 0.100 \\
    $29.7^\circ$ & $1.53$ & G\"ottingen & measured & 0.330 & 12/12 & -0.048 & 0.098 & +0.185 & -0.045 & -- & -- \\
    $29.7^\circ$ & $1.53$ & G\"ottingen & calibrated & 0.106 & 12/12 & +0.041 & 0.095 & +0.264 & -0.024 & -- & -- \\
    \midrule
    $0^\circ$ & $7.2$ & G\"ottingen & measured & 0.330 & 17/17 & -0.097 & 0.097 & -0.097 & -0.098 & -0.013 & 0.013 \\
    $0^\circ$ & $7.2$ & G\"ottingen & calibrated & 0.106 & 17/17 & +0.077 & 0.077 & +0.078 & +0.076 & -0.013 & 0.013 \\
    $2.5^\circ$ & $7.2$ & G\"ottingen & measured & 0.330 & 3/3 & -0.093 & 0.102 & -0.150 & -0.049 & +0.006 & 0.021 \\
    $2.5^\circ$ & $7.2$ & G\"ottingen & calibrated & 0.106 & 3/3 & +0.082 & 0.096 & +0.016 & +0.132 & +0.006 & 0.021 \\
    $5^\circ$ & $6.49$ & G\"ottingen & measured & 0.330 & 4/4 & -0.098 & 0.177 & -0.292 & +0.007 & -0.024 & 0.051 \\
    $5^\circ$ & $6.49$ & G\"ottingen & calibrated & 0.106 & 4/4 & +0.107 & 0.201 & -0.088 & +0.246 & -0.024 & 0.051 \\
    $10^\circ$ & $6.56$ & G\"ottingen & measured & 0.330 & 5/5 & +0.365 & 0.440 & -0.105 & +0.375 & +0.073 & 0.098 \\
    $10^\circ$ & $6.56$ & G\"ottingen & calibrated & 0.106 & 5/5 & +0.421 & 0.463 & +0.068 & +0.387 & +0.073 & 0.098 \\
    $10^\circ$ & $6.56$ & ONERA F1 & calibrated & 0.161 & 5/5 & +0.399 & 0.475 & +0.002 & +0.549 & +0.088 & 0.134 \\
    $10^\circ$ & $6.56$ & ONERA F1 & measured & 0.100 & 5/5 & +0.420 & 0.486 & +0.051 & +0.549 & +0.088 & 0.134 \\
    \bottomrule
  \end{tabular}
\end{table}

\clearpage
\begin{figure}[H]\centering
\includegraphics[width=0.86\textwidth]{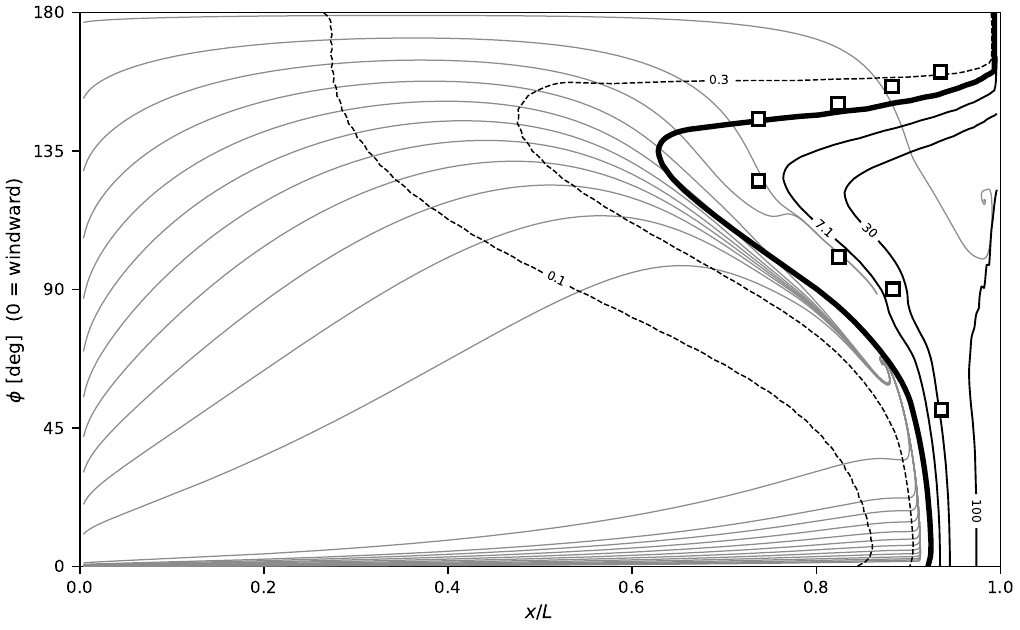}
\caption{Surface at the measured-seed condition, unrolled into $(x/L,\phi)$, $\phi=0$ windward; line art. Grey skin-friction lines integrated from the wall shear vector itself -- the oil-flow analogue, not the inviscid streamlines of Stock Figs.~14--17, and their convergence is the separation signature -- under the $\max_n\chi$ contour family (sub-unity dashed, $\chi=1$ heavy, $\chi=c_{v1}$ and above solid, labeled), $\max_n\chi$ taken on a wall-normal segment shot from each surface point. White-filled squares: measured DFVLR transition. One page per measured condition. $\alpha=5^\circ$, $Re_L=1.52\times10^6$.}\label{f:ov_a5lo_gmeas}
\end{figure}

\begin{figure}[H]\centering
\includegraphics[width=0.86\textwidth]{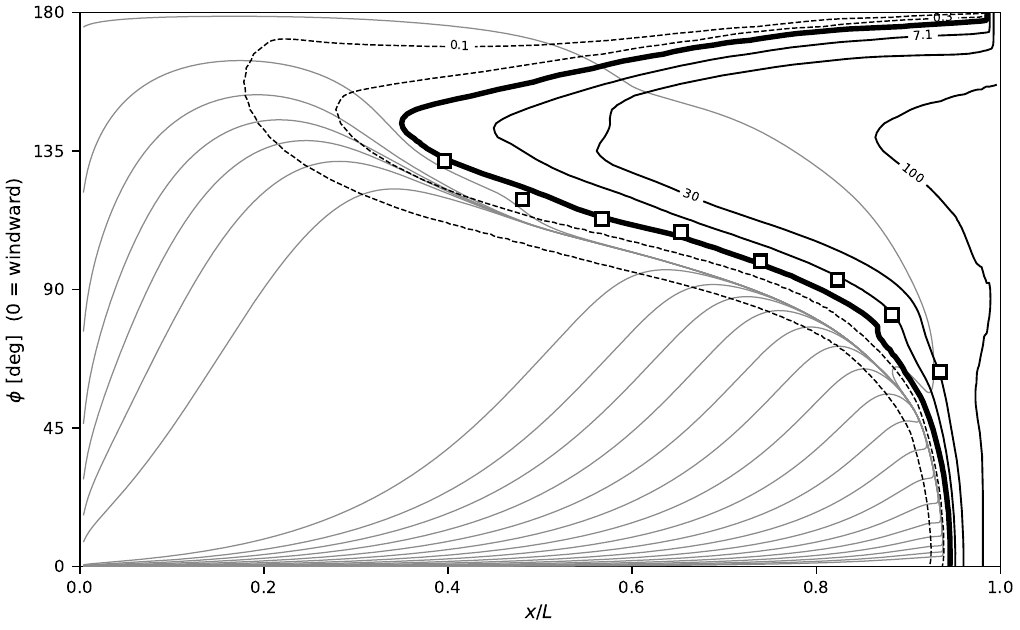}
\figcont{$\alpha=10^\circ$, $Re_L=1.52\times10^6$.}
\end{figure}

\begin{figure}[H]\centering
\includegraphics[width=0.86\textwidth]{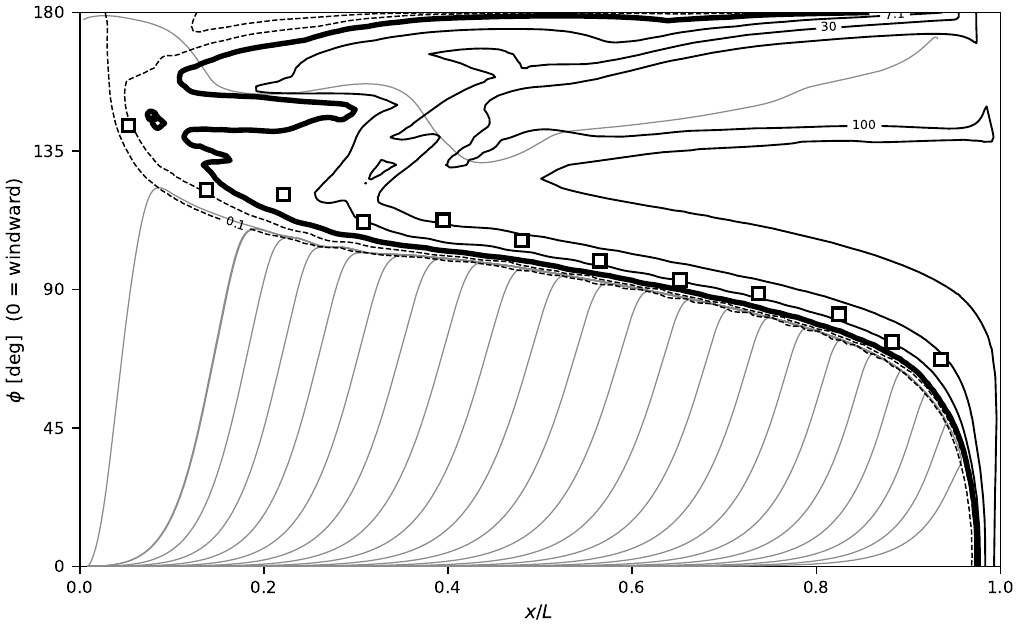}
\figcont{$\alpha=29.7^\circ$, $Re_L=1.53\times10^6$; the friction lines converge onto the open free-vortex-layer separation over most of the leeward half, and the twelve measured stations lie on that line rather than on a transition front.}
\end{figure}

\begin{figure}[H]\centering
\includegraphics[width=0.86\textwidth]{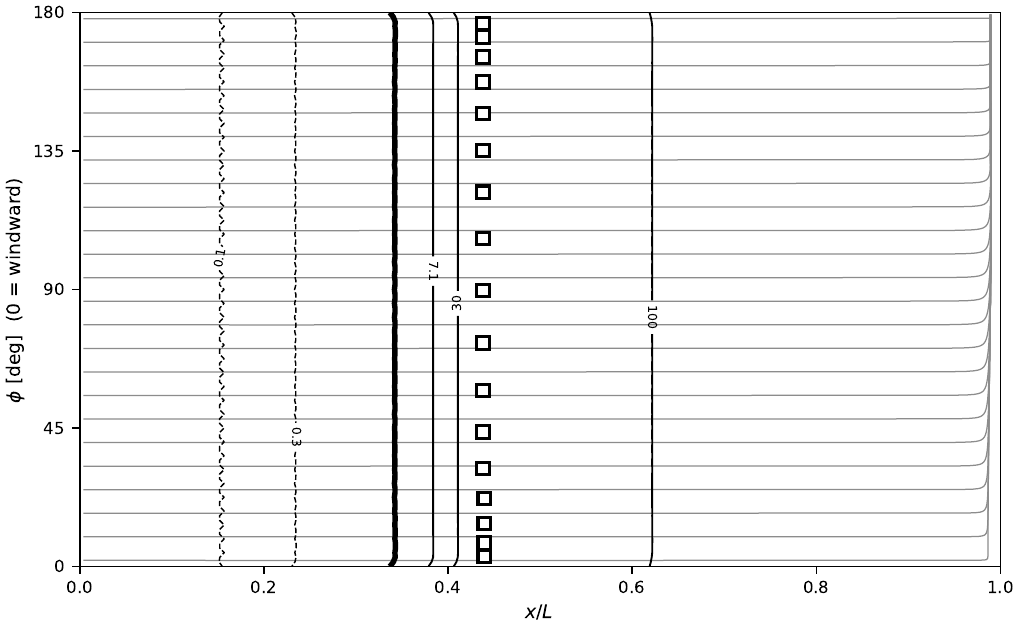}
\figcont{$\alpha=0^\circ$, $Re_L=7.2\times10^6$.}
\end{figure}

\begin{figure}[H]\centering
\includegraphics[width=0.86\textwidth]{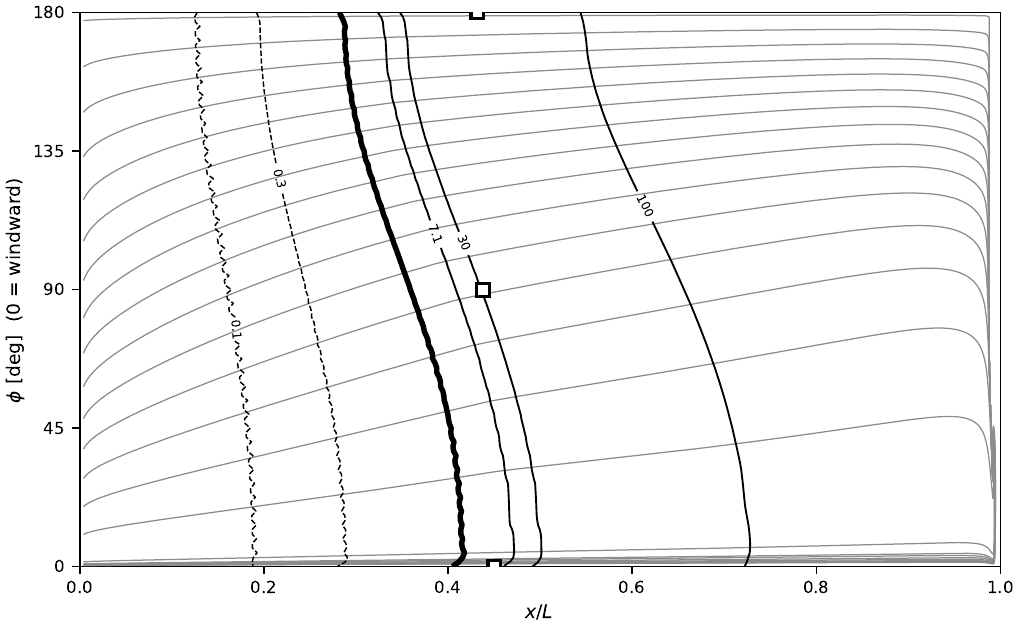}
\figcont{$\alpha=2.5^\circ$, $Re_L=7.2\times10^6$.}
\end{figure}

\begin{figure}[H]\centering
\includegraphics[width=0.86\textwidth]{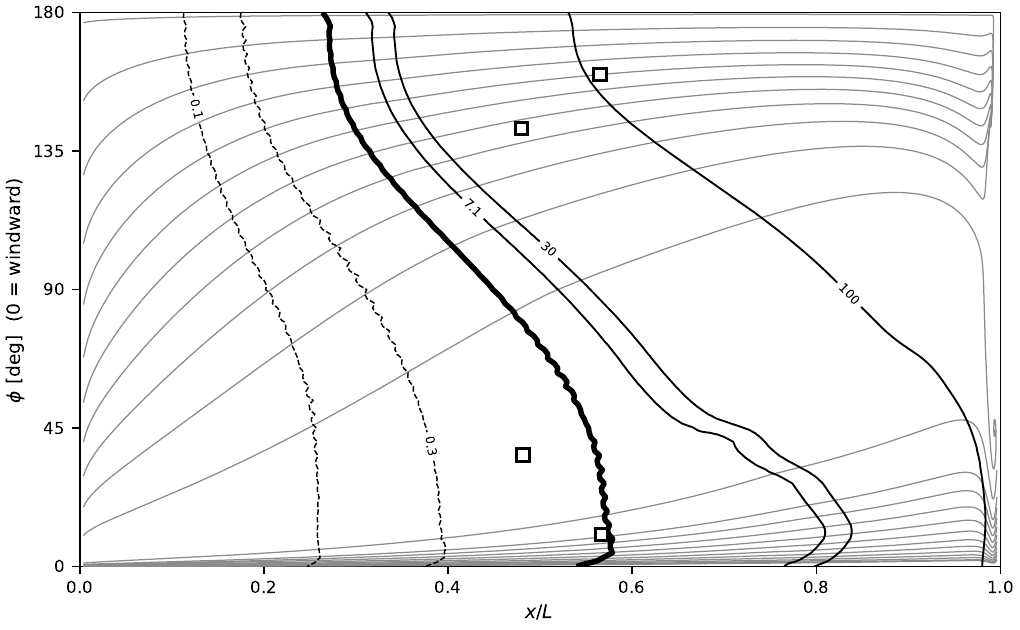}
\figcont{$\alpha=5^\circ$, $Re_L=6.49\times10^6$.}
\end{figure}

\begin{figure}[H]\centering
\includegraphics[width=0.86\textwidth]{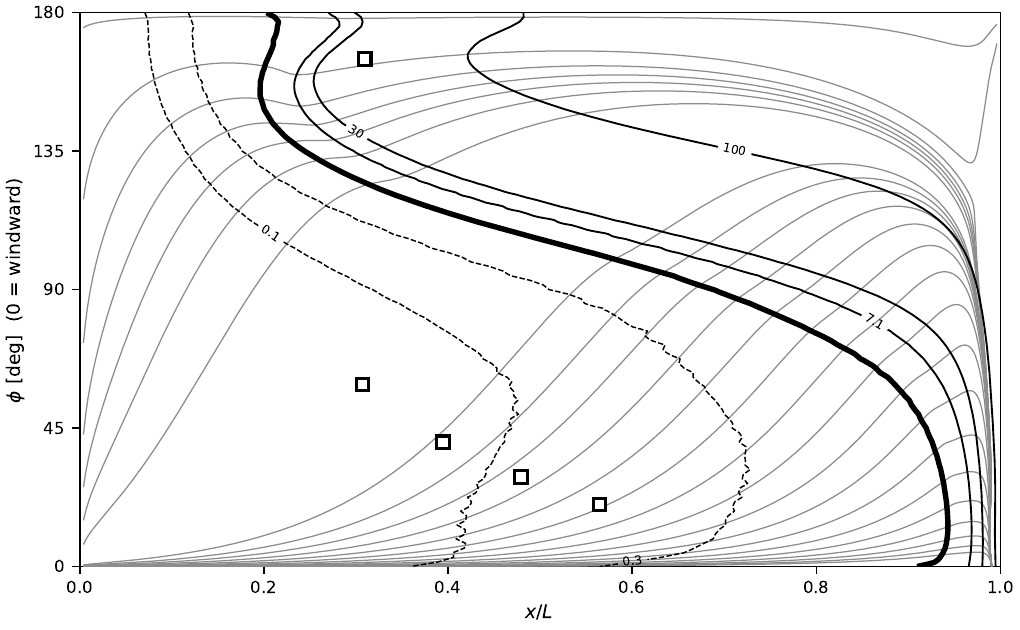}
\figcont{$\alpha=10^\circ$, $Re_L=6.56\times10^6$, G\"ottingen.}
\end{figure}

\begin{figure}[H]\centering
\includegraphics[width=0.86\textwidth]{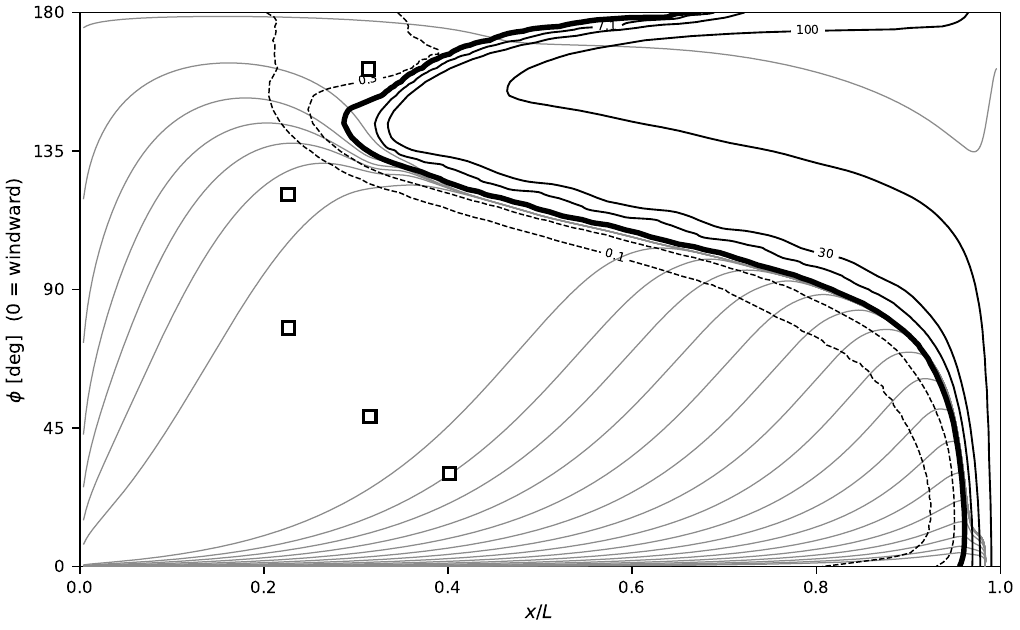}
\figcont{$\alpha=10^\circ$, $Re_L=6.56\times10^6$, ONERA F1.}
\end{figure}

\clearpage

\section{Remarks on the operating envelope}

The following follow from the kernel's algebra evaluated on exact
profiles and on exact perturbations of the computed cases, with no
further computation; each is reproduced by
\texttt{repro/analytic/envelope\_bounds.py}.

\begin{enumerate}\itemsep2pt
\item Rigid rotation is read as amplifying. Against a co-rotating wall the
  velocity relative to the nearest-wall point is $\Omega d$ and the absolute
  vorticity $2\Omega$, so $\hat Y/\hat X\!=\!2$, $\hat Z\!=\!0$ and
  $\hat\Omega\hat I\!=\!0.40$: a solid-body rotation transitions.
\item Frame rotation moves both indicators: the absolute vorticity gives
  $\hat Y\!\to\!\hat Y+2\Omega d$ and the wall-relative velocity
  $\hat X\!\to\!\hat X+\Omega d$. The rate moves by $18.3\,(\ell/L)$, that
  is by $O(Re_\theta/Re_L)$; at $\Omega\!=\!U_\infty/L$ the campaign's own
  stations give
  \begin{center}\small
  \begin{tabular}{lccc}
    \toprule
    case & $Re_L$ & station $x/L$ & $\delta(\hat\Omega\hat I)/\hat\Omega\hat I$ \\
    \midrule
    NLF(1)-0416, $\alpha=-8^\circ$ & $4\times10^6$ & $0.560$ & $+0.7\%$ \\
    NLF(1)-0416, $\alpha=0^\circ$  & $4\times10^6$ & $0.389$ & $+0.6\%$ \\
    NLF(1)-0416, $\alpha=15^\circ$ & $4\times10^6$ & $0.003$ & $+0.0\%$ \\
    Eppler 387, $\alpha=0^\circ$   & $2\times10^5$ & $0.522$ & $+3.0\%$ \\
    Eppler 387, $\alpha=7^\circ$   & $2\times10^5$ & $0.388$ & $+2.6\%$ \\
    two-element, flap bubble        & $10^6$        & $0.850$ & $+1.7\%$ \\
    Daedalus, $\eta=0.10$          & $5\times10^5$ & $0.450$ & $+1.7\%$ \\
    6:1 spheroid, $\alpha=10^\circ$ & $1.5\times10^6$ & $0.653$ & $+1.2\%$ \\
    cylinder, $Re_D=10^5$           & $10^5$        & $0.711$ & $+4.9\%$ \\
    cylinder, $Re_D=10^7$           & $10^7$        & $0.175$ & $+0.2\%$ \\
    \bottomrule
  \end{tabular}
  \end{center}
  The shift scales as $\ell/L$, i.e.\ as $1/\sqrt{Re_L}$ at a fixed station,
  so only the low-Reynolds cases feel a rotating frame at all, and the sign is
  always destabilizing.
\item $\hat I$ is a near-cancellation only on mildly loaded layers:
  $|Y\!-\!X\!-\!Z|/(|X|\!+\!|Y|\!+\!|Z|)$ at the rate-setting station is
  $0.04$ at Falkner--Skan $\beta\!=\!+0.5$, $0.08$ at $\beta\!=\!0$, $0.12$ at
  $\beta\!=\!-0.10$ and $0.29$ at separation, so the $\ell/R\!\sim\!10^{-3}$
  terms are visible in the rate on attached layers and not near separation.
\item Streamwise wall curvature enters the curvature indicator through
  the metric term $\delta Z=\pm(d/2R)Y$ and is therefore already present
  in every computed case, contributing $-0.15\%$ of the rate on the
  NLF(1)-0416 ($\alpha\!=\!0^\circ$, $Re_c\!=\!4\times10^6$, $x/c\!=\!0.389$)
  and $-0.32\%$ on the Eppler 387 ($\alpha\!=\!0^\circ$,
  $Re_c\!=\!2\times10^5$, $x/c\!=\!0.522$).
\item Transverse curvature is the term that grows on slender bodies:
  $\ell/r_0$ scales with the fineness ratio at fixed $Re_L$ and station, so
  the $-2.9$ to $-8.2\%$ over the eight spheroid stations ($\alpha=10^\circ$,
  $Re_L=1.5\times10^6$, $x/L=0.396$--$0.934$) becomes $-4.7$ to $-12.6\%$ at
  10:1, $-8.9$ to $-20.8\%$ at 20:1 and $-15.6$ to $-28.5\%$ at 40:1,
  saturating near $-30\%$ once $\ell\!\sim\!r_0$.
\item The curvature indicator has two candidate realizations:
  $\nabla^2\mathbf u\!\cdot\!\hat{\mathbf u}$, the one used here
  (Eqs.~\eqref{w:gram} and~\eqref{w:ratio}), and
  $\hat{\mathbf n}\!\cdot\!\nabla|\boldsymbol\omega|$. For streamwise
  curvature the two carry the same leading term and the numbers above are
  unambiguous; for the transverse curvature of a body of revolution they do
  not, so the $-2.9$ to $-6.3\%$ computed at the spheroid stations is a
  realization uncertainty rather than a bias.
\item On a concave wall the metric term changes sign and destabilizes, so
  the model is not blind to centrifugal geometry, but its response is
  linear in $\delta/R$ where the instability is governed by the
  G\"ortler number $G=Re_\delta\sqrt{\delta/R}$ (Saric, \emph{Annu.\ Rev.\
  Fluid Mech.}\ \textbf{26}, 379--409, 1994) --- the right sign under the
  wrong scaling.
\item On the exact swept-Hiemenz profile the kernel amplifies at $82\%$ of
  the Blasius rate and switches on at $\bar R\!=\!603$ against the
  linear-stability critical value near $583$, unfitted, though the
  agreement is on the threshold and not on the mechanism.
\item The asymptotic suction layer is read as stable at every Reynolds
  number against a true critical $Re_{\delta^*}\!\approx\!5.4\times10^4$,
  so the model will over-credit suction. Here $\delta^*\!=\!\nu/|v_w|$
  exactly, so $Re_{\delta^*}\!=\!U_\infty/|v_w|$ and the sweep
  $10^2$--$10^6$ is a suction sweep $|v_w|/U_\infty\!=\!10^{-2}$--$10^{-6}$;
  $P\!\le\!0$ is set by the profile shape, which is suction-independent.
  The same evaluation shows that the positive amplifying coordinate excluded
  at a no-slip line does not appear under transpiration of that sign.
\item Because SA's destruction scales as $(\tilde\nu/d)^2$ and is
  negligible where $d$ is large, a wake transports its $\tilde\nu$ without
  decaying, so a surface ingesting an upstream wake---a blade row directly
  in the turbulent wake of another, as much as the two-element section
  here---should be expected to
  transition at impingement rather than by amplification.
\item The assembled model is continuous everywhere and differentiable
  almost everywhere, the exceptions being the clips of the amplifying
  coordinate at $P\!=\!0$ and $P\!=\!1$, the handover switch at
  $\chi\!=\!1$, and the gated maximum at handover, all of which admit the
  soft forms the onset threshold already uses.
\end{enumerate}

\clearpage

\appendix

\section{Appendix: the two-source rate --- inviscid and viscous branches}
\label{app:twosource}

\begin{figure}[tp]\centering
\includegraphics[width=0.99\textwidth]{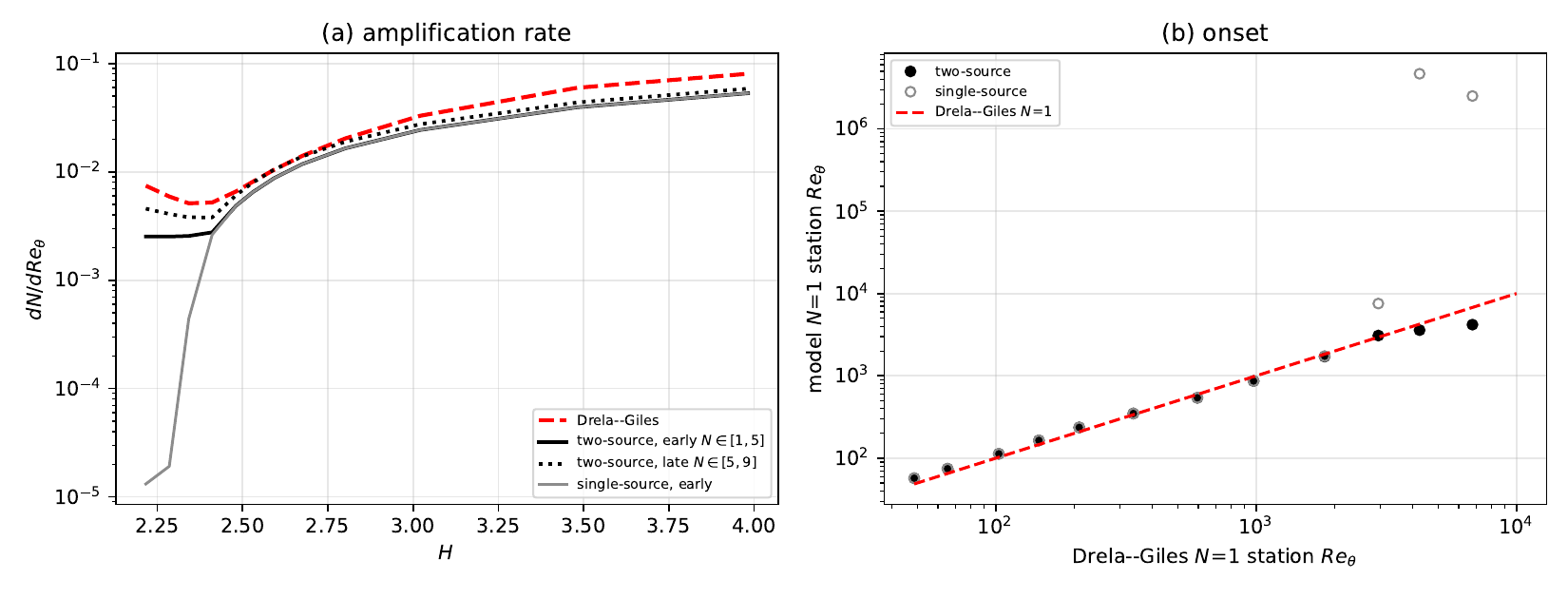}
\caption{Two-source rate against Drela--Giles across the Falkner--Skan
family, $\beta=+1.0$ to $-0.1988$; analytic march, no CFD.
(a)~rate as early ($N\!\in\![1,5]$) and late ($N\!\in\![5,9]$) secants, with
the single-source rate. (b)~marched $N\!=\!1$ station against the
Drela--Giles $N\!=\!1$ station $Re_{\theta0}+1/(dN/dRe_\theta)$; the diagonal
is exact agreement.}\label{fig:tscalibrate}
\end{figure}

The rate uses the single even coordinate $\hat\Omega\hat I$, which carries
both inviscid--inflectional and viscous Tollmien--Schlichting content
because the two co-vary across the Falkner--Skan family. They separate on
the parabola, where $\hat I\equiv0$: plane Poiseuille returns zero rate at
every Reynolds number against a true $Re_c=5772$ (Orszag 1971), and on the
favorable branch the booked amplification falls an order of magnitude below
the Drela--Giles envelope by $\beta\!=\!0.35$ (Fig.~\ref{fig:tscalibrate}).
Blasius carries no inflection, so the missing branch is the main mechanism,
not an exotic one.

A second coordinate separates them:
\begin{equation}
  P_{\mathrm{c}} \;=\; \hat\Omega\,\big\langle -\hat Z\big\rangle_+,
  \qquad \hat Z = Z/U_\mathrm{loc},
  \label{eq:ts_pcurv}
\end{equation}
positive wherever $u''<0$, clipped at an adverse wall, vanishing like $d$ at
the wall and zero in the free stream. Each branch carries its own rate and
its own onset, and the two \emph{sources} are combined:
\begin{gather}
  P_{\mathrm{AI}} = \omega\,\tilde\nu\;
  \operatorname{softmax}_2\!\Big[\,
     a_{\max}\,P_I\;\sigma_{Re}\big(Re_\Omega/Re_\Omega^{\mathrm c,i}\big),\;\;
     a_{\mathrm{visc}}\,P_{\mathrm{c}}\;\sigma_{Re}\big(Re_\Omega/Re_\Omega^{\mathrm c,v}\big)
  \Big],
  \label{eq:ts_twosource}\\[3pt]
  Re_\Omega^{\mathrm c,i} = k\operatorname{softmin}_2\!\Big(C,\,A+B\,P_I^{-2}\Big),
  \qquad
  Re_\Omega^{\mathrm c,v} = A + B_{\mathrm c}\,P_{\mathrm{c}}^{-2},
  \qquad
  \operatorname{softmax}_2(x,y)=\sqrt{x^2+y^2}.
  \notag
\end{gather}
The inviscid gate is unchanged, ceiling included; the viscous branch shares
its floor $A$ and adds one coefficient $B_{\mathrm c}$. Constants:
$a_{\max}\!=\!0.19$, $a_{\mathrm{visc}}\!=\!0.0276$, $A\!=\!124.6$,
$B\!=\!1.424$, $B_{\mathrm c}\!=\!130$, $C\!=\!1851.2$.

The gates are separate, not soft-minimized into one. On Blasius the two
thresholds are $358$ and $2598$; sharing the lower switches the viscous rate
on early, worth $+7.4\%$ on the Blasius anchor. With them separate the
construction reproduces the single-source rate to within $0.7\%$ over the
calibrated family (Table~\ref{tab:tsfamily}), so $a_{\max}$, $k$ and the
airfoil campaign are untouched. Dropping the ceiling $C$ sends
$\beta\!=\!+0.20$ to $0.63\times$ its single-source value.

$a_{\mathrm{visc}}$ is set on the favorable Falkner--Skan family against the
Drela--Giles envelope. Plane Poiseuille then fixes $B_{\mathrm c}$
independently and without refitting, since $P_I\equiv0$ on a parabola:
requiring the frozen-profile eigenvalue of Eq.~\eqref{eq:ts_twosource} to
change sign at $Re_c=5772$ gives Table~\ref{tab:tschannel}. At
$a_{\mathrm{visc}}=0.0276$ the channel asks $B_{\mathrm c}=91$ against the
family's $130$, and the family's constants place the channel neutral point
at $7409$, $1.28\times$ the true value. The channel is a calibration and
verification device only; its observed transition is subcritical and
finite-amplitude.

Three limits. $P_{\mathrm c}$ is largest at a stagnation line, where the
filled attachment profile keeps $\langle-\hat Z\rangle_+$ finite while
$P_I\!\le\!0$ holds the single-source kernel off the nose geometrically, so
the branch would feed amplification into the one region the present
formulation leaves inert; adopting it requires either showing its
stagnation-line rate stays below onset at leading-edge Reynolds numbers or
pairing it with a nose treatment. The viscous rate
$a_{\mathrm{visc}}P_{\mathrm{c}}$ is Reynolds-independent, inherited from
the inviscid branch. And the branches decompose the rate, not the physics:
the viscous gate on Blasius opens only near $Re_\theta\!\approx\!1190$, so
Blasius stays carried by $P_I$. The branch is not part of the model
validated here.

\begin{figure}[tp]\centering
\includegraphics[width=0.99\textwidth]{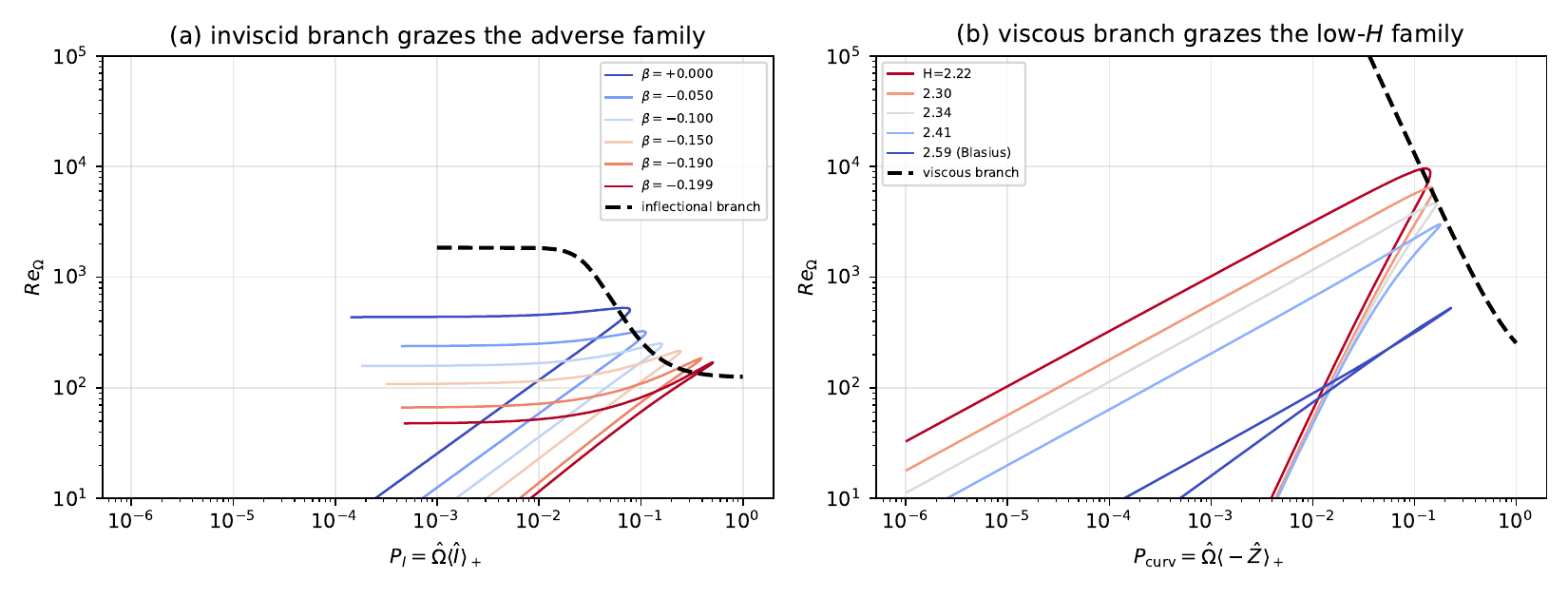}
\caption{Each onset branch against its own family, at each profile's
Drela--Giles critical $Re_{\theta0}(H)$. (a)~inflectional branch
$A+B/P_I^2$, adverse family. (b)~curvature branch
$A+B_{\mathrm c}/P_{\mathrm c}^2$, low-$H$ members, in the coordinate
$P_{\mathrm c}$ of Eq.~\eqref{eq:ts_pcurv}.}\label{fig:tsgraze}
\end{figure}

\begin{figure}[tp]\centering
\includegraphics[width=0.72\textwidth]{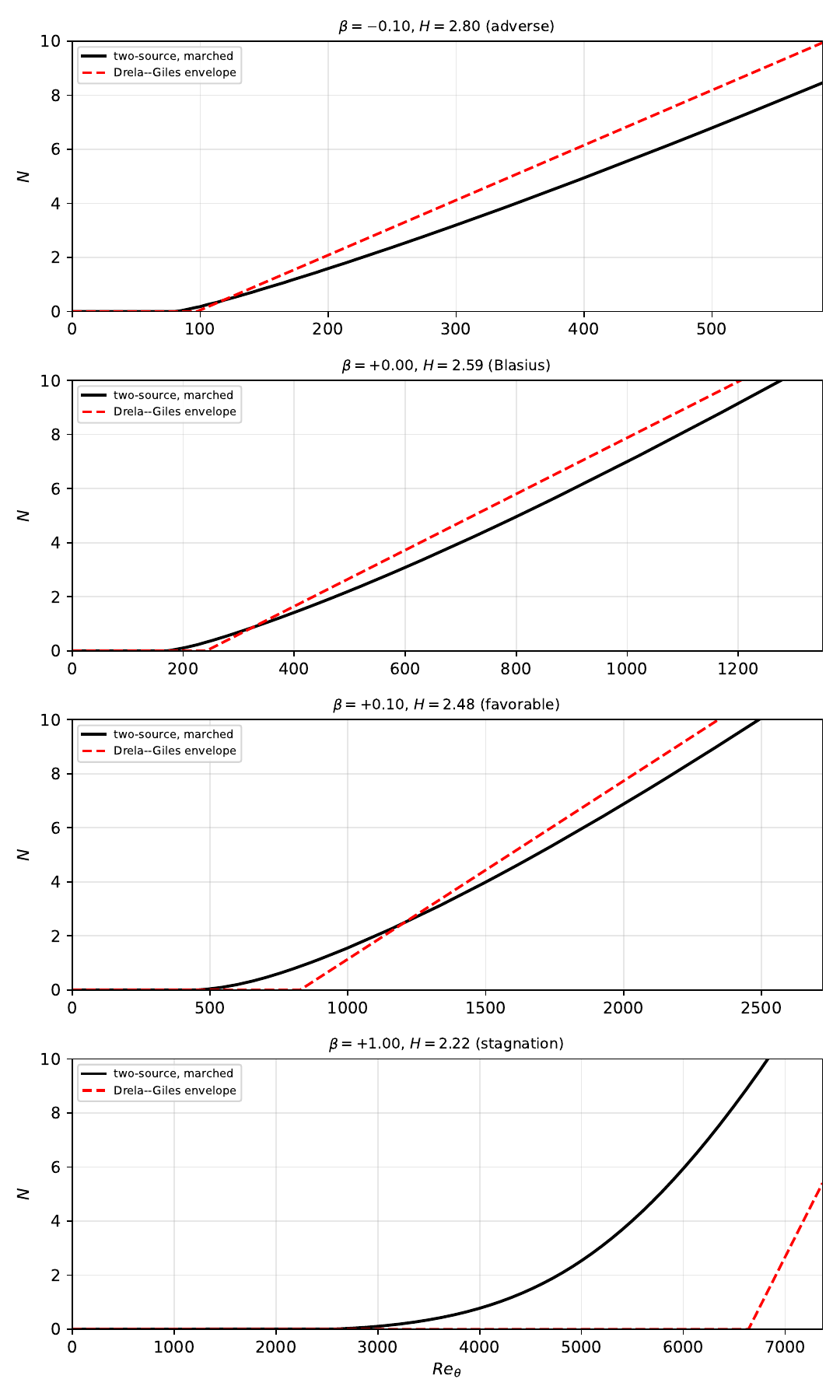}
\caption{Disturbance transport under Eq.~\eqref{eq:ts_twosource} on four
wedges---adverse, Blasius, favorable, and stagnation
($\beta\!=\!+1$)---against each profile's Drela--Giles envelope. On the
stagnation wedge the branch ignites at
$Re_\theta\!\approx\!2.5\times10^3$ against $Re_{\theta0}\!=\!6.5\times10^3$.
Same instrument and same $c_{\nu,\mathrm{ai}}=1/6$ as the model's own
calibration.}\label{fig:tstransport}
\end{figure}

\begin{figure}[tp]\centering
\includegraphics[width=0.99\textwidth]{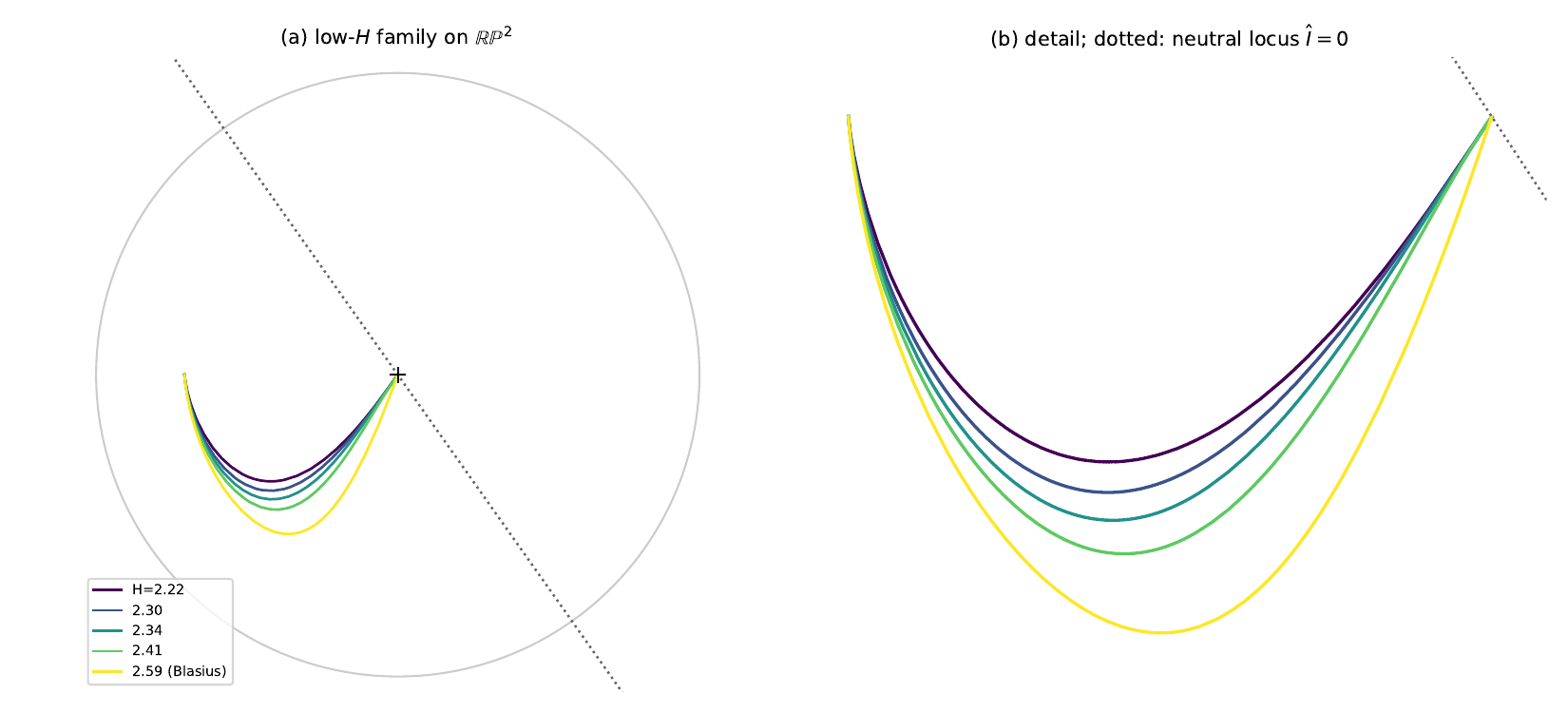}
\caption{Low-$H$ Falkner--Skan family on the $\mathbb{RP}^2$ indicator
sphere, $H\!\approx\!2.22$ to Blasius $H\!=\!2.59$, with (b) near-wall
detail. The family bunches against the neutral parabola locus where $\hat I$
is too small to separate members whose Drela--Giles critical Reynolds
numbers differ by nearly a factor of two; $P_{\mathrm c}$ separates
them.}\label{fig:tssphere}
\end{figure}

\begin{table}[tp]
  \centering\small
  \caption{Peak dimensionless source $a\,\sigma_{Re}$ on Falkner--Skan
  wedges: single-source, and Eq.~\eqref{eq:ts_twosource} at
  $(a_{\mathrm{visc}},B_{\mathrm c})=(0.0276,130)$.}
  \label{tab:tsfamily}
  \begin{tabular}{rrccc}
    \toprule
    $\beta$ & $Re_\theta$ & single-source & two-source & ratio\\
    \midrule
    $0$        & $500$  & $0.01485$ & $0.01485$ & $1.000$\\
    $0$        & $1000$ & $0.01485$ & $0.01487$ & $1.001$\\
    $-0.10$    & $400$  & $0.03082$ & $0.03083$ & $1.000$\\
    $-0.1988$  & $300$  & $0.09692$ & $0.09764$ & $1.007$\\
    $+0.10$    & $1000$ & $0.00695$ & $0.00696$ & $1.000$\\
    $+0.20$    & $2000$ & $0.00281$ & $0.00282$ & $1.004$\\
    $+0.35$    & $4000$ & $0.00033$ & $0.00405$ & $12.4$\\
    \bottomrule
  \end{tabular}
\end{table}

\begin{table}[tp]
  \centering\small
  \caption{Plane-channel check: the $B_{\mathrm c}$ placing the neutral point
  on $Re_c=5772$, and the neutral point obtained at the
  Falkner--Skan-anchored $B_{\mathrm c}=130$.}
  \label{tab:tschannel}
  \begin{tabular}{ccc}
    \toprule
    $a_{\mathrm{visc}}$ & $B_{\mathrm c}$ for $Re_c\!=\!5772$ & $Re_c$ at $B_{\mathrm c}\!=\!130$\\
    \midrule
    $0.0200$ & $81.0$  & $8095$\\
    $0.0276$ & $91.0$  & $7409$\\
    $0.0350$ & $99.9$  & $6916$\\
    $0.0600$ & $127.9$ & $5834$\\
    \bottomrule
  \end{tabular}
\end{table}

\clearpage
\thispagestyle{empty}
\vspace*{\fill}
\begin{center}
  \begin{minipage}{0.6\textwidth}\centering
  \begin{CJK}{UTF8}{bsmi}天下萬物生於有，有生於無。\end{CJK}\\[6pt]
  \emph{The myriad things are born of being;\\
  being is born of non-being.}\\[6pt]
  ---\emph{Daodejing}, ch.~40
  \end{minipage}
\end{center}
\vspace*{\fill}

\end{document}